\pdfoutput=1
\documentclass[11pt,onecolumn,draftclsnofoot]{IEEEtran}
\usepackage{changepage}
\usepackage{algorithm}
\usepackage{array}
\usepackage{textcomp}
\usepackage{stfloats}
\usepackage{url}
\usepackage{verbatim}
\usepackage{cite}
\usepackage{algpseudocode}
\usepackage{tabularx}
\usepackage{graphicx}
\usepackage{caption}
\usepackage{subcaption}
\usepackage{placeins}
\usepackage{fancyhdr}
\usepackage{lastpage}
\usepackage{multirow}
\usepackage{xcolor} % For font color settings
\usepackage{amsmath, amsfonts}
\usepackage{pdflscape}
\usepackage{amssymb}
\usepackage{threeparttable} % For tables with notes
\usepackage{orcidlink}  % For adding ORCID identifiers
\usepackage{makecell}   % For setting table line thickness
\usepackage{booktabs}
\usepackage{afterpage}
\usepackage{tocloft}

\date{}

\begin{document}

\title{
	Technical Report: Towards Spatial Feature  \vspace{-0.5em} \newline Regularization in Deep-Learning-Based  \vspace{-0.5em} \newline Array-SAR~Reconstruction 
}

\author{
	Yu~Ren\orcidlink{0000-0002-8582-680X}\textsuperscript{*}\footnote{\textsuperscript{*} These authors contributed equally.}, 
	 Xu~Zhan\orcidlink{0000-0003-2816-9791}\textsuperscript{*,$\dagger$},   
	 Yunqiao~Hu\orcidlink{0000-0001-6153-2105}, 	
	 Xiangdong~Ma\orcidlink{0000-0002-2834-2163}, 
	 Liang~Liu\orcidlink{0000-0002-2834-2163},	
	 \newline Mou~Wang\orcidlink{0000-0003-3462-3989}, 
	 Jun~Shi\orcidlink{0000-0001-7676-8380},  
	 Shunjun~Wei\orcidlink{0000-0001-8091-9540},	
	 Tianjiao~Zeng\orcidlink{0000-0002-6780-6100}\textsuperscript{$\dagger$}\footnote{$\dagger$ Correspondence: zhanxu@std.uestc.edu.cn, tzeng@uestc.edu.cn, xlzhang@uestc.edu.cn.}, 
	 and~Xiaoling~Zhang\orcidlink{0000-0003-2343-3055}\textsuperscript{$\dagger$}
}
\maketitle

\vspace{-3em}
\begin{abstract}

	Array synthetic aperture radar (Array-SAR), also known as tomographic SAR (TomoSAR), has demonstrated significant potential for high-quality 3D mapping, particularly in urban areas. While deep learning (DL) methods have recently shown strengths in both precision and efficiency for reconstruction, most existing studies rely on pixel-by-pixel reconstruction without fully leveraging prominent spatial features, such as building structures. This limitation often leads to reconstruction artifacts, including holes in building surfaces or fragmented edges. Spatial feature regularization, a concept long proven effective in traditional reconstruction methods, has not yet been comprehensively explored within the context of DL-based approaches. To address this gap, our group has conducted a preliminary and exploratory study, focusing on integrating spatial feature regularization into DL-based Array-SAR reconstruction.

This study addresses several key questions: What are the relevant spatial features in the context of urban-area mapping using Array-SAR? How can these features be effectively described, modeled, and regularized? More importantly, how can these regularized features be incorporated into the reconstruction process using DL networks?

To investigate these questions, the study is divided into five phases: spatial feature description and modeling, spatial feature regularization, feature-enhanced network design, method evaluation, and discussions. Spatial features have been analyzed and summarized as sharp edges and regular geometric shapes commonly found in urban scenes. To model these features, a general intra-slice and inter-slice strategy has been proposed, utilizing 2D slices as basic reconstruction units and fusing them into a complete 3D scene using two distinct approaches: parallel and serial fusion. To regularize these features, two novel computational frameworks—iterative reconstruction with enhancement and light reconstruction and enhancement—have been designed. These frameworks incorporate spatial feature modules into DL networks, leading to the construction and evaluation of four specialized reconstruction networks. A comprehensive evaluation framework has been developed, combining a newly self-constructed urban building simulation dataset with two publicly available datasets. Six diverse tests, ranging from close-point resolution to complex urban landscape challenges, have been conducted to validate the proposed methods.

Evaluation results reveal that the spatial feature regularization frameworks significantly enhance the reconstruction process. Specifically, the proposed methods achieve higher reconstruction precision, retrieve more complete and continuous building structures (e.g., sharp edges and surfaces), and exhibit improved robustness by reducing outliers and noise in the reconstructed scenes. The findings demonstrate the effectiveness of the proposed designs in advancing DL-based Array-SAR reconstruction for urban 3D mapping, providing a robust foundation for future exploration and applications.
	
\end{abstract}


\section*{\textbf{Note}}

1. In this technical report, Yu Ren and Xu Zhan contributed equally to the additional methodologies, experiments, analyses, and content preparation. The authorship order reflects their equal contributions and the expanded scope of this work. This report builds upon the findings presented in the paper titled \textit{"Exploring Spatial Feature Regularization in Deep-Learning-Based TomoSAR Reconstruction: A Preliminary Study and Performance Analysis,"} authored by our group and accepted for publication in IEEE Transactions on Geoscience and Remote Sensing (DOI: 10.1109/TGRS.2024.3521278)\footnote{© 20xx IEEE. Personal use of this material is permitted. Permission from IEEE must be obtained for all other uses, in any current or future media, including reprinting/republishing this material for advertising or promotional purposes, creating new collective works, for resale or redistribution to servers or lists, or reuse of any copyrighted component of this work in other works.}. \textit{Correspondence: zhanxu@std.uestc.edu.cn, tzeng@uestc.edu.cn, xlzhang@uestc.edu.cn.}

2. Yu Ren, Xu Zhan, Yunqiao Hu, Xiangdong Ma, Liang Liu, Mou Wang, Jun Shi, Shunjun Wei, and Xiaoling Zhang are with the School of Information and Communication Engineering, University of Electronic Science and Technology of China, Chengdu 611731, China. Tianjiao Zeng is with the School of Aeronautics and Astronautics, University of Electronic Science and Technology of China, Chengdu 611731, China.

3. This work was supported in part by the National Natural Science Foundation of China under Grants 62471113, 62305049, and 62371104, in part by Aeronautical Science Foundation of China under Grants 1A2024Z071080005, and in part by Sichuan Science and Technology Program 2024NSFSC0479.

\tableofcontents

\clearpage
\section{Introduction}
Array synthetic aperture radar (Array-SAR), or tomographic SAR (tomoSAR), has become a hot topic in research over the last two decades~\cite{reigber2000first, zhu2010very, wang2023mada}. What sets it apart is its ability to create high-resolution 3D maps of large areas using multi-view observations along the elevation direction. The tomoSAR research field has grown immensely during this time. It started with conceptual design and modeling, developing corresponding reconstruction algorithms, and system development studies. Now, it's reached a point where it's being thoroughly tested and used in practical applications on various platforms.

In terms of practical applications, researchers have been busy uncovering tomoSAR's potential in many remote sensing areas. This includes monitoring natural resources like glaciers~\cite{tebaldini2016imaging} and forests~\cite{blomberg2020evaluating}, and uses in urban area modeling and surveillance~\cite{rambour2020interferometric}. TomoSAR's adaptability and reliability have been proven by its successful use in a wide range of SAR systems. These range from ground-based setups ~\cite{chai2019deformation,dinh2013ground}, to unmanned aerial vehicle SAR platforms ~\cite{wang2022first, wang2023uav}, and even large airborne~\cite{jiao2020urban, han2021efficient} and spaceborne SAR systems~\cite{ge2018spaceborne, feng2022elevation}. For example, in 2020, Shi and his team used the TanDem-X spaceborne system to map the entire city of Munich, Germany, with impressive accuracy~\cite{shi2020sar}. Then in 2022, Xu, Qiu, Jiao, and their team used an airborne array-SAR system to map large areas in the Chinese cities of Emei and Yuncheng, achieving similar levels of performance~\cite{jiao2020urban}. These examples just go to show the versatility and practicality of tomoSAR in various geographical areas.

\subsection{TomoSAR reconstruction meets deep learning}

It's worth noting that the team at the German Aerospace Center has recently revealed a plan to develop a global Level of Detail 1 (LoD-1) urban model using TanDem-X data~\cite{zhu2022global}. This ambitious project is fundamentally driven by the tomoSAR technique. As tomoSAR continues to mature, it's evident that the technology has progressed beyond its early development stages and is ready for broader application in both scientific and practical domains.

As large-area mapping using tomoSAR becomes increasingly prevalent and practically feasible, it introduces the significant challenge of managing substantial computational workloads. There is a pressing need for highly efficient processing of tomoSAR data, encompassing tasks across the entire spectrum—from the initial processing of raw collected echoes to the generation of the final geo-coded 3D model. The typical processing chain of tomoSAR comprises three primary steps: pre-processing (including co-registration, de-ramping, and possible noise filtering), reconstruction (the central step involving the conversion of echoes into a coarse 3D mapping result), and post-processing (involving geometrical transformation, scatter extraction, surface extraction, and more) ~\cite{rambour2020interferometric}. Among these processing steps, reconstruction emerges as the most crucial, demanding the majority of computational resources. In essence, optimizing the efficiency of the entire processing workflow depends substantially on enhancing the efficiency of the reconstruction step.

With this perspective in mind, recent developments in deep learning (DL) have played a significant role in the field over the past couple of years. Researchers within the community have begun integrating DL techniques with tomoSAR processing, leveraging DL's efficiency and precision~\cite{wu2020super, wang2021tomosar, wang2023atasi, wang2023mada, qian2022basis, qian2022gamma}. These efforts have primarily centered on the reconstruction step of tomoSAR.

TomoSAR reconstruction, typically regarded as a classic linear inverse problem, involves solving an optimization task. This is conventionally accomplished using iterative processes that leverage the first-order gradient descent method. The effectiveness of the reconstruction process is inherently linked to the step-by-step path tracked from the starting to the ending solution. A more efficient path is characterized by its brevity, requiring fewer steps. Traditionally, the construction of such an efficient path has relied on manual interventions, drawing from practical experiences. However, this approach proves less efficient, particularly when addressing the demands of large-scale area reconstruction.

In stark contrast, DL techniques offer distinct advantages in addressing these challenges. Specifically, the current preliminary work leverages DL networks to solve the linear inverse problem, primarily based on a common framework known as "deep unrolling" or "deep unfolding"~\cite{chen2022learning, monga2021algorithm}. In this approach, the traditional iterative solving process is unrolled (unfolded) through a DL neural network to emulate for a much higher and efficient solution. More details regarding these preliminary works will be reviewed in the next section. These endeavors have exhibited promising outcomes in terms of both efficiency improvement and precision enhancement. Thanks to DL, the traditionally lengthy process, often requiring hundreds of iterations, has now been streamlined to typically fewer than 10 iterations, while maintaining or even surpassing the same level of precision~\cite{qian2022gamma}.

\subsection{Motivations}

Building upon recent achievements in DL-based tomoSAR reconstruction, this study explores spatial feature regularization in DL-based tomoSAR ~\cite{rambour2019introducing}. This direction aligns with the historical evolution of reconstruction methods~\cite{rambour2019introducing, aghababaee2019regularization, zhu2015joint, yao2022tomographic, jiao2023preliminary, han2023geometric, jiao2020urban}, particularly the transition from independent azimuth-range unit reconstructions to simultaneous processing of adjacent units—a shift that enhances spatial structure mapping in urban areas.

From a signal dimensionality perspective, tomoSAR reconstruction reconstructs a stack of multi-view 2D range-azimuth images into a 3D scattering distribution of the observed scene, represented as 3D images in the range-azimuth-elevation coordinate system. While all signals in this process have three-dimensional spatial attributes, current DL-based tomoSAR reconstruction methods typically perform unit-by-unit reconstruction independently~\cite{wang2023atasi, wang2023mada, qian2022basis, qian2022gamma}. Processing the original signal by breaking down the 3D tensor into separate 1D signals (fibers) risks losing natural correlations within the signal, especially those related to building structures, potentially compromising reconstruction precision.

However, as noted in the review article, urban areas predominantly feature buildings with characteristic vertical walls and flat rooftops~\cite{rambour2020interferometric}. By leveraging these structural attributes through geometric constraints, we can ensure reconstruction results conform to a more compact geometric manifold. This approach not only yields higher precision with fewer outliers but also produces smoother, more uniform building surfaces with reduced irregularities. Furthermore, introducing these priors into the signal processing chain may reduce system costs and complexity by replacing complex and expensive baseline sampling resources~\cite{qiu2024microwave3d, qiu2024advances}.

Building upon these insights and addressing the limitations identified in current approaches, our research proposes a novel direction in DL-based tomoSAR reconstruction. We focus specifically on incorporating spatial feature regularization into the reconstruction framework, leveraging the inherent geometric patterns and structural consistencies present in urban environments. This approach not only builds upon the computational efficiency advantages of current DL-based methods but also addresses their key limitation: the lack of spatial context integration in the reconstruction process.

Our primary objective is twofold: first, to achieve enhanced reconstruction precision for urban structures, particularly building features such as vertical walls and flat rooftops; and second, despite the potential increased computational complexity of these features, to maintain the same level of efficiency that makes DL-based approaches attractive for large-scale applications. Through the integration of spatial feature regularization, we aim to develop a more robust reconstruction framework that better captures urban geometric complexities while maintaining the speed advantages of current DL-based methods.

To meet our objectives, we have conducted a preliminary study focused on two core questions as follows:
\begin{enumerate}
    \item \textit{Spatial feature description and modeling:} What are spatial features within the specific context of urban-area mapping using the tomoSAR technique? How can we establish corresponding models related to spatial features?
    \item \textit{Spatial feature regularization and reconstruction:} Most critically, how can we incorporate these modeled features—often referred to as regularization in the optimization context—into our tomoSAR reconstruction process, particularly within the DL network?
\end{enumerate}

\subsection{Results and contributions}

Below, we present a simplified overview of the study process and its results. More detailed information is provided in the subsequent sections.

\vspace{0.5em}
\textit{C.1 Spatial feature description and modeling}
\vspace{0.5em}

To set the stage for our research, we provide detailed descriptions of spatial features in urban area mapping. Our in-depth analysis has identified two key aspects: sharp edges and regular geometric shapes.

Building on this understanding, we focus on spatial feature modeling. Even before the advent of deep learning (DL) techniques, the research community recognized the importance of spatial features. We begin with traditional manual methods—which we term "shallow modeling"—to distinguish them from our more advanced techniques. Our modeling approach has two perspectives. First, we consider the modeling dimensions, ranging from 1D fibers and 2D slices to 3D tensors. Second, we examine their mathematical representations, which are essential for feature regularization. After a thorough review of existing research, we select 3D total variation (3D-TV) norm-based modeling for its comprehensive coverage and suitability for large-scale reconstructions.

Traditional modeling methods are largely manual and have limited regularization strength. The emergence of DL, with its robust feature extraction capabilities, offers potential enhancement to this regularization strength. This insight leads us to explore spatial feature modeling within the DL framework.

Starting with slices as our basic processing unit, we examine two methods for assembling 3D tensors. This investigation has produced two unique modules: the Bi-Parallel-UNet Fusion Module and the Sequential-UNet-LSTM Fusion Module. Both modules employ a two-stage process. In the first stage, we arrange individual slices side-by-side in a specific direction to capture intra-slice details. The second stage enhances slice correlations through either assembly from a different orthogonal direction or refinement of local adjacent slices.

The Bi-Parallel-UNet Fusion Module employs a two-path approach, using parallel streams of U-Net structures to process two-dimensional slices from orthogonal directions. It combines results from both paths to improve inter-slice relationships through fusion of differently processed slices. The Sequential-UNet-LSTM Fusion Module, in contrast, uses a sequential architecture. It begins with a U-Net module for intra-slice enhancement before using an LSTM network to establish spatial correlations between adjacent slices.

\vspace{0.5em}
\textit{C.2 Spatial feature regularization and reconstruction} 
\vspace{0.5em}

After designing the spatial feature modeling modules, we combine them to create spatial feature regularization and design corresponding reconstruction methods. Analyzing the classical inverse process for reconstruction—which underlies both traditional and deep learning (DL) methods—reveals two distinct computational frameworks: 1) iterative reconstruction with enhancement, and 2) light reconstruction and enhancement.

The first framework alternates between data regularization (where the reconstruction result adheres to the system's forward measurement process) and feature regularization (which applies priors concerning the imaging scene's features, including sparse and spatial features). The second framework balances precision and computational load through a light iteration process. This process combines data and feature regularization but focuses only on sparse features, with spatial feature regularization occurring afterward without additional iterations.

Building on these frameworks and leveraging the model-driven advantages of current deep learning reconstruction's deep unrolling framework, we develop four reconstruction neural networks. These networks combine our spatial feature regularization modules and computational frameworks to optimize reconstruction precision and efficiency. The networks—tomo-IRENet-3DTV, tomo-IRENet-U, tomo LRENet-biU, and tomo-LRENet-LSTM—implement different approaches. The first two utilize the iterative reconstruction with enhancement framework, paired with 3D-TV norm-based and single U-Net based regularization modules respectively. The latter two employ the light reconstruction and enhancement framework, incorporating the Bi-Parallel-UNet Fusion Module and Sequential-UNet-LSTM Fusion Module. We select these four networks for comprehensive performance evaluation.

\vspace{0.5em}
\textit{C.3 Evaluation and results} 
\vspace{0.5em}

We have developed a comprehensive evaluation framework to assess the designed reconstruction networks across multiple aspects: evaluation data, methods, perspectives, objects, and metrics. The evaluation uses both simulated data and two sets of publicly measured data. Due to current DL-based tomographic reconstruction networks' limitation to 1D processing, we have designed a new simulation approach specifically for urban buildings.

We have assessed four new reconstruction networks alongside three representative reconstruction methods—a FISTA-based method, SLIMMER, and tomo-IRENet-Raw (\cite{han2021efficient, zhu2010tomographic}). While the first two use traditional iterative optimization, the last follows current DL methodology (\cite{wang2023atasi, wang2023mada, qian2022basis, qian2022gamma}).

The evaluation examines both quality and efficiency. For urban area mapping quality, we focus on four key aspects: resolving overlaid scatterers, scatterer reconstruction precision, minimizing residual noise and clutter, and maintaining spatial structure integrity.

To thoroughly assess these aspects, we have designed six different imaging tests ranging from 1D to 3D cases, incorporating varying levels of building complexity—from simple to complex structures, single to multiple buildings, and simple to diverse urban areas. We have selected seven metrics to quantitatively evaluate 3D image reconstruction accuracy, 3D point cloud geometric spatial structure accuracy, and reconstruction efficiency.

The results have demonstrated that we have achieved our study objectives. Our spatial feature regularization approach, along with the designed modeling modules and reconstruction networks, successfully preserves building spatial structures. Importantly, the reconstruction efficiency remains the same level of to current neural network standards, confirming its viability for large-scale mapping.

\begin{figure}[h]
    \centering
    \includegraphics[width=1.1\textwidth]{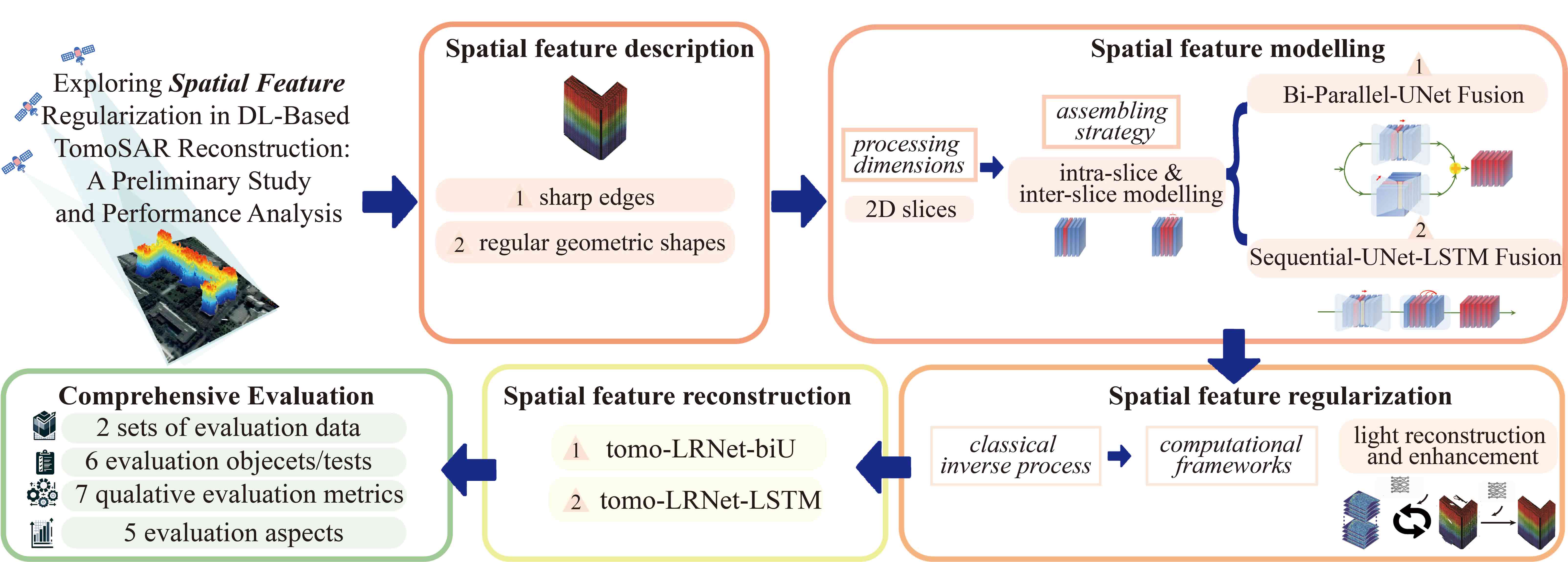}
    \caption{Simplified illustration of the study.}
    \label{fig_1}
\end{figure}

To aid in understanding the study, a simplified illustration is incorporated as in Fig.~\ref{fig_1}.

\subsection{Notions}

As we place emphasis on the comprehensive dimensionality of tomoSAR reconstruction-related data, we view this data in the form of tensors. To facilitate the readers' understanding, we provide an introduction to basic tensor concepts and notions~\cite{liu2022tensor} before delving into the detailed sections that follow.

\textit{Tensor:} In our study, the 3D scene image (representing reconstruction results without additional post-processing) and their corresponding 3D echo (a stack of 2D images from different observation angles) are both third-order tensors. We denote tensors with calligraphic letters, e.g., $\mathcal{X}$. Unless specified otherwise, the scene 3D image is denoted as $\mathcal{X} \in \mathbb{C}^{N_z \times N_x \times N_y}$, where $N_x$, $N_y$, and $N_z$ represent the number of pixel units along the range, azimuth, and elevation directions of the scene 3D image, respectively. The 3D echo is denoted as $\mathcal{Y} \in \mathbb{C}^{N_e \times N_r \times N_a}$, where $N_r$, $N_a$, and $N_e$ correspond to the number of pixel units along the range, azimuth, and elevation directions in the echo 3D image, respectively.

\textit{Vector:} For vectors, we use boldface lowercase letters, e.g., $\mathbf{x}$. A "fiber" in a tensor refers to a column vector within it, denoted as $\mathcal{X}(:,j,k)$. In our study, a 1D echo signal from one range-azimuth pixel unit corresponds to a 1D echo fiber, represented as $\mathbf{y} \in \mathbb{C}^{N_e \times 1}$. Similarly, the corresponding 1D scene signal corresponds to a 1D scene fiber, denoted as $\mathbf{x} \in \mathbb{C}^{N_z \times 1}$.

\begin{figure}[h]
    \centering
    \includegraphics[width=1.0\linewidth]{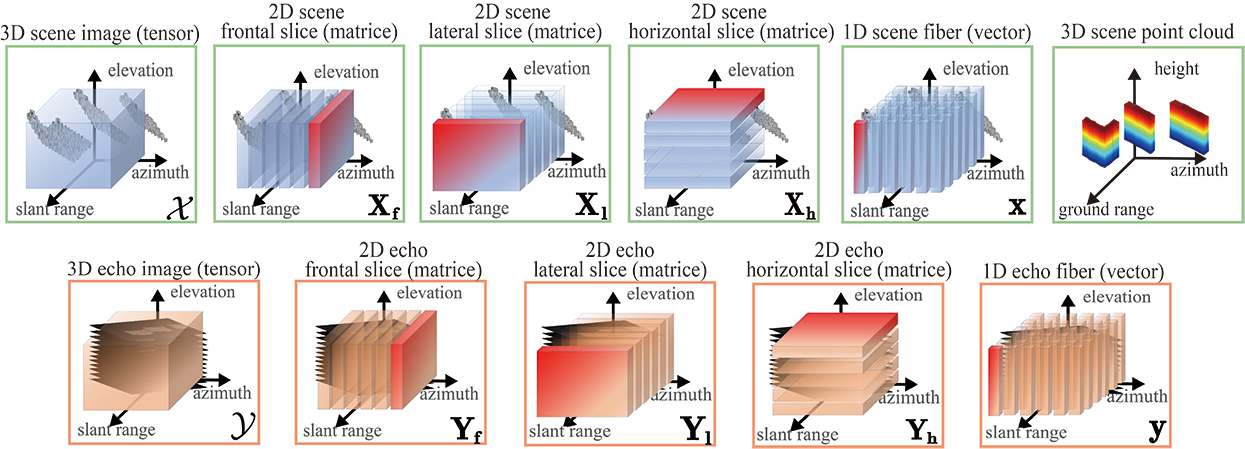} %
    \caption{Simplified illustration of notions in the study.}
    \label{fig_2}
\end{figure}

\textit{Matrice:} Matrices are represented using boldface uppercase letters, such as $\mathbf{X}$. Within a tensor, a "slice" refers to a matrix. Depending on different dimensions, there are three distinct types of slices: frontal, lateral, and horizontal slices, denoted as $\mathcal{X}(:,:,k)$, $\mathcal{X}(:,j,:)$, and $\mathcal{X}(i,:,:)$, respectively. In our study, a 2D echo signal along the elevation-range dimensions corresponds to a 2D echo frontal slice, which is represented as $\mathbf{Y_f} \in \mathbb{C}^{N_e \times N_r}$. The corresponding 2D scene signal along these dimensions corresponds to a 2D scene frontal slice, denoted as $\mathbf{X_f} \in \mathbb{C}^{N_z \times N_x}$. Similarly, we have 2D echo lateral slices (elevation-azimuth dimensions) $\mathbf{Y_l} \in \mathbb{C}^{N_e \times N_a}$ and 2D scene lateral slices $\mathbf{X_l} \in \mathbb{C}^{N_z \times N_y}$. Additionally, there are 2D echo horizontal slices (azimuth-range dimensions) $\mathbf{Y_h} \in \mathbb{C}^{N_a \times N_r}$ and 2D scene horizontal slices $\mathbf{X_h} \in \mathbb{C}^{N_y \times N_x}$ based on similar notational rules.

For clarity, we provide an illustration of these notions, as shown in Fig.~\ref{fig_2}.

\subsection{Organization}

The remaining sections of this report are organized as follows: Section II provides the preliminaries and background. Section III introduces the first phase of the study: spatial feature description and modeling. Section IV presents the second phase: spatial feature regularization. Section V discusses the third phase: spatial-feature-enhanced network design. Section VI details the fourth phase: method evaluation. Section VII delves into the discussion. 

\section{Preliminaries and Background}

This section begins with a brief overview of the generalized reconstruction model commonly used in tomoSAR reconstruction, which forms the foundation of our study. We then examine how reconstruction methods have evolved, comparing approaches both before and after the advent of deep learning.

\subsection{Mathematical model for TomoSAR reconstruction}

\begin{figure}[h]
    \centering
    \includegraphics[width=0.3\linewidth]{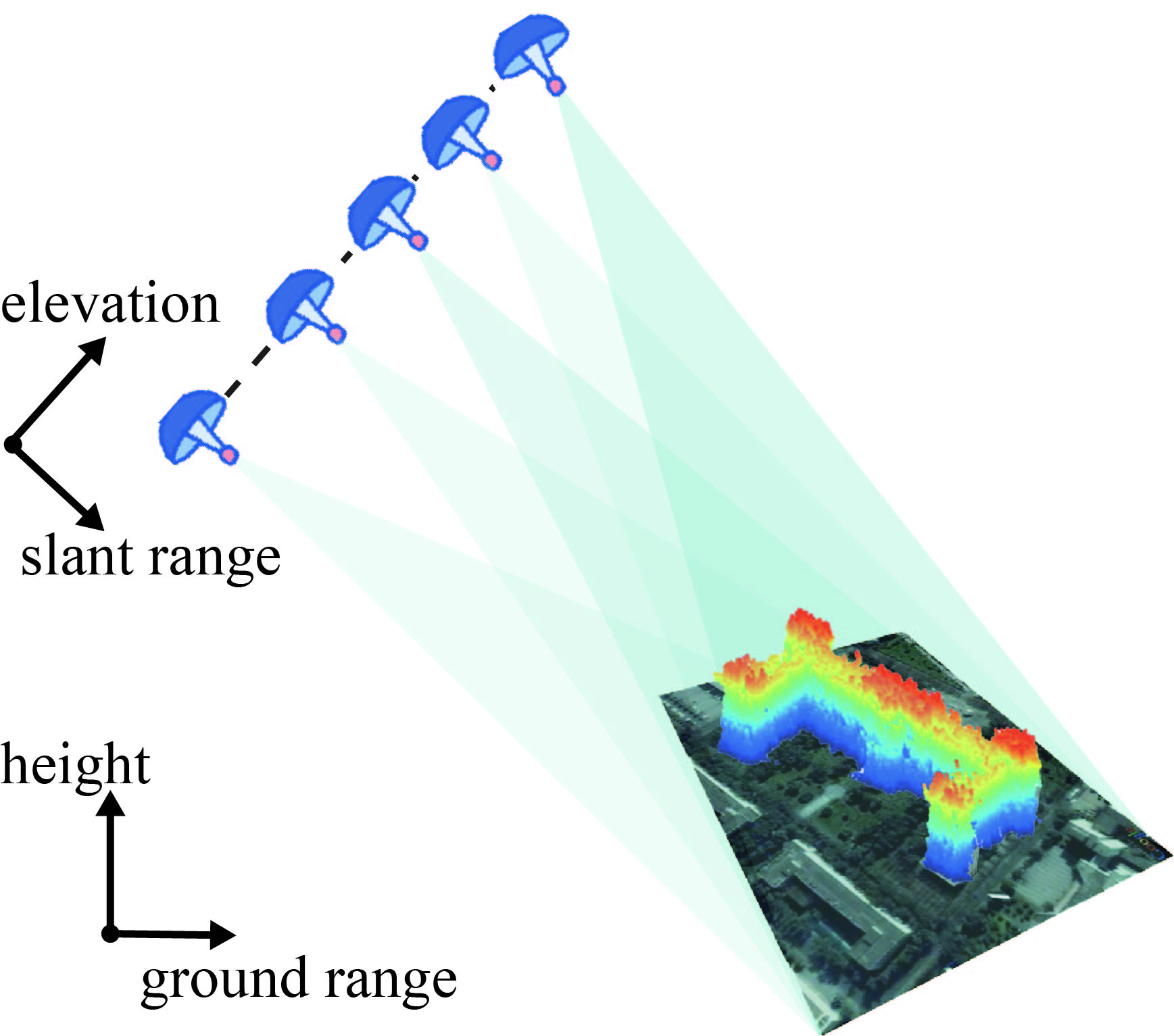} 
    \caption{TomoSAR observation geometrical relationship.}
    \label{fig_3}
\end{figure}

In a tomoSAR system, multiple observations are made from various angles in the elevation dimension, as shown in Fig.~\ref{fig_3}. 

Each observation produces a 2D image of the mapped scene in the slant-range/azimuth dimensions, following the common SAR principle. These 2D observations are stacked along the elevation direction to form the 3D echo image of the scene. These distinct observations create a baseline in the elevation dimension, allowing for the separation of overlaid scatters in the 2D image, ultimately resulting in the generation of a 3D scene image through reconstruction. The relationship between one echo fiber $\mathbf{y}$ and its corresponding scene fiber $\mathbf{x}$ in the 3D images is typically represented as follows\footnote{In this context, we assume that the echo has undergone preprocessing, including co-registration and de-ramping, to simplify the subsequent analysis~\cite{wang2022tomographic}.}.
\begin{equation}
	\label{equation_1}
	\centering
	\mathbf{y}=\mathbf{A}\mathbf{x}+\mathbf{n}
\end{equation}
Here $\mathbf{A} \in \mathbb{C}^{N_e \times N_z}$ represents the system forward measurement matrix, which takes the form of a discrete Fourier Transform (DFT) matrix. Additionally, $\mathbf{n} \in \mathbb{C}^{N_e \times 1}$ represents the measurement noise, typically assumed to follow a complex Gaussian distribution. 

\subsection{Reconstruction methods predating DL}

In essence, tomoSAR reconstruction involves solving the linear equation (\ref{equation_1}) based on the given measurements $\mathbf{y}$. To begin, we provide a brief overview of methods employed prior to the incorporation of DL into the field. 

During the initial phases, since the forward measurement process is the form of Discrete Fourier Transform (DFT), reconstruction was conventionally carried out by applying the inverse Discrete Fourier Transform (IDFT) straightforwardly. Viewed from an optimization perspective, this method is rooted in the principle of ensuring consistency through a least square error (LSE) approach between the measurements and the reconstructed result. Consequently, to achieve a high level of reconstruction precision, a significant number of measurements is required, which results in an extensive baseline with evenly spaced samplings. Meeting these requirements can be quite challenging in tomoSAR systems ~\cite{xiaolan2021sarmv3d, xiaolan2022key,qiu2024advances}. Typically, the available baseline length is limited, and the sampling distance is often uneven. These non-ideal factors can lead to limited resolution in the reconstruction results and can introduce interferences such as sidelobes and clutters. The process of solving a linear equation can also be seen as an inverse problem. However, due to the limited baseline length and uneven sampling distance, this often becomes an ill-posed problem within the context of tomoSAR systems. 

To enhance reconstruction precision and robustness against interferences, a common approach is to introduce additional regularization terms beyond the earlier mentioned regularization term~\cite{wang2022tomographic, shi2019nonlocal, yu2022efficient, wei2015novel, zhu2014superresolving, zhu2010tomographic}. These terms help impose more constraints on the solution process, guiding it towards an optimal solution. The former consistency regularization is often referred to as data regularization, while the latter consistency regularization terms are commonly known as prior regularizations. As the name suggests, these prior regularizations describe assumptions about the features of the result (scene to be mapped), resulting in a reconstruction that adheres better to its original prior distribution. 

In accordance with this philosophy, tomoSAR reconstruction methods have evolved into sparsity-based types over the last decade. A common assumption is that within the mapped scene, only a few strong scatterers truly exist. In other words, the scene image exhibits sparsity in the spatial domain, a feature that can be analytically regularized using the $l_1$ norm as a distance metric. Consequently, the reconstruction process corresponding to solving the optimization problem in  (\ref{equation_2}) ~\cite{zhu2010tomographic} is as follows.
\begin{equation}
\label{equation_2}
\hat{\mathbf{x}}=\mathop{\arg\min_{\mathbf{x}}}\left\{\frac{1}{2}\|\mathbf{y-Ax}\|_2^2 + \lambda\|\mathbf{x}\|_1\right\}
\end{equation}
Here, $\lambda$ serves as a hyper-parameter that characterizes the regularization strength of this prior term. The above optimization problem is normally solved through first-order proximal gradient algorithms~\cite{parikh2014proximal}, like iterative shrinkage threshold algorithm (ISTA)~\cite{parikh2014proximal, beck2009fast}, alternating direction method of multipliers (ADMM)~\cite{parikh2014proximal, wang20223}, etc. These algorithms typically involve iterations containing two essential steps in (\ref{equation_3}) and (\ref{equation_4}), respectively.
\begin{equation}
\label{equation_3}
\mathbf{v}^{(j+1)}=\mathbf{x}^{(j)} + \beta\mathbf{A}^H \left(\mathbf{y}-\mathbf{A}\mathbf{x}^{(j)}\right)
\end{equation}
\vspace{-0.75cm} 
\begin{equation}
\label{equation_4}
\mathbf{x}^{(j+1)}= \mathop{prox}(\mathbf{v}^{(j+1)};\beta,\lambda)
\end{equation}

The two aforementioned steps can be conceptualized as gradient descent and proximal denoising, respectively. \textit{1}) Gradient descent with a stepsize of $\beta$, which corresponds to the data regularization term. \textit{2}) Proximal denoising through soft-thresholding $\mathop{prox}(\cdot)$ with a threshold related to both $\beta$ and $\lambda$, which projects the solution to the desired signal feature space and corresponds to the prior regularization term~\cite{yuan2021snapshot, hastie2015statistical}. The iterative process involving these two main steps helps significantly reduce interferences like sidelobes, clutters, and noises. It also enhances the resolution of scatters and relaxes the requirements on baseline sampling, allowing for fewer samplings with sparser distributions. These advantages make it one of the preferred methods for tomoSAR reconstruction.

\subsection{Reconstruction methods in the DL area}

However, the above solving framework of traditional methods presents two notable disadvantages. Firstly, it requires the manual tuning of hyperparameters, which significantly influences both the accuracy and efficiency of the reconstruction process. Secondly, compared to the traditional Fourier-based reconstruction, the iterative nature of the optimization approach tends to be more time-consuming.

In recent years, the development of deep learning has given rise to a new methodology called learning to optimize~\cite{chen2022learning}. This approach holds great promise in tackling the aforementioned challenges. As the name implies, learning to optimize combines the benefits of data-driven learning with traditional analytic optimization iterations. In essence, by taking inspiration from the original iteration steps, a learnable architecture is constructed. This leads to higher-quality reconstruction results achieved with significantly fewer iterations and automatically learned hyperparameters.

Expanding on the learning to optimize methodology, a technique known as deep unfolding, or deep unrolling, has emerged~\cite{monga2021algorithm}. In deep unfolding, the optimization algorithm is unfolded or unrolled and integrated into the structure of a neural network. This process involves representing the iterative steps of the optimization algorithm as sequential layers within the network architecture. Additionally, to ensure computational efficiency, the unfolded optimization algorithm is often truncated to a significantly smaller fixed number of iterations. By incorporating the optimization algorithm into a neural network framework, deep unfolding enables end-to-end training and optimization, leveraging the learning power of deep learning techniques to enhance the performance of solving inverse problems.

In the field of TomoSAR imaging, ISTA is one of the most commonly unrolled algorithms~\cite{wang2023atasi, wang2023mada, qian2022basis, qian2022gamma}. Specifically, the Learned ISTA (LISTA) network framework has gained significant popularity and preference. Within the framework, key components such as the forward measurement matrix, gradient descent stepsize, and thresholding value are adaptively learned. Building upon the LISTA framework, Qian proposed the $\gamma$-net ~\cite{qian2022gamma}, which represents the first application of its kind in the field of TomoSAR imaging. The proposed network shows promising performances in terms of computation efficiency, super-resolution power, and estimation accuracy. 

Subsequent works have been conducted later, focusing on the framework of unfolding ISTA, to improve performance. For instance, \textit{1}) Wang introduced the ATASI-Net~\cite{wang2023atasi}, which focuses on addressing a large number of training parameters in the forward measurement matrix. This approach utilizes the analytic ISTA (AISTA) algorithm to determine the matrix offline, reducing training complexity and enhancing efficiency. \textit{2}) To tackle the problem of target loss, innovative techniques have been proposed. In the case of ATASI-Net, an adaptive element-wise threshold is obtained using a log-sum penalty function instead of the traditional $l_1$ norm. This modification helps preserve target information during the reconstruction process. Qian's another recent work introduced the concept of gated units inspired by recurrent neural networks, enhancing the ability to capture temporal dependencies and preserve targets~\cite{qian2022basis}). \textit{3}) Furthermore, the issue of mismatched forward measurement matrices due to spatial variations has been addressed. MAda-Net~\cite{wang2023mada}, for instance, introduces the measured response with spatial variety, allowing the network to compensate for the mismatch and enhance reconstruction accuracy. These advancements underscore the ongoing efforts to more accurate reconstructions.

\begin{figure}[t]
    \centering
    \begin{subfigure}[b]{0.24\linewidth}
        \includegraphics[width=\linewidth]{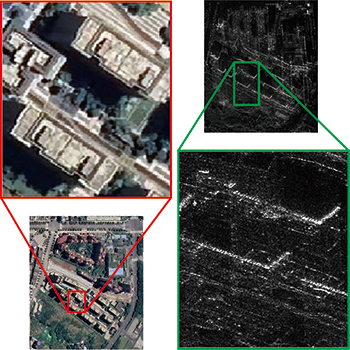}
        \caption{}
        \label{fig:sub1}
    \end{subfigure}
    \begin{subfigure}[b]{0.24\linewidth}
        \includegraphics[width=\linewidth]{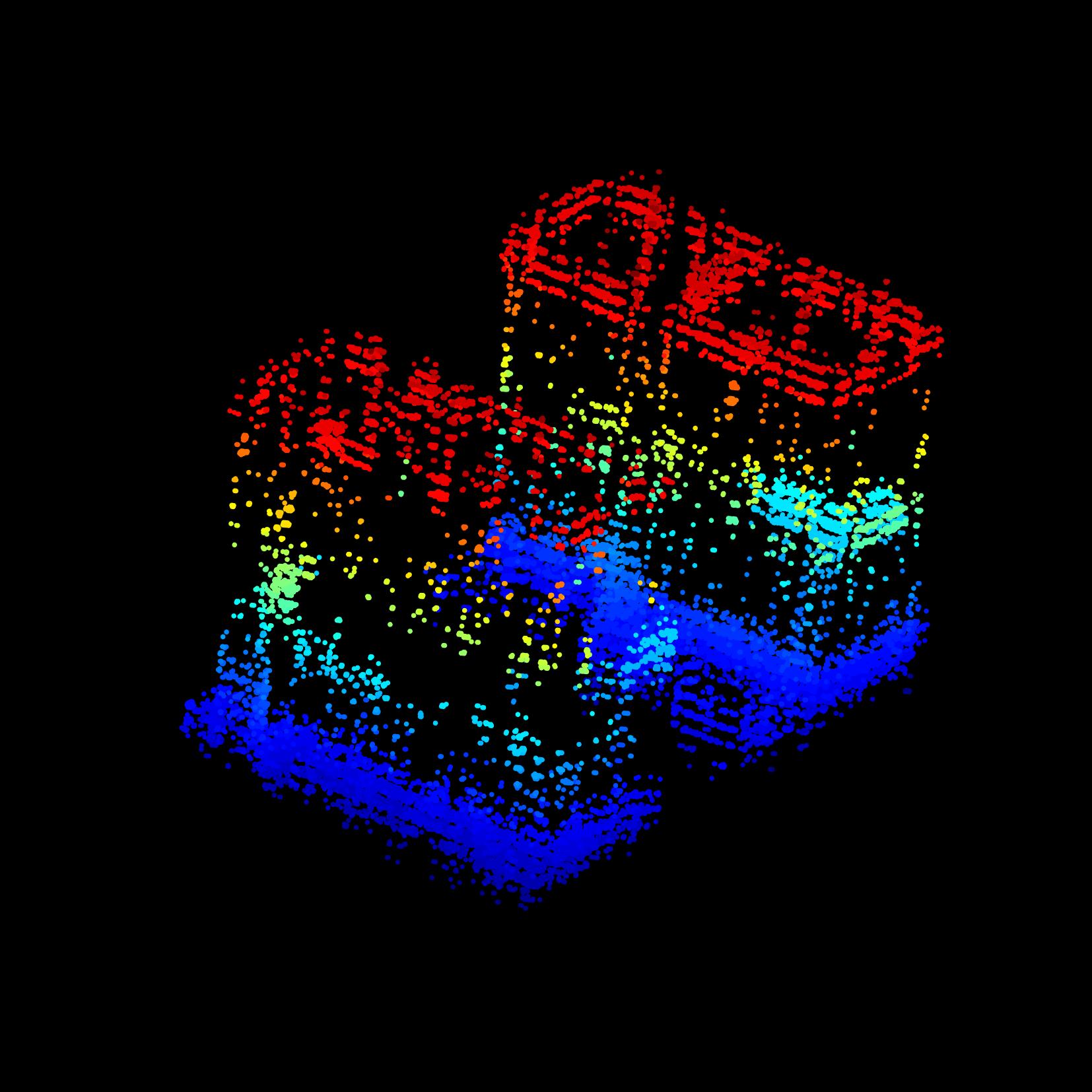}
        \caption{}
        \label{fig:sub2}
    \end{subfigure}
    \begin{subfigure}[b]{0.24\linewidth}
        \includegraphics[width=\linewidth]{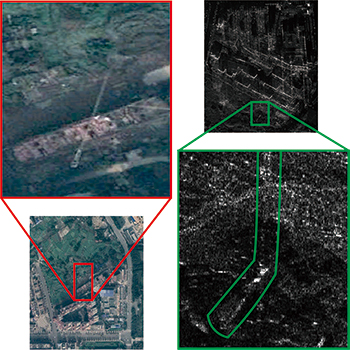}
        \caption{}
        \label{fig:sub3}
    \end{subfigure}
    \begin{subfigure}[b]{0.24\linewidth}
        \includegraphics[width=\linewidth]{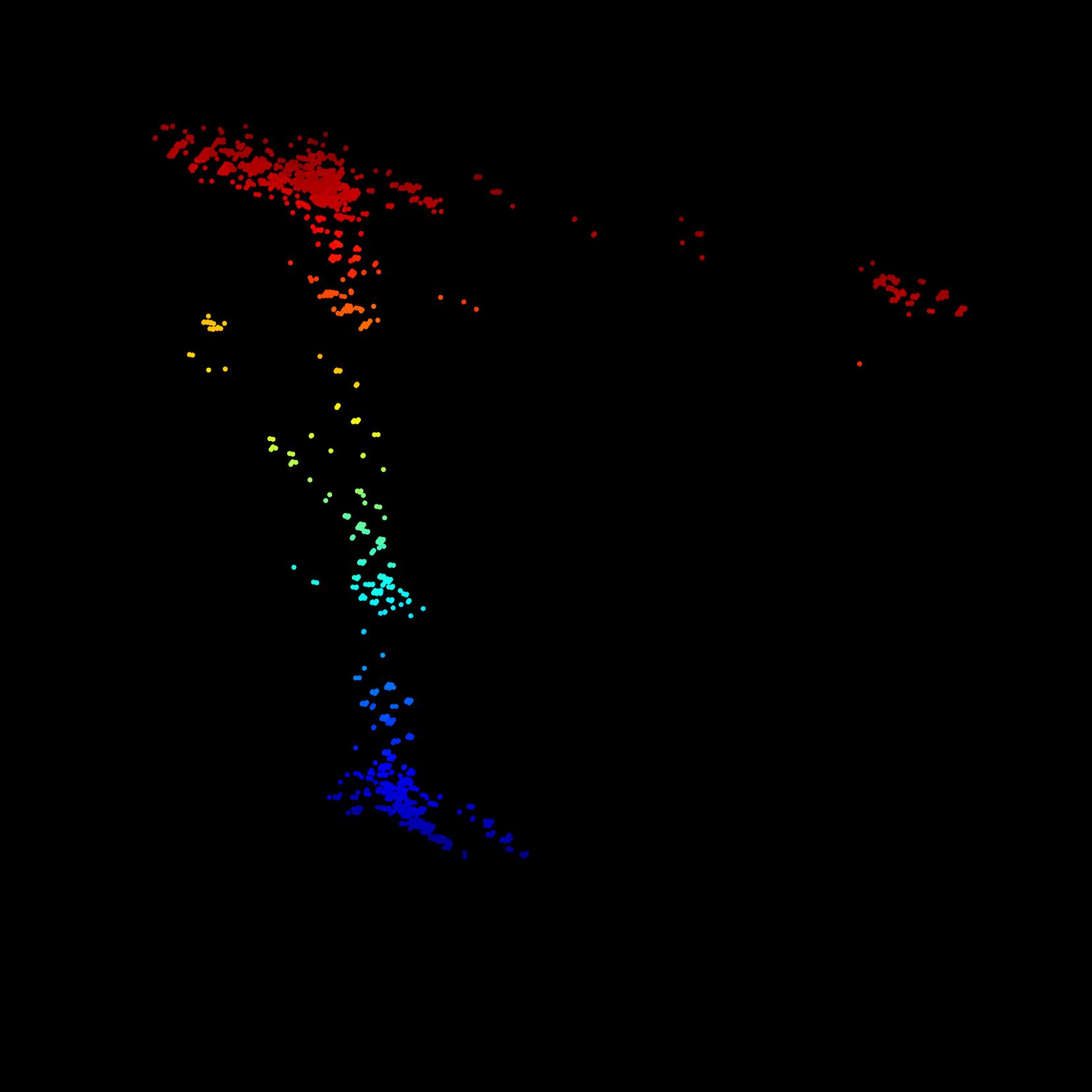}
        \caption{}
        \label{fig:sub4}
    \end{subfigure}
    \caption{Examples of spatial feature loss. (a) Optical image and 2D SAR image of two regular buildings in the scene. (b) Corresponding reconstruction results, showing holes in the surface of the buildings. (c) Optical image and 2D SAR image of a crane in the scene. (d) Corresponding reconstruction results, revealing a fragmented crane arm. \protect\footnotemark}
    \label{fig_4}
\end{figure}

However, it's worth emphasizing that in the recent works mentioned above, each scene fiber is processed independently, and the final reconstructed scene is constructed by piecing together these individually reconstructed fibers. Indeed, this approach has practical utility, but it's crucial to recognize that processing fibers independently can disrupt the intrinsic spatial relationships between fibers that are often found in targets such as urban buildings. Such disturbances to spatial features can manifest as gaps in building surfaces or fragmented edges. Typical examples are illustrated in Fig.~\ref{fig_4}. Based on this observation, we conducted the following study to delve deeper, focusing on the spatial feature regularization problem in DL-based TomoSAR, intending to solve this issue for better reconstruction results in urban-area mapping.

\footnotetext{These reconstructed results are obtained by the ISTA-unfolded neural network on the public Emei dataset~\cite{xiaolan2021sarmv3d, xiaolan2022key}.}

\section{Phase one: Spatial Feature Description \& Modeling}

Starting from this section, we officially delve into the first phase of our study, focusing on spatial feature description and modeling. This phase serves as the foundation for the subsequent phases. Through spatial feature description and modeling, we aim to gain a more specific and practical understanding closely related to the context of tomoSAR mapping objects, particularly in urban areas. In this section, we first analyze classical urban area objects to reveal their spatial features. Then, we analyze traditional methods for modeling spatial features, which typically involve manual design of such priors. In contrast to these conventional approaches and inspired by the spirit of deep learning, we propose utilizing deep learning modules to extract these priors automatically. For ease of comprehension, we name these two types of modeling as shallow and deep modeling, respectively.

\subsection{Spatial feature description}

In the context of urban area mapping, most of the objects within the area and of particular interest for further applications, such as urban planning, are man-made structures like buildings, bridges, and overpasses. The shape pattern is a distinct identifier for each object to be 3D mapped. It typically comprises two essential factors: structure and geometry~\cite{yang2022dsg}. To be more specific, the structure of a man-made object defines how its components are arranged to create the overall shape. For instance, in the case of a building, the structural elements include walls, windows, and a roof. These components combine in a specific way to form the building's shape. And, the geometry of an object refers to its size and the spatial relationships between its components. In the context of urban mapping, this means the building's height, width, and length, as well as the angles between surfaces, such as the roof and walls. 

In the case of objects under microwave illumination and imaging using tomoSAR systems, two significant aspects become apparent:  

\textit{1}) \textit{Sharp edges:} Radar waves interact with the edges and corners of man-made structures, leading to strong radar reflections. These interactions create distinct, high-contrast edges in radar images. For example, the edges of buildings appear as sharp lines.

\begin{figure}[t]
    \centering
    \begin{subfigure}[b]{0.24\linewidth}
        \includegraphics[width=\linewidth]{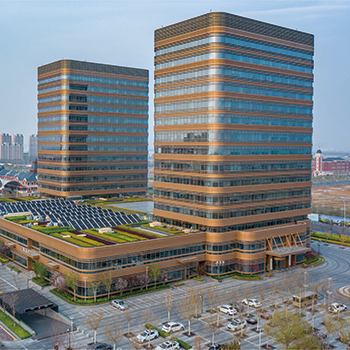}
        \caption{}
        \label{fig:sub1}
    \end{subfigure}
    \begin{subfigure}[b]{0.24\linewidth}
        \includegraphics[width=\linewidth]{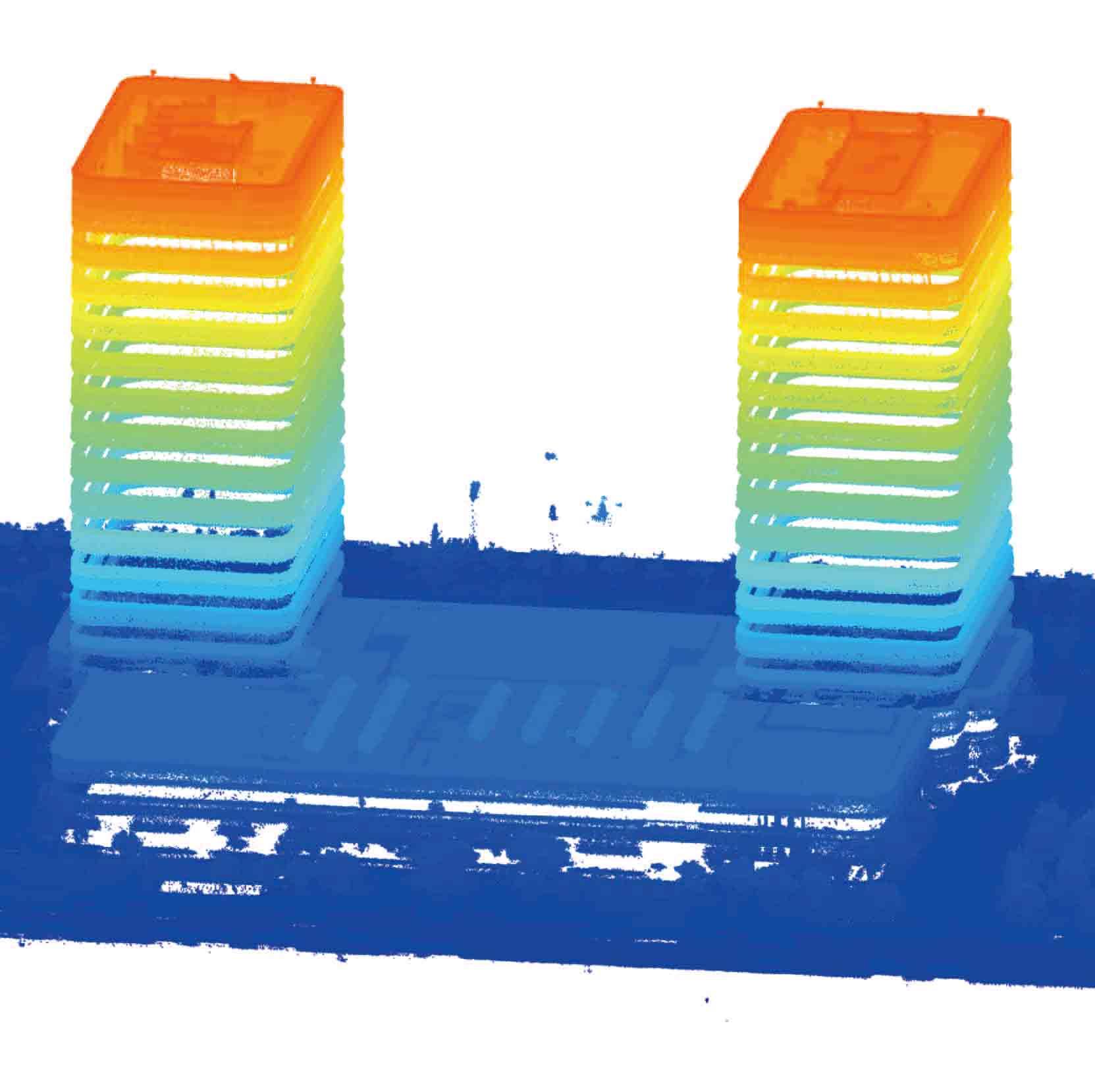}
        \caption{}
        \label{fig:sub2}
    \end{subfigure}
    \begin{subfigure}[b]{0.24\linewidth}
        \includegraphics[width=\linewidth]{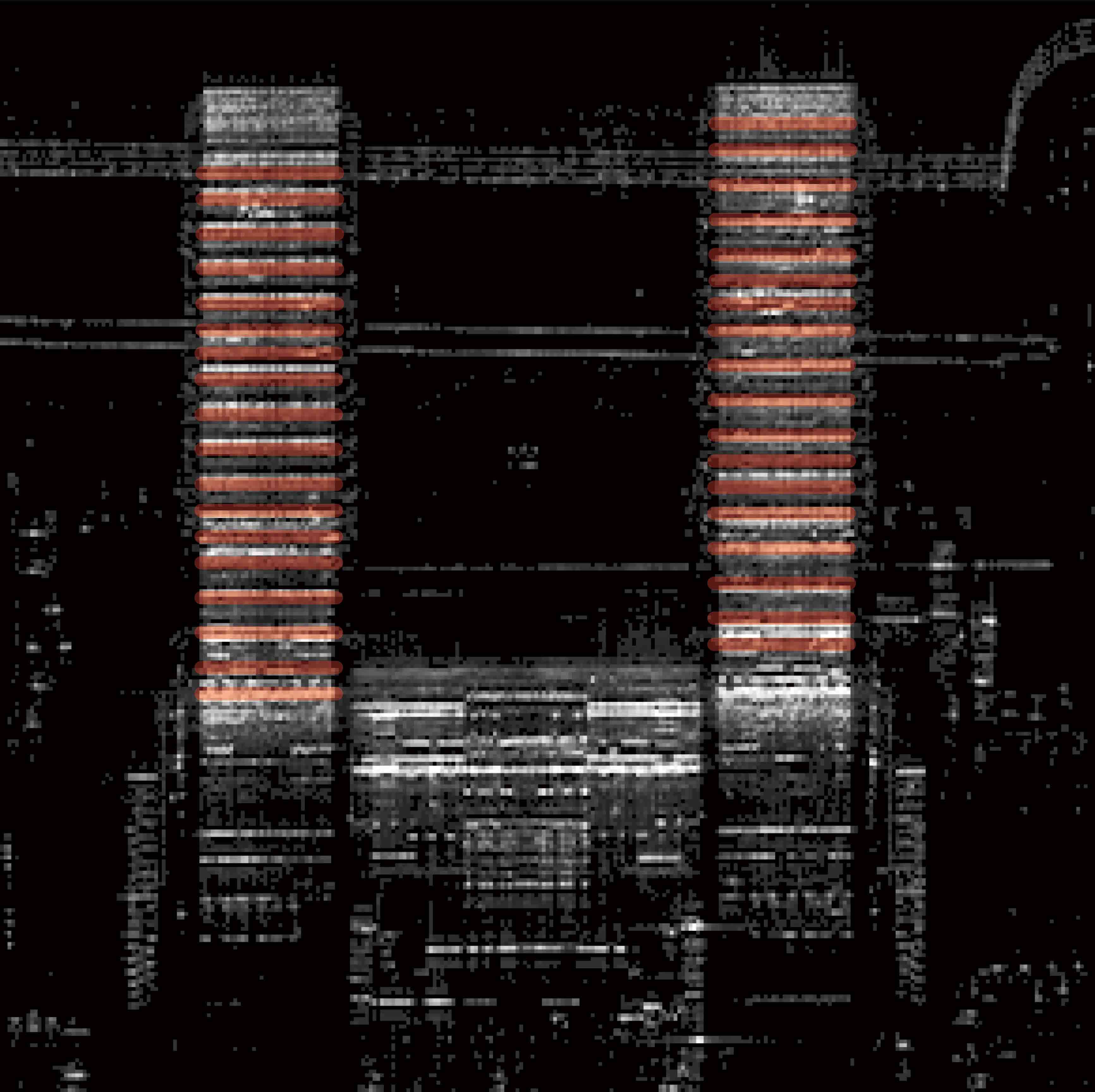}
        \caption{}
        \label{fig:sub3}
    \end{subfigure}
    \begin{subfigure}[b]{0.24\linewidth}
        \includegraphics[width=\linewidth]{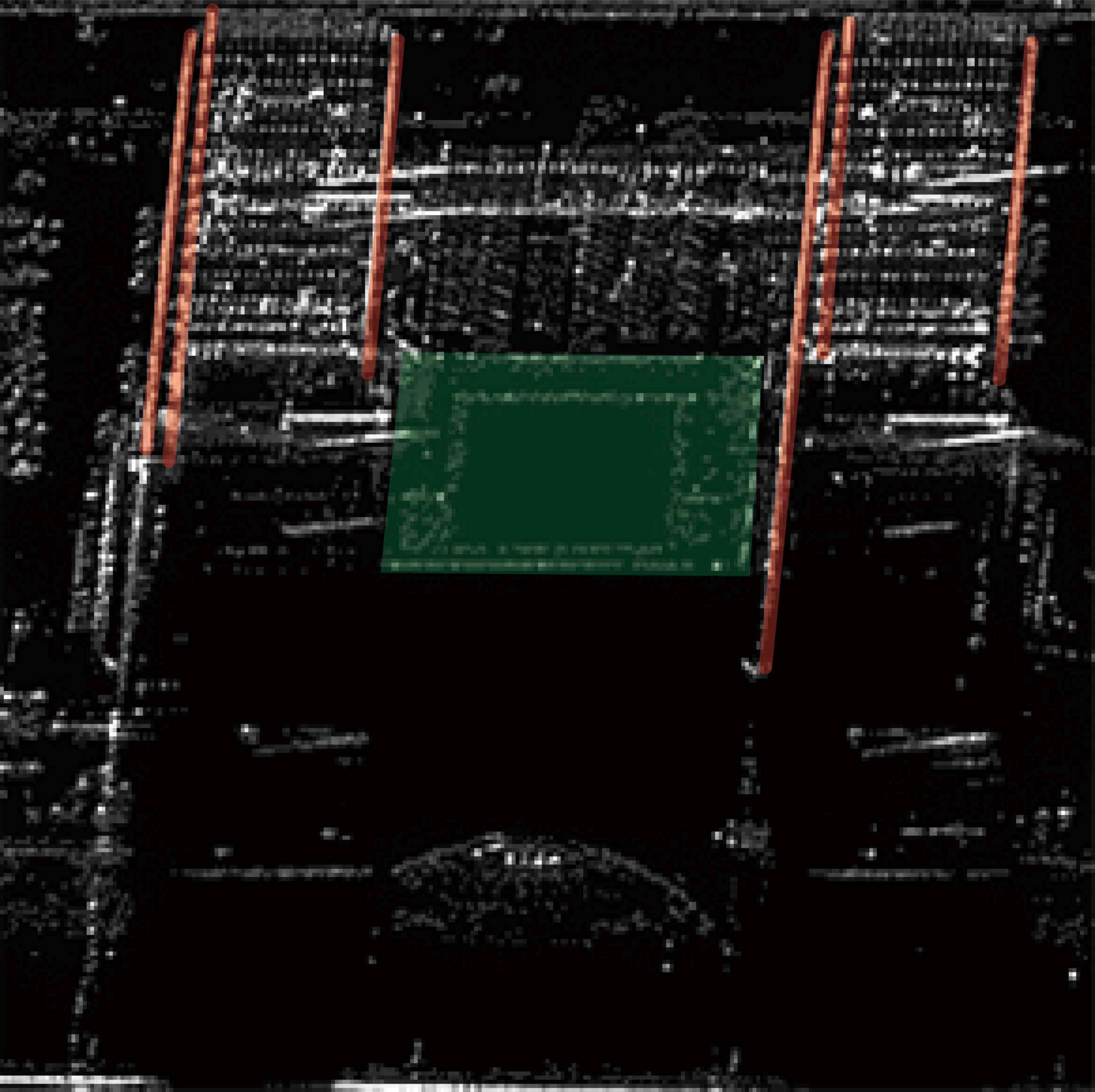}
        \caption{}
        \label{fig:sub4}
    \end{subfigure}
    \caption{Examples of spatial features in radar images. (a) Optical image of the scene. (b) 3D Lidar point cloud of the scene. (c) Radar image of the scene from one angle, with line structures marked in red. (d) Radar image of the scene from another angle, with line structures marked in red, and a regular geometric shape highlighted with a green box. \protect\footnotemark}
    \label{fig_5}
\end{figure}

 \textit{2}) \textit{Regular geometric shapes:} Buildings, roads, bridges, and other man-made structures often have regular geometric shapes. Buildings, for instance, often have rectangular or quadrilateral forms. Due to distinctive shape patterns that allow for easy differentiation from other categories, such as trees and bushes, through visual intuition, even in microwave radar images, as illustrated in Fig.~\ref{fig_5}. 

Such observations align with findings in previous tomoSAR studies that have utilized non-deep learning reconstruction methods~\cite{rambour2019introducing, aghababaee2019regularization, zhu2015joint, yao2022tomographic, jiao2023preliminary, han2023geometric, jiao2020urban}. These features have also shared a same critical role in various alternative approaches to urban area 3D mapping. For instance, there is a burgeoning trend in the field involving the use of deep neural networks to directly infer height information from a single 2D SAR image, showcasing the versatile applications of these features~\cite{recla2022deep, sun2022large, chen20193d}.

Unlike the commonly adopted sparsity feature of scatters found in (\ref{equation_2}), which predominantly concerns the scatter intensity relationships and is modeled in the intensity domain, the features we've discussed earlier delve into the spatial relationships and positions of scatters in the spatial domain. They capture information about how different scatters are positioned in relation to one another, their orientations, shapes, and overall spatial layout within the scene. 

\footnotetext{The LiDAR point cloud and radar images are obtained from~\cite{xiaolan2021sarmv3d, xiaolan2022key}.}

These features are distinct from the more commonly adopted intensity features, and we refer to them as spatial features. Viewed from another geometric manifold perspective, spatial features introduce constraints on the potential support set in the reconstruction result~\cite{jiao2023preliminary}. For instance, scatters on the surface or edges of objects should have similar heights, forming clusters, or they should exhibit gradual and continuous variations instead of sudden changes, as opposed to clutter or noise scatters that are independent of each other. 

\subsection{Spatial feature shallow modeling}

After obtaining a more specific and concrete interpretation of spatial features in the context of urban area mapping, the next crucial step is to find ways to model these features, just as intensity features are usually modeled through the intensity $l_1$ norm in (\ref{equation_2}). This modeling of spatial features is essential for harnessing their potential to improve reconstruction precision and guide the optimization process toward more accurate results.

Given the prior work in the area that has explored spatial feature regularization and demonstrated its effectiveness, we primarily draw inspiration from their ideas to establish the foundation for our study's modeling. Underlying these approaches is a shared fundamental concept: the utilization of distance metrics to evaluate the similarities between individual scatters and their neighboring counterparts. A higher value of distance typically indicates greater dissimilarity. A lower distance value between scatters indicates that they are more similar and likely belong to the same component.~\cite{aghababaee2019regularization}. After implementing such regularization, the subsequent reconstruction (optimization) processes will therefore be affected by the additional loss introduced by the corresponding distance value.

The metrics employed in prior studies can be categorized into two types: stochastic, as demonstrated by Jiao et al. ~\cite{jiao2020urban}, and deterministic, as seen in the works of Rambour et al.~\cite{rambour2019introducing}, Aghababaee et al.~\cite{aghababaee2019regularization}, Zhu et al.~\cite{zhu2015joint}, Yao et al.~\cite{yao2022tomographic}, Jiao et al.~\cite{jiao2023preliminary}, and Han et al.~\cite{han2023geometric}. In light of the established framework in  (\ref{equation_2}) that we aim to build upon, which is deterministic and also widely adopted by current deep learning methods, we opt for this form of regularization due to its seamless integration into the existing framework. 

In the domain of deterministic modeling, numerous approaches are closely linked to the height of scatters, assuming that adjacent scatters exhibit similar or limited height variances. These approaches can be applied to sets of fibers or individual fibers. The measurement of similarities can be approached from various perspectives. For example, works like those of Aghababaee et al.~\cite{aghababaee2019regularization} and Zhu et al.~\cite{zhu2015joint} make use of the $l_1$ norm to quantify the differences between a scatter fiber and its adjacent ones, or employ the $l_{1,2}$ norm for fiber cluster, respectively. On the other hand, in the research conducted by Zhao et al.~\cite{yao2022tomographic}, the $\mathrm{Shattern}\!-\!\mathrm{P}$ norm is applied to model the low-rank characteristics of fiber clusters. These modeling approaches are summarized as follows, respectively.
\begin{equation}
\label{equation_5}
g_1(\mathbf{x}) = \sum_{q} \left\lVert \mathbf{x - x_q} \right\rVert_1
\end{equation}
\vspace{-0.75cm} 
\begin{equation}
\label{equation_6}
g_2(\mathbf{X})=\|\mathbf{X}\|_{1,2}
\end{equation}
\vspace{-0.75cm} 
\begin{equation}
\label{equation_7}
g_3(\mathbf{X}) = \left\lVert \mathbf{X} \right\rVert_p
\end{equation}
Here, $\mathbf{x_q}$ represents the adjacent fibers surrounding $\mathbf{x}$. The first two processes use a single fiber, while the last process uses fiber clusters.

The three approaches, as evident from equations (\ref{equation_5}) to (\ref{equation_7}), primarily enforce regularization on a per-fiber basis. In contrast to the aforementioned options, the 3D Total Variation (3D-TV) norm used in the work of Rambour et al.~\cite{rambour2019introducing} views the entire scene tensor as the target of the regularization. Processing slices in different orientations, rather than focusing solely on individual fibers, may offer a more flexible and potentially scalable approach for large-area mapping. This method has the potential to better integrate spatial information across varying perspectives, which could be advantageous for handling complex scenes. Traditional approaches, including those that rely on image processing techniques such as clustering, classification, fitting, and even morphological operations, have shown effectiveness in specific scenarios~\cite{guo2024urban, li2024morphology, ma2024homor, wang2024efficient, han2023geometric, jiao2023preliminary}. However, these methods often involve manual intervention or parameter tuning, which can be difficult to integrate seamlessly into automated deep learning frameworks. Such dependencies may hinder their application in large-scale mapping tasks, where automation, scalability, and efficiency are critical. 

The 3D-TV norm directly models the entire 3D scene tensor and is defined as follows:
\begin{equation}
\label{equation_8}
g_{TV}(\mathcal{X}) = \left\lVert \mathbf{D_h}(\mathcal{X}) \right\rVert_1 + \left\lVert \mathbf{D_l}(\mathcal{X}) \right\rVert_1 + \left\lVert \mathbf{D_f}(\mathcal{X}) \right\rVert_1
\end{equation}

where $\mathbf{D_h}$, $\mathbf{D_l}$, and $\mathbf{D_f}$ indicate differential operator on the horizontal, lateral and frontal dimension, respectively. 

As shown in (\ref{equation_8}), in contrast to the earlier approaches presented in (\ref{equation_5}) to (\ref{equation_7}), the fundamental processing elements have shifted from individual fibers to slices of varying dimensions. Building upon our earlier analysis of spatial feature description, two critical aspects have been identified: sharp edges and regular geometric shapes. These features are particularly susceptible to various forms of noise, including outliers and clutter, which can degrade reconstruction quality.

To address these challenges, the 3D-TV norm plays a pivotal role. It is renowned for its ability to promote sparsity in the gradient domain, effectively reducing noise while preserving the sharpness of edges—one of the key spatial features. For regular geometric shapes, which inherently exhibit structural smoothness, the 3D TV norm further proves advantageous by minimizing deviations from this smoothness. This property helps maintain the regularity and integrity of geometric structures, preventing artifacts and distortions during the reconstruction process.

The effectiveness of the 3D TV norm has been demonstrated across a wide range of applications in image processing. Notably, it has been applied successfully in domains such as array-SAR~\cite{wang2022ctv}, hyperspectral image restoration~\cite{peng2020enhanced}, and snapshot compressive imaging~\cite{yuan2021snapshot}, among others~\cite{boominathan2020phlatcam}. These examples underscore the versatility and robustness of the 3D TV norm in addressing the critical aspects of noise reduction and spatial feature preservation in complex imaging tasks.

\vspace{1em}
\hrule
\vspace{1em}
\subsubsection*{Note} 

In addition to the approaches mentioned earlier for capturing spatial structural features, several novel methods have been explored in recent works related to TomoSAR and similar fields~\cite{yao2022tomographic, wu2023learning, pu2021ortp, lin2023deep, lu2021accurate}. These methods provide alternative perspectives for describing spatial features. One such perspective involves modeling high-similarity features through a low-rank property, which assumes that the scene tensor is redundant~\cite{yao2022tomographic, wu2023learning, pu2021ortp}. This redundancy is enhanced by promoting sparsity in the transformed domain, typically achieved through singular value decomposition (SVD). Another notable approach focuses on capturing directive features using the shearlet transform~\cite{lin2023deep, lu2021accurate}. The shearlet transform emphasizes sparsity in the feature domain, making it particularly effective for detecting anisotropic structures like edges while suppressing isotropic elements such as noise and clutter.

While each of these methods has unique advantages and nuances, they share a common goal: to identify a transform domain that enforces feature regularization through sparsity, thereby distinguishing meaningful spatial structures from other elements in the image. However, when embedding regularization into deep learning (DL) frameworks, practical considerations such as precision, efficiency, and storage requirements play a crucial role. Among the various options available, it is important to evaluate the computational burden associated with the forward and inverse transform operators. For example, SVD and the shearlet transform, while powerful, can introduce higher computational overhead compared to simpler operators such as the differential operator used in total variation methods.

In this context, the 3D TV norm has been chosen as the preferred approach for our study. It strikes a practical balance between computational efficiency and feature modeling effectiveness. By focusing on shallow modeling of spatial features, the 3D TV norm aligns well with our research objectives, offering a lightweight yet robust solution that addresses the challenges of precision, efficiency, and scalability in large-scale tomographic reconstruction tasks.
\vspace{1em}
\hrule
\vspace{1em}

\subsection{Spatial feature deep modeling}

In the previous section, we have reviewed potential solutions for spatial feature modeling based on current literature and related works in other imaging modalities. A key observation from this review is that these solutions largely rely on manual designs, which are guided by prior knowledge and expert recognition of spatial structures.

Before exploring spatial feature deep modeling, it is important to revisit the developments in sparse reconstruction, as it serves as the foundational theory behind TomoSAR reconstruction. Sparse reconstruction traditionally represents spatial features in a specifically designed domain, often through the construction of a sparse coding dictionary. This approach allows spatial features to be expressed as a linear combination of only a few atoms (basis elements) in the dictionary, while non-ideal factors such as noise and clutter remain unrepresented~\cite{kang2020learning}. The effectiveness of this approach depends heavily on how well the dictionary is designed, as a more suitable dictionary enables a sparser and more efficient representation of the features.

To address the limitations of manually designed dictionaries, dictionary learning techniques have been introduced as an enhancement. These methods aim to learn a dictionary tailored to the data by training on small batches of clean images (or training samples) through optimization algorithms such as the alternating direction method of multipliers (ADMM)~\cite{kang2020learning}. This adaptive approach provides a significant improvement over manual designs, offering greater flexibility and accuracy in modeling spatial features. However, it remains a semi-manual process that relies on predefined assumptions and limited training data.

Building on these advancements, our work explores the potential of deep learning to move beyond traditional dictionary learning. By leveraging neural networks, we aim to automatically learn complex and expressive representations directly from larger datasets, bypassing the need for manually designed or semi-optimized dictionaries. This shift represents a significant evolution in spatial feature modeling, allowing for greater precision, adaptability, and scalability. To clearly differentiate this new approach from earlier methods, we refer to it as "deep modeling," while retaining the term "shallow modeling" for manual and dictionary-based techniques.

This progression—from manual designs to dictionary learning and finally to deep modeling—reflects a natural transition in the field toward more automated and efficient approaches. By building on the strengths of prior methods while addressing their limitations, deep modeling opens up new possibilities for advancing spatial feature representation in TomoSAR .

\newpage
\vspace{0.5em}
\textit{C.1 Processing dimensions}
\vspace{0.5em}

When employing deep neural networks for spatial feature modeling, it is essential to carefully consider the fundamental elements the network will process, as this directly influences the network’s architecture. A straightforward approach might involve using the entire scene tensor as input. While 3D neural networks have shown success in other imaging modalities—such as video snapshot compressive imaging~\cite{sun2022video} and hyperspectral image restoration~\cite{ma2023learning}—the unique characteristics of radar images, particularly their inherent sparsity, present challenges for such an approach. Unlike optical images, radar images are significantly sparser, which can hinder the effectiveness of 3D tensor-based processing. Furthermore, employing 3D neural networks introduces computational challenges due to the heavy demands of 3D operators, such as convolution and pooling. These operations can be resource-intensive, increasing computational costs and potentially limiting scalability. This conflicts with our primary objective: developing a TomoSAR reconstruction method that achieves higher precision while maintaining the efficiency required for large-scale urban mapping.

Another consideration is the availability of suitable training data. Constructing a sufficiently large and diverse dataset of 3D training samples for radar images can be a challenging task, adding another layer of complexity to the adoption of 3D neural networks. Given these constraints, we lean toward lighter and more practical solutions that align with the strategies used in spatial feature shallow modeling.

To address these challenges, we focus on using slices of the scene as the fundamental elements for deep neural network processing. This approach offers several advantages. First, it significantly reduces computational demands, as working with 2D slices is inherently less resource-intensive than processing full 3D tensors. Second, it allows us to leverage the extensive research and mature tools developed for 2D image processing in the computer vision community. These tools have demonstrated remarkable effectiveness and robustness across a wide range of applications, making them particularly well-suited for our purposes. By adopting this strategy, we aim to strike a balance between precision, efficiency, and feasibility, ensuring that our approach aligns with the requirements of large-scale urban mapping while fully capitalizing on the strengths of deep learning for 2D image processing.

\begin{figure}[h]
    \centering
    \includegraphics[width=0.35\linewidth]{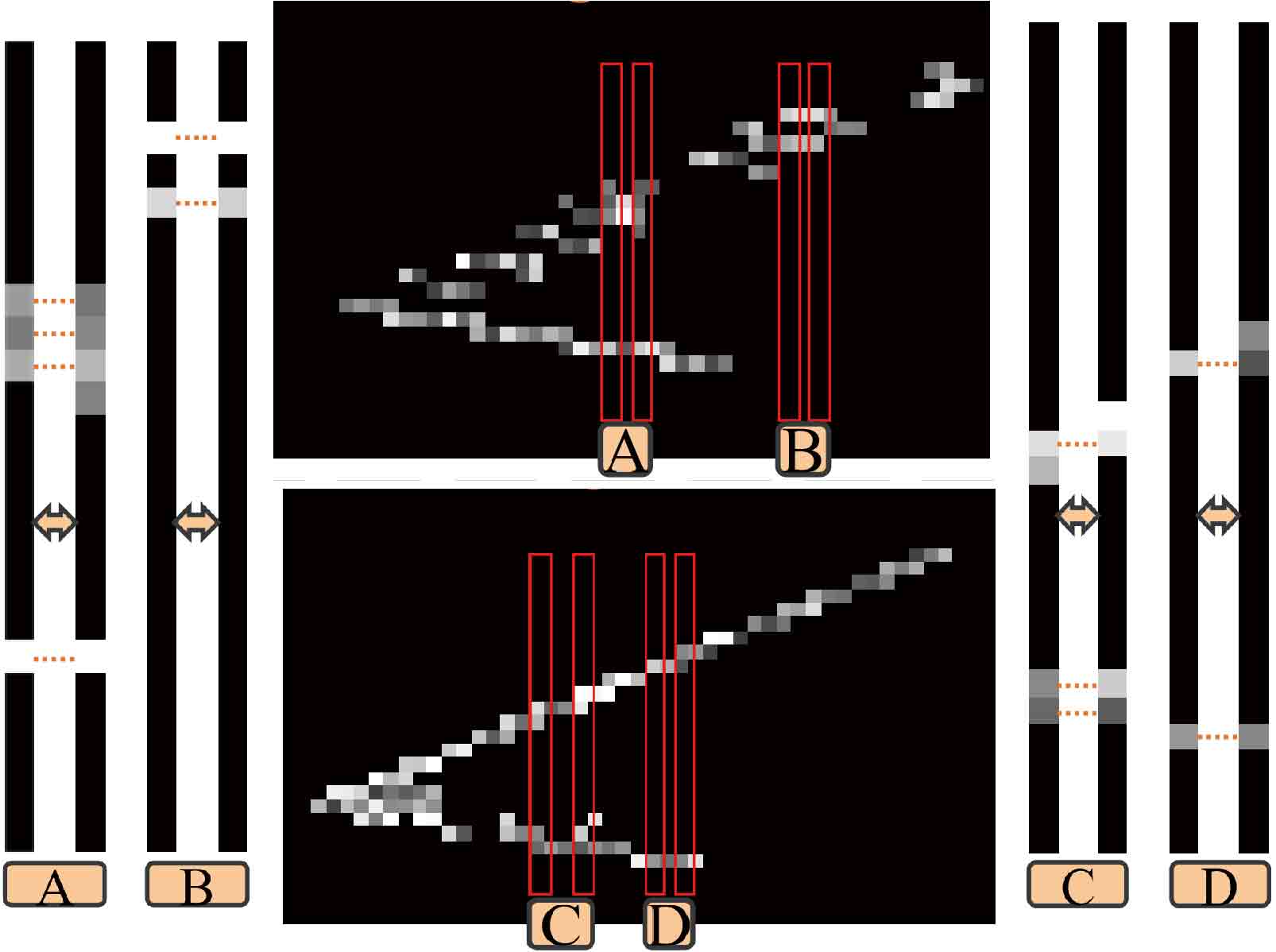} % 使用 \linewidth 来指定图的宽度为单栏宽度
    \caption{Scene slices illustrate specific spatial features. There are two slices along the slant-range and elevation dimensions. Four sets of adjacent scene fibers are randomly selected, highlighted with red rectangles, and labeled as A, B, C, and D. These slices display edges distinctly and reveal a high correlation between adjacent fibers.\protect\footnotemark}
    \label{figure_six}
\end{figure}

A fundamental characteristic of these 2D scene slices further supports this decision. Each slice, which is essentially a 2D image constructed from adjacent scene fibers, naturally preserves spatial structures within its plane. This is in stark contrast to the much simpler 1D scene fiber, where such spatial relationships are absent. For better understanding, Fig.~\ref{figure_six} illustrates this concept, showcasing two frontal slices of the scene. The figure highlights four adjacent fibers, enclosed in red rectangles, and demonstrates their high correlation. Additionally, it reveals the presence of clear edges within the slices, emphasizing the ability of 2D slices to retain meaningful spatial features. This inherent characteristic makes 2D slices a more effective and efficient choice for feature modeling and reconstruction compared to alternative approaches.

\footnotetext{The two slices are obtained through the simulation method introduced latter.}

Next, to effectively reveal the spatial features within each slice, we employ the widely recognized encoder-decoder structure of deep neural networks. This structure has proven its effectiveness across various applications, including computer vision~\cite{ji2021cnn, badrinarayanan2017segnet}, reconstruction tasks for different imaging modalities~\cite{yao2020enhanced, haggstrom2019deeppet, sun2022hyperspectral, monakhova2019learned}, and more recently, SAR imaging tasks~\cite{pu2021deep, wei2022carnet}. The principle behind this structure is compelling: it encodes input data into a compressed format, akin to condensing a large amount of information into a smaller, more manageable representation, and subsequently decodes this compact form back into the original space. This process ensures that the critical features of the data are carefully preserved while reducing redundancy.

\begin{figure}[h]
    \centering
    \includegraphics[width=0.5\linewidth]{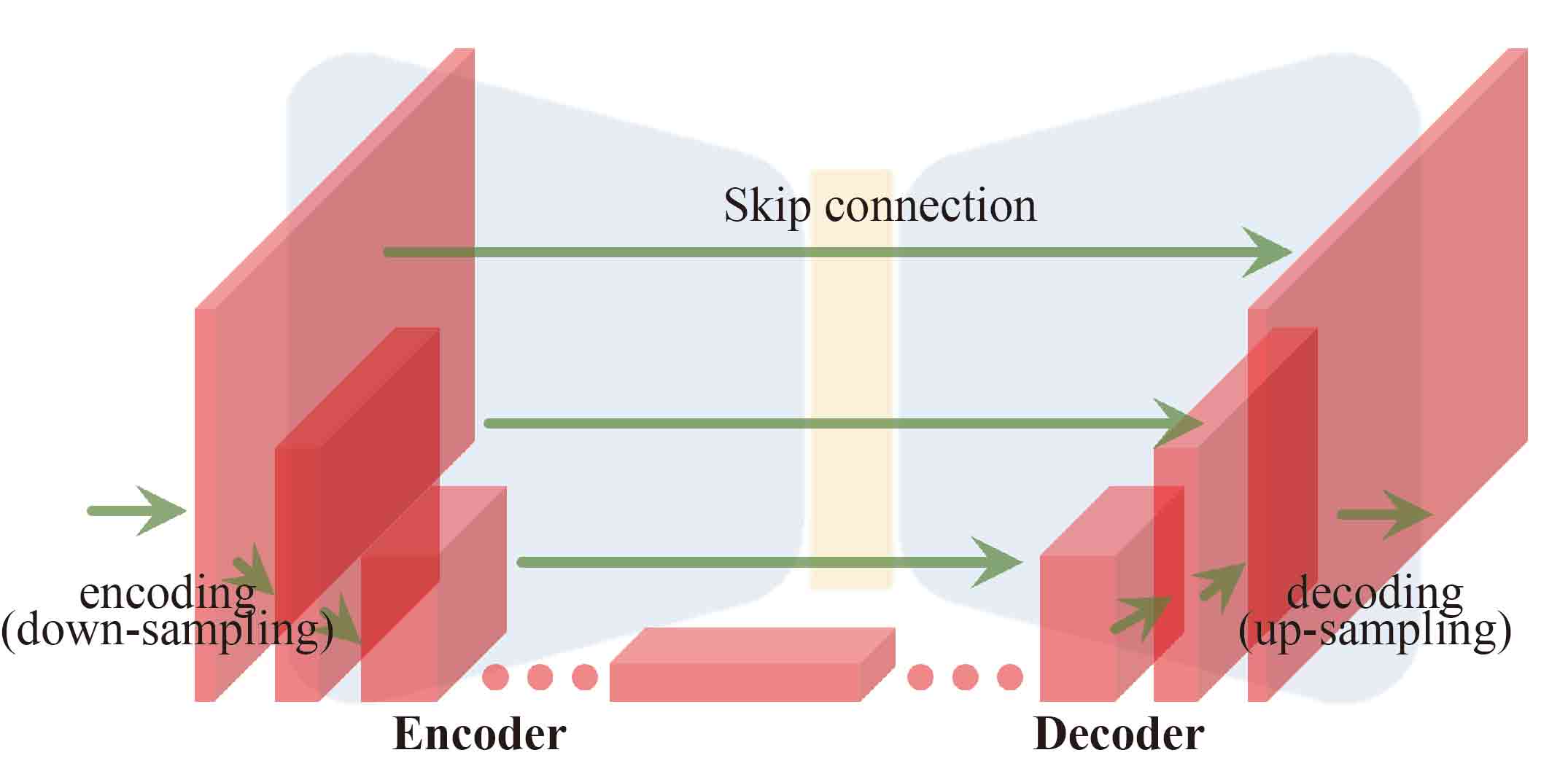} 
    \caption{Simplified illustration of the U-Net structure. The structure is known for its encoder-decoder framework. A raw image is repeatedly encoded (downsampled) and then decoded (upsampled) back to its original size. It also features skip connections between the same levels.}
    \label{fig_7}
\end{figure}

Within this encoder-decoder framework, we adopt the U-Net structure as the specific model for our work, as illustrated in Fig.~\ref{fig_7}. U-Net enhances the traditional encoder-decoder framework with its characteristic 'U'-shaped architecture and the addition of skip connections, making it particularly effective for spatial feature extraction and reconstruction tasks. The model encodes the input image into compact representations and subsequently decodes these back to the original size, with skip connections enabling efficient integration of features across different levels. These enhancements allow the U-Net to strike a balance between capturing detailed spatial information and reconstructing accurate outputs. 

The following outlines the specific features of the U-Net model and their relevance to our study:

\begin{itemize}
    \item \textit{Effective Architecture:}
	The U-Net’s characteristic 'U'-shaped architecture consists of two main components: an encoder (contracting path) and a decoder (expanding path). The encoder reduces the spatial dimensions of the input image while increasing its feature depth, effectively capturing critical data representations. The decoder then restores the spatial dimensions using the information encoded in the contracting path, ensuring accurate reconstruction of spatial features. This dual-path design allows for efficient representation learning while maintaining a strong focus on spatial detail.
    \item \textit{Preservation of Spatial Information:}
    A key strength of the U-Net is its ability to capture spatial information at multiple scales throughout the encoding process. This spatial information, which holds crucial image features, is efficiently integrated into the decoding process to rebuild the input data. This characteristic is particularly valuable for handling spatial relationships and layouts in 2D slices, which are essential for accurate feature modeling.
    \item \textit{Feature Fusion through Skip-Connections:}
    The U-Net architecture incorporates skip connections that link corresponding layers in the encoder and decoder paths. These connections enable the fusion of fine-grained, detail-oriented information from earlier layers with high-level abstract features from deeper layers. This feature fusion across scales enhances the model’s ability to extract and reconstruct spatial features, providing a more comprehensive representation of the input data.
    \item \textit{Efficiency in Training:}
    U-Net models are known for their efficiency in training, requiring fewer training samples compared to other deep neural network architectures. This efficiency makes them especially practical for our task, given the limited availability of training samples for radar-based applications.
\end{itemize}

These features make the U-Net model an attractive and well-suited choice for revealing spatial features in 2D slices. By leveraging this architecture, we can effectively capture and reconstruct the critical spatial structures within each slice while maintaining computational efficiency. 

\newpage
\vspace{0.5em}
\textit{C.2 Modeling strategies}
\vspace{0.5em}

Given that we are not utilizing the entire tensor of the 3D scene, it becomes crucial to determine an effective way to reconstruct the original tensor using 2D scene slices. The previous processing primarily focused on individual slices, leveraging the U-Net model to extract and emphasize their spatial characteristics. This approach revealed the inherent spatial structures within each slice. However, such a method may fall short in capturing spatial features that span across multiple slices. To address this limitation, we must consider broader spatial structures that extend beyond a single slice, enabling the optimal utilization of spatial features across multiple slices. In this study, we refer to this comprehensive strategy as intra-slice and inter-slice modeling.

While the intra-slice modeling aspect has already been discussed, we now turn our attention to inter-slice modeling. To effectively capture inter-slice spatial relationships, we propose two distinct deep modeling modules: the Bi-Parallel-UNet Fusion and the Sequential-UNet-LSTM Fusion. These two approaches are designed to address the challenge of integrating information across multiple slices, each with a specific focus and methodology. 

\vspace{1em}
\textit{1) Bi-Parallel-UNet Fusion}

We begin by introducing the most straightforward strategy for constructing the 3D scene tensor. This method involves assembling 2D slices of a specific type, such as frontal slices, along the slant-range and elevation dimensions, similar to assembling a jigsaw puzzle. When these slices are arranged at various azimuth locations, they collectively form the 3D scene tensor along the azimuth direction. This approach leverages the inherent spatial structure present within the frontal slices to reconstruct the broader scene. However, while this method is conceptually simple, it relies heavily on the quality and resolution of the frontal slices alone, which may limit the ability to fully capture features spanning across different dimensions.

An equally valid and complementary alternative is to use lateral slices, which lie in the elevation-azimuth plane, at different slant-range locations. This approach allows the construction of the 3D scene tensor in the range direction instead. Unlike frontal slices, lateral slices capture spatial features from a different perspective, emphasizing variations and structures along the range dimension. By combining both perspectives, the reconstruction process can better capture the complete spatial structure of the 3D scene, ensuring a more accurate and robust representation.

\begin{figure}[h]
    \centering
    \includegraphics[width=0.35\linewidth]{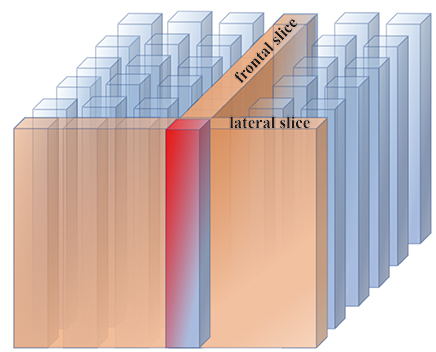} 
    \caption{A scene tensor can be constructed by assembling scene slices obtained from different directions. Each scene fiber within this tensor is shared by scene slices that correspond to various directions. This sharing of scene fibers indicates the correlation between different slices.}
    \label{fig_8} 
\end{figure}

A critical observation is that a single scene fiber appears in at least two slice types—for example, the same fiber may be present in both a frontal slice and a lateral slice. This is illustrated in Fig.~\ref{fig_8}, where the overlap of fibers between slices is shown. 

These overlaps indicate a strong interaction and correlation between the different slice types. Such correlations are valuable as they provide complementary information that can be used to enhance the reconstruction process. For example, features that are unclear or partially represented in one slice type may be more distinctly visible in another, and combining these perspectives ensures a more complete capture of the scene's spatial features.

To fully exploit these correlations, we have designed the Bi-Parallel-U-Net Fusion module, which consists of three key operators: $\mathcal{E}_\mathbf{l}(\cdot)$, $\mathcal{E}_\mathbf{f}(\cdot)$, and $\mathcal{M}(\cdot)$. These operators are defined to model spatial features within slices and fuse them across slice types. Specifically, $\mathcal{E}_\mathbf{l}(\cdot)$ encodes and extracts spatial features from lateral slices (elevation-azimuth dimension), while $\mathcal{E}_\mathbf{f}(\cdot)$ handles spatial feature extraction from frontal slices (slant-range and elevation dimension). The fusion operator $\mathcal{M}(\cdot)$ integrates the features from both slice types.

\begin{figure}[h]
    \centering
    \includegraphics[width=0.5\linewidth]{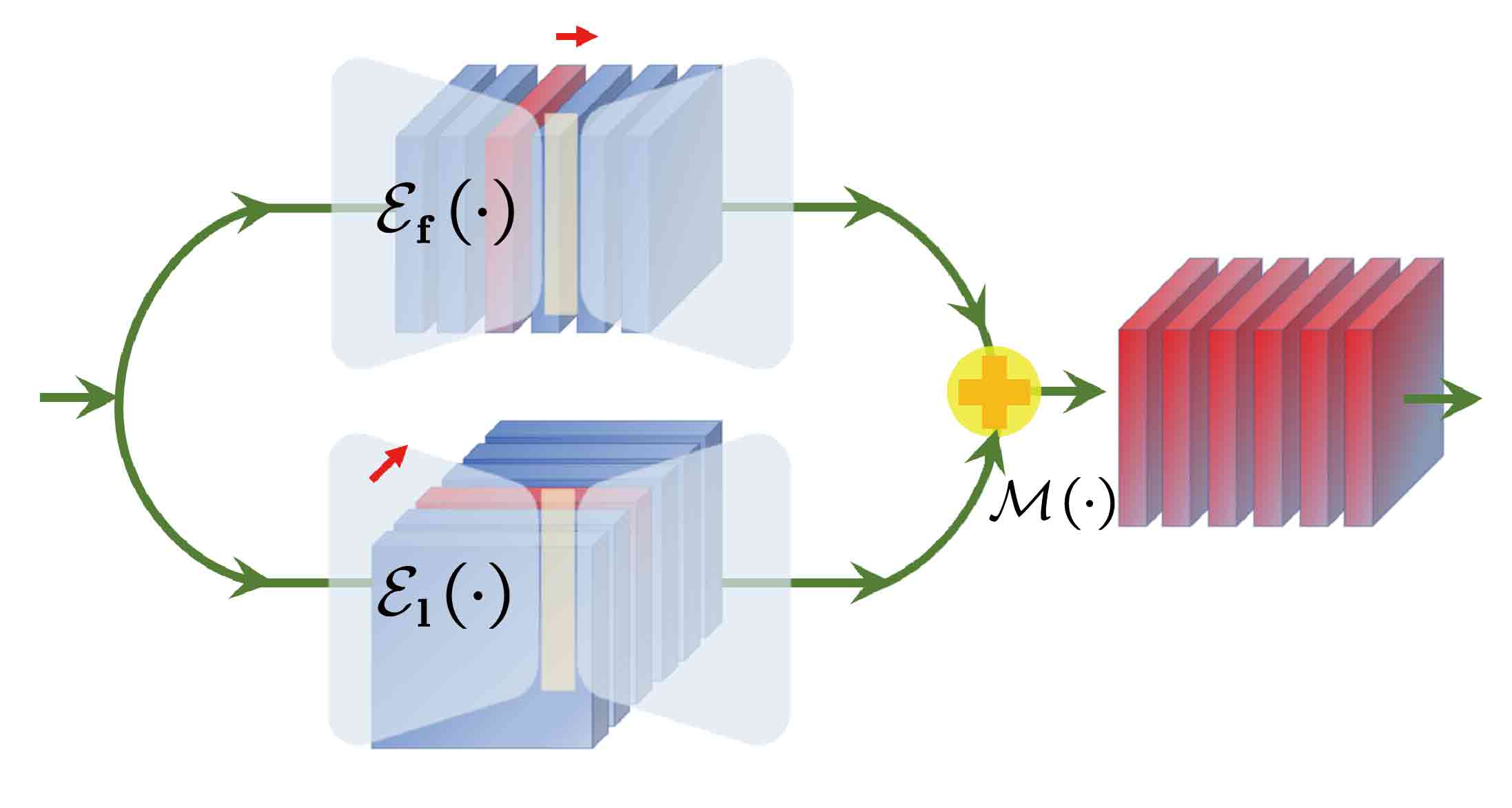} 
    \caption{Simplified illustration of Bi-Parallel-UNet Fusion module's structure. It comprises two parallel U-Net paths. These paths model spatial features within slices from different directions. Additionally, it includes a combination to model spatial features across various directions.}
    \label{fig_9}
\end{figure}

After processing the slices with $\mathcal{E}_\mathbf{l}(\cdot)$ and $\mathcal{E}_\mathbf{f}(\cdot)$, two separate 3D scene tensors are generated—one representing the lateral slices and the other representing the frontal slices. These tensors encapsulate distinct spatial features unique to their respective slice types. The fusion operator $\mathcal{M}(\cdot)$ is then applied to combine these tensors using a max operation. The simplified structure of this module is shown in Fig.~\ref{fig_9}.

\textit{2) Sequential-UNet-LSTM}
\vspace{1em}

\begin{figure}[h]
    \centering
    \includegraphics[width=0.6\linewidth]{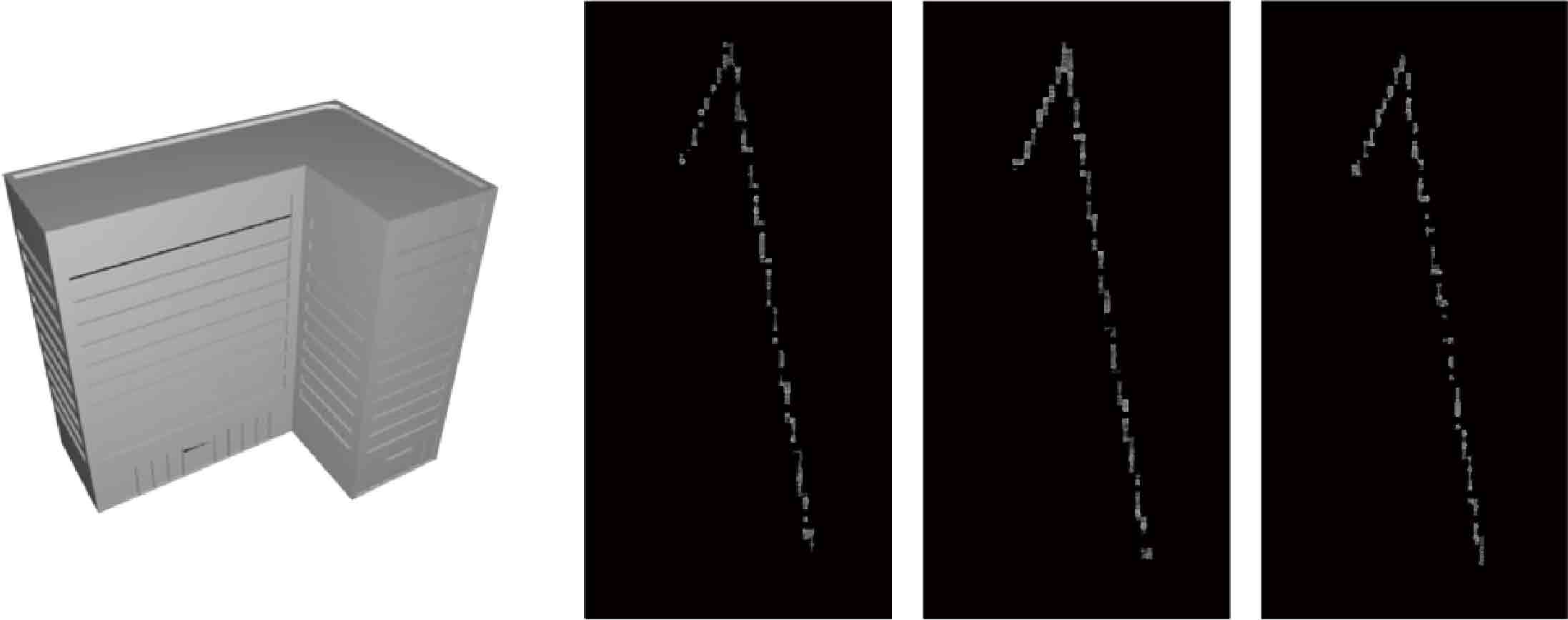}
    \caption{Correlations between adjacent slices: Showcases a building model
	and three of its five adjacent slices.\protect\footnotemark}
    \label{fig_10}
\end{figure}

In the previous module, we leveraged the correlation between slices by recognizing that a single fiber is shared across slices of varying dimensions. Building on this foundation, this module shifts focus to a new perspective: the local correlation between adjacent slices, as illustrated in Fig.~\ref{fig_10}. Adjacent slices exhibit a high degree of correlation, being highly repetitive and sharing similar features as part of the same structure.

\footnotetext{Slices are obtained through the simulation method introduced latter.}

\begin{figure}[h]
    \centering
    \includegraphics[width=0.7\linewidth]{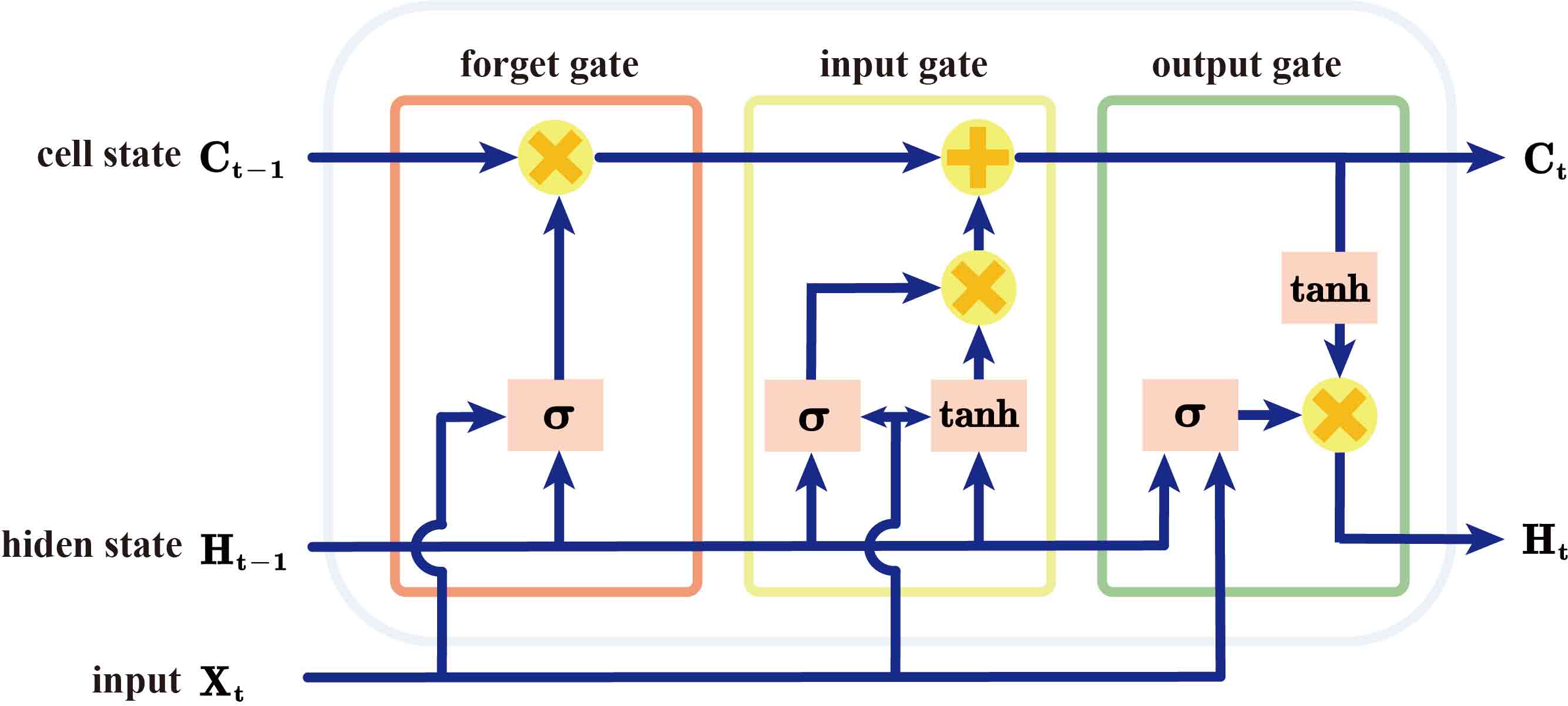} 
    \caption{Illustration of one basic LSTM structure. It maintains long-term memory through the cell state and short-term memory through the hidden state. This is achieved using several gate mechanisms: the forget gate, the input gate, and the output gate. The forget gate discards irrelevant information, the input gate updates new information, and the output gate releases outdated information.}
    \label{fig_11}
\end{figure}

To leverage these correlations, we adopt the Long Short-Term Memory (LSTM) architecture. LSTM, a variant of recurrent neural networks (RNNs), is particularly well-suited for capturing long-term dependencies in sequential data. Fig.~\ref{fig_11} illustrates the basic structure of an LSTM, highlighting its key components and features.  

There are two core concepts in LSTM's information state: cell state and hidden state. 
\begin{itemize}
\item \textit{Cell state:} The cell state, acting as the structure’s long-term memory, runs through the cascaded LSTM structures. It carries information across many steps, allowing the network to remember and use long-term dependencies in the data.
\item \textit{Hidden state:} The hidden state, functioning as the structure’s short-term memory, is the output of each LSTM module for each step. It carries information to the next step in the cascaded structures, influenced by the previous hidden state, the current input, and the cell state.
\end{itemize}

The LSTM structure operates a series of gates to manage information flow, including the forget gate, input gate, and output gate. 

\begin{itemize}
    \item \textit{Forget gate:} The forget gate determines what information should be discarded from the cell state, i.e., what prior information is no longer relevant. This is achieved by processing the current input and the previous hidden state through a sigmoid function.
    \item \textit{Input gate:} The input gate decides what new information should be added to the cell state. It first determines the update values by passing the current input and the previous hidden state through a sigmoid function. Next, it creates a vector of candinate new values for the state using a tanh function. The cell state is then updated by multiplying the update values by these candidate values.
    \item \textit{Output gate:} The output gate dictates the next hidden state. It uses a sigmoid function on the current input and the previous hidden state to decide what parts of the updated cell state should be output. Then, it applies a tanh function to the updated cell state and multiplies it by the sigmoid function's output to determine the final output.
\end{itemize}

This sophisticated mechanism enables LSTMs to capture both short-term and long-term dependencies effectively, making them particularly well-suited for sequential data. Recent studies, such as Qian~\cite{qian2022basis}, have leveraged this capability by modeling temporary fiber results from the tomoSAR iterative process as time-series data. In Qian’s approach, LSTMs reintroduce long-term information during iterations, significantly reducing target loss.

Our research, however, diverges from Qian’s by focusing on spatial rather than temporal dependencies. Specifically, we aim to model the correlations among adjacent slices during scene reconstruction, and for this purpose, we adopt a variant of LSTM known as convolutional LSTM (convLSTM). ConvLSTM has been shown to be more effective in image processing tasks~\cite{NIPS2015_07563a3f}, as it replaces the traditional fully connected operations in LSTMs with convolutional operations. This modification makes convLSTM better suited for capturing spatial correlations within image data.

\begin{figure}[h]
    \centering
    \includegraphics[width=0.7\linewidth]{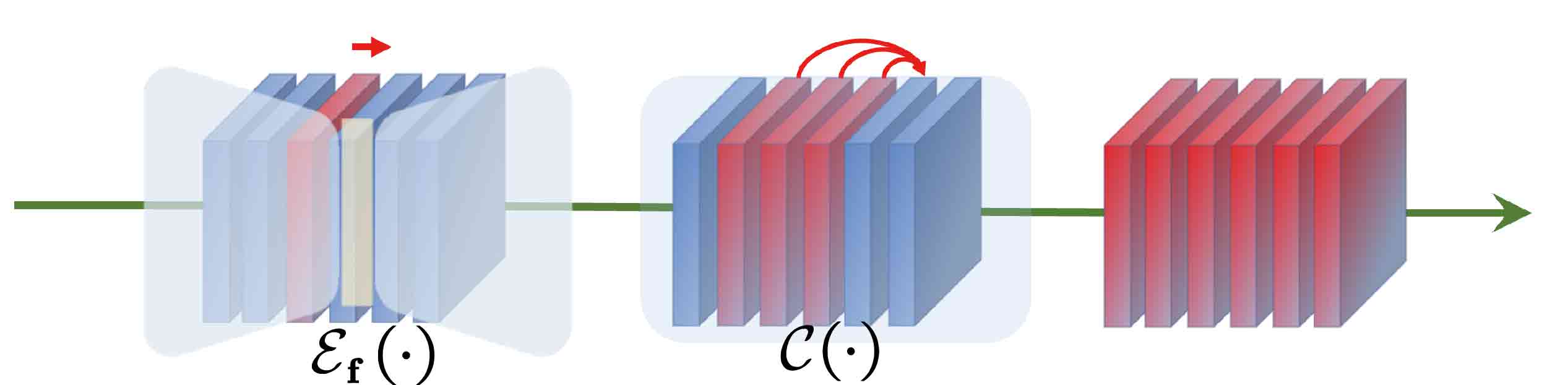} %
    \caption{Simplified illustration of Sequential-UNet-LSTM Fusion module's structure. It consists of a sequential combination of a U-Net based structure for intra-slice modeling and an LSTM based structure for inter-slice modeling.}
    \label{fig_12}
\end{figure}

Based on this concept, we have developed the Sequential-UNet-LSTM Fusion module, which is designed to model inter-slice correlations during the reconstruction process. A simplified structure of this module is presented in Fig.~\ref{fig_12}. To achieve this, we first apply an operator, denoted as $\mathcal{E}_\mathbf{f}\left(\cdot\right)$, to each slice in one direction. This step models the spatial features within each slice and, from a global perspective, assembles these slices to form the initial scene tensor. Subsequently, from a local perspective, we introduce a correlation operator, $\mathcal{C\left(\cdot\right)}$, based on the LSTM structure to capture spatial dependencies among adjacent slices effectively.

\section{Phase two: Spatial Feature Regularization}

In Section III of our research, we have successfully completed the initial phase, which was dedicated to an in-depth exploration of spatial features. This phase involved a comprehensive analysis of their characteristics and culminated in the development of specialized modeling modules, each addressing key aspects of spatial representation. These modules form the essential foundation for advancing the reconstruction process.

With the initial phase concluded, we now transition into the second phase, which focuses on the integration of these spatial features into the reconstruction process. The primary aim of this phase is to develop the computational framework necessary to support and enhance the reconstruction process. Having already built the core reconstruction module in the previous phase, the current study is centered on assembling and refining the framework to ensure it can effectively utilize the spatial features identified and modeled earlier.

\subsection{Iterative reconstruction with enhancement}
\vspace{0.5em}
\textit{A.1 Shallow modeling regularization}
\vspace{0.5em}

In our study, we adopt the "deep unrolling" methodology to structure our computational framework~\cite{monga2021algorithm}. This method provides more clarity than techniques such as the deep end-to-end (E2E) CNN~\cite{yuan2021snapshot}\footnote{The E2E CNN is to learn (an approximation of) the inverse process of the forward model}. By combining optimization and deep learning, it results in a model that is more interpretable and physics-aware \cite{datcu2023explainable}. The methodology comprises three main steps: 1) Define the optimization problem. 2) Formulate the iteration steps. 3) Network mapping. 

\vspace{1em}
\textit{1) Define the optimization problem}

The first step in using this methodology is defining the optimization problem that our computational framework aims to solve. Among the modeling modules, the 3D-TV-based shallow modeling is the simplest to implement. Following the principles of multi-task learning~\cite{yang2021structure}, we incorporate the 3D-TV norm into the initial optimization problem in (\ref{equation_2}). This introduces a new regularization term, creating a new task to enhance spatial features. As a result, we have a modified optimization problem, which can be explained as follows.
\begin{equation}
\label{equation_9}
\hat{\mathcal{X}} = \mathop{\arg\min_{\mathcal{X}}} \Bigg\{\frac{1}{2} \| \mathcal{Y} - \mathcal{A}(\mathcal{X}) \|_F^2 
 + \lambda_{1} \| \mathcal{X} \|_1 + \lambda_{2} g_{TV}(\mathcal{X}) \Bigg\}
\end{equation}

Here, $\mathcal{A}\left(\cdot\right)$ denotes the system's forward measurement operator that maps the scene image $\mathcal{X}$ to the echo image $\mathcal{Y}$. $\| \cdot \|_F$ indicates the Frobinus norm of a tensor. $\lambda_1$ and $\lambda_2$ are weight parameters used to balance the regularization terms.

\textit{2) Formulate the iteration steps}

The solution process involves iterations of gradient descent and proximal projection. Here are the steps for a single round of iteration.
\begin{equation}
\label{equation_10}
    \mathcal{Z}^{(k)} = \mathcal{X}^{(k)} - \alpha \nabla f(\mathcal{X}^{(k)})
\end{equation}
\vspace{-0.75cm} 
\begin{equation}
\label{equation_11}
    \mathcal{X}^{(k+1)} = \text{prox}_{\alpha \lambda_1 \|\cdot\|_1 + \alpha \lambda_2 g_{TV}}\left(\mathcal{Z}^{\left(k\right)}\right)
\end{equation}

Here, $f(\mathcal{X}^{(k)})=\frac{1}{2}\|\mathcal{Y}-\mathcal{A}\left(\mathcal{X}^{\left(k\right)}\right)\|_F^2$ , $\alpha$ is the step size for gradient descent. The term $\text{prox}_{\alpha \lambda_1 \|\cdot\|_1 + \alpha \lambda_2 g_{TV}}\left(\cdot\right)$  represents the proximal operator, employed to project the data point onto the data subspace's manifold. 

\textit{3) Network mapping}

Within the "deep unrolling" framework, we construct basic modules to mimic a single iteration. By connecting multiple modules, we replicate the full iterative process. This approach enables us to create a deep neural network capable of performing tomographic SAR reconstruction, effectively implementing the first proposed strategy of incorporating spatial feature modeling into regularization. In this regularization framework, the modeling strategy is manually guided, while the regularization strategy leverages deep learning.

\vspace{0.5em}
\textit{A.2 Deep modeling regularization}
\vspace{0.5em}

Next, our focus shifts to regularizing the other modeling modules previously designed, which are based on deep learning. As mentioned earlier, we will continue to employ a model-driven regulation approach, favored for its interpretability and awareness of physical process.

To refine this approach, a deeper exploration of the iterative resolution process mentioned earlier is necessary. This exploration is expected to yield valuable insights to design computation framework to regularize other modules, leading to several observations:

\textit{1. Data Features Perspective:} The iteration process generally comprises two main steps: the gradient step and the proximal mapping steps. These steps align with the system's forward measurement process, the scene's sparse feature, and its spatial feature, respectively. Broadly, the first step relates to the system's feature, influenced by system parameters and the system-scene geometry. Thus, the iterative process can be viewed as alternating between system feature consistency and scene feature consistency.

\textit{2. Data Space Perspective:} We deal with two primary spaces: the scene image space and the measured data (or echo) space. The operator $\mathcal{A}^\dagger$, being the adjoint of the system forward measurement operator  $\mathcal{A}$, essentially performs a preliminary back-projection, or rough imaging. Throughout the iteration, data transitions from the scene image space to the measured data space and back again.

\textit{3. Data Distribution Perspective:} The iteration process is essentially about posterior estimation of the scene image based on measured data. In line with Bayesian theorem, this estimation hinges on two aspects: the prior distribution of the scene image and the likelihood of the measured data given the scene image. These correspond to the gradient step and the proximal mapping steps, respectively. The gradient step projects the data point onto the manifold of the likelihood distribution. And the proximal mapping step project the data point into the manifold of the prior distribution

From these observations, it is evident that the reconstruction process, supported by a physical model, alternates between initial imaging and subsequent image enhancement. In this context, we name the previous framework as iterative reconstruction with enhancement. And, since the system-related aspects are largely defined by the physical model that is fixed, the precision of reconstruction is significantly influenced by the image enhancement phase. Naturally, we can substitute the image enhancement section with previously designed modules. Based on (\ref{equation_10}) and (\ref{equation_11}), we can achieve a more flexible computational framework.
\begin{equation}
\label{equation_12}
    \mathcal{Z}^{(k)} = \mathcal{X}^{(k)} - \alpha \nabla f(\mathcal{X}^{(k)})
\end{equation}
\vspace{-0.75cm}
\begin{equation}
\label{equation_13}
    \mathcal{X}^{(k+1)} = \mathcal{E}\left(\mathcal{Z}^{\left(k\right)}\right)
\end{equation}

In this case, we use $\mathcal{E}\left(\cdot\right)$ to represent an image enhancement operator. Within this flexible computational framework, we integrate the traditional iterative computational framework and the designed deep modeling modules. For the first proposed shallow module (the 3DTV-based norm), it aligns with the proximal operator in  (\ref{equation_11}). For other deep modeling modules, it aligns with specifically designed deep regularization modules. 

The flexible computational framework we've introduced can be viewed from another angle, especially when considering the proximal mapping problem as outlined in (\ref{equation_11}). In this context, the image enhancement phase is similar to a denoising process, particularly when assuming the presence of Gaussian noise. Here, the designed deep modules can be seen as an advanced image denoiser. This interpretation closely aligns with the recent promising technique known as "Plug and Play (PnP)" in the computational imaging community ~\cite{yuan2021snapshot}. The PnP framework replaces the proximal mapping step in optimization-based algorithms with a deep denoiser. These concepts have shown promising results in other imaging modalities, such as snapshot compressive imaging~\cite{yuan2021snapshot}. They've also been effectively introduced into SAR imaging recently\cite{wang20223, oral2024plug, wang2024array}. However, there's a difference between our framework and the PnP framework. In the PnP framework, the deep denoiser is pretrained and independent of the entire iterations, while other computation steps still follow traditional manual methods. In contrast, we make all the computation steps in our framework trainable.

\begin{figure}[h]
    \centering
    \includegraphics[width=0.5\linewidth]{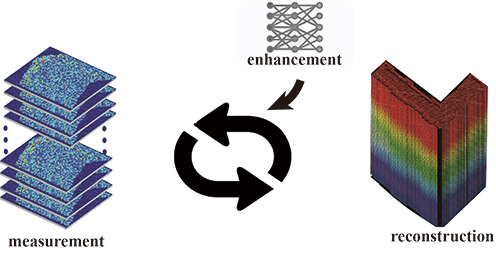} 
    \caption{Illustration of the computational framework of iterative reconstruction with enhancement.The reconstruction process, supported by a physical model, alternates between initial imaging and subsequent image enhancement to regularize scene's various features.}
    \label{figure_13}
\end{figure}

In summary, we provide an illustration in Fig.~\ref{figure_13} for better comprehension of this computational framework.

\subsection{Light reconstruction and enhancement}

The computational framework we have designed for iterative reconstruction with enhancement successfully integrates all earlier designed shallow or deep spatial regularization modules. And it does this while remaining within the confines of a deep unfolding framework that is both physically-aware and interpretable.

The adopted deep unfolding framework has been empirically proven to reduce the traditional solving time from hundreds of iterations to less than ten~\cite{qian2022gamma, wang2022ctv}, thus enhancing the reconstruction efficiency. Most of these cases have been validated using a manually modeled regularization module designed to regulate the sparsity feature, specifically the $l_1$ norm~\cite{qian2022basis,qian2022gamma,wang2023atasi,wang2023mada}. This module processes calculations on a pixel-by-pixel basis in the raw image space, leading to minimal computational. However, our more flexible extensions, especially when replaced with the designed deep modules, might increase both computational and storage requirements. Given this increased complexity in the computational framework, we aim to consider an alternative strategy that provides a balanced framework in terms of reconstruction precision and efficiency. 

The regularization module is implemented in the second stage of each iteration, making the increase in complexity directly proportional to the number of times the reconstruction process is iterated. The straightforward idea is to minimize its usage, as it might not be necessary to run this module from start to end. This idea aligns with the empirical findings that even in traditional iterative processes, the reconstructed result quickly approaches the final output after a few initial iterations. The majority of remaining iterations focus on enhancement, dealing with residual noise, and refining details. 

Following this vallina idea, we design a new computational framework, namely the light reconstruction and enhancement framework. This strategy simplifies the process by first executing a light reconstruction as a warm start, then enhancing the results using the designed regularization module for enhancement. This two-step like framework allows us to leverage the benefits of both models, ensuring efficient computation and high-precision output. 

First, we perform a quick, preliminary reconstruction using the traditional manually $l_1$-norm regularized for sparsity feature. Then, we enhance the reconstructed image using a more advanced model that can capture complex spatial features. And this framework can be mathematically expressed as:

\textit{1. Light Reconstruction Stage:} The initial phase involves a lighter, iterative reconstruction process:
\begin{equation}
\label{equation_14}
\mathcal{Z}^{(k+1)} = \mathcal{X}^{(k)} - \alpha \nabla_{\mathcal{X}^{\left(k\right)}}{\left(\frac{1}{2} \| \mathcal{Y} - \mathcal{A}\left(\mathcal{X}^{(k)}\right) \|_F^2\right)}
\end{equation}
\vspace{-0.75cm}
\begin{equation}
\label{equation_15}
\mathcal{X}^{(k+1)} = \text{prox}_ {\alpha \lambda_1 \|\cdot\|_1}\left(\mathcal{Z}^{(k+1)}\right)\
\end{equation}

where $\text{prox}_ {\alpha \lambda_1 \|\cdot\|_1}\left(\cdot\right)$ indicates the proximal mapping operator for the $l_1$ norm of the sparse feature.

\textit{2. Enhancement Stage:} In the subsequent phase, we apply a sophisticated spatial feature regularization module to the initial estimate:
\begin{equation}
\label{equation_16}
\hat{\mathcal{X}}^{'} = \mathcal{E}(\hat{\mathcal{X}})\end{equation}

Here, $\mathcal{E}(\cdot)$ represents the spatial feature regularization module, enhancing the preliminary output from the light reconstruction stage.

\begin{figure}[h]
    \centering
    \includegraphics[width=0.6\linewidth]{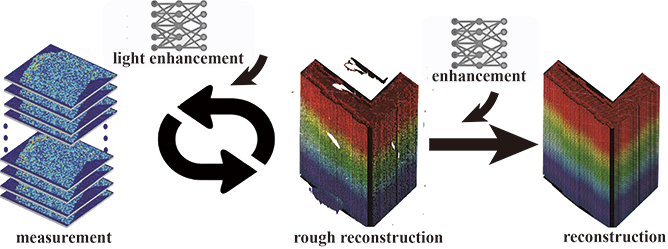} 
    \caption{The reconstruction process is divided into two stages. Initially, the light reconstruction only regularizes the sparsity feature. After this stage, the spatial feature is enhanced once.}
    \label{figure_14}
\end{figure}

This two-stag framework effectively leverages the benefits of both the traditional $l_1$-norm-based approach for initial reconstruction and the advanced model for capturing complex spatial features in the enhancement stage, ensuring both efficient computation and high-precision reconstruction result. We also provide an illustration in Fig. \ref{figure_14} for better comprehension of this new framework. 

\section{Phase three: Spatial-Feature-Enhanced Network design}

Building upon our previous work, we have developed two distinct computational frameworks to integrate our shallow and deep regularization modules. The first framework, which we call the iterative reconstruction with enhancement framework, follows the deep unrolling methodology. It begins with a shallow regularization module and incorporates a versatile enhancement step that can work with both shallow and deep regularization approaches.

Our second framework, the light reconstruction and enhancement framework, strikes a balance between efficiency and precision. It first performs a streamlined iterative reconstruction using lightweight sparse feature regularization, followed by an enhancement phase that employs more sophisticated spatial feature regularization.

Having developed these frameworks and regularization modules, we have focused on creating complete tomoSAR reconstruction networks. While various combinations of modules and frameworks have been possible, we have prioritized achieving high reconstruction precision while maintaining computational efficiency comparable to existing networks. This approach has led us to develop four distinct networks for evaluation: the tomo-IRENet-TV utilizing shallow 3D-TV modules and tomo-IRENet-U incorporating partial deep modules for the iterative reconstruction with enhancement framework, and the tomo-LRENet-biU and tomo-LRENet-LSTM for the light reconstruction and enhancement framework. We will detail each of these networks in detail in the following sections.

\subsection{tomo-IRENet-TV}

tomo-IRENet-TV uses the iterative reconstruction with enhancement framework for computational reconstruction, and the 3D-TV norm for spatial feature regularization. To build the network, we first need to derive the specific formulae for equations (\ref{equation_12}) and (\ref{equation_13}). We use the Split-Bregman approach for this derivation~\cite{wei2022learning}. We provide the complete iteration steps here. Detailed derivations can be found in the Appendix \ref{appendix_a}.

\textit{1. Measurement Consistency:}
\begin{equation}
    \mathcal{Z}^{(k+1)} = \mathcal{X}^{(k)} - \alpha  \mathcal{A}^\dagger\left(\mathcal{A}\left(\mathcal{X}^{(k)}\right) - \mathcal{Y}\right)
    \label{equation_17}
\end{equation}

\textit{2. Feature Consistency:} (for $i = 1, 2, 3$):

\quad$\circ$\text{\ Sparsity Feature} $l_1$ \text{term:}
\begin{adjustwidth}{0.5cm}{0cm} 
    \vspace{-0.5cm}
    \begin{equation}
        \mathbf{X}_i' = \mathbf{X}_i^{(k)} -\tau_1 \left(\mathbf{Q}_i \mathbf{X}_i^{(k)} - \mathbf{p}_i^{(k)}\right)
        \label{equation_18}
    \end{equation}
    \vspace{-0.5cm}
    \begin{equation}
        \mathbf{X}_i^{(k+1)} = \text{prox}_{\lambda_1 \tau_1}\left(\mathbf{X}_i'\right)
        \label{equation_19}
    \end{equation}
\end{adjustwidth}

\quad $\circ$\text{\ Spatial Feature} \text{TV} \text{term:}
\begin{adjustwidth}{0.5cm}{0cm} % 调整这里的第一个参数来增加缩进
    \vspace{-0.5cm}
    \begin{equation}
        \mathbf{V}_i' = \mathbf{V}_i^{(k)} -\tau_2 \mu \left(\mathbf{V}_i^{(k)} - \mathbf{D}_i\mathbf{X}_i^{(k+1)} - \mathbf{B}_i^{(k)}\right)
        \label{equation_20}
    \end{equation}
    \vspace{-0.5cm}    
    \begin{equation}
        \mathbf{V}_i^{(k+1)} = \text{prox}_{\lambda_2 \tau_2}\left(\mathbf{V}_i'\right)
        \label{equation_21}
    \end{equation}
\end{adjustwidth}

\quad $\circ$\text{\ Dual Variables} $\mathbf{B}_i^{(k)}$ \text{Update:}
\begin{adjustwidth}{0.5cm}{0cm} % 调整这里的第一个参数来增加缩进
    \vspace{-0.5cm}
    \begin{equation} 
        \mathbf{B}_i^{(k+1)} = \mathbf{B}_i^{(k)} + \mathbf{D}_i\mathbf{X}_i^{(k+1)} - \mathbf{V}_i^{(k+1)}
        \label{equation_22} 
    \end{equation}
\end{adjustwidth}

\quad $\circ$ $\mathcal{X}^{(k)}$ \text{Update:}
\begin{adjustwidth}{0.5cm}{0cm} % 调整这里的第一个参数来增加缩进
    \vspace{-0.5cm}
    \begin{equation} 
        \mathcal{X}^{(k+1)} = \frac{1}{3} 
        \Big\{
        \text{unfold}\left(\mathbf{X}_1^{(k+1)}\right) + \text{unfold}\left(\mathbf{X}_2^{(k+1)}\right)
        +\text{unfold}\left(\mathbf{X}_3^{(k+1)}\right) 
        \Big\}
        \label{equation_23}    
    \end{equation}
\end{adjustwidth}

Where $\mathbf{Q}_i = \frac{1}{\alpha}\mathbf{I} + \mu\mathbf{D}_i^H\mathbf{D}_i$ and $\mathbf{p}_i = \frac{1}{\alpha}\mathbf{Z}_i^{(k+1)} + \mu\mathbf{D}_i^H\left(\mathbf{V}_i^{(k)} - \mathbf{B}_i^{(k)}\right)$. Some new variables are introduced here. For the sake of simplicity, we define a few key ones here, while the rest can be found in the Appendix \ref{appendix_a}. The steps involved in the feature consistency are calculated mainly in a matrix-wise manner and are primarily formed through $\text{fold}\left(\cdot\right)$ operator (vectorizing slices of the tensor and stacking them). $\mathbf{D}_i, i=1,2,3$ represents the difference matrix along different dimensions (frontal, lateral and horizontal). $\mathbf{V}_i$ is an alternative variable of $\mathbf{X}_i$ for decoupling. And $\mathbf{B}_i$ is the Bregman dual variable. $\text{unfold}\left(\cdot\right)$ is the adjoint operator of $\text{fold}\left(\cdot\right)$, which reshapes the matrix back into its original tensor form.

We have developed a tomoSAR reconstruction network, named tomo-IRENet-TV, based on the above iteration steps. Specifically, each round of iteration is represented by three modules corresponding to different regularization terms: the system's feature, the scene's sparse feature, and the spatial feature. These modules are assembled into one basic block, and several blocks are then arranged in sequence to form the entire reconstruction network.

\begin{figure}[!h]
    \centering
    \includegraphics[width=0.80\linewidth]{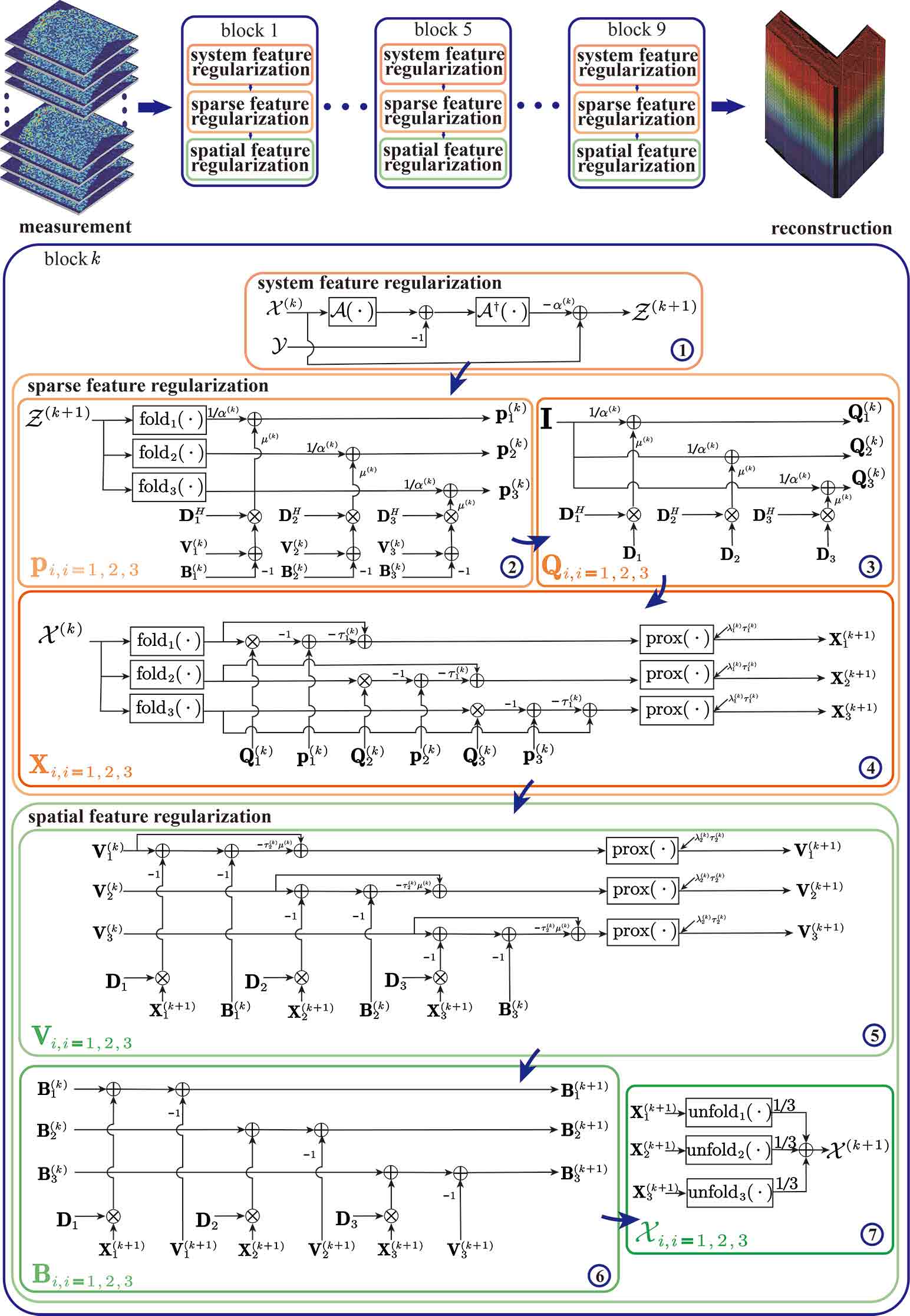}
    \caption{The structure of tomo-IRENet-TV. The structure of tomo-IRENet-TV mimics an iterative reconstruction with enhancement process, consisting of nine blocks. Each block contains three modules designed to perform the regularizations of system's features, the scene's sparse features, and the scene's spatial features, respectively. And the calculation steps basicaly follow (\ref{equation_17}) to (\ref{equation_23}). The hyper-parameters involved are learnable and can vary between different blocks. The input of this network is directly the 3D echo tensor and the output of it is the 3D scene tensor.} 
    \label{figure_15}
\end{figure}

It's important to note that multiple hyperparameters are involved during the iteration process. Traditionally, these are manually set. However, to utilize the power of deep learning, we have made these parameters learnable to enhance the reconstruction estimation. Specifically, all parameters are learnable and can vary in different iteration rounds. In other words, each block can have different hyperparameters. The complete network structure is shown on the following page in Fig.~\ref{figure_15}, due to space constraints.

\newpage
\clearpage

\subsection{tomo-IRENet-U}
The previous tomo-IRENet-TV has been designed to strictly follow the deep unrolling framework. The enhancement part is handled by manually defined enhancement operators. Since we have previously designed deep modules for spatial feature regularization beyond these manual and shallow methods, it seems a natural progression to use these deep modules for enhancement. However, due to the potential computational and storage demands of these modules - particularly when used as extensively as in this iterative reconstruction with enhancement framework - we've chosen to use only a portion of them, specifically the U-Net module, for enhancement. We've named this new network tomo-IRENet-U. It acts as a transition network between the initial one (tomo-IRENet-TV) and subsequent ones that adopt complete deep modules, which offer much greater flexibility.

Tomo-IRENet-U adopts the iterative reconstruction with enhancement framework and utilizes the U-Net module for whole scene feature regularization. The U-Net design integrates both $l_1$ and 3D-TV regularizations effectively. In terms of $l_1$ regularization, U-Net's encoding phase performs adaptive sparse transformation, processing images into sparse representations that highlight key features while reducing redundancy - similar to how $l_1$ regularization selects sparse features. Additionally, U-Net fulfills 3D TV regularization objectives through its multi-layered convolutional structure, which effectively manages multi-dimensional data while maintaining spatial geometry and smoothness throughout the image, aligning with 3D TV's spatial feature preservation goals.

Different from the previous one, this network uses the echo slice as the basic processing unit to match the U-Net module's processing dimensions. The entire reconstruction result is obtained by processing each slice independently and assembling them at the end. Similarly, the network is constructed using cascaded iteration blocks. Within each block, there are two processing modules that ensure measurement consistency and feature consistency respectively.

Here are the steps for one iteration of this network, using a frontal slice as an example.

\textit{1. Measurement Consistency:}
\begin{equation} 
    \mathbf{Z}^{(k+1)} = \mathbf{X_f}^{(k)} - \alpha^{\left(k\right)}\mathbf{A}^H\left(\mathbf{A}\left(\mathbf{X_f}^{\left(k\right)}\right) - \mathbf{Y}\right)
    \label{equation_24}
\end{equation}

In this iteration, the input is $\mathbf{X_f}^{(k)}$ and the corresponding echo slice is $\mathbf{Y_l}$. The variable $\alpha^{\left(k\right)}$ is learnable and can vary across different iterations.

\textit{2. Feature Consistency:}
\begin{equation} 
    \mathbf{X_l}^{(k+1)} = \mathcal{E}_U\left(\mathbf{Z}^{(k+1)}\right)
    \label{equation_25}
\end{equation}

Here, $\mathcal{E}_U\left(\cdot\right)$ represents the enhancement operator through the U-Net module. 

\begin{figure}[!h]
    \centering
    \includegraphics[width=\textwidth]{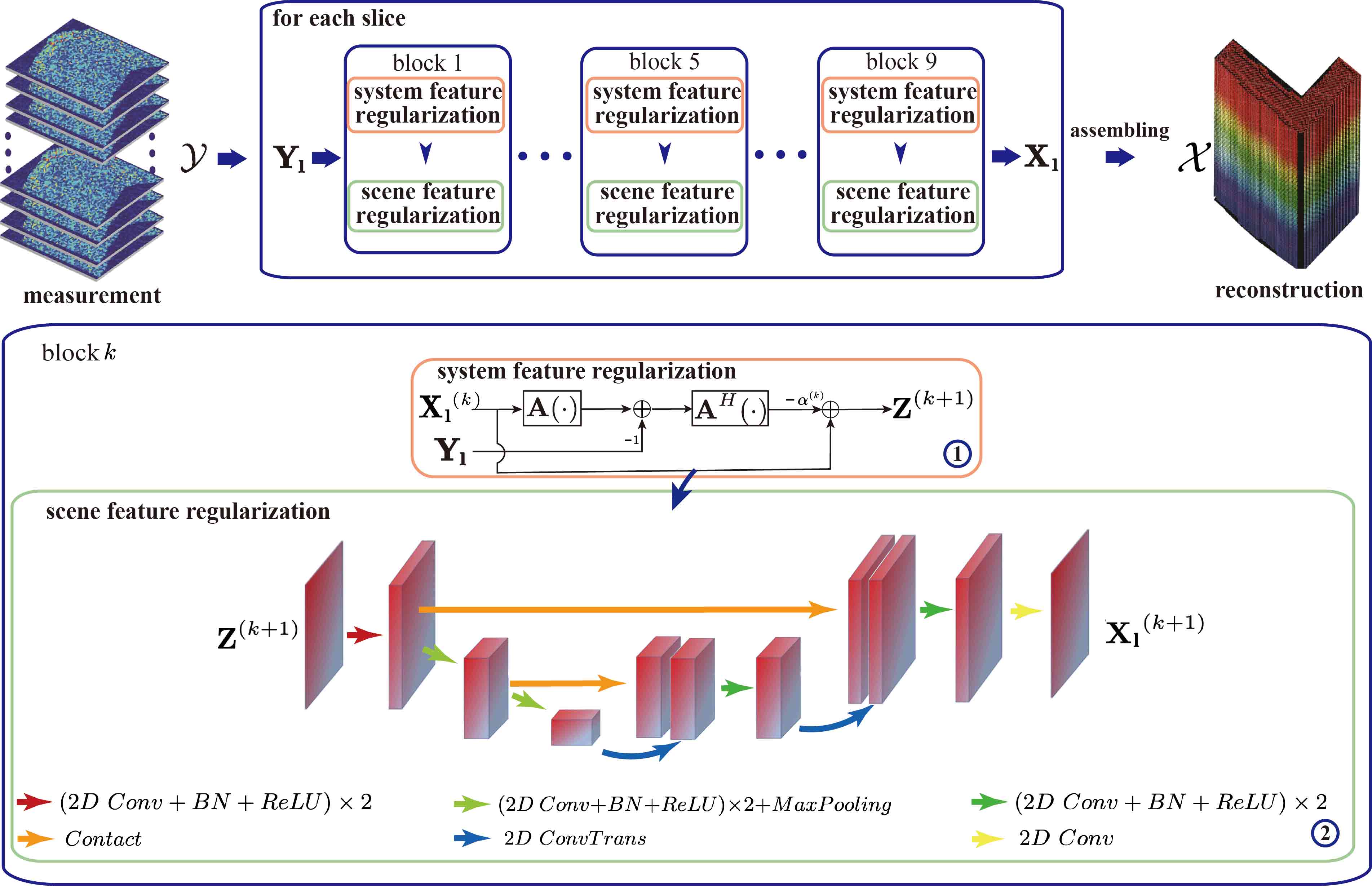}
    \caption{The structure of tomo-IRENet-U is similar to tomo-IRENet-TV. The network uses an iterative reconstruction with enhancement framework. However, the enhancement part is replaced by a subset of the deep modules, a U-Net structure, which serves as the scene feature regularizations. The U-Net structure increases the features of the input first, then downsamples twice and upsamples twice, with a skip-connection at the same level features. At last, features are fused to get the enhanced slice. The network processes the echo tensor one lateral slice at a time and then assembles the processed slices to obtain the final scene tensor.}
    \label{figure_16}
\end{figure}

The network architecture's structure is illustrated in Fig.~\ref{figure_16}. The U-Net module part processes the input in a series of steps. Initially, it begins with two rounds of 2D convolutions, batch normalization, and ReLU activation. Following this, it conducts another two rounds of 2D convolutions, batch normalization and ReLU, before carrying out a max pooling operation. This results in the first set of downsampled features. The downsampling steps are then repeated to produce the second set of downsampled features. Subsequently, the network applies a 2D deconvolution and concatenates this with the first set of downsampled features. The resulting concatenated features are fused through two rounds of 2D deconvolutions, batch normalization and ReLU activation, leading to the first set of upsampled features. These upsampling steps are repeated to generate the second set of upsampled features. Finally, the network applies a 2D convolution to produce the final output.

\subsection{tomo-LRENet-biU and tomo-LRENet-LSTM} 

The first two networks use an iterative reconstruction with enhancement framework, while the last two use light reconstruction and enhancement frameworks, sharing the same light reconstruction component but different spatial feature regularization modules. The light reconstruction component refers to a network that performs only system and scene sparse feature regularizations. Since we reconstruct slice-wise rather than fiber-wise, the $l_1$ norm-based sparse feature regularization network needs these iteration step modifications:

\textit{1. Measurement consistency: }
\begin{equation} 
    \mathbf{Z}^{(k+1)} = \mathbf{X_l}^{(k)} - \alpha^{\left(k\right)}  \mathbf{A}^H\left(\mathbf{A}\left(\mathbf{X_l}^{\left(k\right)}\right) - \mathbf{Y}\right)
    \label{equation_26}
\end{equation}

In this iteration, the input is $\mathbf{X_f}^{(k)}$ and the corresponding echo slice is $\mathbf{Y_f}$. The variable $\alpha^{\left(k\right)}$ is learnable and can vary across different iterations.

\textit{2. Sparse feature consistency:}
\begin{equation} 
\mathbf{X_l}^{(k+1)} = \mathop{sign}\left(\mathbf{Z}^{(k)}\right)\odot \mathop{max}\left({\left|\mathbf{Z}^{(k)}\right|-\theta^{(k)}},\mathbf{0}\right)
\label{equation_27}
\end{equation}

Here, $\mathop{sign}\left(\cdot\right)$ represents the symbol function. $\theta^{(k)}$ is also learable and can vary across different iterations. In comparison with the fiber-wise approaches, the difference lies in that the threshold is related to the entire slice, not just one fiber.

With these iterative steps, we can similarly construct the light reconstruction part of the network using cascaded iteration blocks. Within one iteration block, the network performs the two steps mentioned above.

\begin{figure*}[t]
        \centering
        \begin{subfigure}[b]{\linewidth}
            \caption*{{\fontsize{8pt}{10pt}\selectfont\centering \qquad\qquad tomo-LRENet-biU}}
            \includegraphics[width=\linewidth]{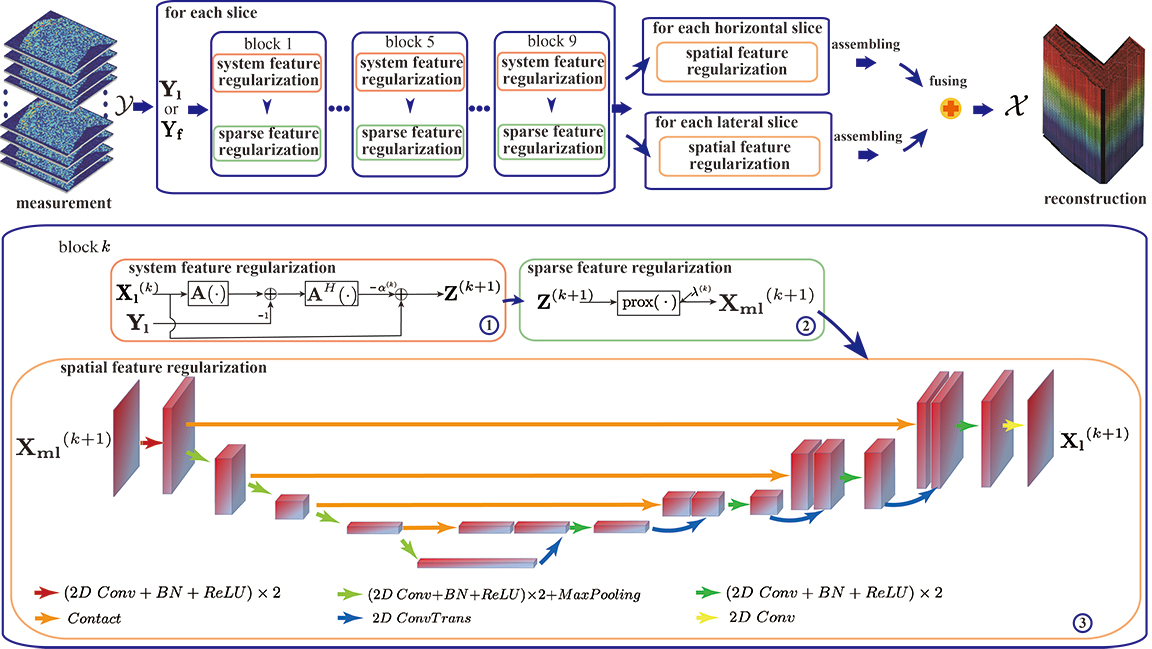}
            \caption{}
            \label{figure_17_1}
        \end{subfigure}
        \vspace{11pt}
        \begin{subfigure}[b]{\linewidth}
            \caption*{{\fontsize{8pt}{10pt}\selectfont\centering\qquad\qquad tomo-LRENet-LSTM}}
            \includegraphics[width=\linewidth]{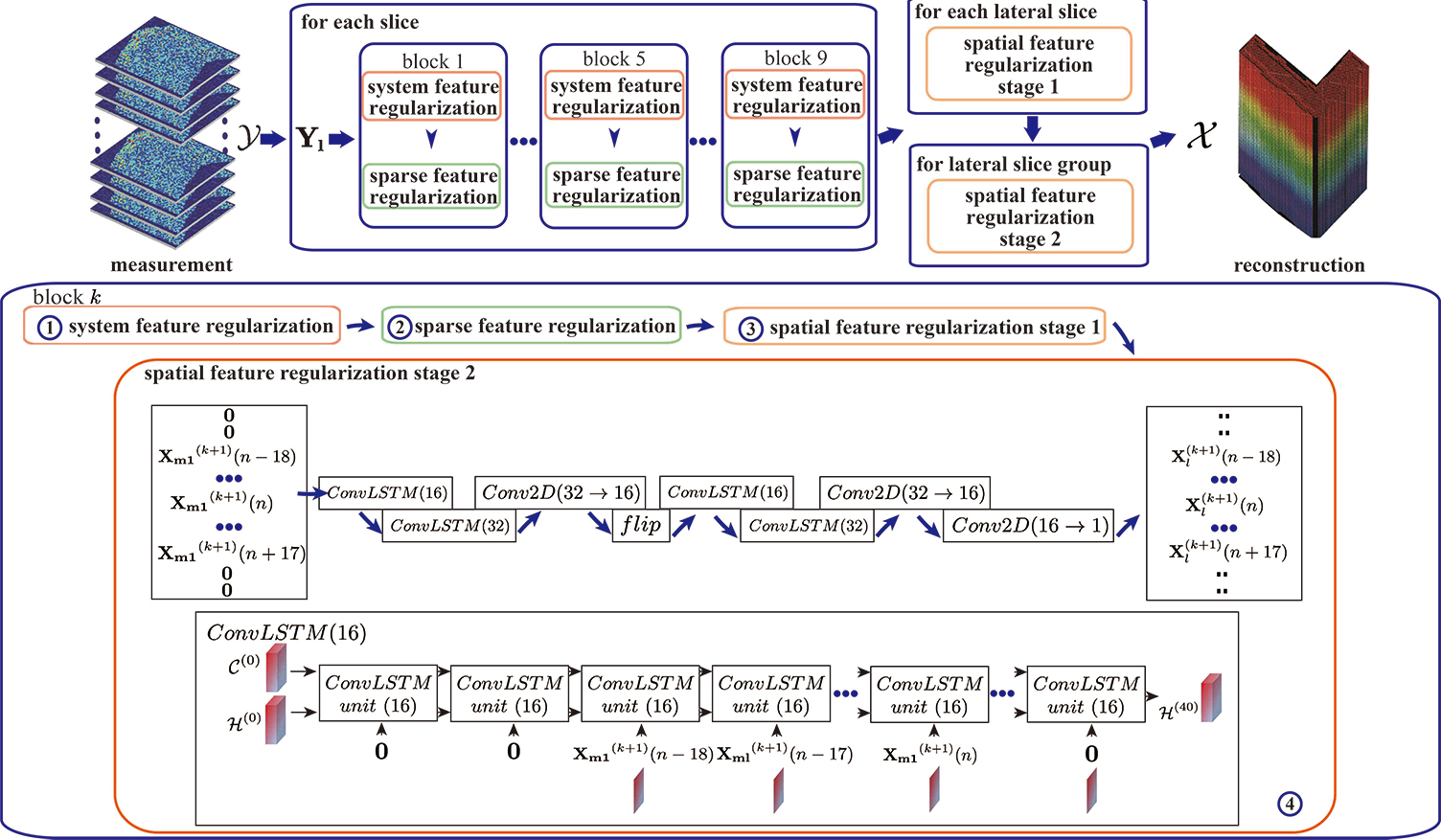}
            \caption{}
            \label{figure_17_2}
        \end{subfigure}
        \caption{Structure of reconstruction networks: (a) tomo-LRENet-biU, (b) tomo-LRENet-LSTM. The second network shares the same structure as the first, except for the LSTM related modules. The common part is simplified due to limited space.}
        \label{figure_17}
    \end{figure*}

For the enhancement part, we use two deep modules: Bi-Parallel-UNet Fusion and Sequential-UNet-LSTM. These form two distinct networks, named tomo-LRENet-biU and tomo-LRENet-LSTM respectively. Their structures are depicted in Fig.~\ref{figure_17_1} and Fig.~\ref{figure_17_2} respectively. 

\newpage

\FloatBarrier

\FloatBarrier

\vspace{1em}
The U-Net structures in both networks are similar to that of tomo-IRENet-U. However, since they don't require iteration, the instances of upsampling and downsampling have increased from two to four times. Tomo-LRENet-biU individually processes lateral and frontal slices, assembling two scene tensors. These tensors are then merged through pixel-wise maximum operations. Tomo-LRENet-LSTM's spatial feature regularization part has two stages: slices are processed individually in the first stage, followed by grouping multiple slices and processing them through the LSTM-based structure. The final scene tensor is formed by assembling all processed slices. Specifically, it processes lateral slices individually in the first stage, followed by grouping multiple slices for processing through an LSTM-based structure. The final scene tensor is assembled from all processed slices. Specifically, a sequence of 36 slices, with 2 zero-padded slices at the front and end, is processed together, with each slice representing one time-step in the LSTM. The structure comprises 4 core ConvLSTMs: the first two process the slices from beginning to end, and the last two from end to beginning. Initially, the slices are processed into 16 features per slice in the first ConvLSTM, then into 32 features in the second. The feature count is reduced to 16 via 2D convolution. The group is then flipped and processed with the last two ConvLSTMs. Finally, features are fused into one through 2D convolution, and the final reconstructed tensor is assembled by combining different groups and discarding the padded slices.

The choice of the number of slices affects the performance. A smaller number of slices may only capture limited sections of a building, missing correlations between different parts of the structure. On the other hand, a larger number of slices might include irrelevant regions, such as surrounding clutter or empty space, introducing redundancy and reducing both efficiency and the clarity of the reconstructed results. Based on empirical analysis, we have found that selecting between 28 and 36 slices generally yields satisfactory outcomes. Therefore, the current setting of 36 slices is chosen to balance these factors and ensure robust performance across various scenarios\footnote{Future work will explore incorporating non-local correlations beyond adjacent slices to capture more comprehensive structural information across the entire scene, thereby allowing more non-redundant slices to be processed.}.

\section{Phase four: Method Evaluation}

Up to this point, we have obtained four unique reconstruction networks from the early stages of our study. In the succeeding phase, we plan to perform thorough evaluations to measure these networks' performance. We initiate by discussing the data used for evaluation, followed by an introduction to the selected evaluation methods. Subsequently, we introduce a framework designed for a comprehensive, multi-faceted analysis. The evaluation results will be presented in the end.

\subsection{Evaluation data}

We use both simulated and measured data to evaluate different methods, adhering to the standard approach. We focus heavily on deep learning-based reconstruction methods that need supervised training to determine optimal parameters. In this case, simulated data refers to training data, not just testing data. It's important to note that current deep learning-based reconstruction methods are fiber-wise~\cite{wang2023atasi, wang2023mada, qian2022basis, qian2022gamma}, which doesn't suit our slice-wise or tensor-wise study. Therefore, we've designed a new approach for simulating data. We'll detail this method first and then introduce the measured data we use.

\vspace{0.5em}
\textit{A.1 Simulated data}
\vspace{0.5em}

To effectively train neural networks and assess their performance, a dataset with numerous instances of input data paired with their corresponding ground truth is crucial. However, in SAR imaging, obtaining the ground truth of the imaging scene is not feasible. Approaches like electromagnetic computation can approximate the ground truth but at prohibitively high costs, especially in large-scale remote sensing scenarios like urban areas. As a result, meeting both the size and diversity requirements for the dataset becomes challenging.

\begin{figure}[h]
    \centering
    \includegraphics[width=0.6\linewidth]{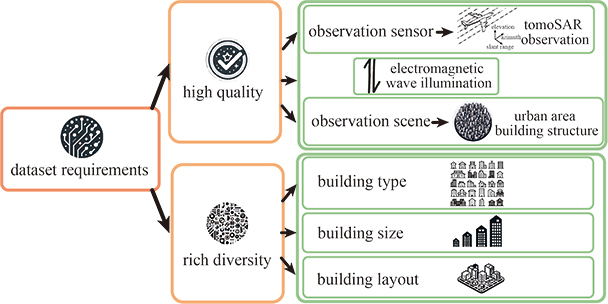}
    \caption{Dataset requirements for the study.}
    \label{figure_19}
\end{figure}

Therefore, in accordance with existing literatures in the field of SAR imaging,~\cite{wang2023atasi, wang2023mada, qian2022basis, qian2022gamma, sun2022large, chen20193d,wang2022ctv, wu2023learning, pu2021deep,  wei2022carnet, wei2022learning}, a simulation dataset is utilized in this study. Specifically, in the context of this study, which focuses on spatial feature-based urban area reconstruction, the simulation dataset must satisfy the following features from the perspectives of quality and diversity, as shown in Fig. \ref{figure_19}. 

\begin{itemize}
    \item \textit{Quality:}
    The dataset should exhibit high fidelity by accurately capturing the forward relationship from the observation scene to SAR's collected echoes, incorporating characteristics from both the SAR sensor and the observation scene. This entails revealing the spatial structures of buildings in the urban area from the observation scene aspect, incorporating the SAR imaging coordinate system from the SAR sensor aspect, and considering the electromagnetic wave illumination and occlusion of different parts of the buildings from the SAR-scene relationship aspect.
    \item \textit{Diversity:}
    The dataset should cover a wide range of variabilities to ensure the effectiveness of the reconstruction networks across different settings. It is essential to include various building types in urban areas, as well as diverse spatial distributions in the 3D space. This encompasses capturing the variability in building shapes, sizes and orientations.
\end{itemize}

However, it is worth noting that the current literature primarily emphasizes pixel-wise reconstruction, which leads to inadequate simulation flowcharts and resulting datasets in capturing the desired features mentioned earlier. For instance, in the mentioned study~\cite{qian2022gamma}, the elevation profiles of each range-azimuth pixel are independently obtained, and the positions of scatters within each profile are randomly set. Consequently, the random positioning of scatters in 3D space fails to preserve the spatial structures of buildings, making this method unsuitable for our study, which focuses on spatial feature-based urban area reconstruction. 

Beyond the spatial feature aspect, we also emphasize the core issue of TomoSAR reconstruction—the layover problem. In other words, our dataset should also exhibit diversity in terms of layover scenarios.

\begin{figure}[h]
    \centering
    \begin{subfigure}[b]{0.24\linewidth}
        \includegraphics[width=1\linewidth]{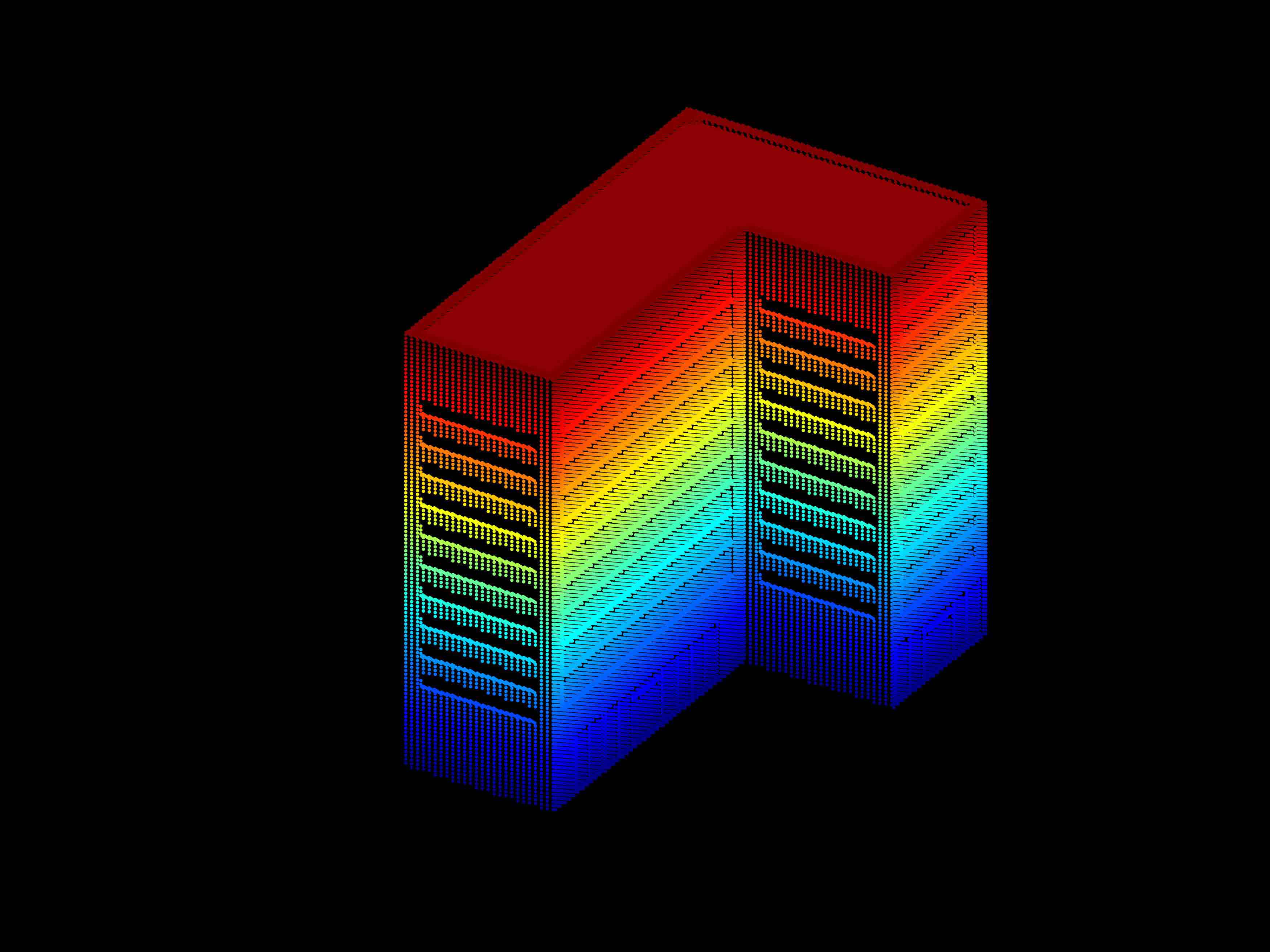}
        \caption{}
        \label{figure21_1}
    \end{subfigure}
    \begin{subfigure}[b]{0.24\linewidth}
        \includegraphics[width=1\linewidth]{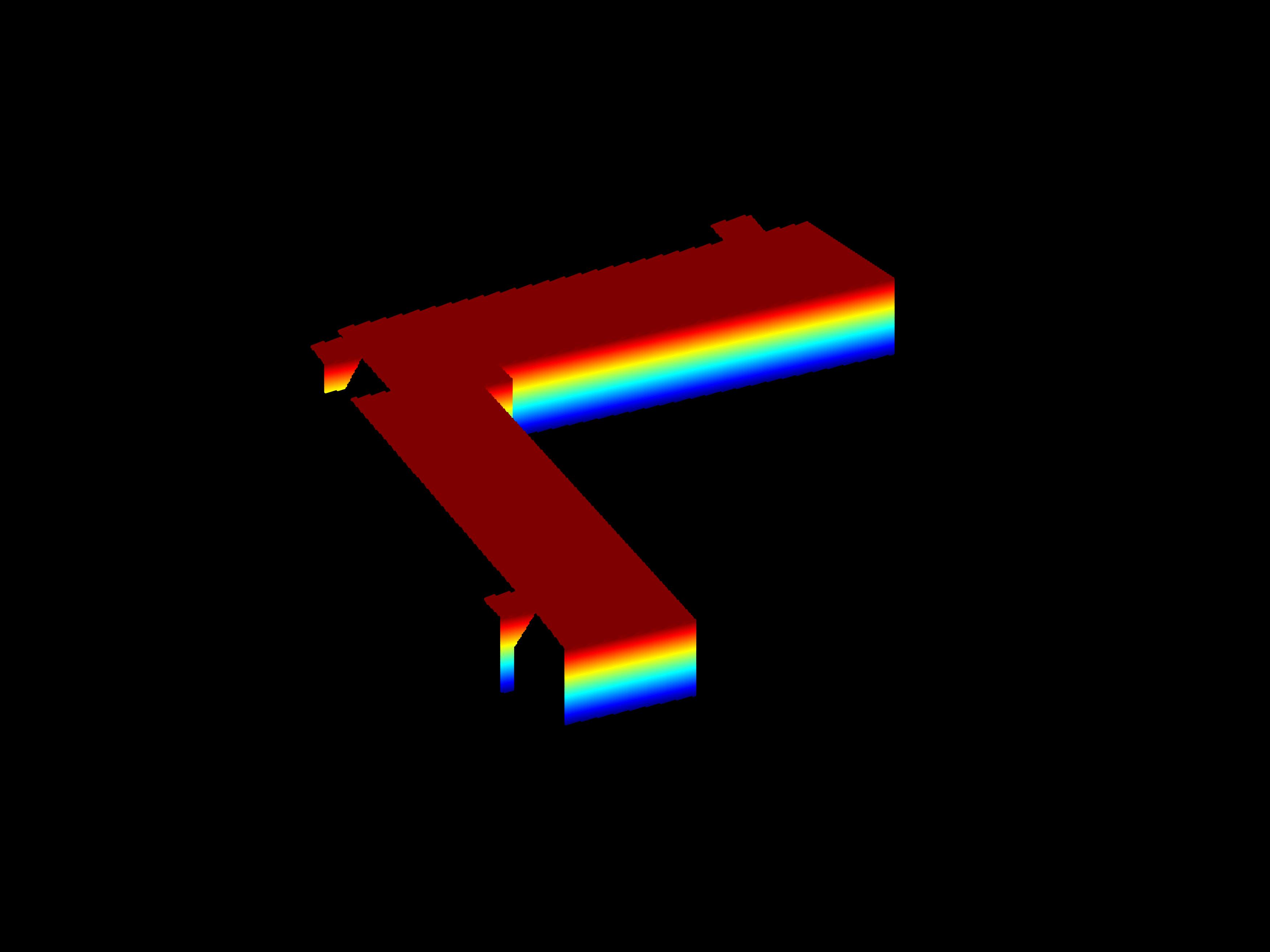}
        \caption{}
        \label{figure21_2}
    \end{subfigure}
    \begin{subfigure}[b]{0.24\linewidth}
        \includegraphics[width=1\linewidth]{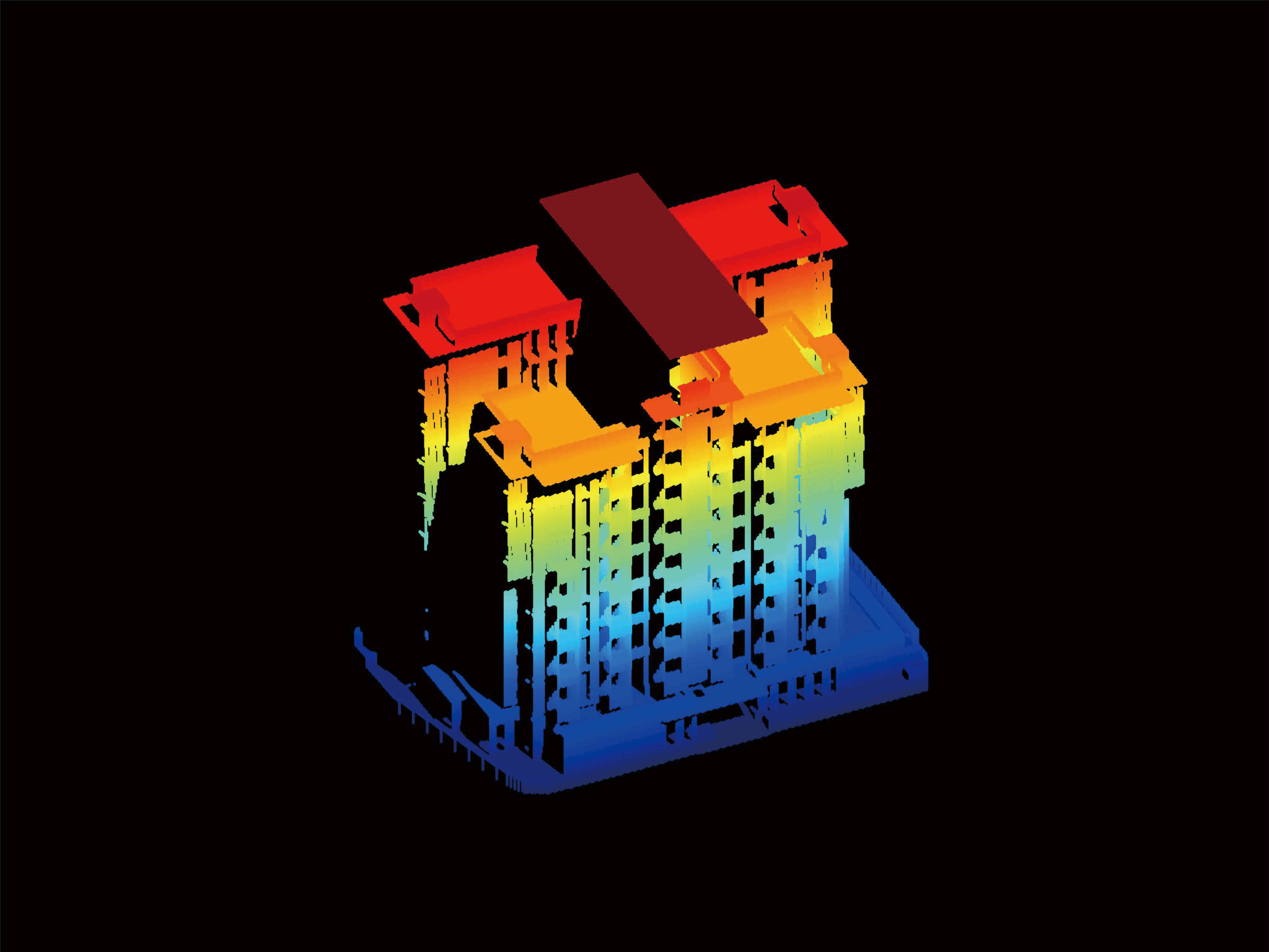}
        \caption{}
        \label{figure21_3}
    \end{subfigure}
    \begin{subfigure}[b]{0.24\linewidth}
        \includegraphics[width=1\linewidth]{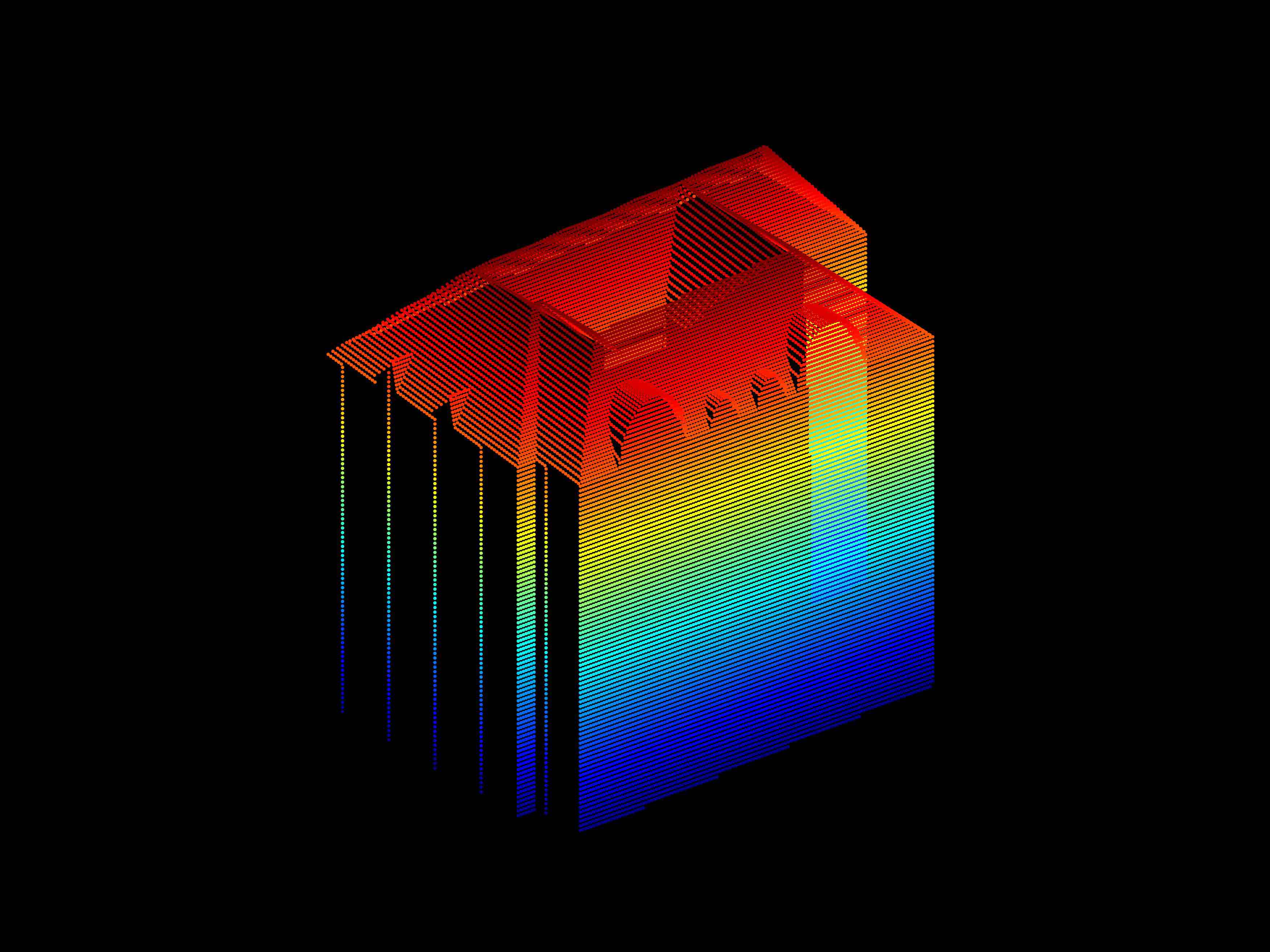}
        \caption{}
        \label{figure21_4}
    \end{subfigure}
    \caption{Building model examples: (a) High-rise vertical buildings, (b) Flat buildings, (c) Complex buildings, (d) Other structures.}
    \label{figure_21}
\end{figure}

To address this diversity challenge, our simulation dataset is built upon multiple building models, some of which are illustrated in Fig.~\ref{figure_21}. We establish a fixed global imaging plane. For each imaging scene, we randomly select a single building model, adjusting its scale and position within the plane. Using open-source ray-tracing software, we generate high-quality ground truth data in the form of frontal and lateral slices. We then produce measured data through a forward measurement process. We repeat this process to generate diverse imaging scenes, each featuring a different building model.

\begin{figure}[h]
    \centering
    \includegraphics[width=0.6\linewidth]{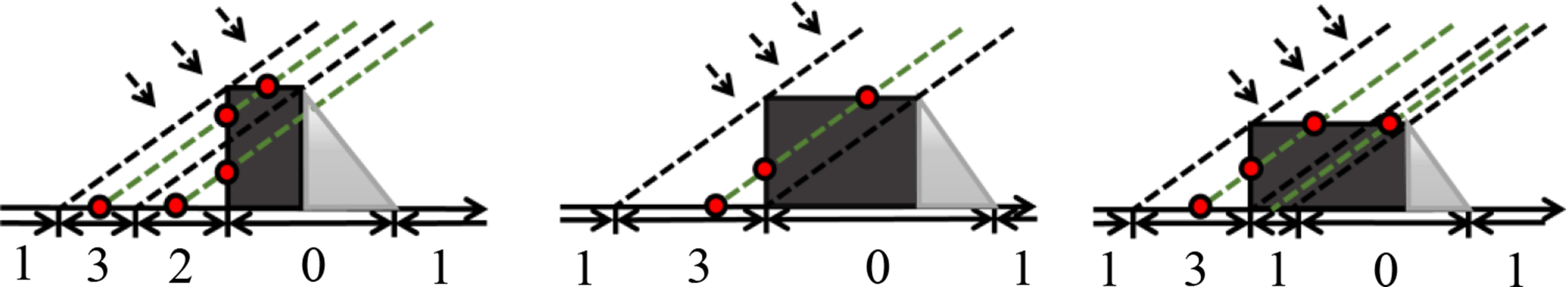}
    \caption{Illustration of layover types generated from a building model. Typical layover scenarios involving 1 to 3 scatterers exist for a single model. From left to right, the building model with varying sizes produces diverse layover spatial distributions. The number below each axis indicates the quantity of layovers present in each scenario.}
    \label{figure_19_insert_20}
\end{figure}

From the fiber-wise reconstruction perspective, the adoption of various building models allows us to represent multiple typical overlay scenarios, such as the rooftop-ground overlay, wall-surface and ground overlay, as well as more complex cases like the rooftop, ground, and wall-surface overlay. Additionally, through random placement and scaling of building models, we introduce variability in scatterer distances and spatial distributions, thus enriching the diversity of layover scenarios, as shown in Fig.~\ref{figure_19_insert_20}.

\begin{figure}[h]
    \centering
    \includegraphics[width=0.6\linewidth]{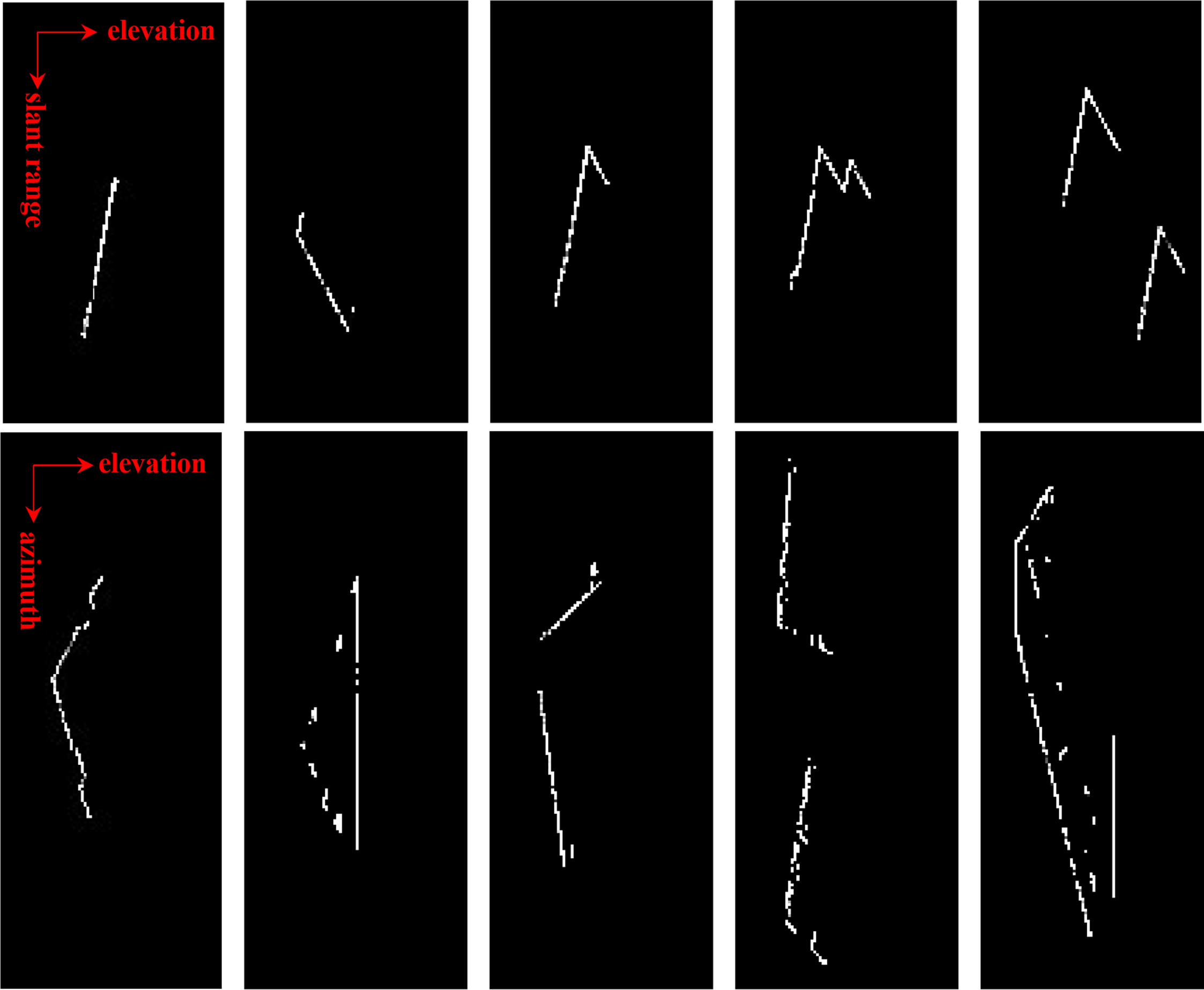}
    \caption{Illustration of layover types generated from multiple models. First row: frontal slices. Second row: lateral slices. Various models generate diverse distributions of line segment patterns.}
    \label{figure_19_insert2_20}
\end{figure}

Taking a step further, from the slice-wise reconstruction perspective, scatterers from different adjacent fibers form line segment structures. The geometry and shape diversity of the building models enhances the diversity of these line segment patterns. We provide examples of frontal and lateral slices from the dataset in Fig.~\ref{figure_19_insert2_20}, showing how the spatial distribution of these line segments varies across different locations, orientations, lengths, etc.

To ensure sufficient diversity, we incorporate 36 unique building models broadly classified into four types: (1) simple high-rise vertical buildings, (2) flat buildings, (3) complex buildings, and (4) other structures. Each building is randomly rescaled and translated in various directions, creating six different scenes per building, resulting in a total of 216 unique scenes. These scenes produce 27,689 frontal slices and 17,942 lateral slices.

Finally, regarding system-related parameters, we primarily follow those from the public Emei dataset \cite{xiaolan2021sarmv3d,xiaolan2022key}, which we adopt as a reference to validate our simulation framework. These can be found in Table~\ref{table_1}. 

\begin{table}[h]
    \caption{System Parameters for Data Simulation}
    \label{table_1}
    \footnotesize
    \centering
    \begin{tabular}{|l|l|}
    \hline
    \textbf{Parameter} & \textbf{Value} \\
    \hline
    Wavelength & 0.031 m \\
    Number of Flyovers & 12 \\
    Reference Slant Range & 2040.3406 m \\
    Reference Incidence Angle & 31.6453$^\circ$ \\
    Reference Height & 1736.9668 m \\
    Signal-to-noise Ratio (SNR) & 5 dB \\
    \hline
    \end{tabular}
\end{table}

\begin{figure}[h]
    \centering
        \includegraphics[width=1.0\linewidth]{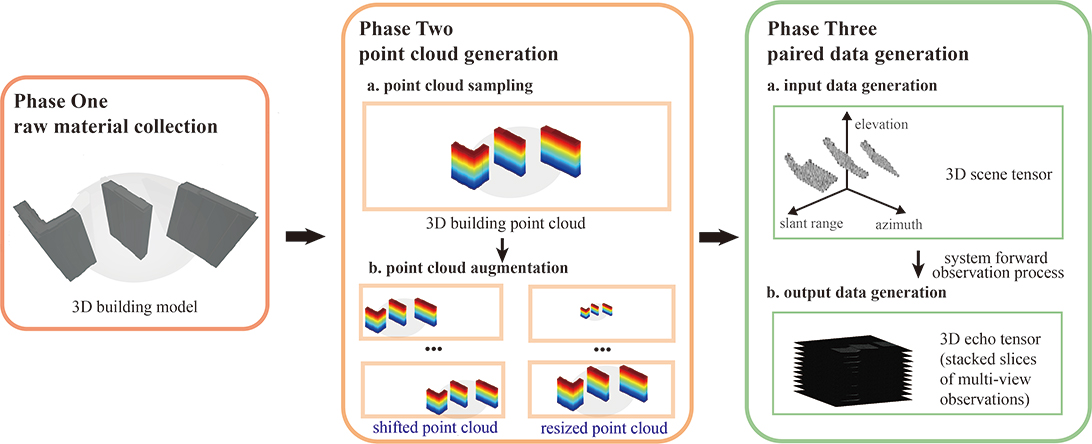}
        \caption{Flowchart of the designed simulation process.}
    \label{figure_20}
\end{figure}

The simulation flowchart\footnote{To provide a clearer understanding of the flowchart, we have included more details and the pseudocode of the simulation process in the Appendix \ref{appendix_b}.} comprises three main steps: raw material collection, point cloud generation, and paired data generation, as illustrated in Fig.~\ref{figure_20}. This flowchart allows us to systematically construct a high-fidelity and diverse dataset that effectively meets the requirements of our study.

\newpage
\textit{A.2 Measured data}
\vspace{0.5em}

To effectively evaluate our methods, we complement simulated data with measured data, tapping into two publicly available 3D datasets within the community~\cite{xiaolan2021sarmv3d,xiaolan2022key}. The utilization of measured data provides a more realistic assessment of our techniques. Our choice is influenced by the availability and characteristics of the datasets in our community. While there are two publicly available 3D datasets, we found that each had its distinct advantages and limitations for our specific research needs.

The first dataset, known as GOTCHA, was released by the Air Force Research Laboratory (AFRL) in the United States~\cite{ge2018spaceborne}. It features X-band data collected via a multi-pass circular SAR modality, encompassing 8 elevation angles and complete azimuth observations. Primarily, it consists of images of civilian vehicles. By selecting specific azimuth observation angles, we can approximate the data to resemble those acquired through tomoSAR modalities. However, given the limited height of the vehicles in GOTCHA, resulting in only a slight degree of layover, this dataset does not ideally match our imaging target (building structures). Therefore, we turn to the second dataset, SARMV3D-1.0, recently released jointly by the Aerospace Information Research Institute of the Chinese Academy of Sciences, Fudan University, and other institutions in China~\cite{xiaolan2021sarmv3d,xiaolan2022key}. This dataset is captured using an array-based multi-baseline InSAR airborne system, offering superior coherence due to single-pass data collection. It effectively mitigates issues like atmospheric and motion phase errors, making it highly suitable for method evaluation.

\begin{figure}[b] % 使用 figure* 环境使图形跨越两栏，'t' 表示顶部
    \centering
    % 第一个小图
    \begin{subfigure}[b]{0.24\linewidth}
        \includegraphics[width=1\linewidth]{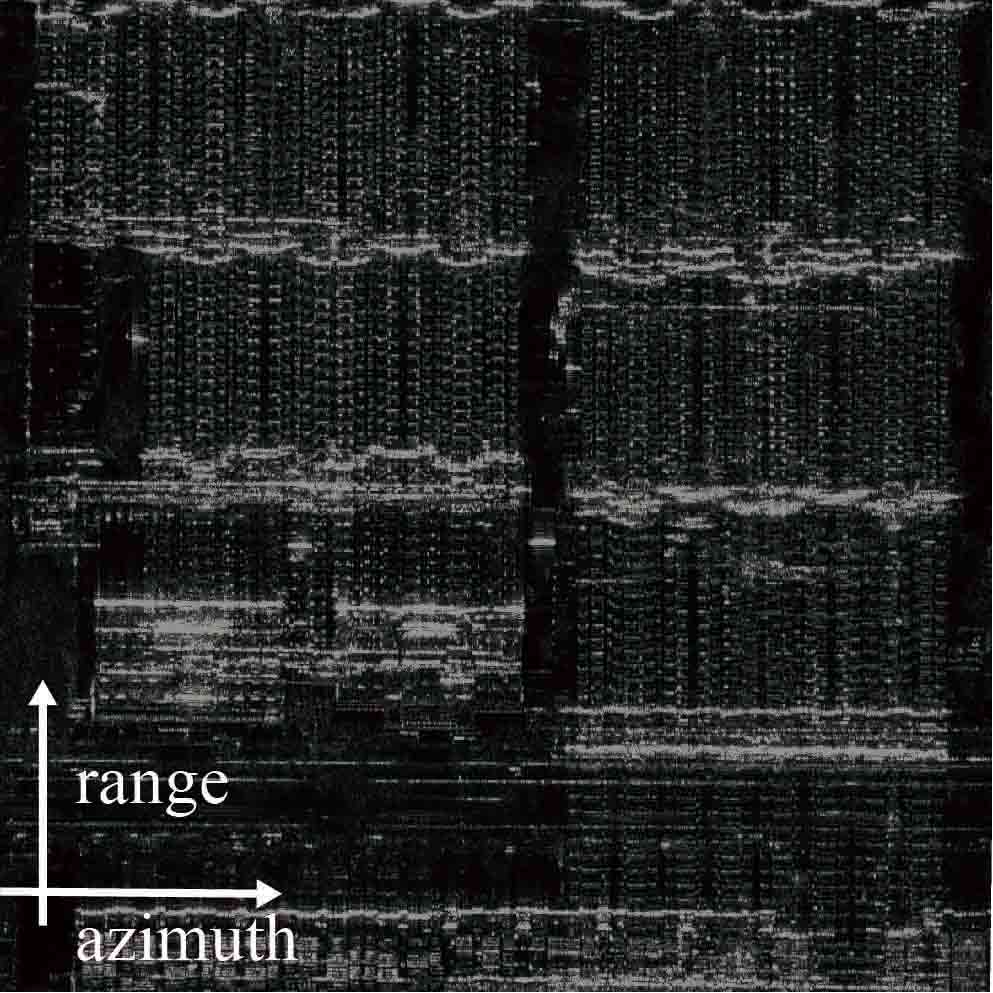}
        \caption{}
        \label{figure22_1}
    \end{subfigure}
    % 第二个小图
    \begin{subfigure}[b]{0.24\linewidth}
        \includegraphics[width=1\linewidth]{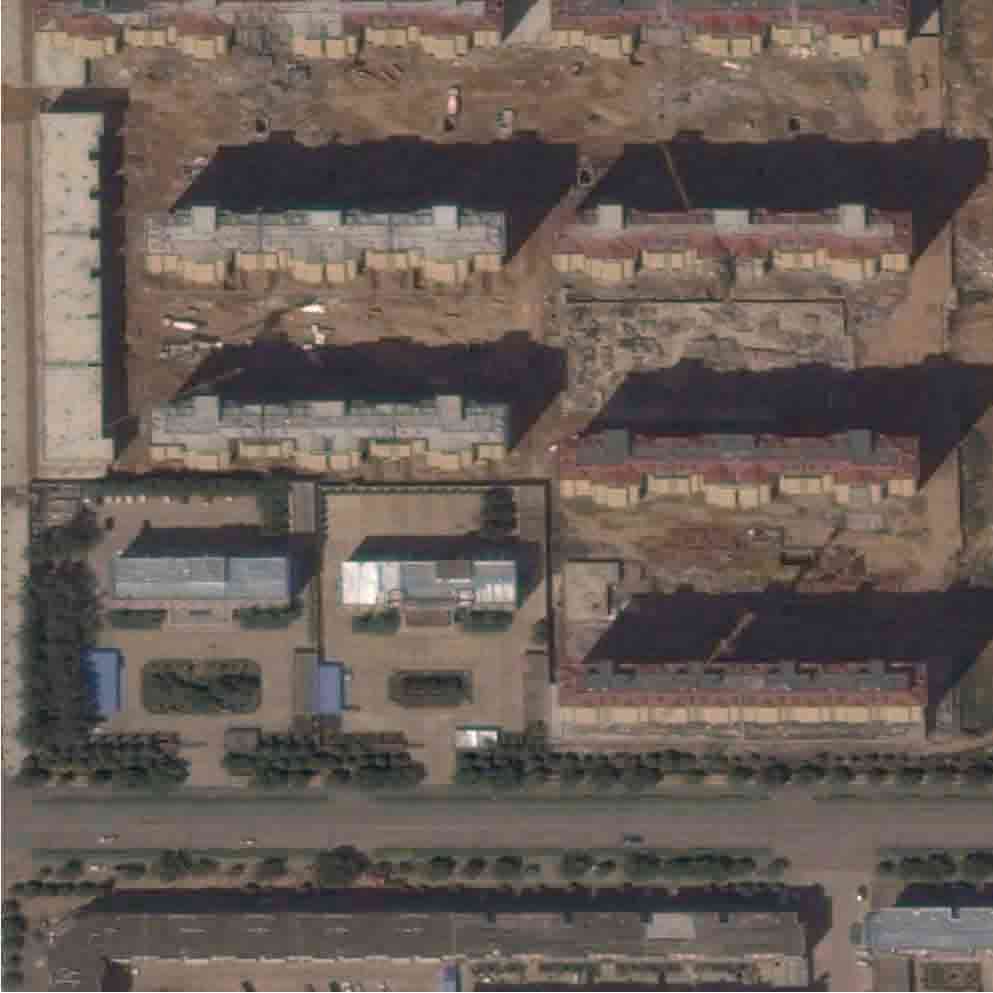}
        \caption{}
        \label{figure22_2}
    \end{subfigure}
    % 第三个小图
    \begin{subfigure}[b]{0.24\linewidth}
        \includegraphics[width=1\linewidth]{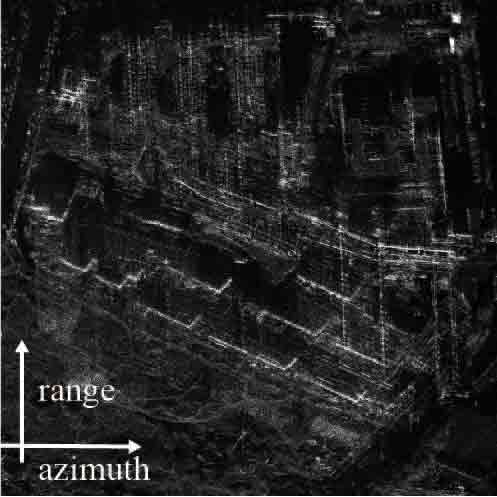}
        \caption{}
        \label{figure22_3}
    \end{subfigure}
    % 第四个小图
    \begin{subfigure}[b]{0.24\linewidth}
        \includegraphics[width=1\linewidth]{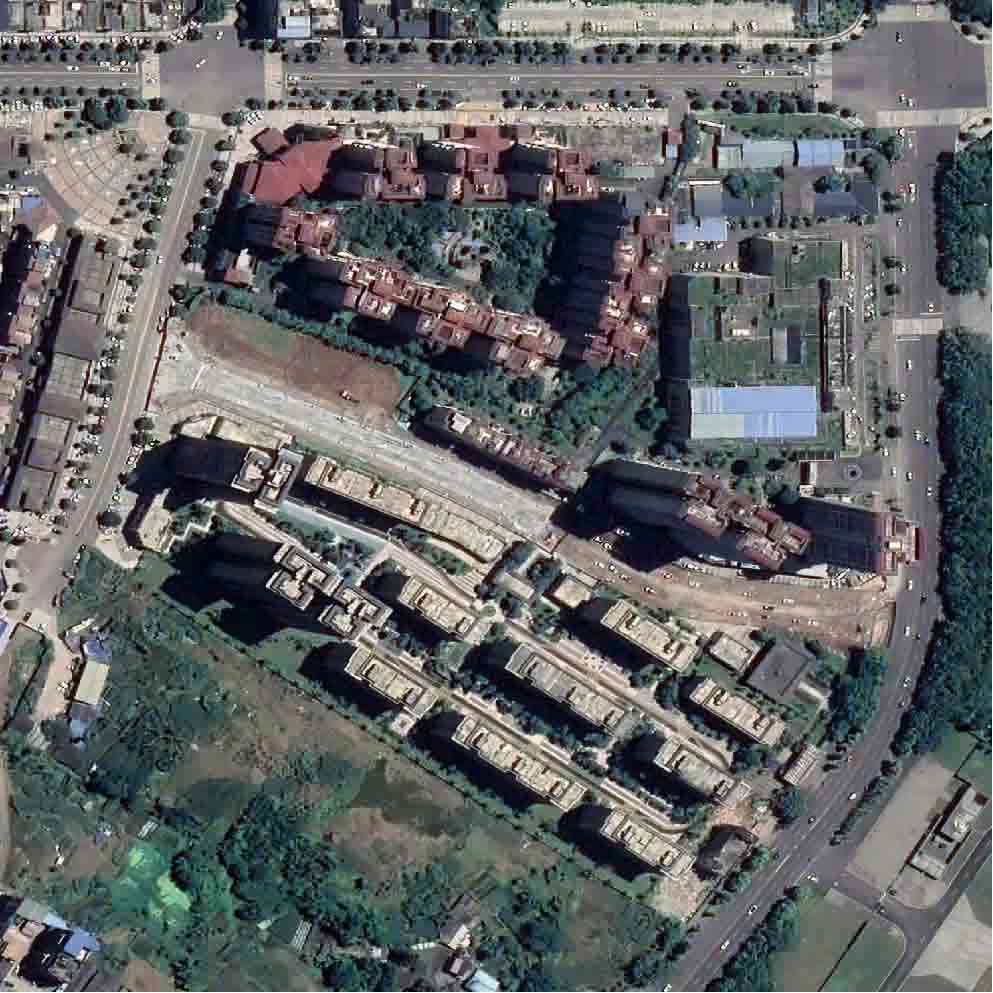}
        \caption{}
        \label{figure22_4}
    \end{subfigure}

    \caption{The evaluation adopted measured data from the SARMV3D-1.0 dataset. This dataset includes two different scenes from Yuncheng, Shanxi, and Emei, Sichuan in China. (a) shows the radar image of the Yuncheng scene, while (b) shows its optical image. (c) presents the radar image of the Emei scene, and (d) shows the corresponding optical image.}
    \label{figure_22}
\end{figure}

SARMV3D-1.0 encompasses two urban scenes from Emei and Yuncheng in China, as illustrated in Fig.~\ref{figure_22}. Each scene is presented with an optical and a 2D radar image. The Yuncheng area in the left part mainly has apartment buildings that are all the same height and face the same way, giving it a neat and organized look. On the other hand, the Emei area in the right part is more varied. It has different kinds of buildings, some are small houses and some are tall buildings, all facing different directions. The Emei area also has some green spaces, making it look more diverse than Yuncheng. The diverse nature of these scenes, captured in different bands (X and Ku) and under various baseline conditions, adds a rich variety to our evaluation process. The key parameters of the data are summarized in Table~\ref{tabel_2}.

\begin{table}[h]
    \caption{Yuncheng and Emei Dataset Parameters}
    \label{tabel_2}
    \footnotesize
    \centering
    \begin{tabular}{|l|c|c|c|}
    \hline
    \textbf{Parameter}       & \textbf{Yuncheng} & \textbf{Emei} & \textbf{Unit} \\
    \hline
    Pixel Number (Range)     & 1800           & 1220       &      \\
    Pixel Number (Azimuth)   & 3600           & 3100       &      \\
    Pixel Size (Range)       & 0.14           & 0.15       & m    \\
    Pixel Size (Azimuth)     & 0.11           & 0.07       & m    \\
    Waveband                 & X              & Ku         &      \\
    Scene Height             & 420            & 595        & m    \\
    Flight Height            & 2156.97        & 1668.62    & m    \\
    Number of Flyovers       & 12             & 8          &      \\
    Range Bandwidth          & 810            & 500        & MHz  \\
    Closest Slant Range      & 1917.72        & 1174.74    & m    \\
    \hline
    \end{tabular}
    \end{table}

	\subsection{Evaluation methods}

	In addition to the four methods (the reconstruction networks) we propose in our study, we have chosen three other methods for evaluation. Of these, two are traditional sparsity-based reconstruction methods solved through different optimization algorithms: the FISTA-based method and SLIMMER~\cite{beck2009fast, zhu2010very,han2021efficient}. The remaining method is a sparsity-based reconstruction method, solved through a deep neural network. This follows the deep unrolling framework for the ISTA solver, a common approach in recent work. The corresponding optimization problem to unroll is the one in (\ref{equation_2}). The resulting network can be interpreted within the framework of iterative reconstruction with enhancement. As the enhancement phase occurs in raw image space, we refer to it as tomo-IRENet-Raw.
	
	All experiments have been conducted on a single computer equipped with a GeForce RTX 2070 with 8GB memory, an Intel i7-10700k, and 32GB of memory.
	
	For the two conventional approaches, we've fine-tuned their parameters using expertise and recommendations from literatures to ensure optimal performance across various scenes. The specifics of the neural network-based methods' implementations are as follows:
	
	\textit{1) tomo-IRENet-Raw:} Adaptive learning is applied to parameters including thresholds and step sizes, set within a network of 9 iterative blocks. The learning rate is configured at 0.0001, with a batch size of 3600 fibers. For the training phase, a pixel-wise $l_2$ loss function is employed, and the network undergoes 500 training iterations using the Adam optimizer.
	
	\textit{2) tomo-IRENet-TV:} This setup includes nine iterative blocks, alearning rate of 0.0001, and handles a single tensor per batch. The training, conducted 45 times, utilizes the pixel-wise $l_2$ loss function, with Adam as the optimizer.
	
	\textit{3) tomo-IRENet-U:} This setup also includes nine iteration blocks and a learning rate of 0.0001, an processes 8 slices per batch. The training, conducted 80 times, uses the pixel-wise $l_2$ loss function, with Adam for optimization. The training dataset comprises 27,689 frontal slices.
	
	\textit{4) tomo-LRENet-biU:} The initial phase of this model, focusing on light reconstruction, includes nine iteration blocks, a learning rate of 0.0001, and a batch size of 32 slices. The training, which uses the pixel-wise $l_2$ loss function, is conducted 80 times with Adam as the optimizer. The dataset for training includes 27,689 frontal slices and 17,942 lateral slices. The training is phased, initially focusing on the light reconstruction part until convergence, followed by joint end-to-end training with the enhancement part.
	
	\textit{5) tomo-LRENet-LSTM:} Mirroring the biU configuration, this setup uses nine iteration blocks for light reconstruction with a learning rate of 0.0001 and a batch size of 32 slices. It undergoes 80 epochs of training using pixel-wise $l_2$ loss, with Adam as the optimizer. The training uses 27,689 frontal slices in the same two-phase manner. The LSTM component's performance is influenced by the series size (number of slices), with larger series generally yielding better correlation modeling. However, due to memory and computational constraints, we choose to use 36 slices per series, adding 2 padding slices at both the beginning and end of the sequence.
	
	\subsection{Evaluation framework}

	To comprehensively assess the performance of various reconstruction methods, we have developed an evaluation framework that considers a range of perspectives. 
	
	\vspace{0.5em}
	\textit{C.1 Evaluation perspectives}
	\vspace{0.5em}
	
	An optimal reconstruction method is expected to deliver high-quality mapping results efficiently. The definition of high-quality encompasses several key aspects:
	
	\begin{itemize}
	
	\item \textit{Resolving Capability of Overlaid Scatterers:} This measures the method's ability to distinguish scatterers that are close in the elevation. A method that can achieve higher resolution with the same baseline configuration is considered to deliver superior results.
	\item \textit{Reconstruction Precision of Scatterers:} This criterion evaluates the accuracy with which individual scatterers are reconstructed, focusing on how closely the reconstructed image matches the actual scene in terms of both scatters’ positions and intensities. The closer the match, the higher the quality of the reconstruction.
	
	\item \textit{Minimization of Residual Noises or Clutters:} Given the typical low signal-to-noise ratio (SNR) challenges of tomoSAR systems, an effective method should demonstrate the capability to significantly reduce residual noise and clutter in the result.
	
	\item \textit{Spatial Structure Integrity:} As our study is particularly focused on urban mapping, maintaining the integrity of the spatial structure in the reconstructed scene is imperative. The method should accurately represent the geometric layout and interrelationships of building structures, which is a crucial aspect of urban area mapping. Compromises in spatial integrity, such as holes in surfaces or discontinuities in lines, are indicative of lower quality reconstruction.
	
	\end{itemize}
	
	\vspace{0.5em}
	\textit{C.2 Evaluation objects}
	\vspace{0.5em}

\begin{figure}[h] % 使用 figure* 环境使图形跨越两栏，'t' 表示顶部
		\centering
		% 第一个小图
		\begin{subfigure}[b]{0.24\linewidth}
			\includegraphics[width=1\linewidth]{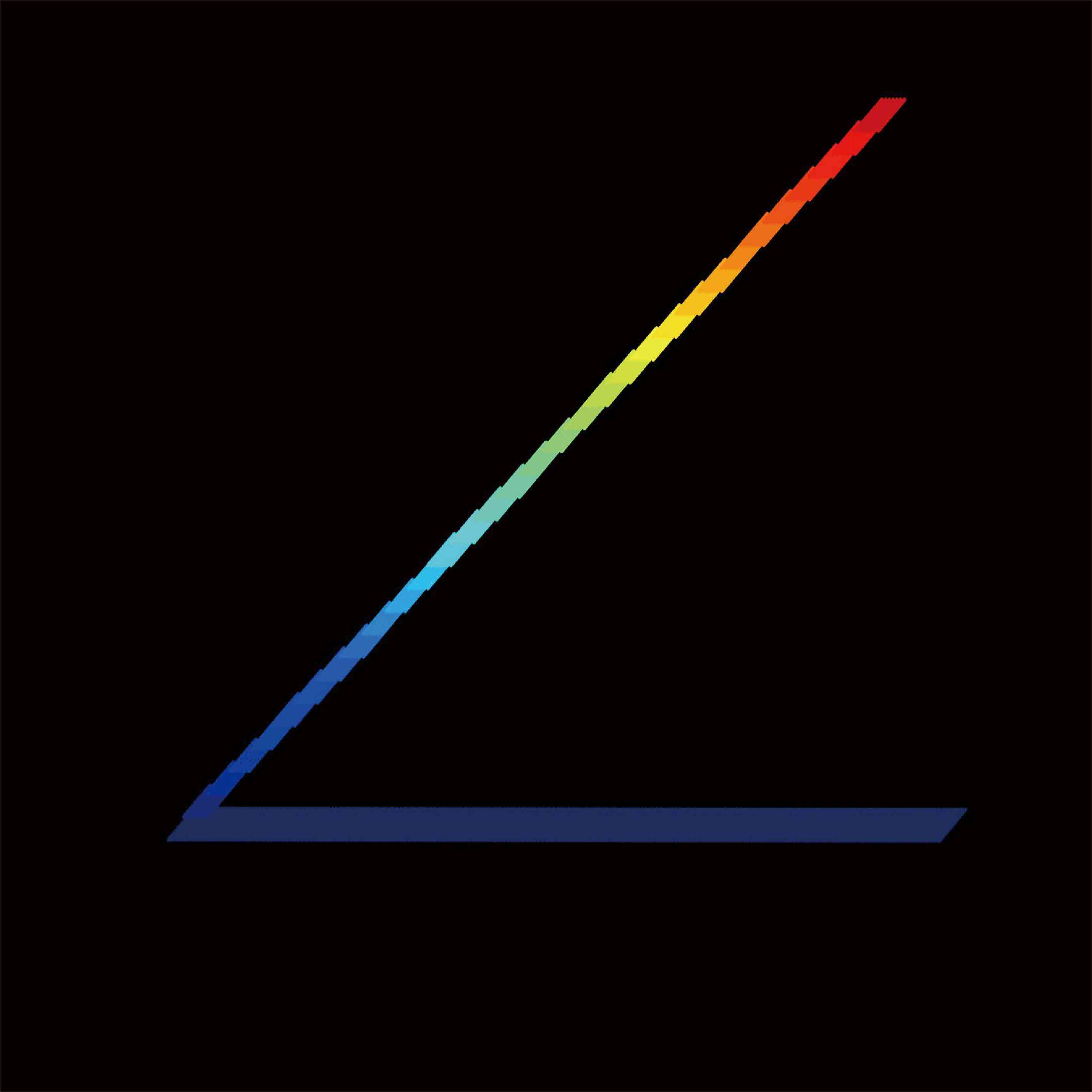}
			\caption{}
			\label{figure23_1}
		\end{subfigure}
		% 第二个小图
		\begin{subfigure}[b]{0.24\linewidth}
			\includegraphics[width=1\linewidth]{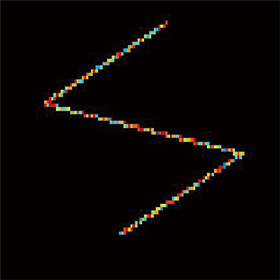}
			\caption{}
			\label{figure23_2}
		\end{subfigure}
		% 第三个小图
		\begin{subfigure}[b]{0.24\linewidth}
			\includegraphics[width=1\linewidth]{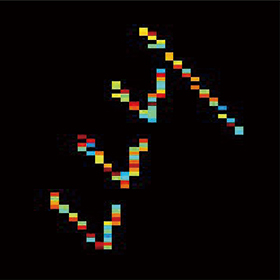}
			\caption{}
			\label{figure23_3}
		\end{subfigure}
		% 第四个小图
		\begin{subfigure}[b]{0.24\linewidth}
			\includegraphics[width=1\linewidth]{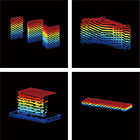}
			\caption{}
			\label{figure23_4}
		\end{subfigure}
	
		\caption{Test objects 1-4. (a) Two closely positioned scatterers with a gradually increasing distance. (b) A one-step structure object. (c) A multi-step structure. (D) Objects of different classical architectural forms.}
		\label{figure_23}
	\end{figure}

	To effectively evaluate the performance of all methods, particularly in relation to the criteria outlined earlier, we have selected a range of test objects. These objects vary in complexity, from simpler individual components to more complex urban scenes. This variety ensures a comprehensive assessment of each method's capability in handling different reconstruction challenges, from individual fibers to full tensors and varying levels of building complexity. The first five test objects are simulated, while the last two are real scenes. The specifications are as follows:

	\begin{itemize}
	
	\item \textit{Test Object 1 (Close Scatterer Resolution Test):} Comprises two closely positioned scatterers within a single building fiber, primarily testing the method's resolving capability. See Fig.~\ref{figure_23}-a.
	
	\item \textit{Test Object 2 (Basic One-Step Structure Reconstruction Test):} A single building slice with a step-like structure, arising from the standard architectural combination of floors, walls, and roofs. This object tests the precision in reconstructing fundamental stepped geometries. See Fig.~\ref{figure_23}-b.

	\item \textit{Test Object 3 (Advanced Multi-Step Structure Reconstruction Test):} Multiple building slices forming a multi-stepped structure, typically found in clusters of closely situated buildings. This object increases complexity to assess the handling of layered geometries. See Fig.~\ref{figure_23}-c.
	
	\item \textit{Test Object 4 (Classical Architectural Forms Test): } Consists of four single buildings, each with a different complexity level, including an L-shaped facade building, a complex planar structure, a low-rise box-type building, and a flat-type building. This object is designed to evaluate performance across simple to complex building structures. See Fig.~\ref{figure_23}-d.
	
	\item \textit{Test object 5 (Uniform Urban Structure Applicability Test):} The Yuncheng scene, an actual urban area with uniform building structures, serving to assess the real-world applicability of the methods.
	
	\item \textit{Test object 7 (Diverse Urban Landscape Challenge):} The Emei scene, characterized by its diverse urban landscape with varying building types and orientations, providing a more challenging test of the methods' efficacy in complex urban environments.
		
	\end{itemize}
	
	\vspace{0.5em}
	\textit{C.3 Evaluation metrics}
	\vspace{0.5em}

	In the evaluation process, we first conduct a qualitative assessment of the results. This involves examining the visual coherence, realism, and overall integrity of the reconstructed scenes. We focus this analysis on the perceptual effectiveness of each method in representing urban environments. 
	
	Following this, we perform a quantitative assessment, measuring specific aspects such as resolution accuracy, reconstruction precision, reconstruction efficiency, and structural fidelity against evaluation metrics. These metrics are categorized into three types, each targeting a different aspect of the reconstruction.
	
	\vspace{0.5em}
	\textit{C.3.1 3D image accuracy metrics}
	\vspace{0.5em}
	
	These metrics assess the accuracy of the reconstructed result compared to the actual scene, focusing on scatters' intensity accuracy and their spatial positioning. 
	
	\begin{itemize}
		\item \textit{Root Mean Square Error}  ($\mathrm{RMSE}$) : RMSE measures the intensity differences between the reconstructed result and the actual scene. A lower RMSE value indicates a higher-quality reconstruction, where the reconstructed intensities closely match the actual ones. It's defined as:
		\begin{equation}
		\label{equation_28}
		\mathrm{RMSE} = \sqrt{\frac{1}{N} \sum_{i=1}^{N} (\hat{\mathcal{X}}_i - \mathcal{X}_i)^2}
		\end{equation}
	
		Here, \( \hat{\mathcal{X}}_i \) and \( \mathcal{X}_i \) represent the pixel intensities in the reconstructed and actual images, respectively, while \( N \) denotes the total number of pixels.
	
		\item \textit{Peak Signal-to-Noise Ratio} ($\mathrm{PSNR}$):  PSNR compares the maximum power of the actual scne to the power of residual noise (difference between reconstructed result and the actual scene). Higher PSNR values indicates higher quality. It's defined as:
		\begin{equation}
		\label{equation_29}
		\mathrm{PSNR} = 20 \cdot \log_{10}\left(\frac{\mathrm{MAX}_{\mathcal{X}}}{\mathrm{RMSE}}\right)
		\end{equation}
	
		Here, \(\mathrm{MAX}_{\mathcal{X}} \) denotes the maximum possible pixel intensity in the actual image \( \mathcal{X} \). The RMSE value is in the denominator, representing the noise level.
		 
	\end{itemize}
	
	\vspace{0.5em}
	\textit{C.3.2 3D point cloud geometric spatial structure metrics} 
	\vspace{0.5em}

	These metrics evaluate the positional accuracy of scatterers within the 3D point cloud reconstruction and their adherence to the geometric spatial structure of the actual environment. Ensuring the reconstructed scene mirrors the actual three-dimensional layout accurately, the metrics include:
	
	\begin{itemize}
		\item \textit{Scatterer Correspondence} ($\mathrm{Precision}$): Also known as Precision, this metric quantifies the proportion of reconstructed points that accurately match the target scatterers in the actual scene. It's defined as:
		\begin{equation}
		\label{equation_30}
		\mathrm{Precision} = \frac{T_p}{N_p}
		\end{equation}
	
		where \( T_p \) represents the count of accurately reconstructed points, and \( N_p \) is the total number of points in the reconstruction. A point is considered accurately reconstructed if its distance to the nearest actual point, calculated as \( d_i = \min \|p_i - p^*\|_2^2 \) (where \( p_i \) is the location of the reconstructed point and \( p^* \) are the locations of actual points), is within a predefined threshold \( \tau_p \).
	
		\item \textit{Reconstruction Completeness} ($\mathrm{Recall}$): Also known as Recall, this metric measures the completeness of the reconstruction by evaluating the proportion of actual scatterers that are accurately captured in the reconstructed model. It's defined as:
		\begin{equation}
		\label{equation_31}
		\mathrm{Recall} = \frac{T_p}{A_p}
		\end{equation}
	
		where \( T_p \) denotes the number of actual scatterers correctly identified in the reconstruction, and \( A_p \) is the total number of actual scatterers in the scene. This metric contrasts with Precision by focusing on capturing all relevant scatterers, thereby measuring the reconstruction's completeness. In other words, precision measures the accuracy of the reconstructed points (how many reconstructed points are true), whereas Recall assesses the completeness of the reconstruction (how many true scatterers are reconstructed).
	
		\item \textit{Average Euclidean Distance} ($D_{pcm}$): Also known as Point Cloud Matching Distance, this metric \(D_{pcm}\) quantifies the average Euclidean distance between each scatterer in the reconstructed result and the closest scatterer in the actual scene. It is defined as:
		\begin{equation}
		\label{equation_32}
		D_{pcm} = \frac{1}{N} \sum_{i=1}^{N} \min_{p^* \in P^*} \|p_i - p^*\|_2
		\end{equation}
	
		where \( p_i \) represents the position of a scatterer in the reconstructed result, and \( P^* \) consists of the scatterer positions in the actual scene. A lower \( D_{pcm} \) value suggests greater spatial accuracy in the reconstruction.
	
		\item \textit{Spatial Distribution Consistency \((V)\):} Also known as Variance, this metric, \( V \), assesses the consistency of the reconstructed scatterers relative to their actual positions by measuring the dispersion of distances between each reconstructed point and its nearest actual point. It's defined as:
		\begin{equation}
		\label{equation_33}
		V = \frac{1}{N} \sum_{i=1}^{N} (d_i - \bar{d})^2
		\end{equation}
	
		where \( d_i \) is the distance between each reconstructed point and the closest actual point, and \( \bar{d} \) is the mean of these distances. A lower \( V \) indicates a more uniform and precise spatial distribution in the reconstruction.
	
	\end{itemize}
	
	\vspace{0.5em}
	\textit{C.3.3 Reconstruction efficiency metric}
	\vspace{0.5em}

	This metric measures the time efficiency of the reconstruction process, assessing how quickly and effectively each method produces the final reconstructed image. Specifically, we adopt Average Reconstruction Time.
		\begin{itemize}
			\item \textit{Average Reconstruction Time \((T_{ag})\):} This metric calculates the mean time taken for the reconstruction process across different methods. A shorter \( T_{ag} \) indicates a more efficient reconstruction process.  
		\end{itemize}
	
\subsection{Evaluation results}

The evaluation results from applying the above framework to the seven methods are outlined below. They are arranged from the simplest, the close scatterer resolution test, to the most complex, the diverse urban landscape challenge.

\vspace{0.5em}
\textit{D.1 Results of test object 1 (close scatterer resolution test)}
\vspace{0.5em}

In this test, we counduct experiments similar to those reported in the literature, focusing on evaluating the resolution capability of various scatterer reconstruction methods~\cite{qian2022gamma, yexian2022comparative, 
zhu2011super}. Specifically, we place two scatterers within a single fiber and adjusted their relative distance to assess the performance of different reconstruction techniques.

We establish a relative reference point by determining the theoretical resolution ($\rho_s$) corresponding to the result achieved through matched-filtering with uniform sampling for a specific baseline length. The distance between the scatterers was systematically varied from 0 to 1.4 times $\rho_s$.

\begin{figure}[t]
    \centering
    % 第一行四张图
    \begin{subfigure}[b]{0.24\linewidth}
        \caption*{{\fontsize{8pt}{10pt}\selectfont\centering \qquad\qquad FISTA-based}}
        \includegraphics[width=\linewidth]{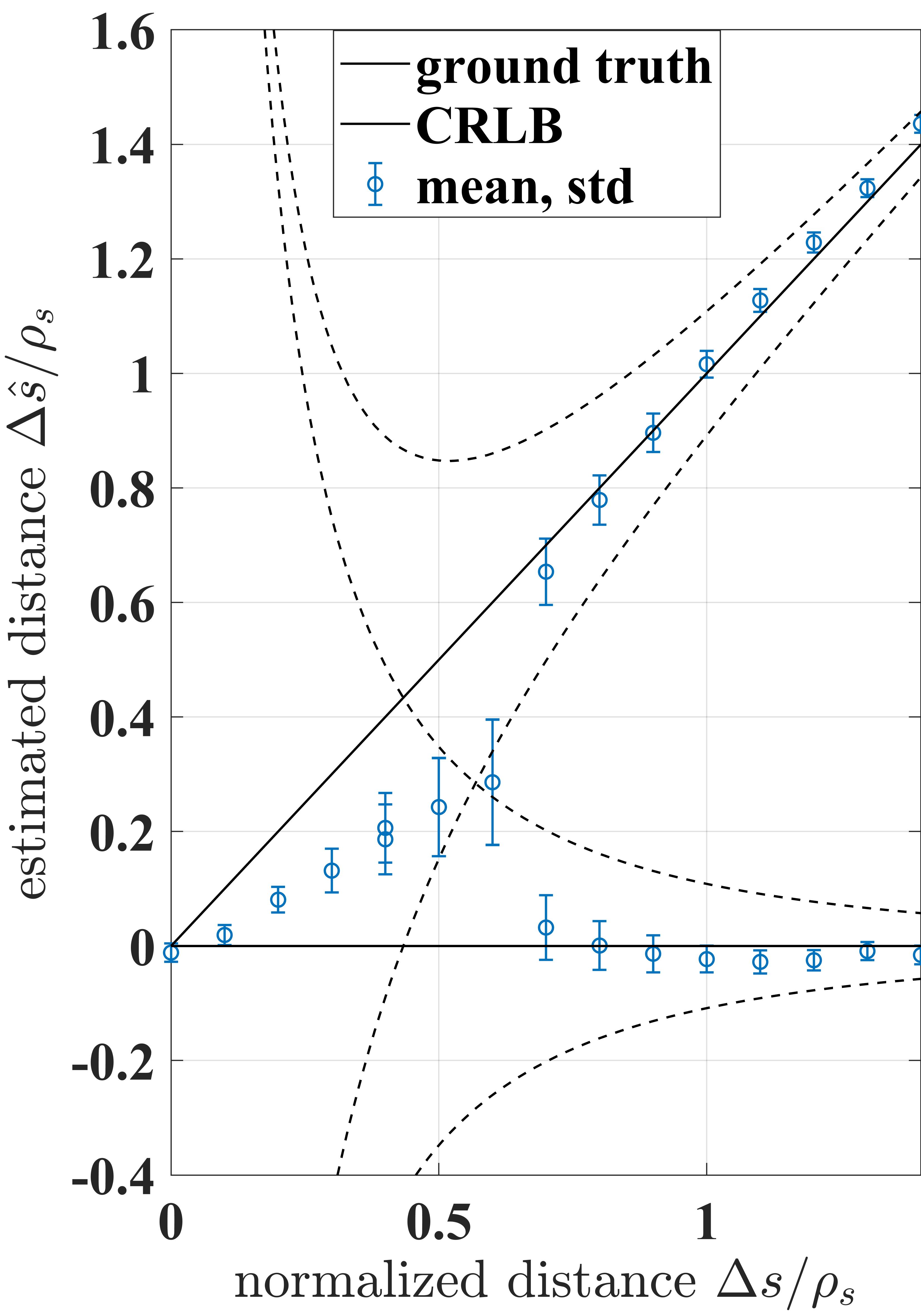}
        \caption{}
        \label{figure24_1}
    \end{subfigure}
    \begin{subfigure}[b]{0.24\linewidth}
        \caption*{{\fontsize{8pt}{10pt}\selectfont\centering\qquad\qquad SLIMMER}}
        \includegraphics[width=\linewidth]{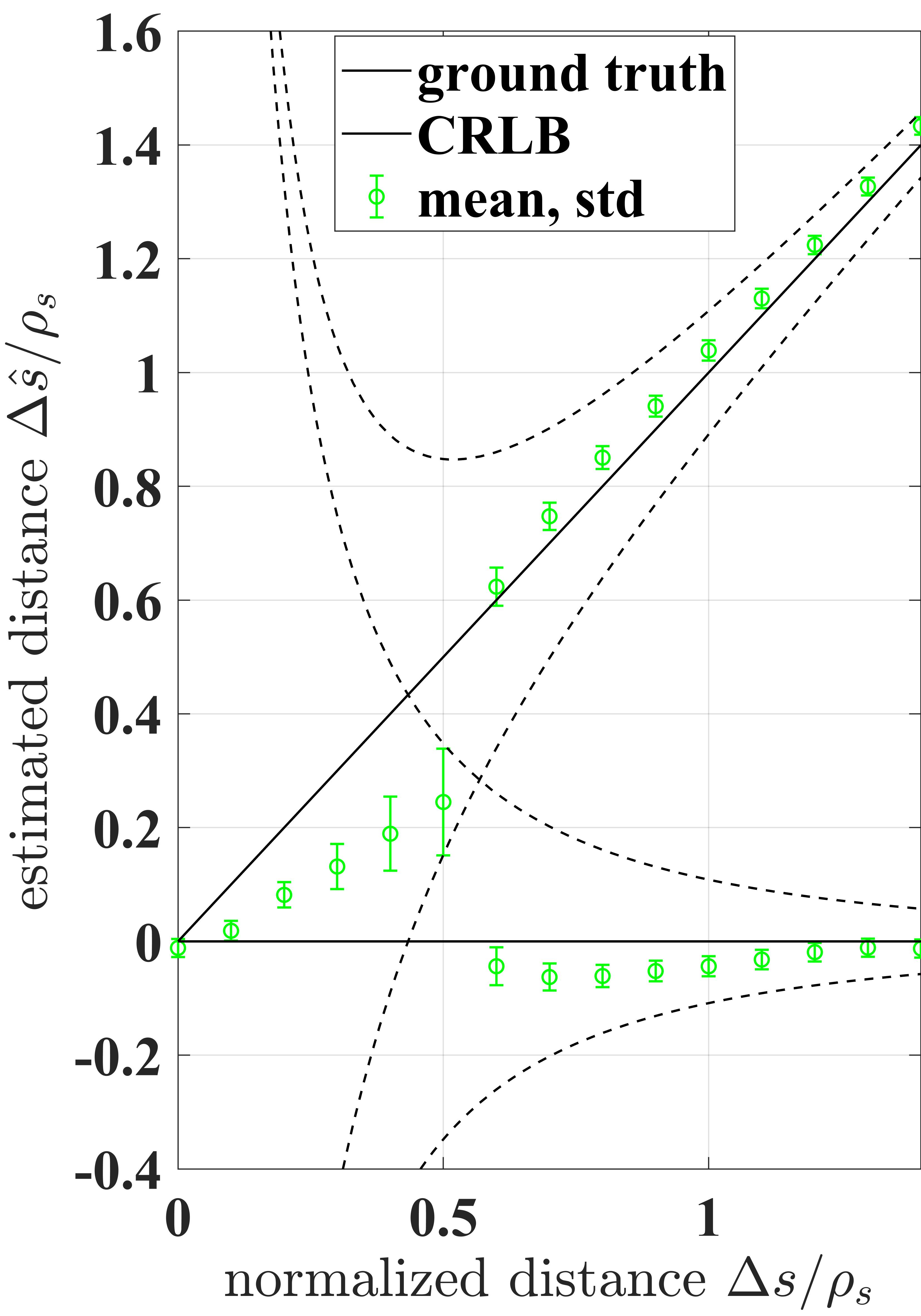}
        \caption{}
        \label{figure24_2}
    \end{subfigure}
    \begin{subfigure}[b]{0.24\linewidth}
        \caption*{{\fontsize{8pt}{10pt}\selectfont\centering\qquad\qquad tomo-IRENet-Raw}}
        \includegraphics[width=\linewidth]{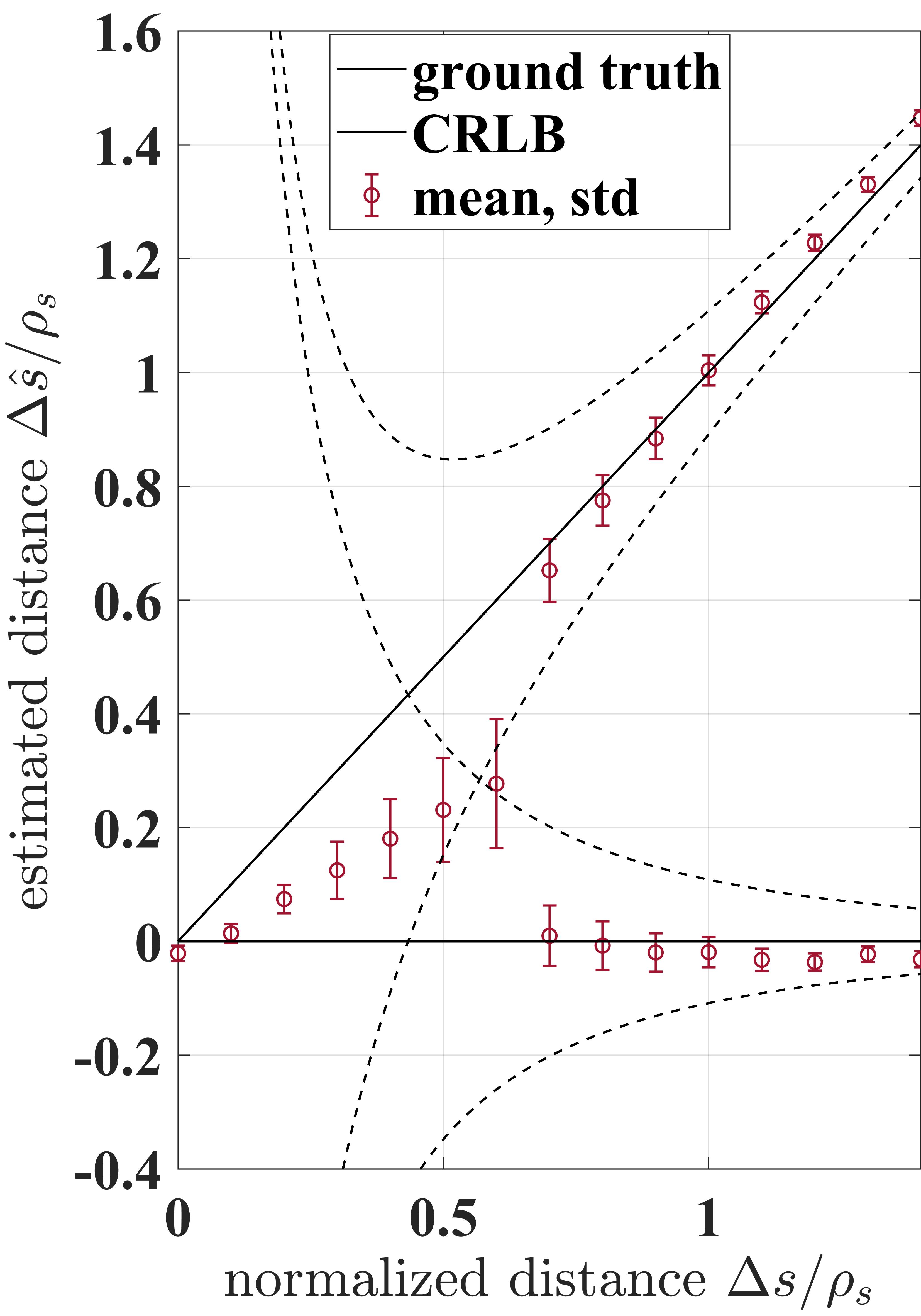}
        \caption{}
        \label{figure24_3}
    \end{subfigure}
    \begin{subfigure}[b]{0.24\linewidth}
        \caption*{{\fontsize{8pt}{10pt}\selectfont\centering\qquad\qquad  tomo-IRENet-TV}}
        \includegraphics[width=\linewidth]{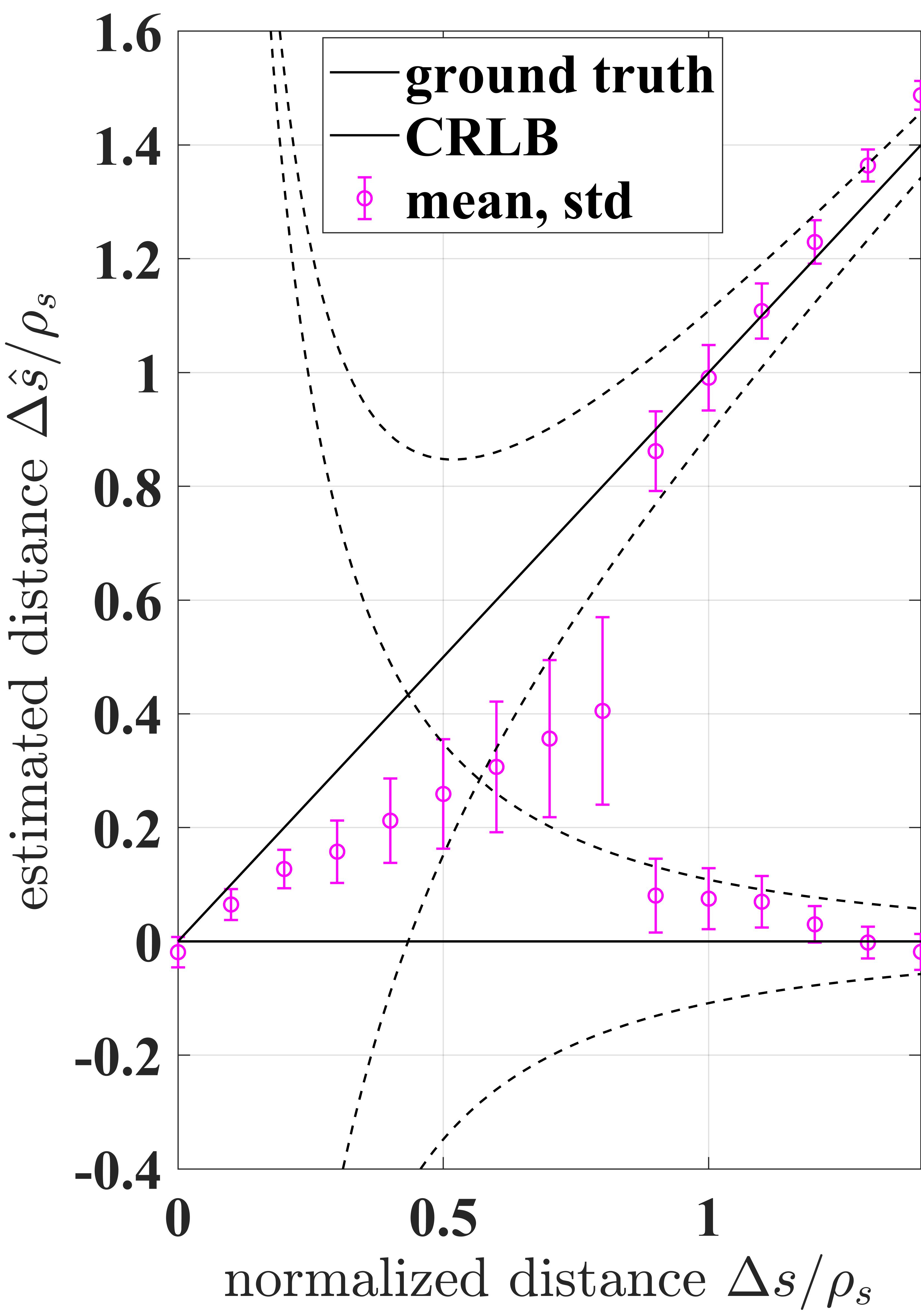}
        \caption{}
        \label{figure24_4}
    \end{subfigure}

    % 第二行三张图
    \begin{subfigure}[b]{0.24\linewidth}
        \caption*{{\fontsize{8pt}{10pt}\selectfont\centering\quad\qquad tomo-IRENet-U}}
        \includegraphics[width=\linewidth]{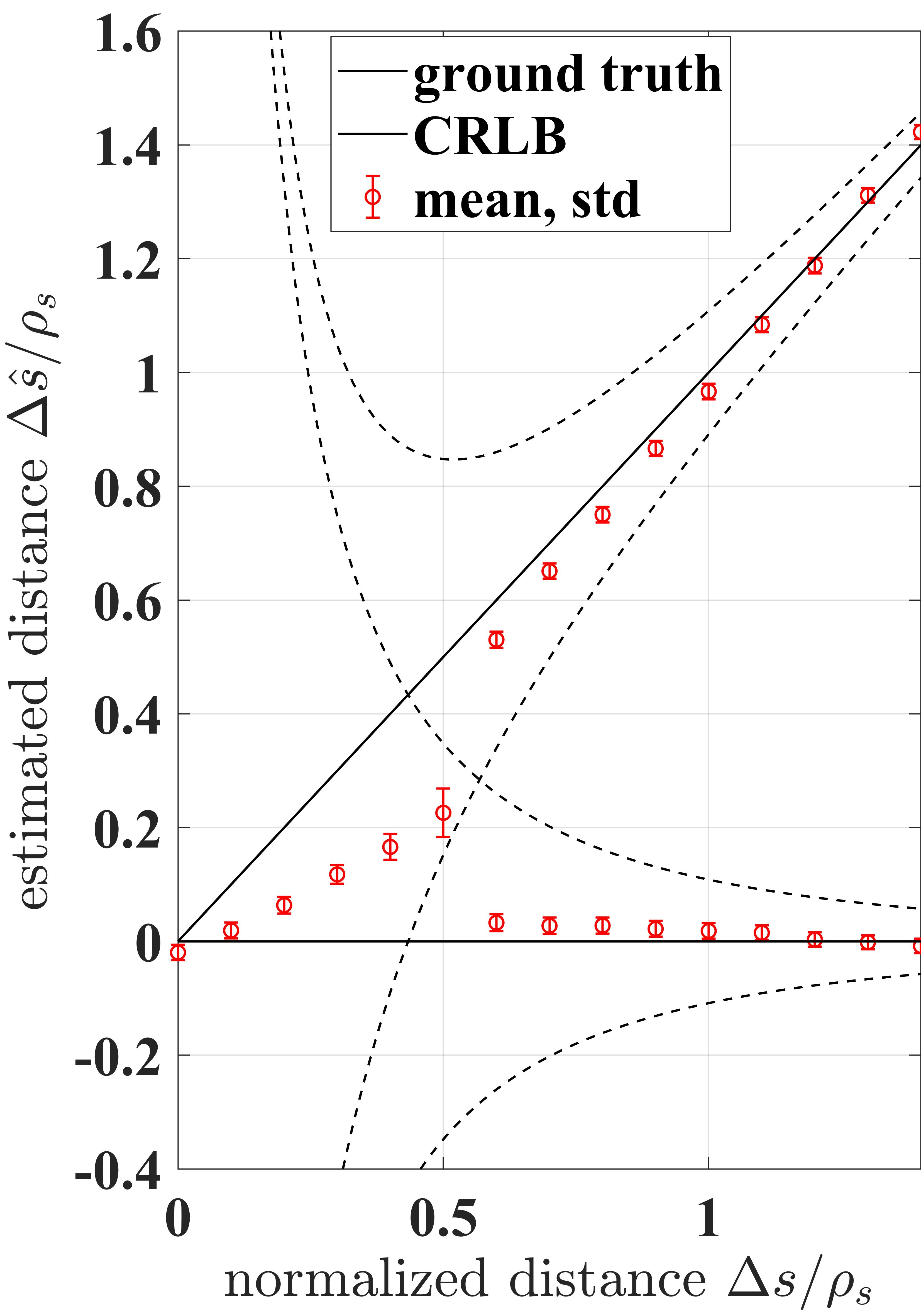}
        \caption{}
        \label{figure24_5}
    \end{subfigure}
    \begin{subfigure}[b]{0.24\linewidth}
        \caption*{{\fontsize{8pt}{10pt}\selectfont\centering\quad\qquad  tomo-LRENet-biU}}
        \includegraphics[width=\linewidth]{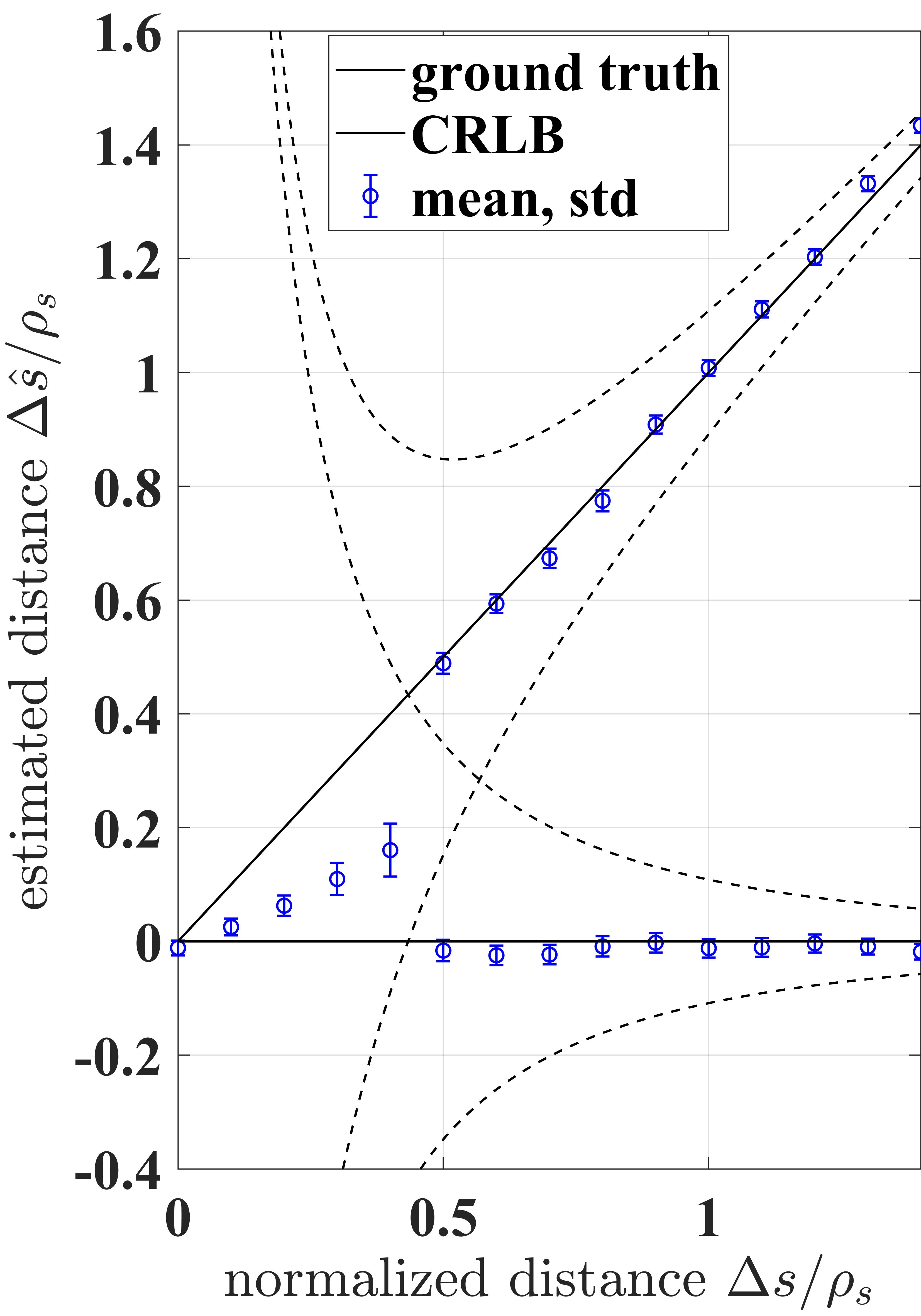}
        \caption{}
        \label{figure24_6}
    \end{subfigure}
    \begin{subfigure}[b]{0.24\linewidth}
        \caption*{{\fontsize{8pt}{10pt}\selectfont\centering \quad\qquad tomo-LRENet-LSTM}}
        \includegraphics[width=\linewidth]{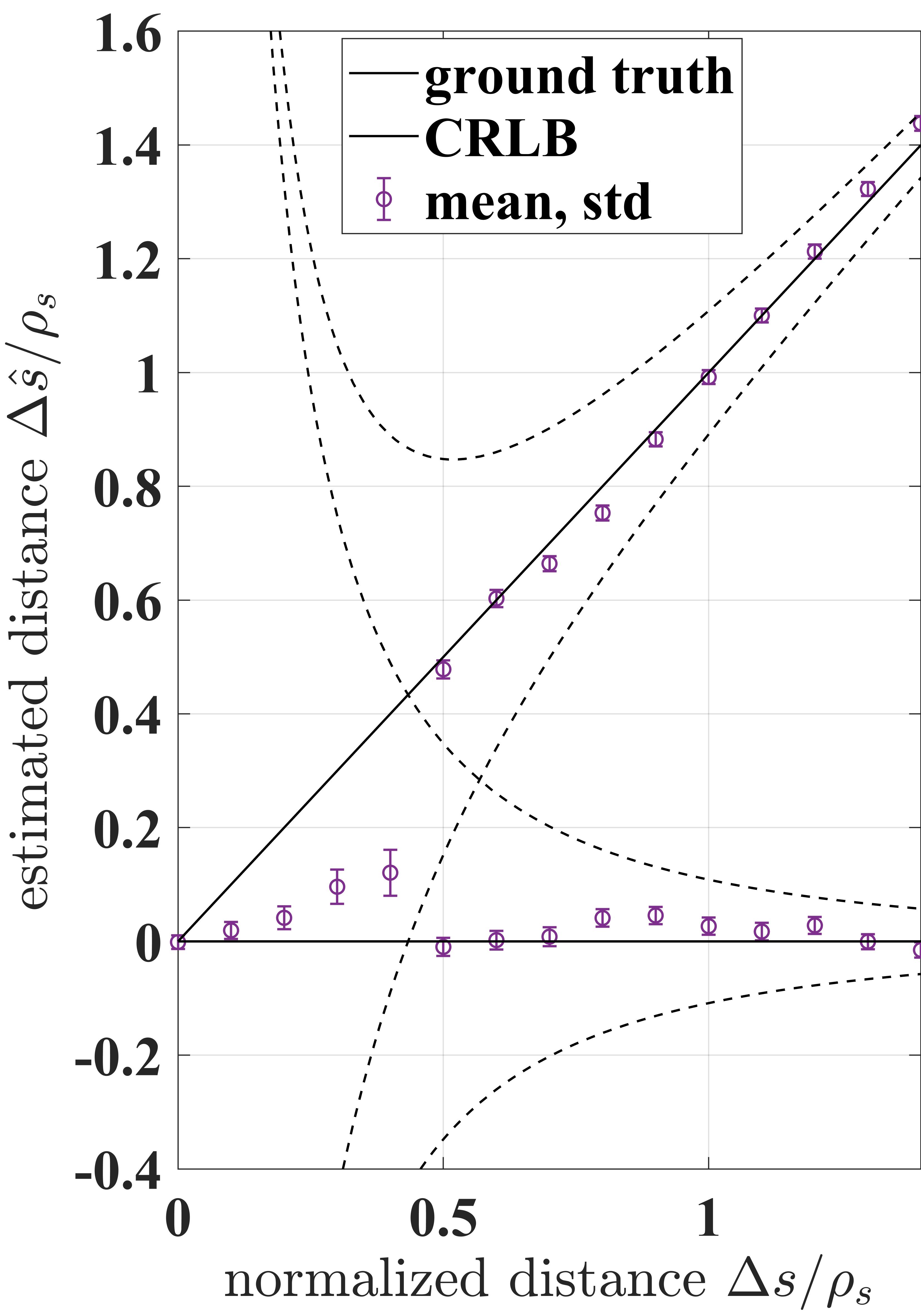}
        \caption{}
        \label{figure24_7}
    \end{subfigure}

    \caption{Results of test1. (a) Result of FISTA-based method. (b) Result of SLIMMER. (c) Result of tomo-IRENet-Raw. (d) Result of tomo-IRENet-TV. (e) Result of tomo-LRENet-U. (f) Result of tomo-LRENet-biU. (g) Result of tomo-LRENet-LSTM.}
    \label{figure_24}
\end{figure}

We have conducted 10,000 Monte Carlo trials under a signal-to-noise ratio (SNR) of 5dB to ensure thorough testing. The results are presented in Fig.~\ref{figure_24}. The solid black line represents the actual positions, while the black dashed curves signify the theoretical lower limit of estimation error (Cramér–Rao bound - CRLB)~\cite{zhu2011super}. The colored circular markers denote the reconstructed positions, with the hollow circle in the center indicating the mean value. The error bars associated with the data points represent the standard variance, indicating the range of uncertainty. Each data point reflects the performance of the method at different levels of normalized separation between scatterers.

Regarding resolving ability, the top performers are tomo-LRENet-biU and tomo-LRENet-LSTM, which can resolve scatters with a normalized relative distance greater than 0.4. Following these are SLIMMER and tomo-IRENet-U, which can handle scatters separated by a distance greater than 0.5. FISTA and tomo-IRENet-Raw are next, handling a distance of 0.6. Tomo-IRENet-TV has the worst performance among the seven methods, only managing a distance of 0.7.

Overall, deep learning methods generally outperform or match the traditional methods, with tomo-IRENet-TV being the exception. Among these, tomo-IRENet-Raw performs at the same level as traditional methods. Networks that consider spatial features, such as tomo-LRENet-biU, tomo-LRENet-LSTM, and tomo-IRENet-U, outperform traditional methods. The adoption of sparsity regularizations in the raw image space boosts the resolving ability for all methods.

The relatively poor performance of tomo-IRENet-TV may be due to the introduction of 3D-TV regularizations, which force a trade-off between two regularization terms. Despite this, it maintains some super-resolving ability, albeit weaker than the others. The use of deep spatial feature modeling, as opposed to shallow spatial feature modeling, appears to alleviate this issue by taking a composite approach rather than a separate one. Except for tomo-IRENet-TV, considering spatial feature regularization in neural networks appears to be beneficial overall.

Interestingly, SLIMMER performs well due to two refinements during reconstruction in the original design, enhancing robustness to outliers but potentially increasing computational time.

Beyond resolving ability, in terms of reconstruction precision and stability, tomo-IRENet-U, tomo-LRENet-biU, and tomo-LRENet-LSTM perform best. FISTA, SLIMMER, and tomo-IRENet-Raw are in the second best category, with tomo-IRENet-Raw slightly ahead. Tomo-IRENet-TV has the worst performance, with a larger error bound indicating weaker stability. This again highlights the challenge of balancing two different regularization terms.

In summary, while the adoption of deep learning for spatial feature regularization does not significantly boost resolving ability, it offers other important benefits. The similar level of resolving ability stems from the methods' reliance on initial 3D reconstruction as their foundation. However, the improvements in variance and stability are significant. This suggests that the methods developed in this study provide more stable tomoSAR reconstruction.

\vspace{0.5em}
\textit{D.2 Results of test object 2 (basic one-step structure reconstruction test)}
\vspace{0.5em}

\begin{figure}[h]
    \centering
    % 第一排的四张图
    \begin{subfigure}[b]{0.24\linewidth}
        \caption*{{\fontsize{8pt}{10pt}\selectfont\centering FISTA-based}}
        \includegraphics[width=\linewidth]{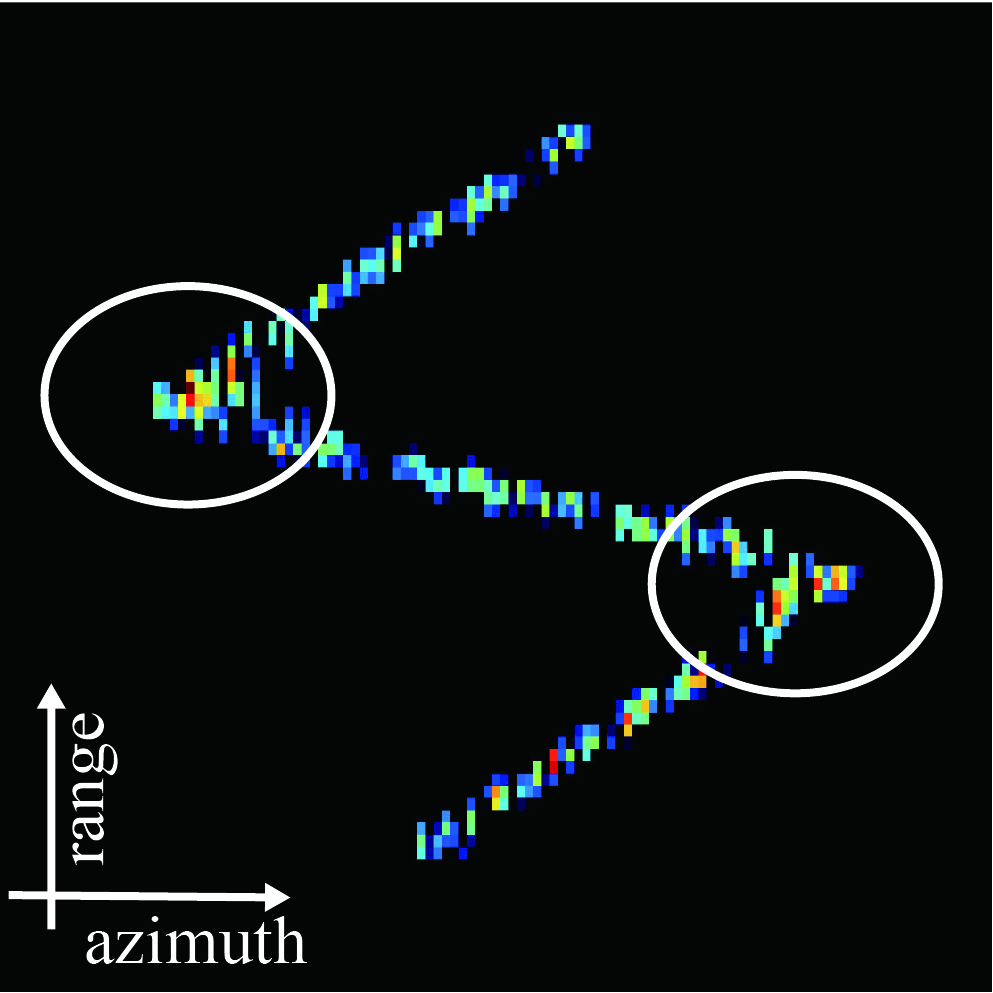}
        \caption{}
        \label{figure25_1}
    \end{subfigure}
    \begin{subfigure}[b]{0.24\linewidth}
        \caption*{{\fontsize{8pt}{10pt}\selectfont\centering SLIMMER}}
        \includegraphics[width=\linewidth]{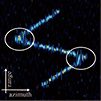}
        \caption{}
        \label{figure25_2}
    \end{subfigure}
    \begin{subfigure}[b]{0.24\linewidth}
        \caption*{{\fontsize{8pt}{10pt}\selectfont\centering tomo-IRENet-Raw}}
        \includegraphics[width=\linewidth]{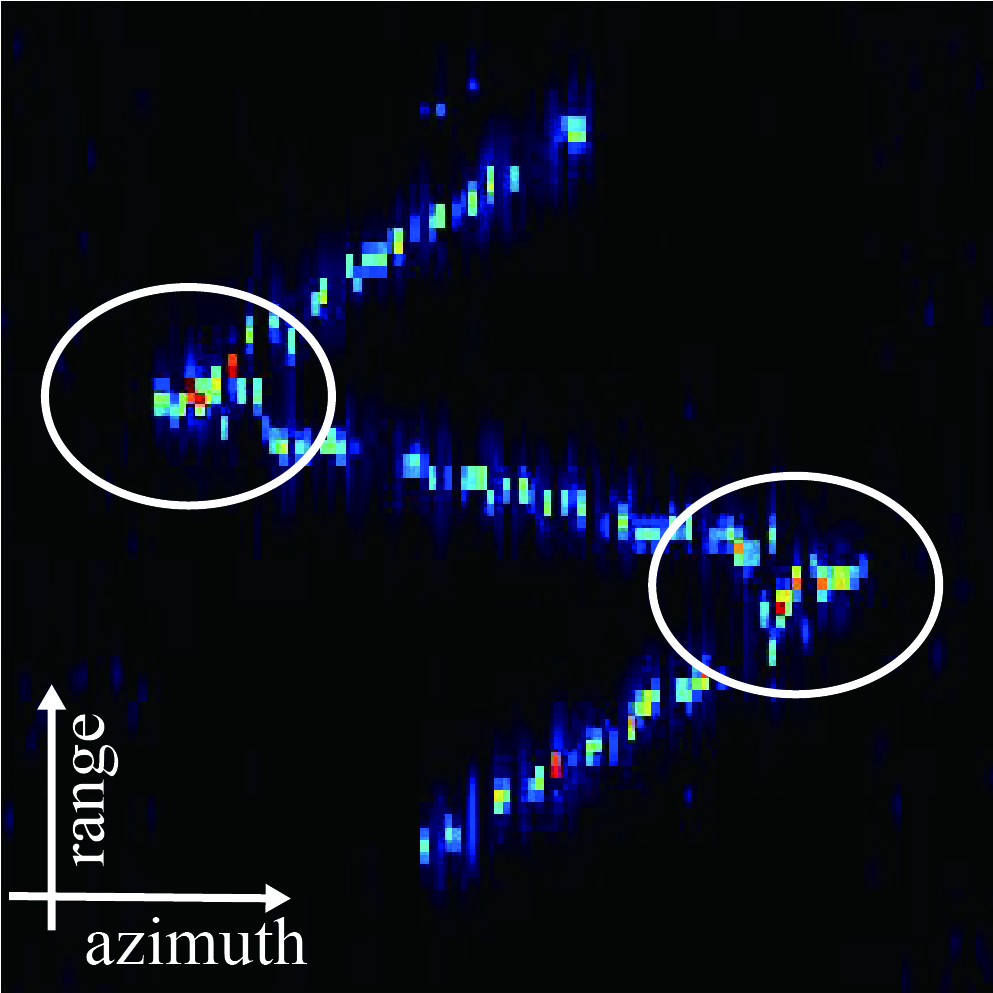}
        \caption{}
        \label{figure25_3}
    \end{subfigure}
    \begin{subfigure}[b]{0.24\linewidth}
        \caption*{{\fontsize{8pt}{10pt}\selectfont\centering tomo-IRENet-TV}}
        \includegraphics[width=\linewidth]{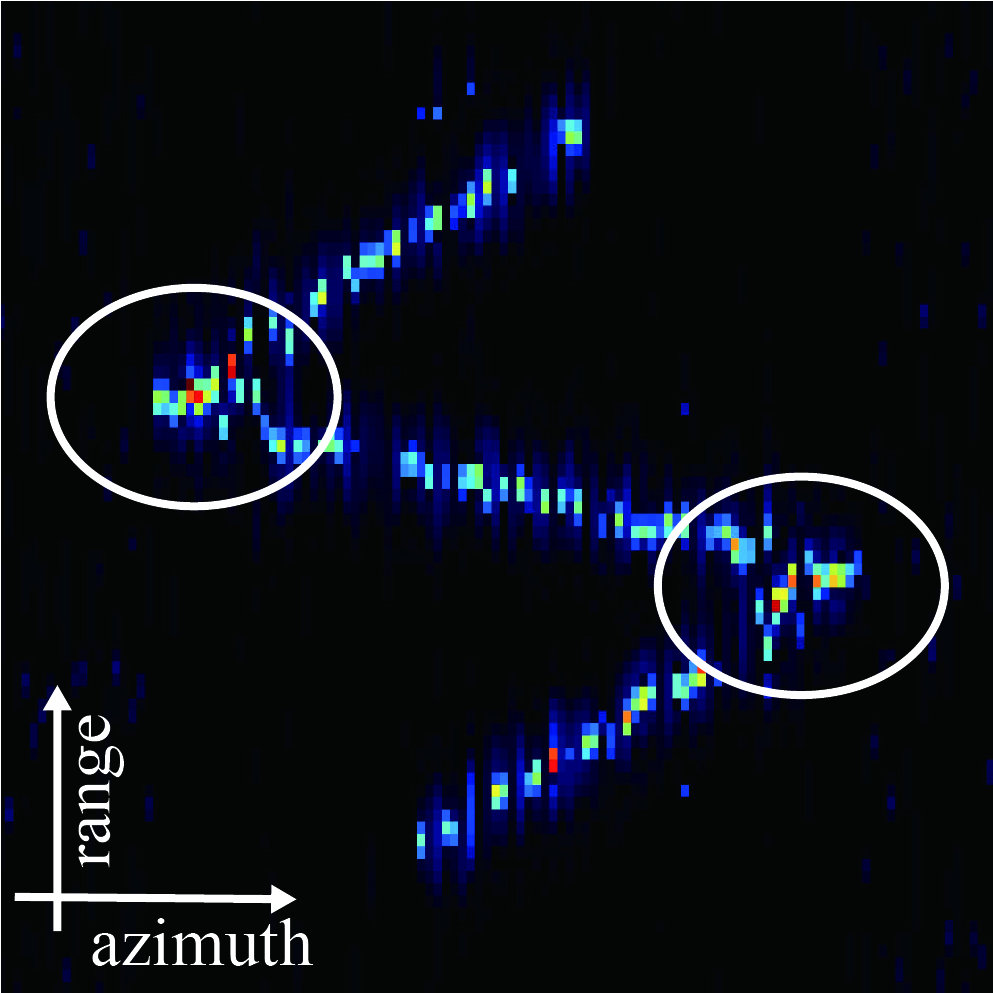}
        \caption{}
        \label{figure25_4}
    \end{subfigure}
    
    % 第二排的四张图
    \begin{subfigure}[b]{0.24\linewidth}
        \caption*{{\fontsize{8pt}{10pt}\selectfont\centering tomo-IRENet-U}}
        \includegraphics[width=\linewidth]{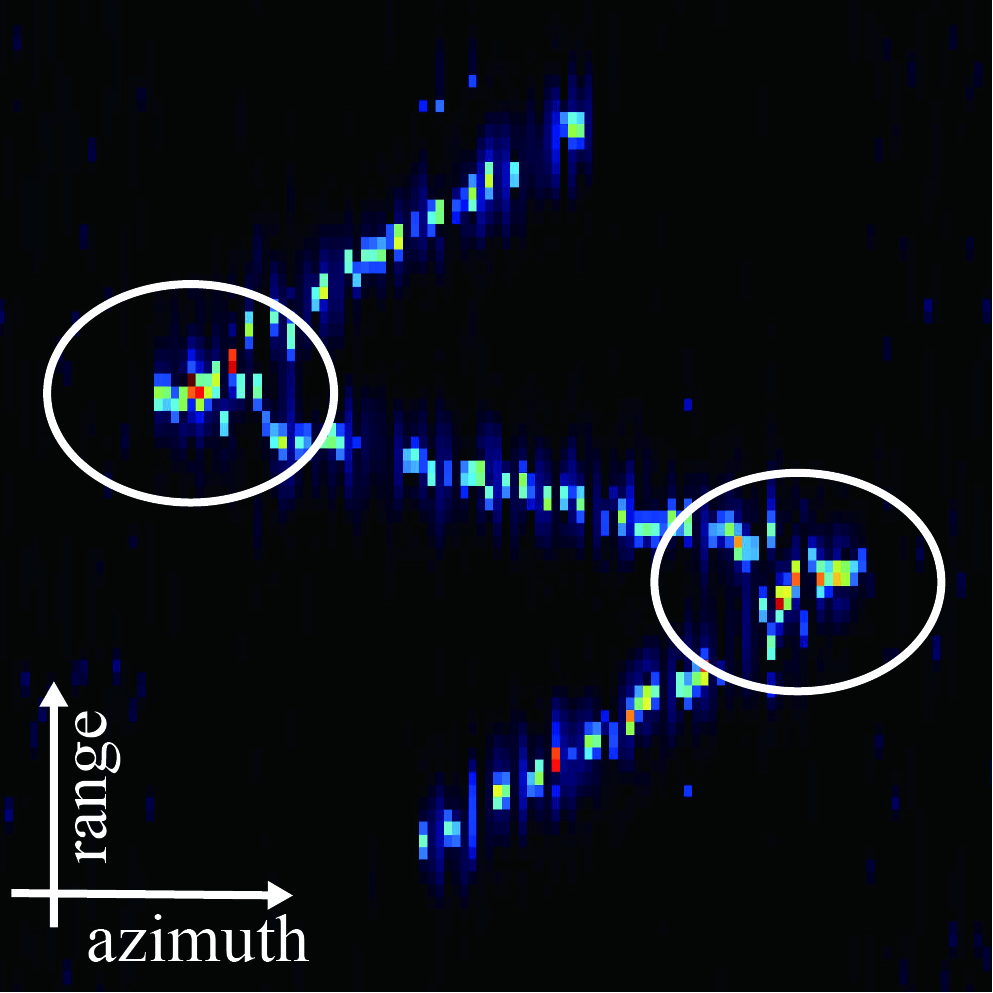}
        \caption{}
        \label{figure25_5}
    \end{subfigure}
    \begin{subfigure}[b]{0.24\linewidth}
        \caption*{{\fontsize{8pt}{10pt}\selectfont\centering tomo-LRENet-biU}}
        \includegraphics[width=\linewidth]{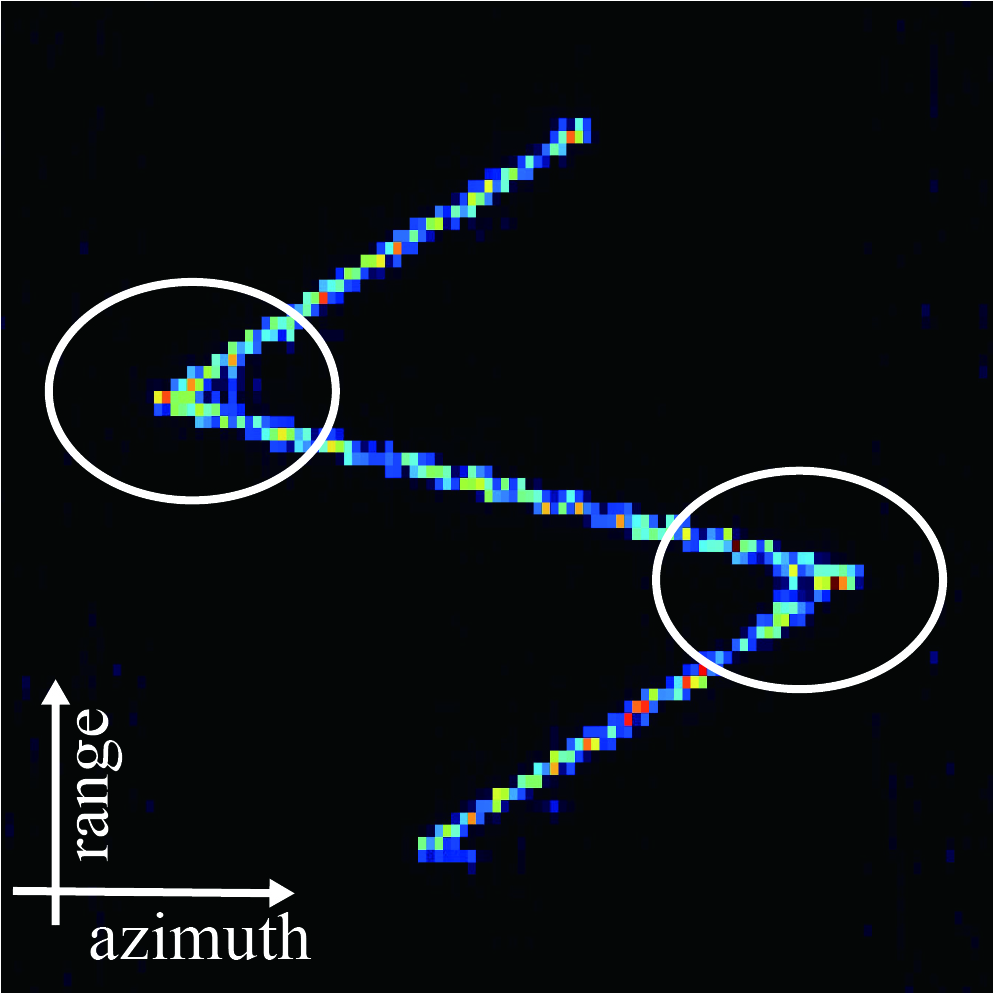}
        \caption{}
        \label{figure25_6}
    \end{subfigure}
    \begin{subfigure}[b]{0.24\linewidth}
        \caption*{{\fontsize{8pt}{10pt}\selectfont\centering tomo-LRENet-LSTM}}
        \includegraphics[width=\linewidth]{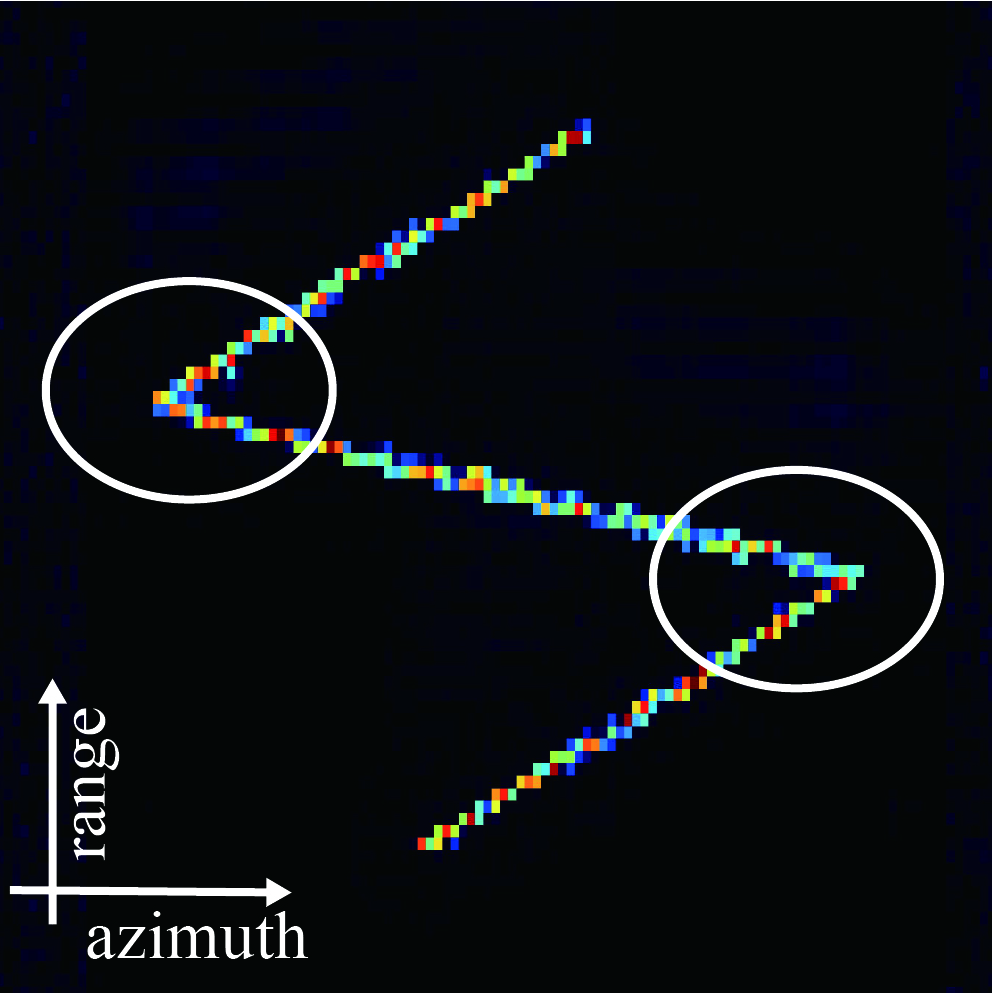}
        \caption{}
        \label{figure25_7}
    \end{subfigure}
    \begin{subfigure}[b]{0.24\linewidth}
        \caption*{{\fontsize{8pt}{10pt}\selectfont\centering ground truth}}
        \includegraphics[width=\linewidth]{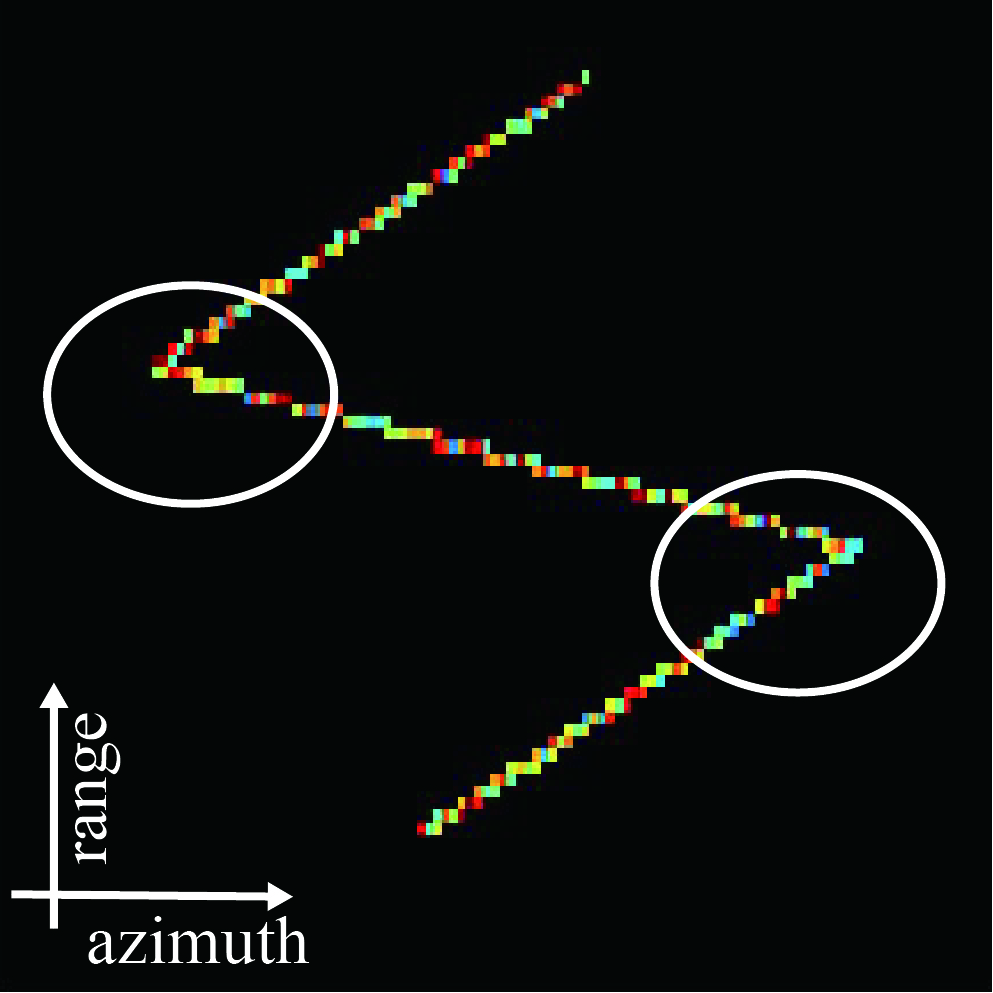}
        \caption{}
        \label{figure25_8}
    \end{subfigure}

    \caption{Results of test2. (a) Result of FISTA-based method. (b) Result of SLIMMER. (c) Result of tomo-IRENet-Raw. (d) Result of tomo-IRENet-TV. (e) Result of tomo-LRENet-U. (f) Result of tomo-LRENet-biU. (g) Result of tomo-LRENet-LSTM. (h) Ground truth.} 
    \label{figure_25}
\end{figure}

In this test, the testing dimension is expanded from being fiber-wise to slice-wise. The structure, a result of the combination of floor, wall facade, and rooftop, resembles a one-step shape in the slant-range/azimuth/elevation coordinate system. This test allows us to observe the condition of spatial structures beyond the resolution of seven methods. The results are illustrated in Fig.~\ref{figure_25}.

For this test object, the challenges of maintaining the spatial structures are evident in two aspects: the three continuous lines and two corners. Both traditional methods and tomo-IRENet-Raw, which only consider the sparse feature, suffer from structural distortions. The lines jitter with broken segments, and there are more outliers spread throughout the results compared to other methods.

Methods that take spatial features into account show less jittering of lines, resulting in a smoother output. Among these, tomo-IRENet-TV performs the worst in terms of resolution and the level of sidelobes. The two methods that use the U-Net module for spatial feature modeling perform at the same level, and their residual outliers are fewer than the others.

The tomo-LRENet-LSTM method performs the best. It produces a clearer corner structure compared to tomo-IRENet-U and tomo-LRENet-biU, both of which have more indistinct corners. Additionally, the line structure is slightly slimmer, and it has the fewest outliers.

An interesting observation emerges when comparing the results of the first test object in Fig.~\ref{figure_24} with this test object's results. The current test, featuring similar corner structures where two scatterers' distances gradually decrease, demonstrates better resolving ability. This improvement may stem from the previous test object being a one-dimensional fiber, where the reconstruction network processed two scatterers with a fixed distance—offering limited spatial features to harness—one at a time, with the corner-structure pattern obtained by assembling different results. In contrast, this test allows the reconstruction network to directly process scatterers at various distances. Compared to the previous situation, the current object incorporates more spatial features for the neural network to leverage, thus achieving a greater boost in resolving ability.

\vspace{0.5em}
\textit{D.3 Results of test object 3 (advanced multi-step structure reconstruction test)}
\vspace{0.5em}

\begin{figure}[h]
    \centering
    % 第一排的四张图
    \begin{subfigure}[b]{0.24\linewidth}
        \caption*{{\fontsize{8pt}{10pt}\selectfont\centering FISTA-based}}
        \includegraphics[width=\linewidth]{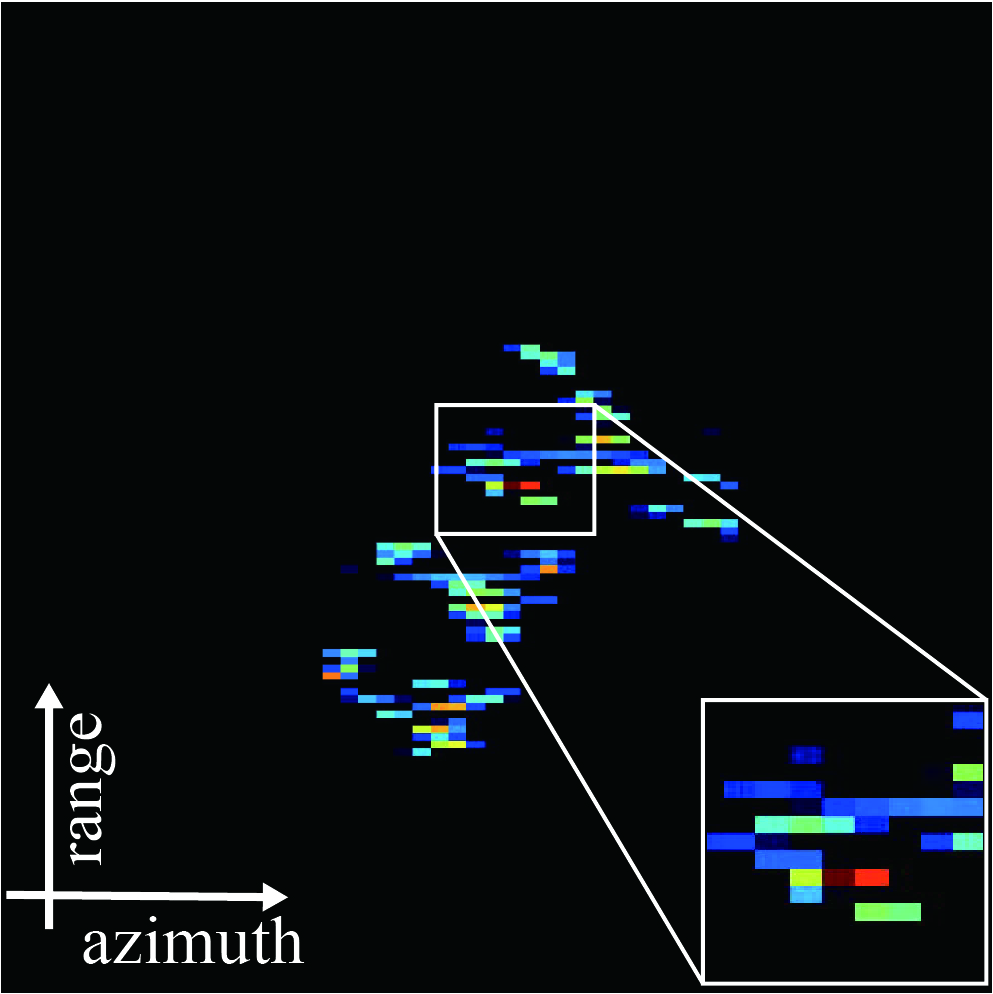}
        \caption{}
        \label{figure26_1}
    \end{subfigure}
    \begin{subfigure}[b]{0.24\linewidth}
        \caption*{{\fontsize{8pt}{10pt}\selectfont\centering SLIMMER}}
        \includegraphics[width=\linewidth]{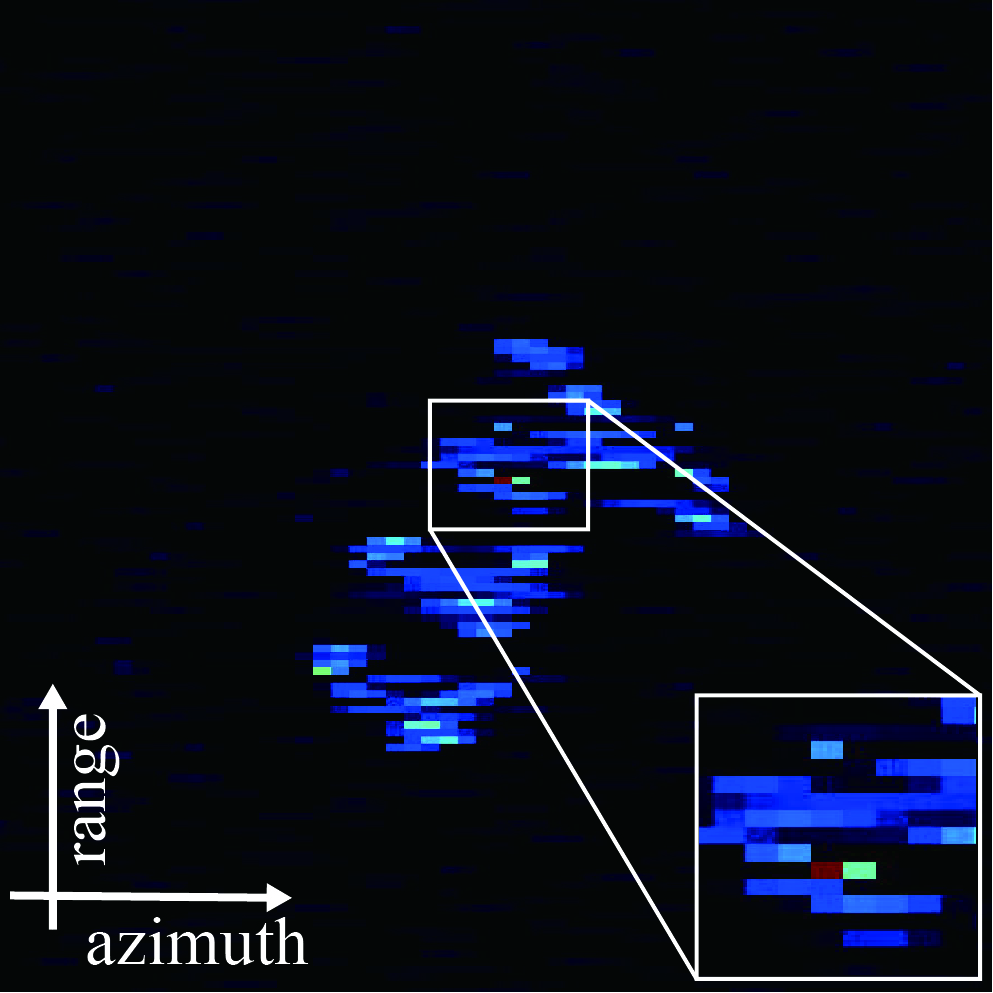}
        \caption{}
        \label{figure26_2}
    \end{subfigure}
    \begin{subfigure}[b]{0.24\linewidth}
        \caption*{{\fontsize{8pt}{10pt}\selectfont\centering tomo-IRENet-Raw}}
        \includegraphics[width=\linewidth]{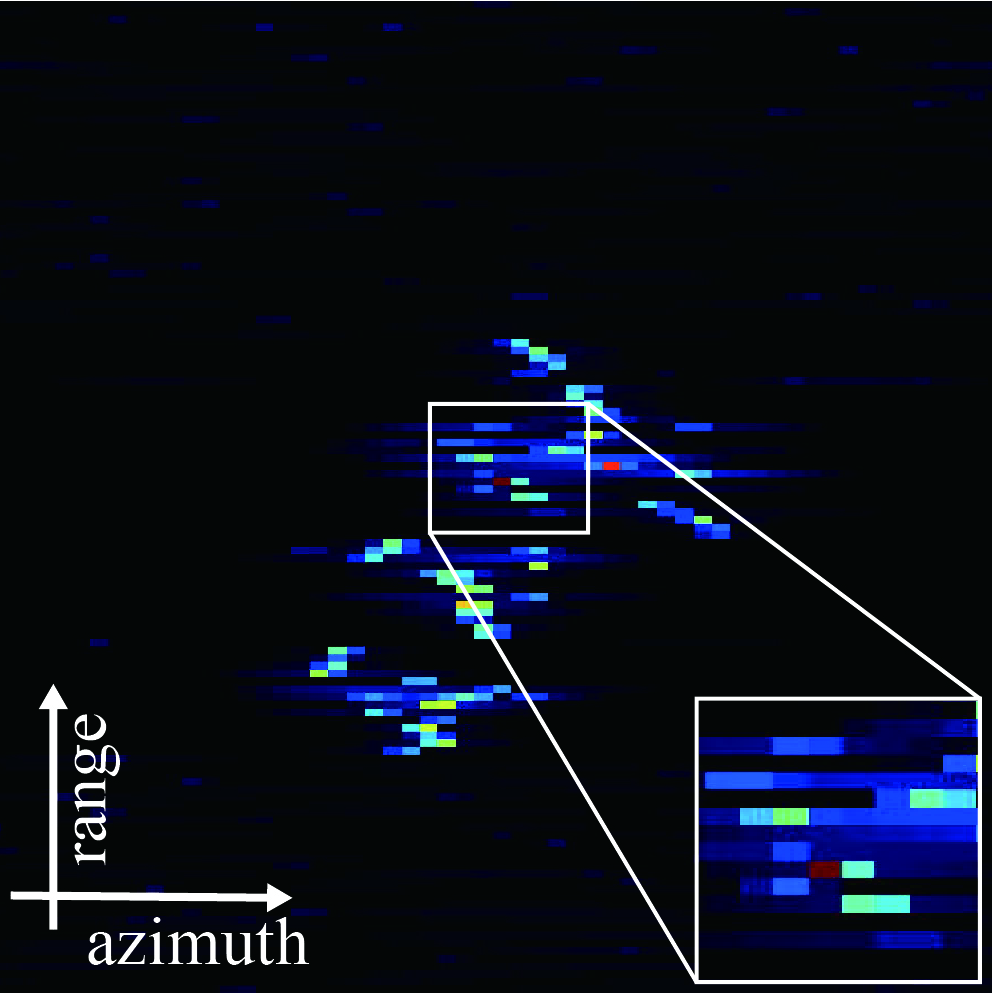}
        \caption{}
        \label{figure26_3}
    \end{subfigure}
    \begin{subfigure}[b]{0.24\linewidth}
        \caption*{{\fontsize{8pt}{10pt}\selectfont\centering tomo-IRENet-TV}}
        \includegraphics[width=\linewidth]{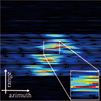}
        \caption{}
        \label{figure26_4}
    \end{subfigure}
    
    % 第二排的四张图
    \begin{subfigure}[b]{0.24\linewidth}
        \caption*{{\fontsize{8pt}{10pt}\selectfont\centering tomo-IRENet-U}}
        \includegraphics[width=\linewidth]{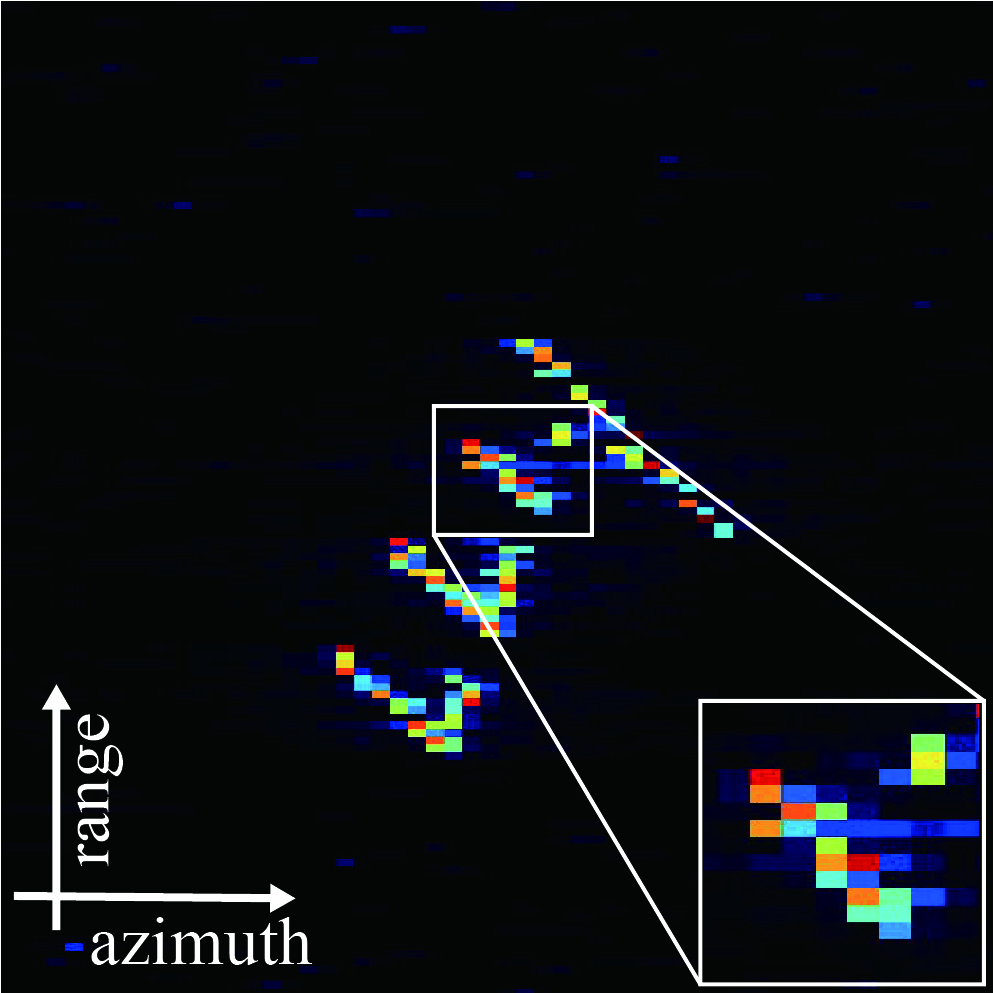}
        \caption{}
        \label{figure26_5}
    \end{subfigure}
    \begin{subfigure}[b]{0.24\linewidth}
        \caption*{{\fontsize{8pt}{10pt}\selectfont\centering tomo-LRENet-biU}}
        \includegraphics[width=\linewidth]{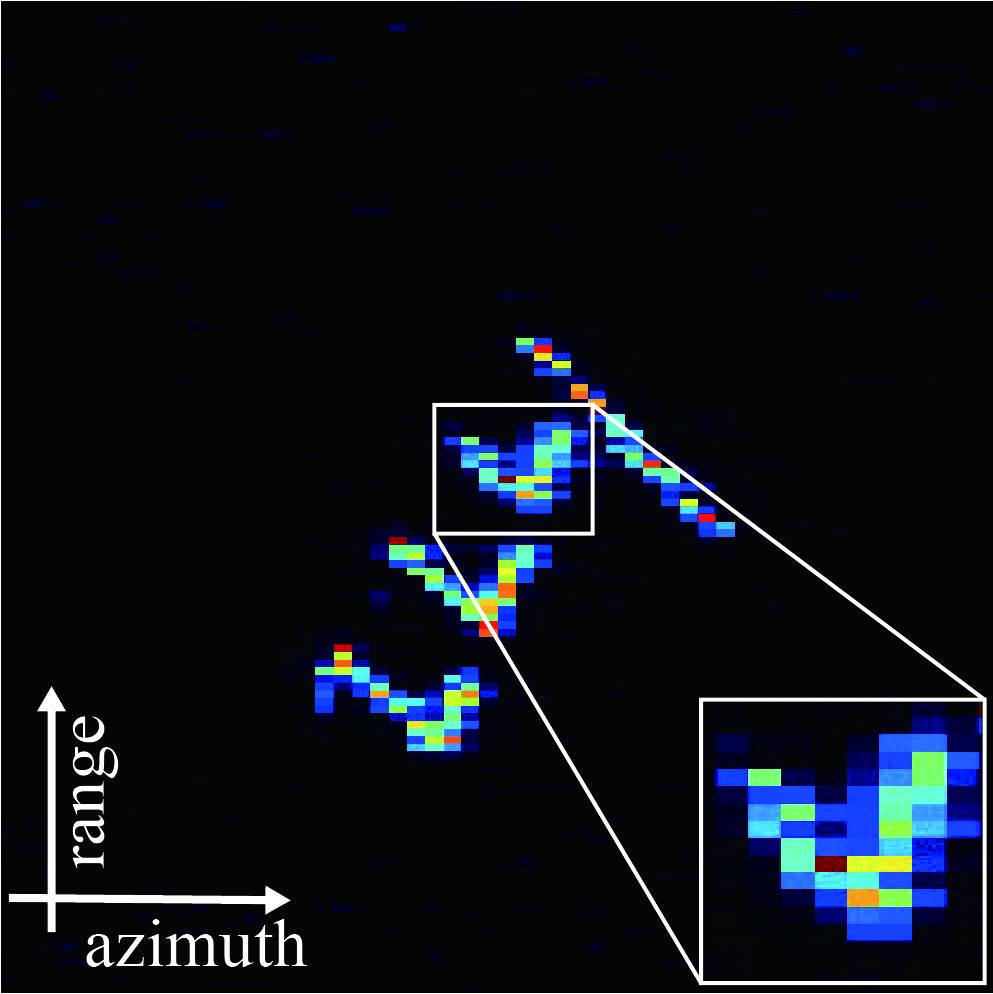}
        \caption{}
        \label{figure26_6}
    \end{subfigure}
    \begin{subfigure}[b]{0.24\linewidth}
        \caption*{{\fontsize{8pt}{10pt}\selectfont\centering tomo-LRENet-LSTM}}
        \includegraphics[width=\linewidth]{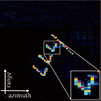}
        \caption{}
        \label{figure26_7}
    \end{subfigure}
    \begin{subfigure}[b]{0.24\linewidth}
        \caption*{{\fontsize{8pt}{10pt}\selectfont\centering ground truth}}
        \includegraphics[width=\linewidth]{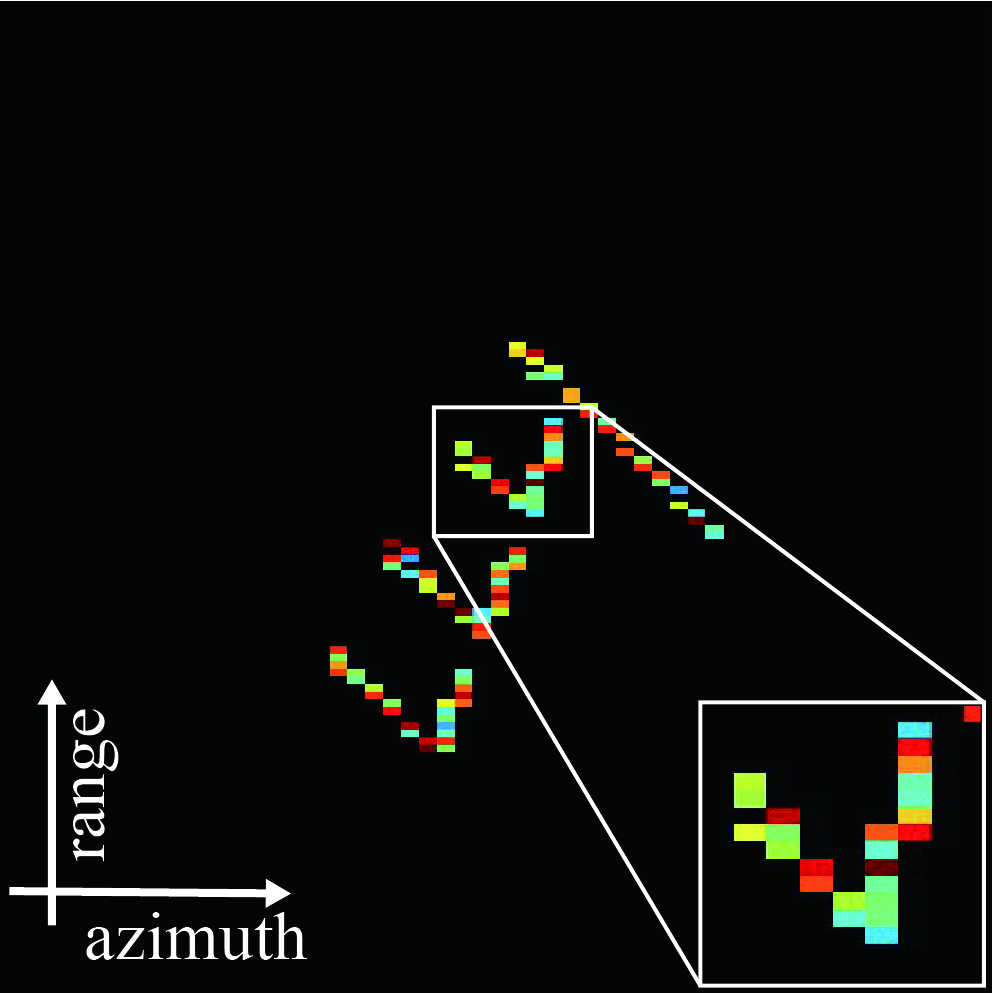}
        \caption{}
        \label{figure26_8}
    \end{subfigure}

    \caption{Results of test3. (a) Result of FISTA-based method. (b) Result of SLIMMER. (c) Result of tomo-IRENet-Raw. (d) Result of tomo-IRENet-TV. (e) Result of tomo-LRENet-U. (f) Result of tomo-LRENet-biU. (g) Result of tomo-LRENet-LSTM. (h) Ground truth.}
    \label{figure_26}
\end{figure}

In this test, we progress from the simple one-step structure reconstruction to a more advanced multi-step scenario. This is particularly relevant for situations involving numerous dense, small buildings. The results are illustrated in Fig.~\ref{figure_26}. 

Generally, the results of this test follow the patterns observed in the previous one-step test. Methods that only consider sparse features fail to reconstruct all the corners. The four methods that also consider spatial features manage to reconstruct them, with the exception of tomo-IRENet-TV due to its lower resolution. Among the remaining three methods, there apparent performance disparities when compared to the prior test that achieved a comparable level of simplicity. Specifically focusing on the top corner, which is distinctly marked with red rectangles, it is clear that tomo-LRENet-LSTM outperforms the rest, followed in close succession by tomo-LRENet-biU, and then tomo-IRENet-U. On a related note, the line smoothness observed in the final reconstructions also mirrors this trend.

\vspace{0.5em}
\textit{D.4 Results of test object 4 (classical architectural forms test)}
\vspace{0.5em}

\begin{figure}[h]
    \centering

    % 第一行的八张图
    \begin{subfigure}[b]{0.11\linewidth}
        \caption*{{\fontsize{8pt}{10pt}\selectfont\centering FISTA-Based
        }}
        \includegraphics[width=\linewidth]{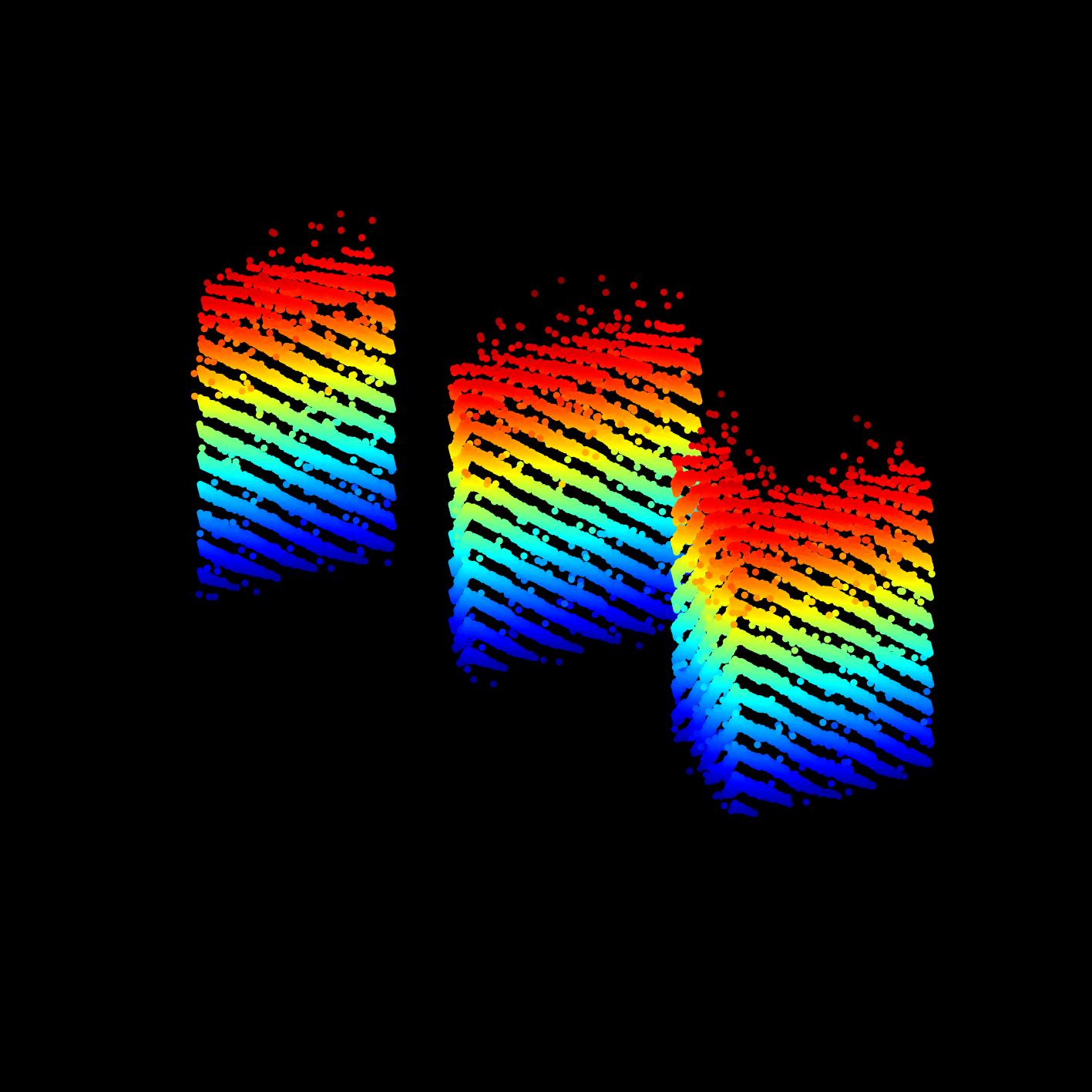}
        \caption*{{\fontsize{8pt}{10pt}\selectfont\centering (a-1)}}
        \label{figure27_a_1}
    \end{subfigure}
    \begin{subfigure}[b]{0.11\linewidth}
        \caption*{{\fontsize{8pt}{10pt}\selectfont\centering SLIMMER}}
        \includegraphics[width=\linewidth]{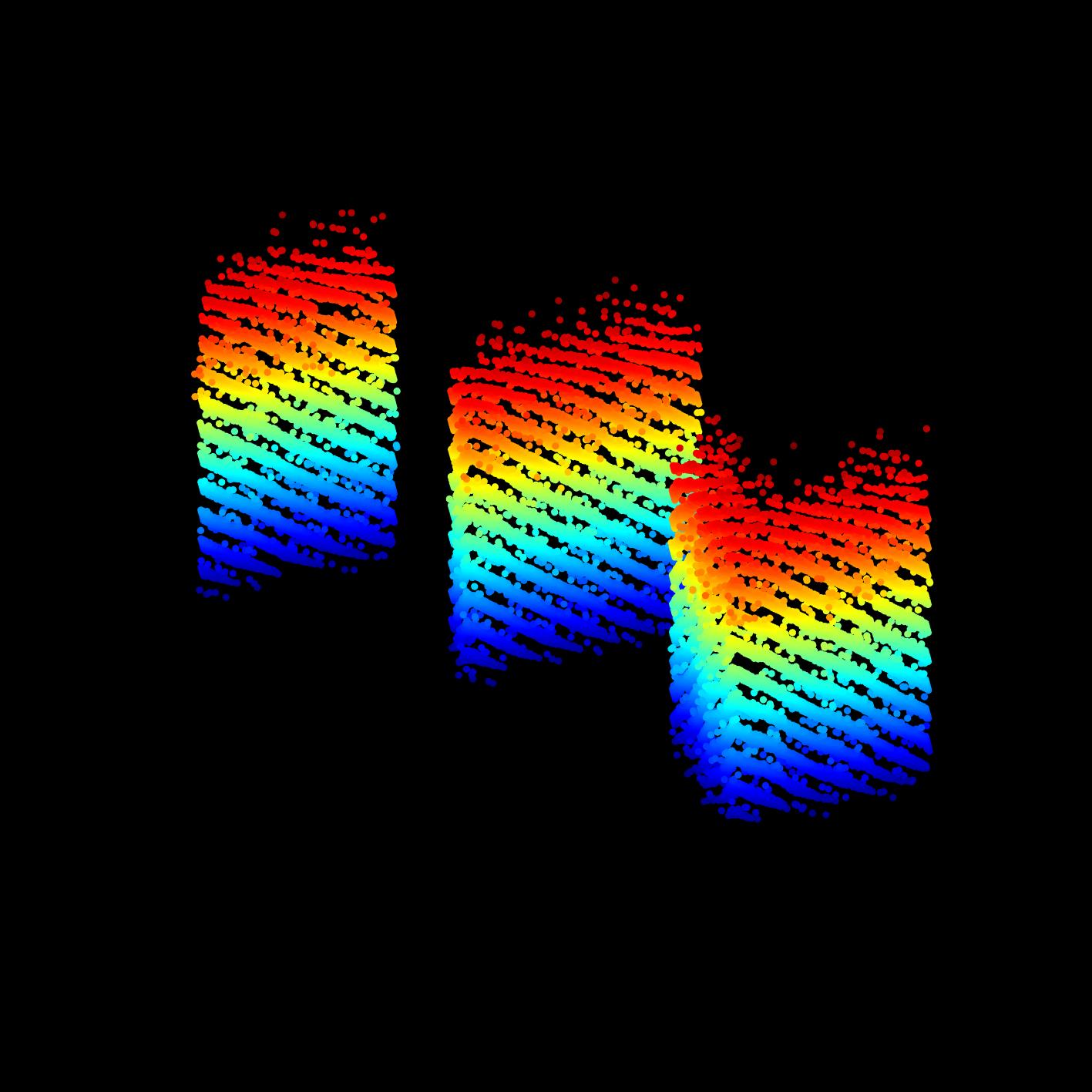}
        \caption*{{\fontsize{8pt}{10pt}\selectfont\centering (a-2)}}
        \label{figure27_a_2}
    \end{subfigure}
    \begin{subfigure}[b]{0.11\linewidth}
        \begin{tabular}{c}
        {\fontsize{8pt}{10pt}\selectfont tomo-} \\
        {\fontsize{8pt}{10pt}\selectfont IRENet-Raw}
        \end{tabular}
        \includegraphics[width=\linewidth]{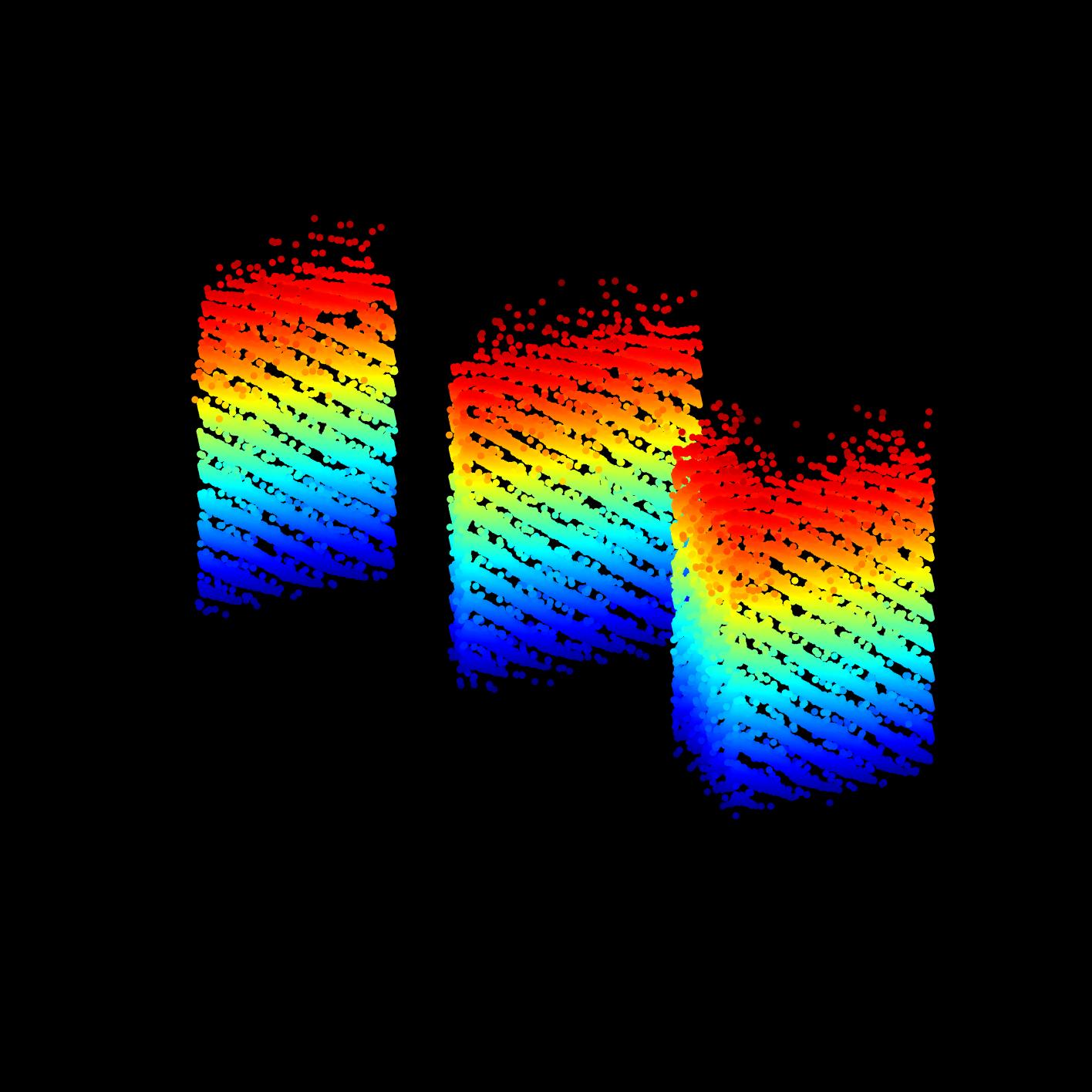}
        \caption*{{\fontsize{8pt}{10pt}\selectfont\centering (a-3)}}
        \label{figure27_a_3}
    \end{subfigure}
    \begin{subfigure}[b]{0.11\linewidth}
        \begin{tabular}{c}
        {\fontsize{8pt}{10pt}\selectfont tomo-} \\
        {\fontsize{8pt}{10pt}\selectfont IRENet-TV}
        \end{tabular}
        \includegraphics[width=\linewidth]{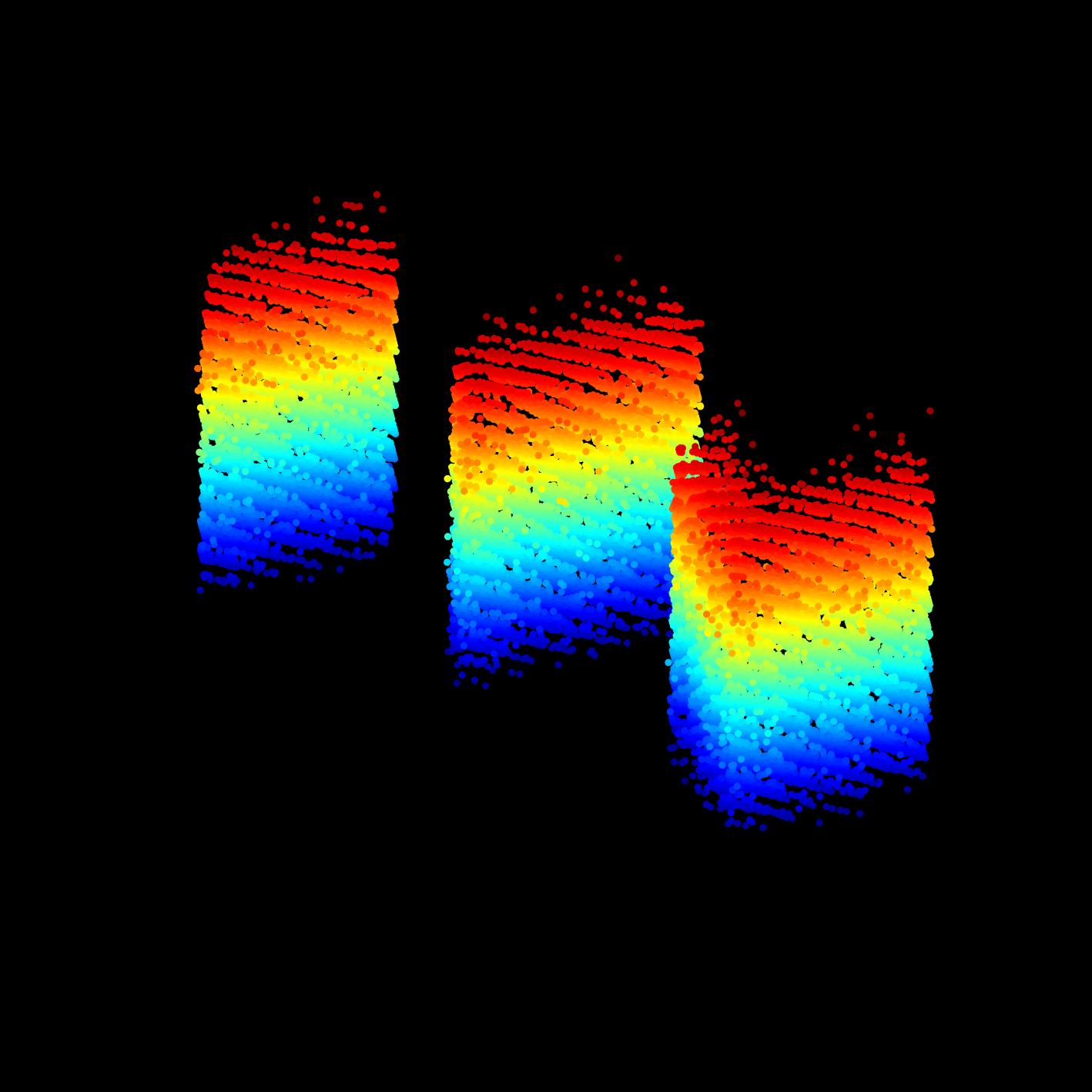}
        \caption*{{\fontsize{8pt}{10pt}\selectfont\centering (a-4)}}
        \label{figure27_a_4}
    \end{subfigure}
    \begin{subfigure}[b]{0.11\linewidth}
        \begin{tabular}{c}
        {\fontsize{8pt}{10pt}\selectfont tomo-} \\
        {\fontsize{8pt}{10pt}\selectfont IRENet-U}
        \end{tabular}
        \includegraphics[width=\linewidth]{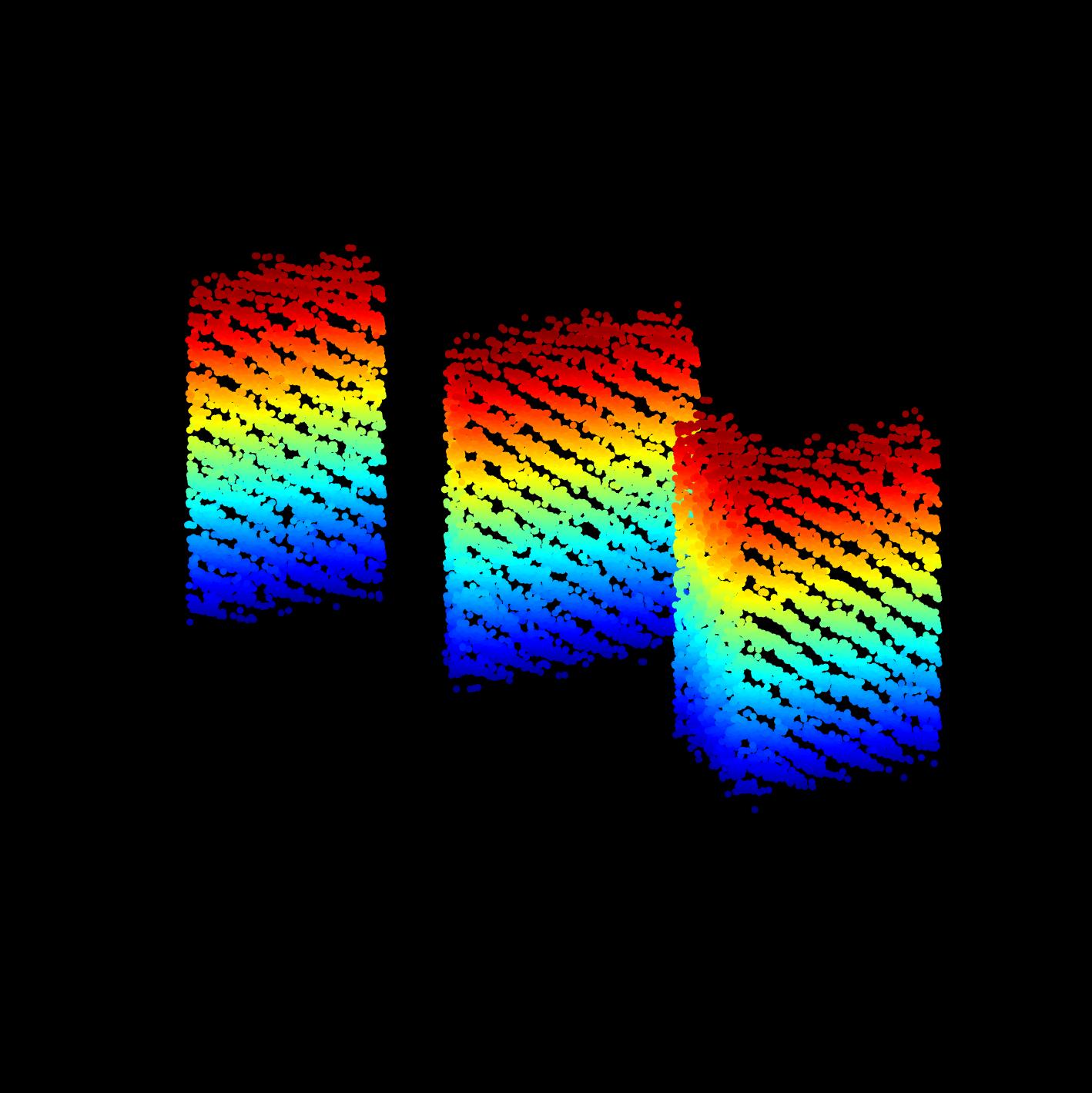}
        \caption*{{\fontsize{8pt}{10pt}\selectfont\centering (a-5)}}
        \label{figure27_a_5}
    \end{subfigure}
    \begin{subfigure}[b]{0.11\linewidth}
        \begin{tabular}{c}
        {\fontsize{8pt}{10pt}\selectfont tomo-} \\
        {\fontsize{8pt}{10pt}\selectfont LRENet-biU}
        \end{tabular}
        \includegraphics[width=\linewidth]{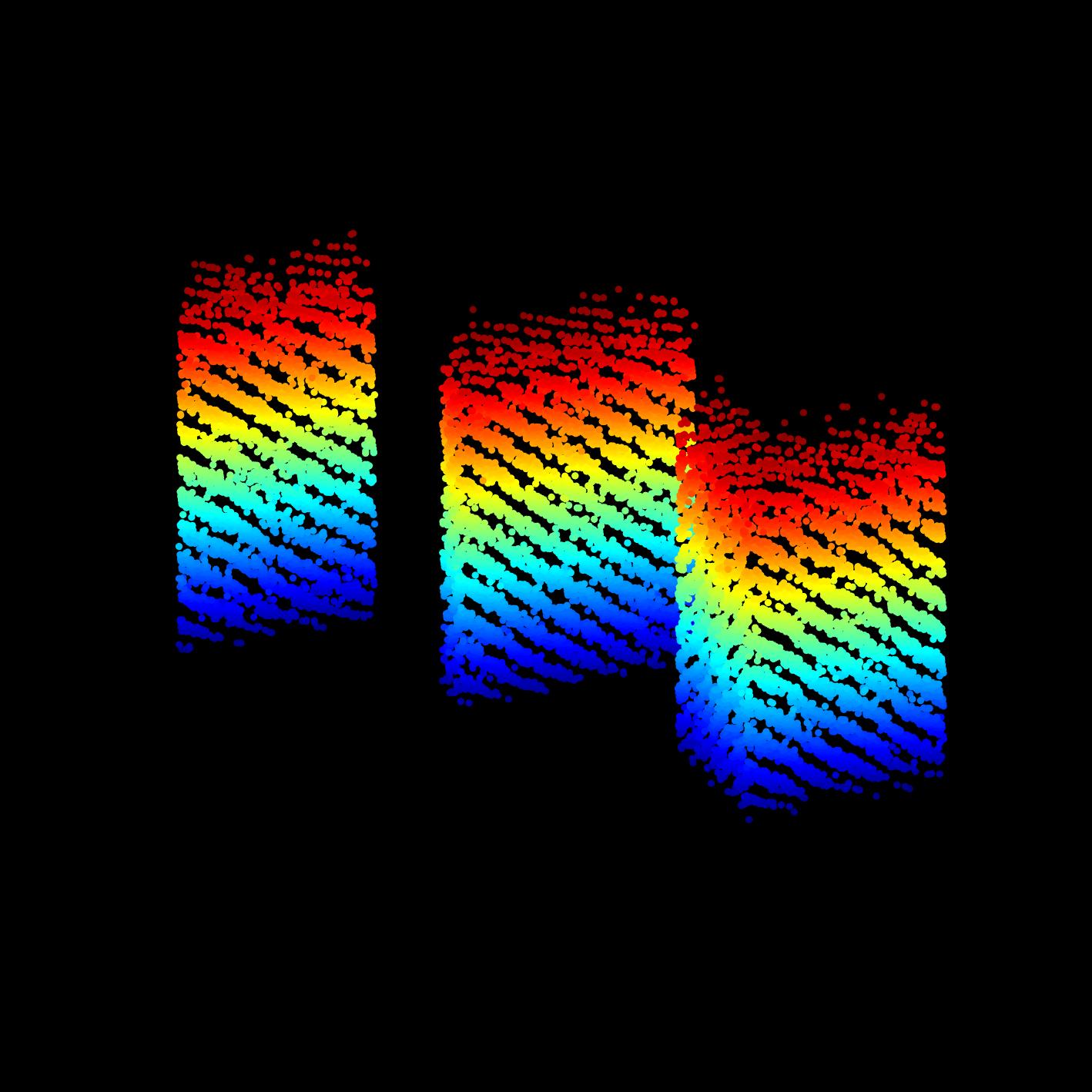}
        \caption*{{\fontsize{8pt}{10pt}\selectfont\centering (a-6)}}
        \label{figure27_a_6}
    \end{subfigure}
    \begin{subfigure}[b]{0.11\linewidth}
        \begin{tabular}{c}
        {\fontsize{8pt}{10pt}\selectfont tomo-} \\
        {\fontsize{8pt}{10pt}\selectfont LRENet-LSTM}
        \end{tabular}
        \includegraphics[width=\linewidth]{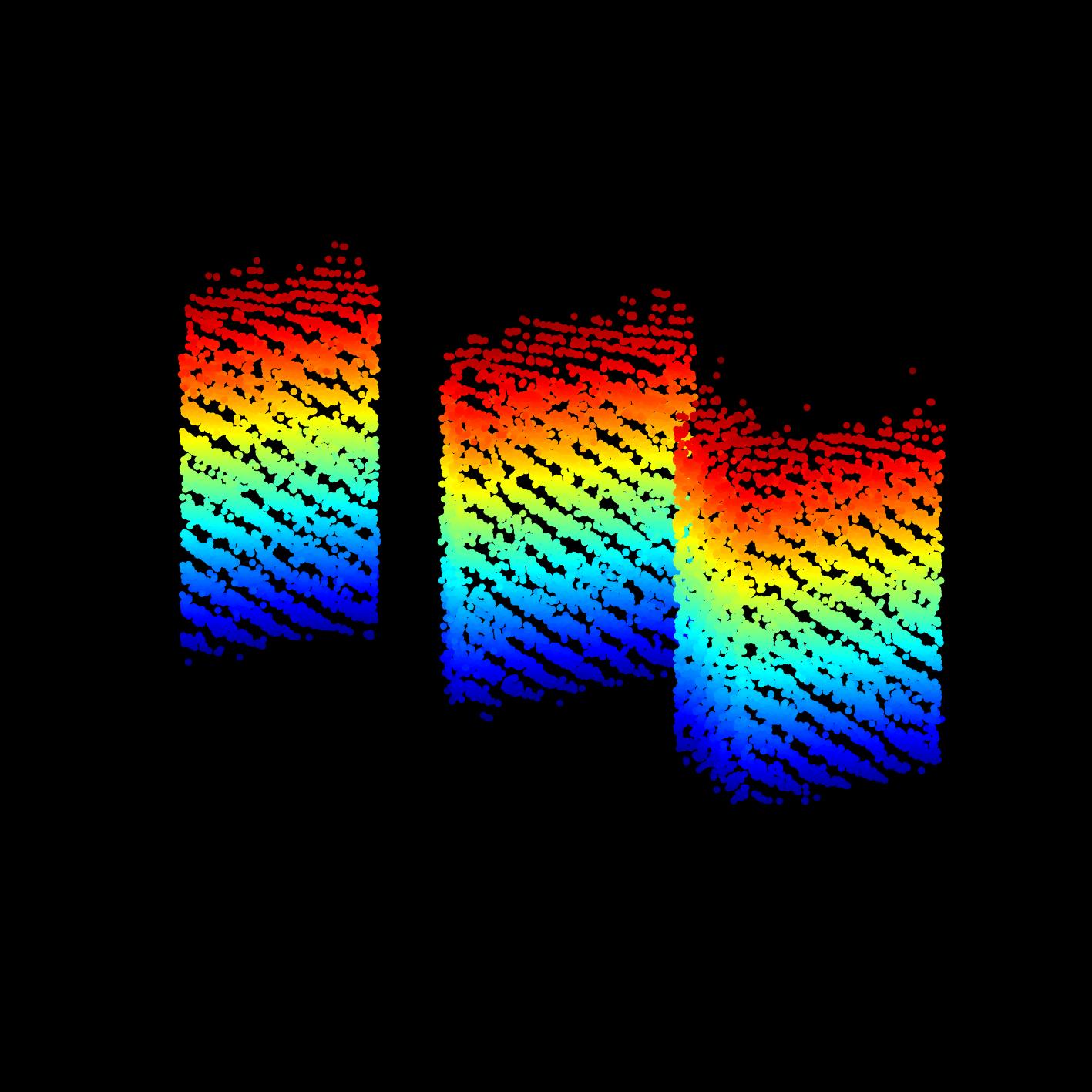}
        \caption*{{\fontsize{8pt}{10pt}\selectfont\centering (a-7)}}
        \label{figure27_a_7}
    \end{subfigure}
    \begin{subfigure}[b]{0.11\linewidth}
        \caption*{{\fontsize{8pt}{10pt}\selectfont\centering ground truth}}
        \includegraphics[width=\linewidth]{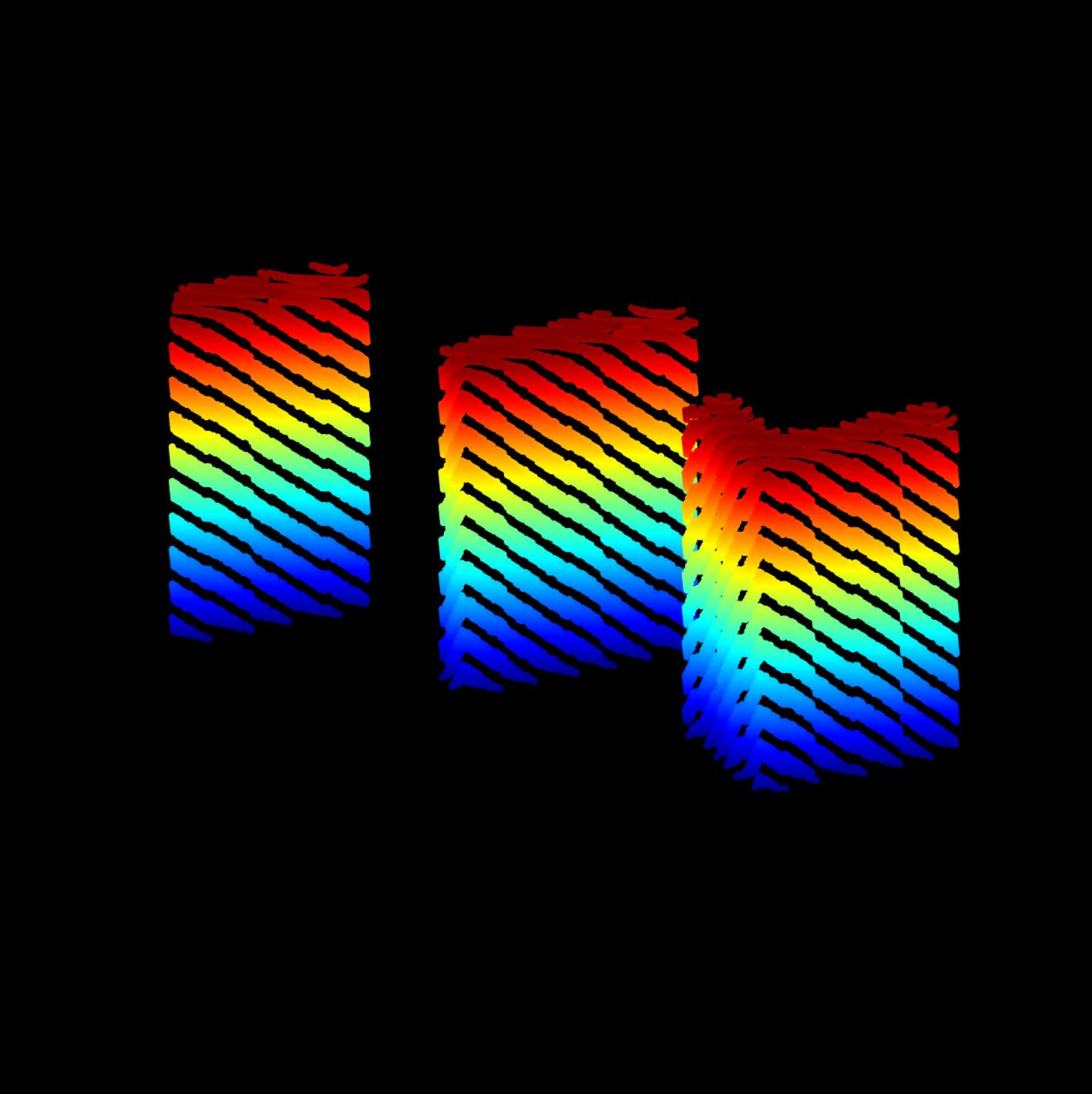}
        \caption*{{\fontsize{8pt}{10pt}\selectfont\centering (a-8)}}
        \label{figure27_a_8}
    \end{subfigure}

    % 第二行的八张图
    \begin{subfigure}[b]{0.11\linewidth}
        \includegraphics[width=\linewidth]{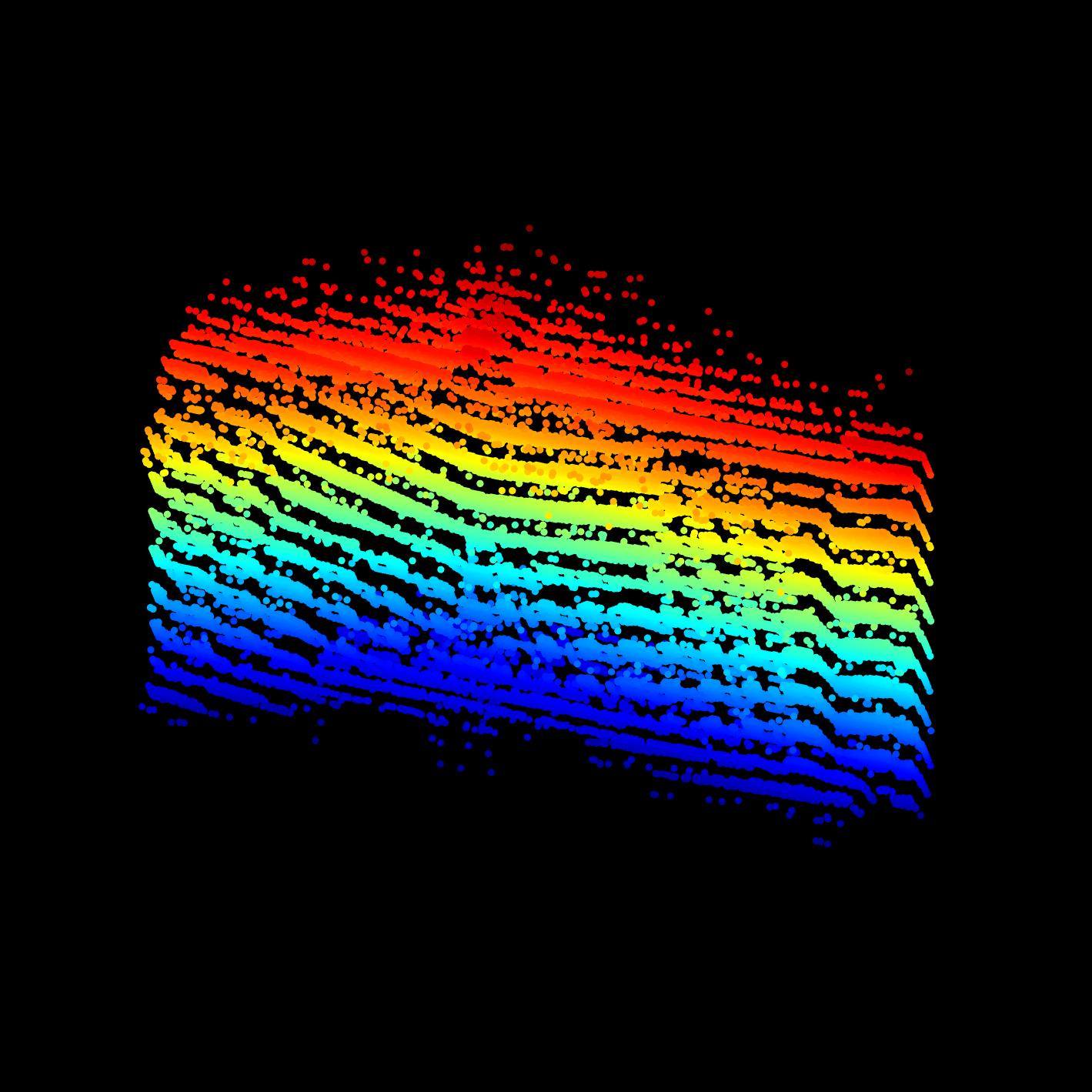}
        \caption*{{\fontsize{8pt}{10pt}\selectfont\centering (b-1)}}
        \label{figure27_b_1}
    \end{subfigure}
    \begin{subfigure}[b]{0.11\linewidth}
        \includegraphics[width=\linewidth]{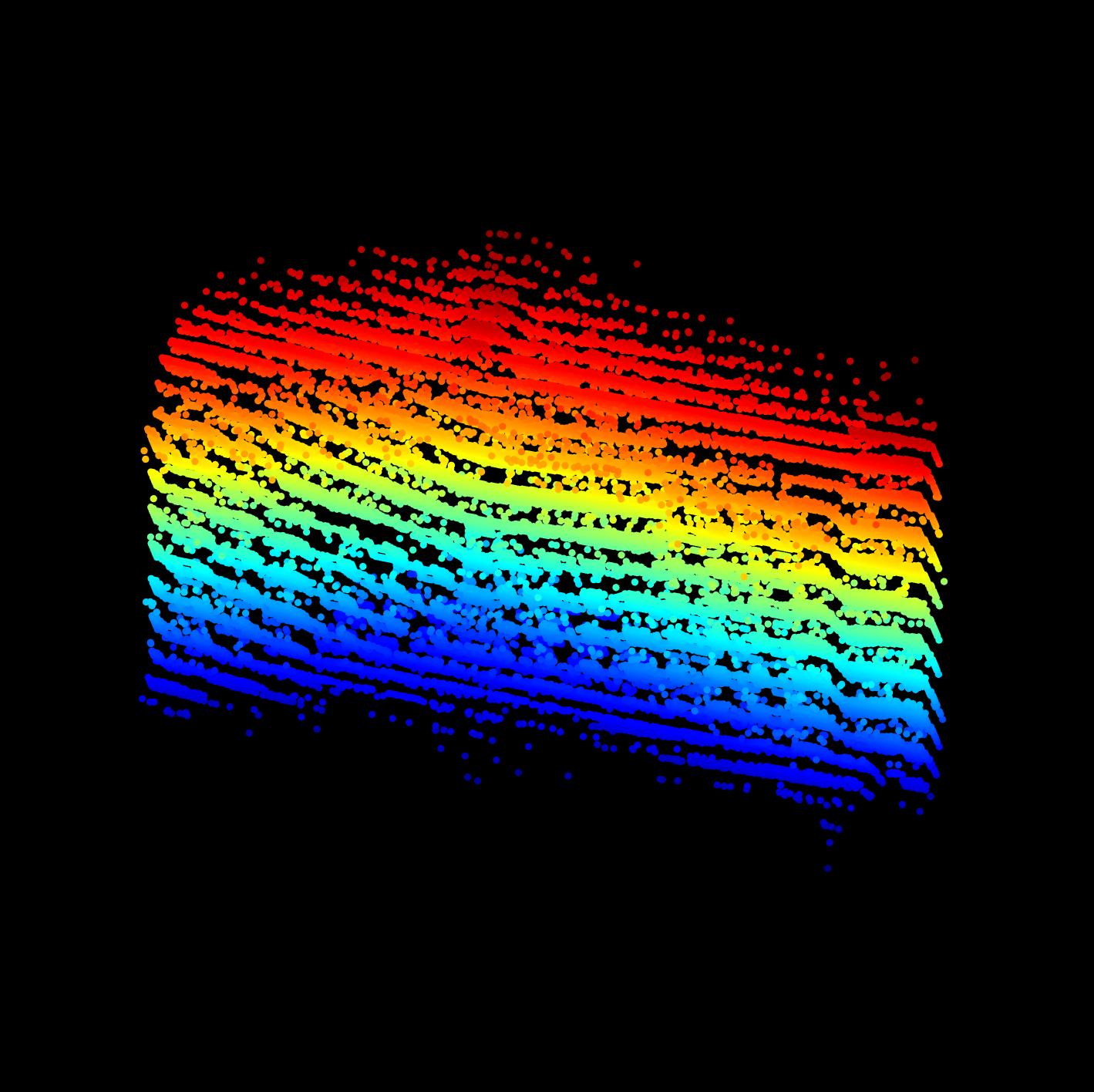}
        \caption*{{\fontsize{8pt}{10pt}\selectfont\centering (b-2)}}
        \label{figure28_b_2}
    \end{subfigure}
    \begin{subfigure}[b]{0.11\linewidth}
        \includegraphics[width=\linewidth]{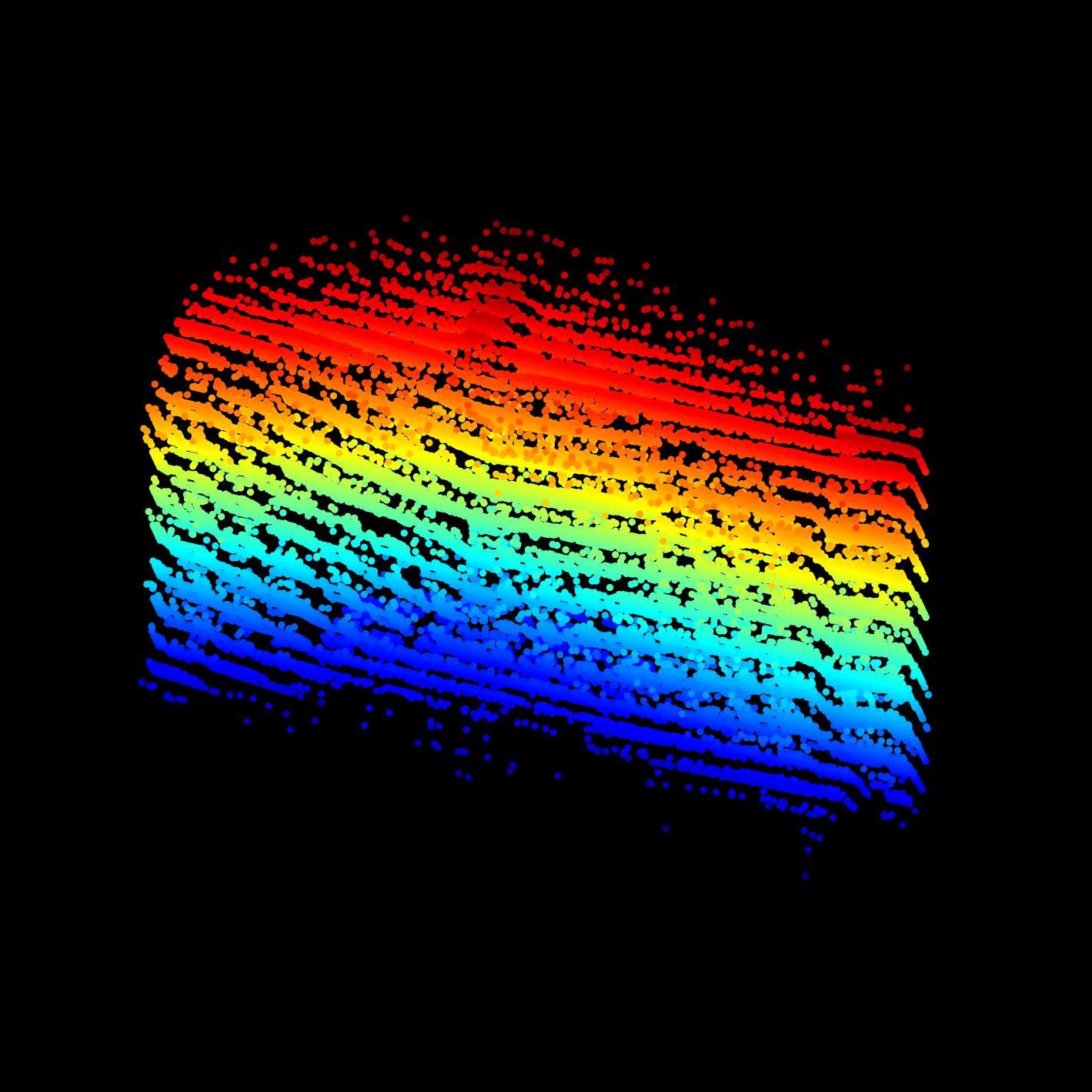}
        \caption*{{\fontsize{8pt}{10pt}\selectfont\centering (b-3)}}
        \label{figure27_b_3}
    \end{subfigure}
    \begin{subfigure}[b]{0.11\linewidth}
        \includegraphics[width=\linewidth]{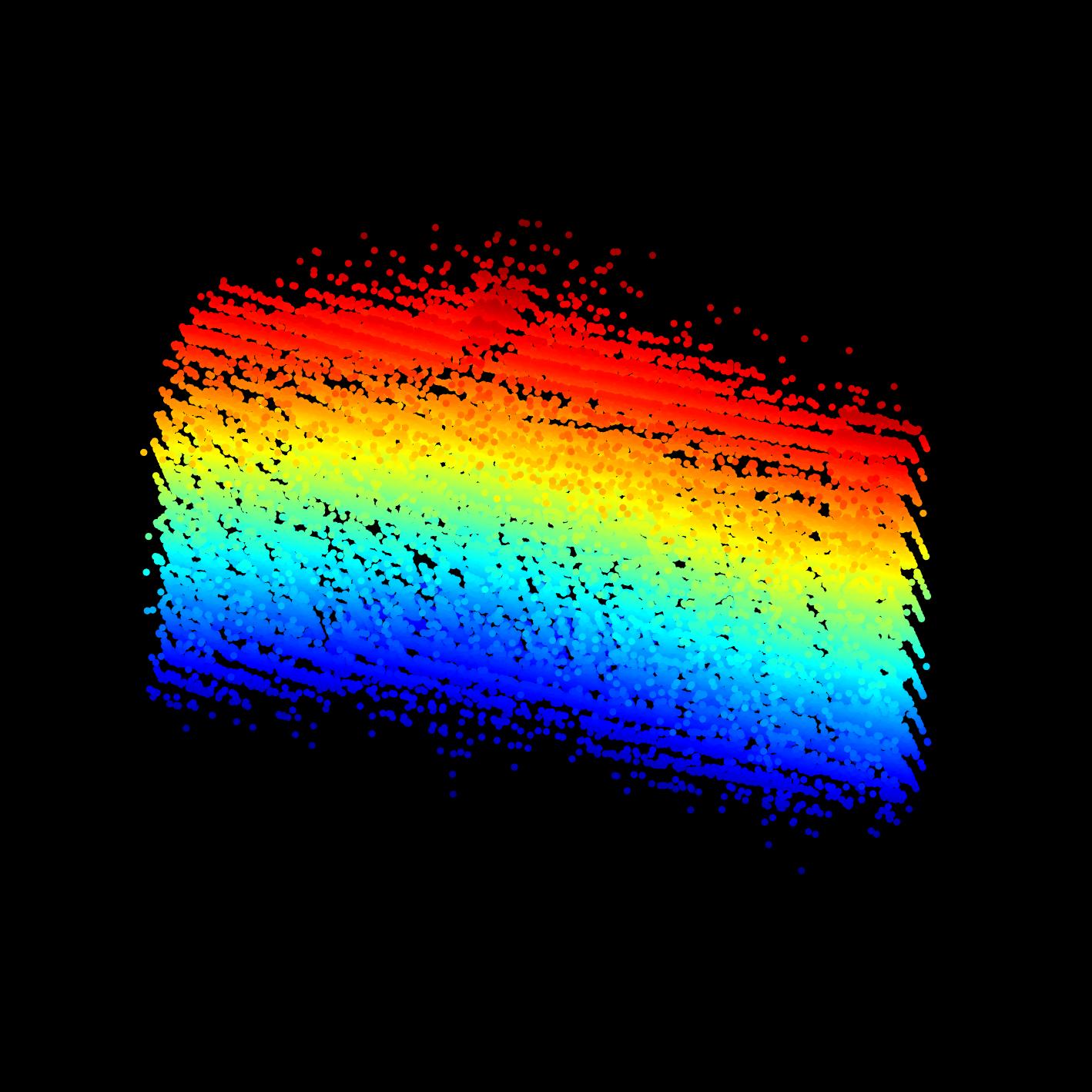}
        \caption*{{\fontsize{8pt}{10pt}\selectfont\centering (b-4)}}
        \label{figure27_b_4}
    \end{subfigure}
    \begin{subfigure}[b]{0.11\linewidth}
        \includegraphics[width=\linewidth]{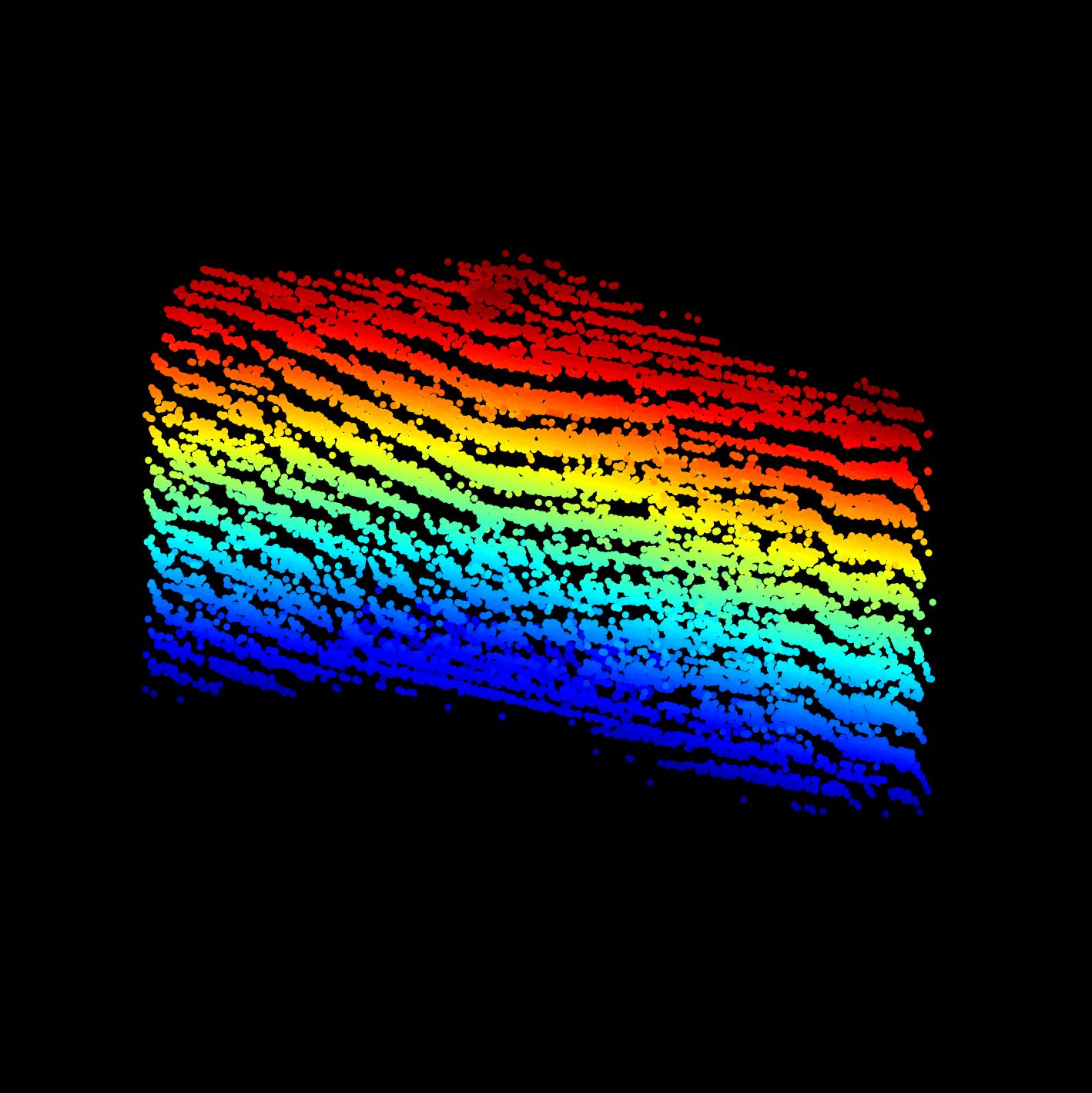}
        \caption*{{\fontsize{8pt}{10pt}\selectfont\centering (b-5)}}
        \label{figure27_b_5}
    \end{subfigure}
    \begin{subfigure}[b]{0.11\linewidth}
        \includegraphics[width=\linewidth]{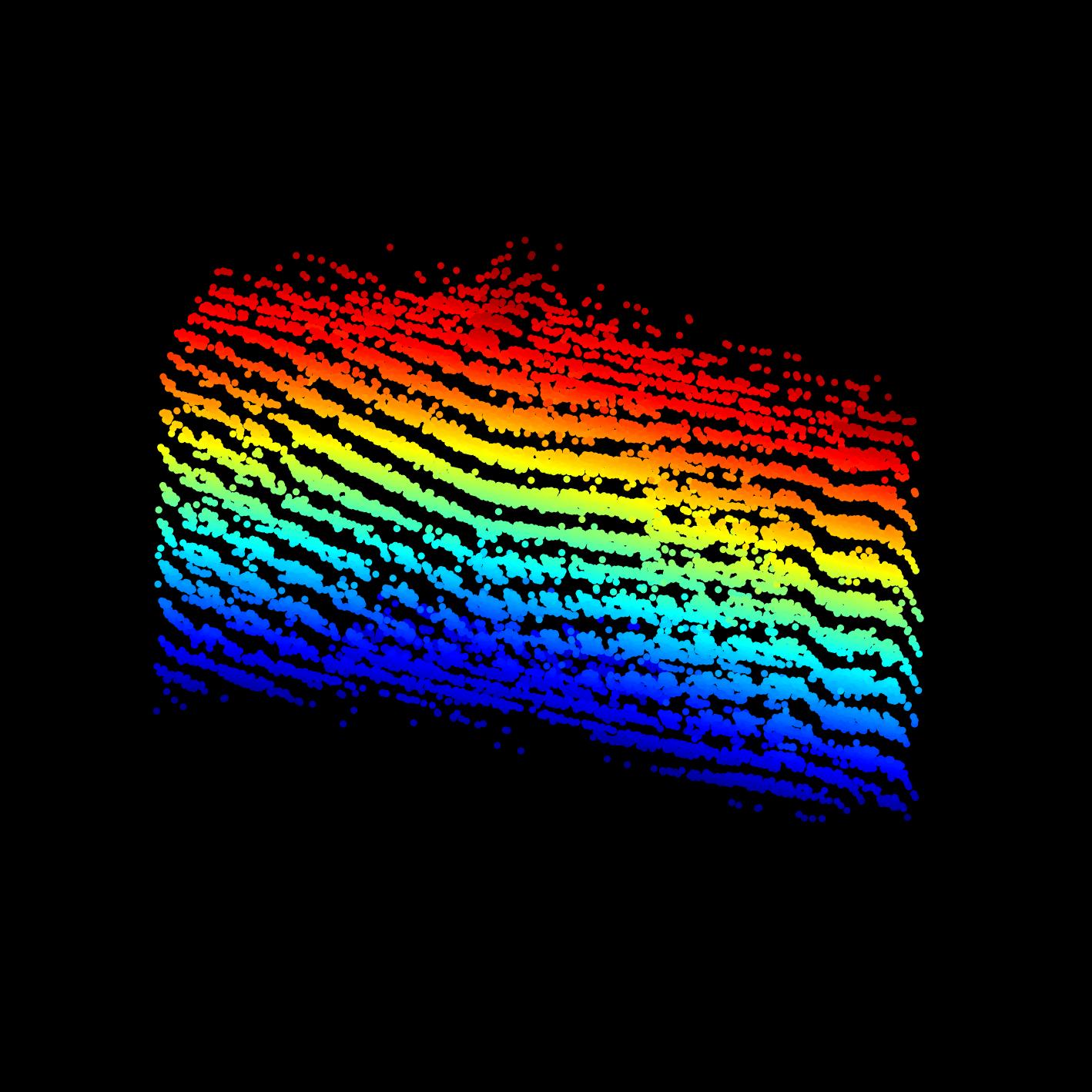}
        \caption*{{\fontsize{8pt}{10pt}\selectfont\centering (b-6)}}
        \label{figure27_b_6}
    \end{subfigure}
    \begin{subfigure}[b]{0.11\linewidth}
        \includegraphics[width=\linewidth]{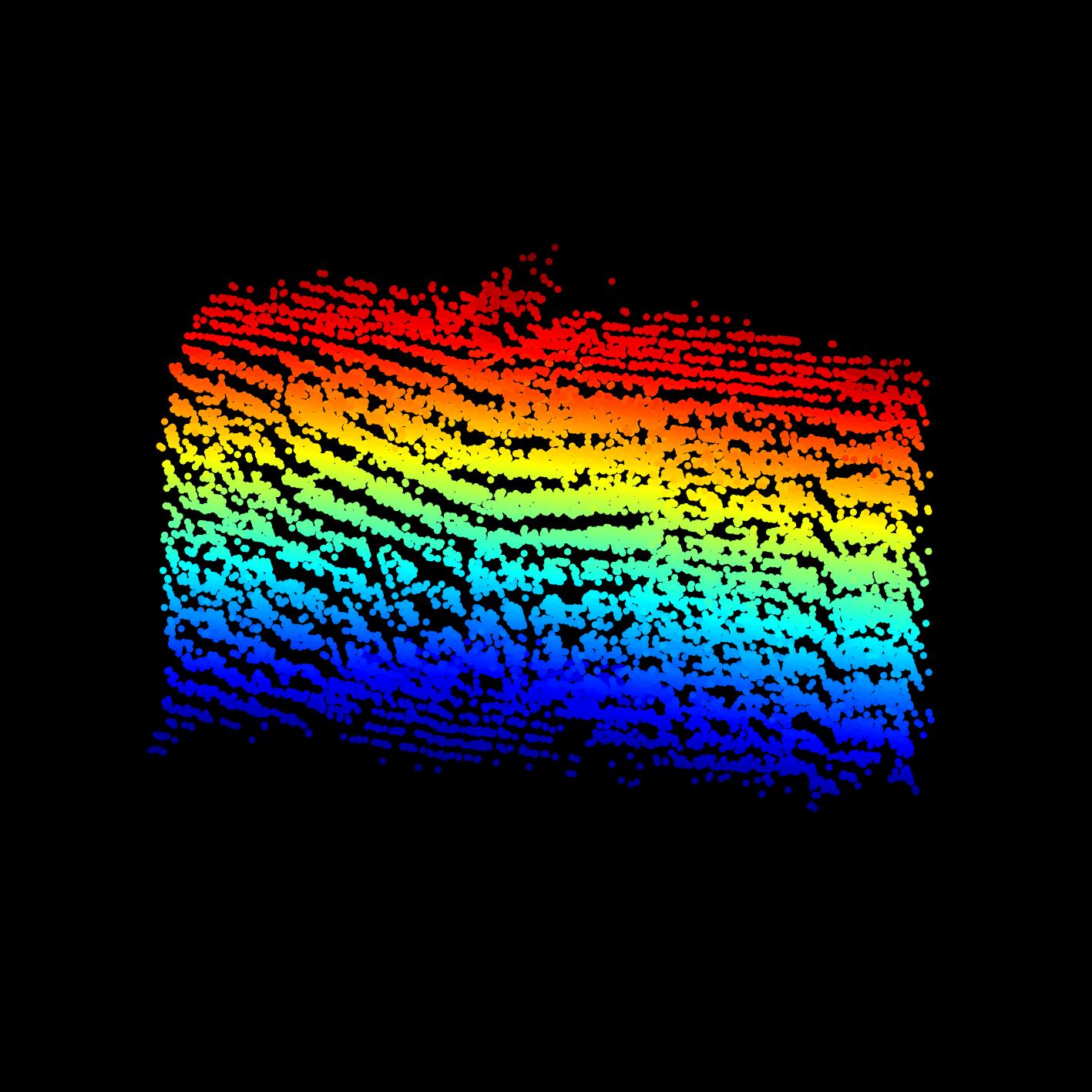}
        \caption*{{\fontsize{8pt}{10pt}\selectfont\centering (b-7)}}
        \label{figure27_b_7}
    \end{subfigure}
    \begin{subfigure}[b]{0.11\linewidth}
        \caption*{{\fontsize{8pt}{10pt}\selectfont\centering }}
        \includegraphics[width=\linewidth]{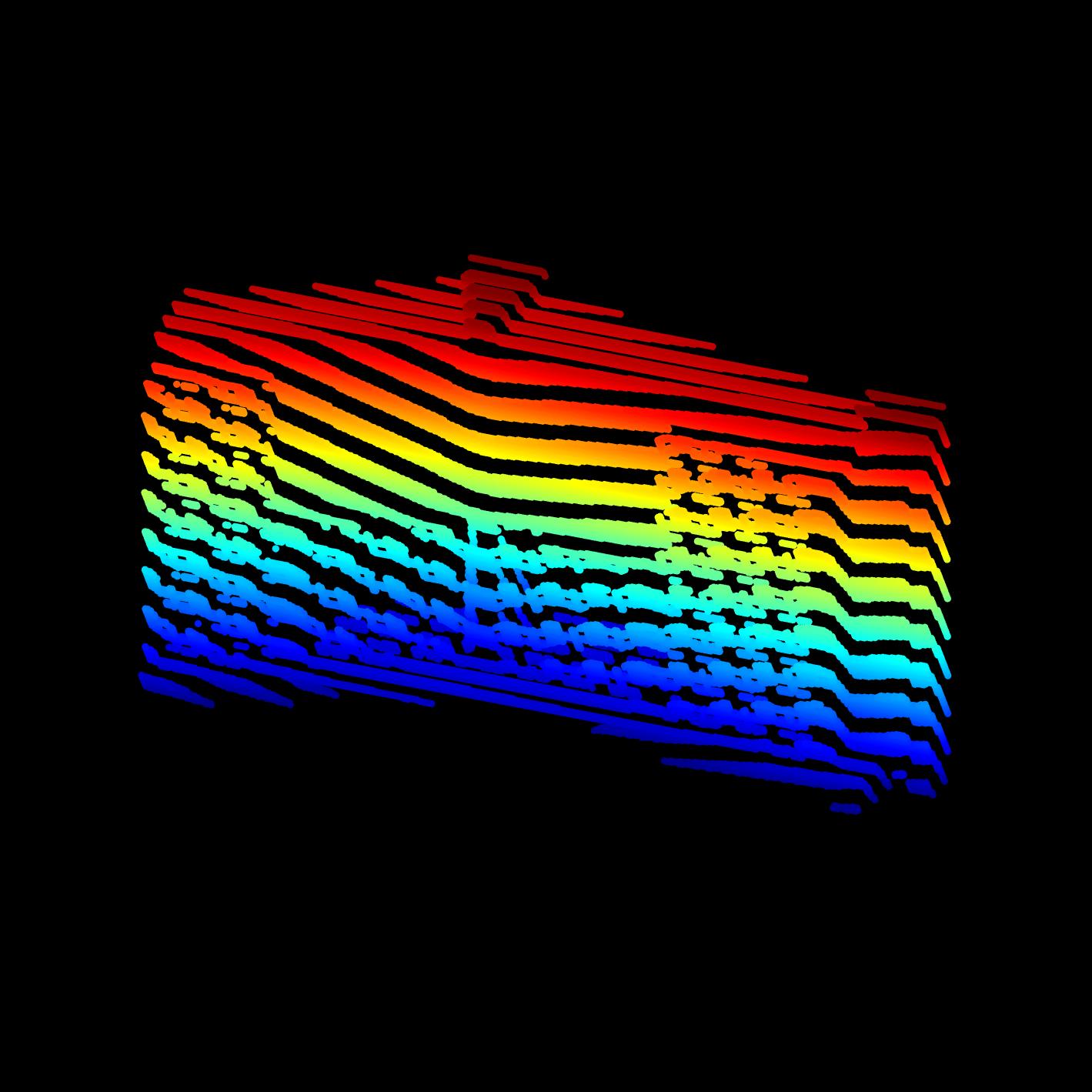}
        \caption*{{\fontsize{8pt}{10pt}\selectfont\centering (b-8)}}
        \label{figure27_b_8}
    \end{subfigure}

    % 第三行的八张图
    \begin{subfigure}[b]{0.11\linewidth}
        \includegraphics[width=\linewidth]{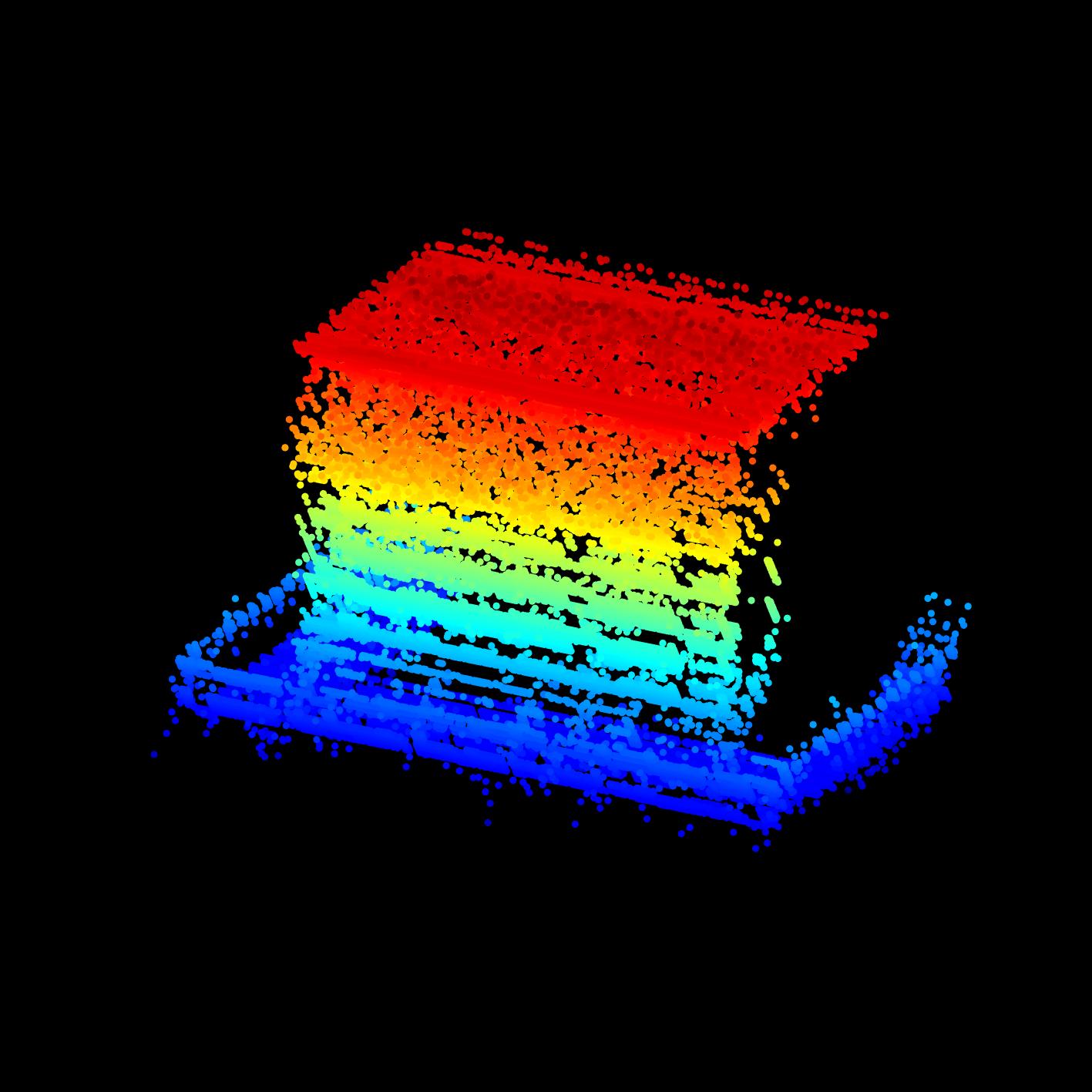}
        \caption*{{\fontsize{8pt}{10pt}\selectfont\centering (c-1)}}
        \label{figure27_c_1}
    \end{subfigure}
    \begin{subfigure}[b]{0.11\linewidth}
        \includegraphics[width=\linewidth]{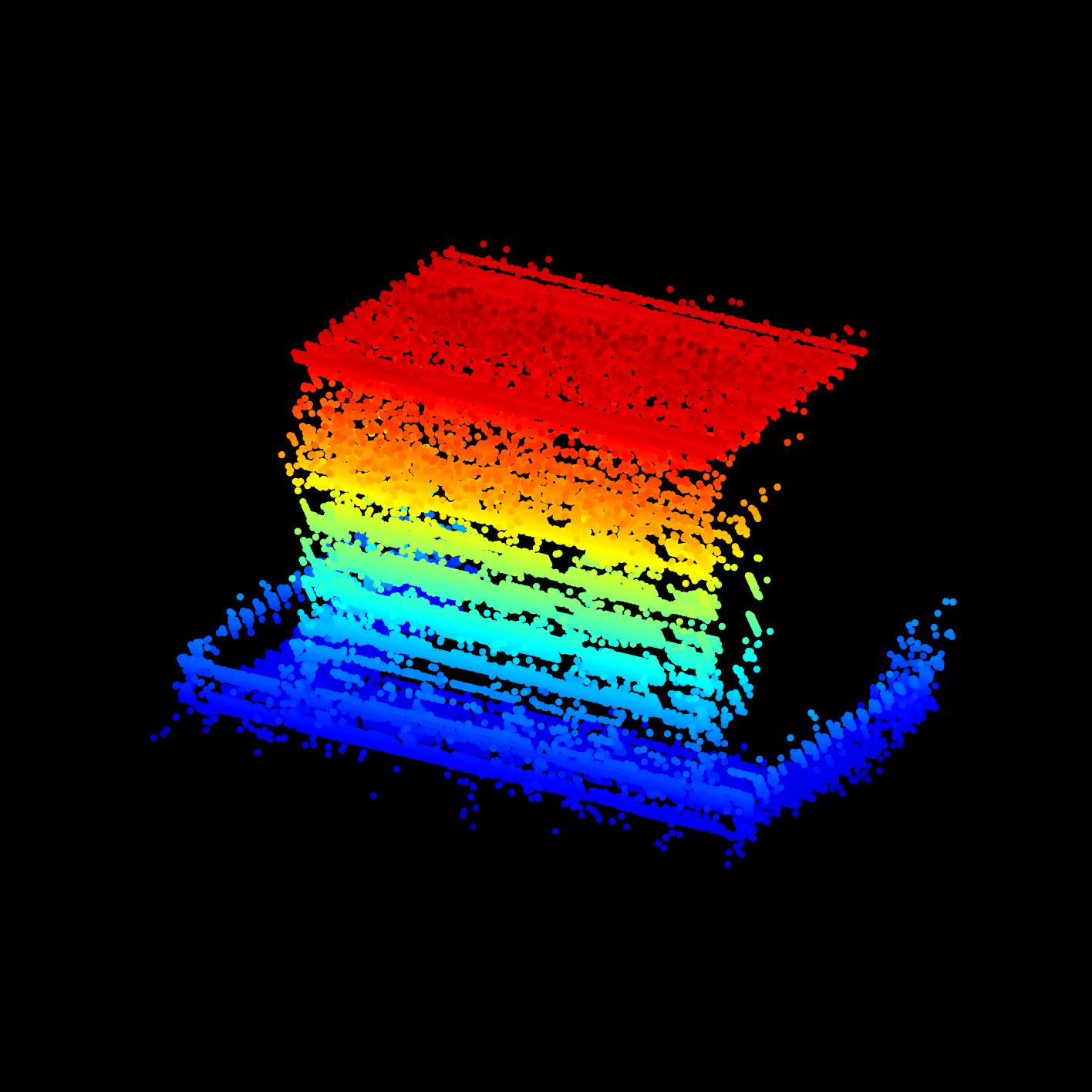}
        \caption*{{\fontsize{8pt}{10pt}\selectfont\centering (c-2)}}
        \label{figure27_c_2}
    \end{subfigure}
    \begin{subfigure}[b]{0.11\linewidth}
        \includegraphics[width=\linewidth]{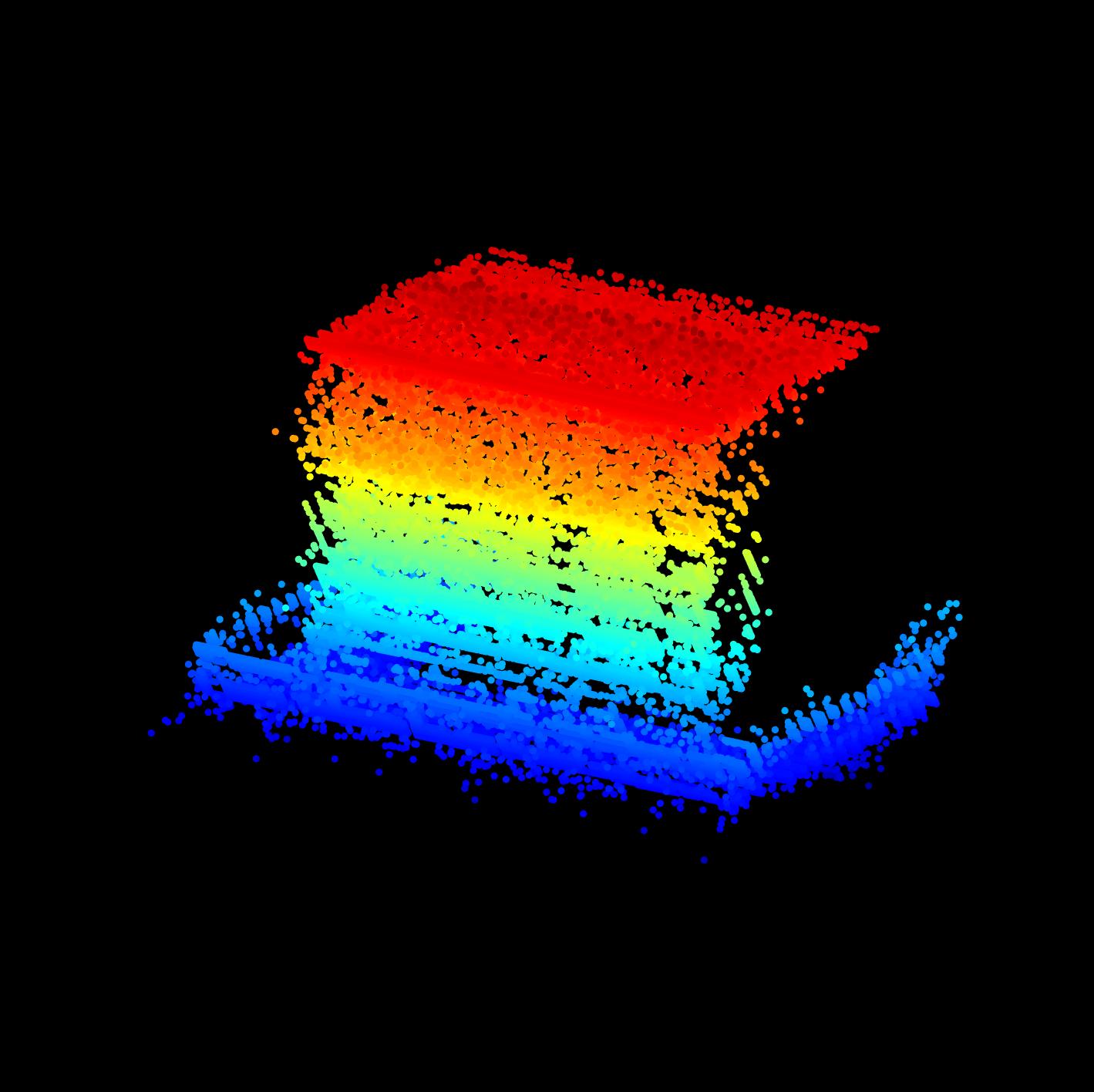}
        \caption*{{\fontsize{8pt}{10pt}\selectfont\centering (c-3)}}
        \label{figure27_c_3}
    \end{subfigure}
    \begin{subfigure}[b]{0.11\linewidth}
        \includegraphics[width=\linewidth]{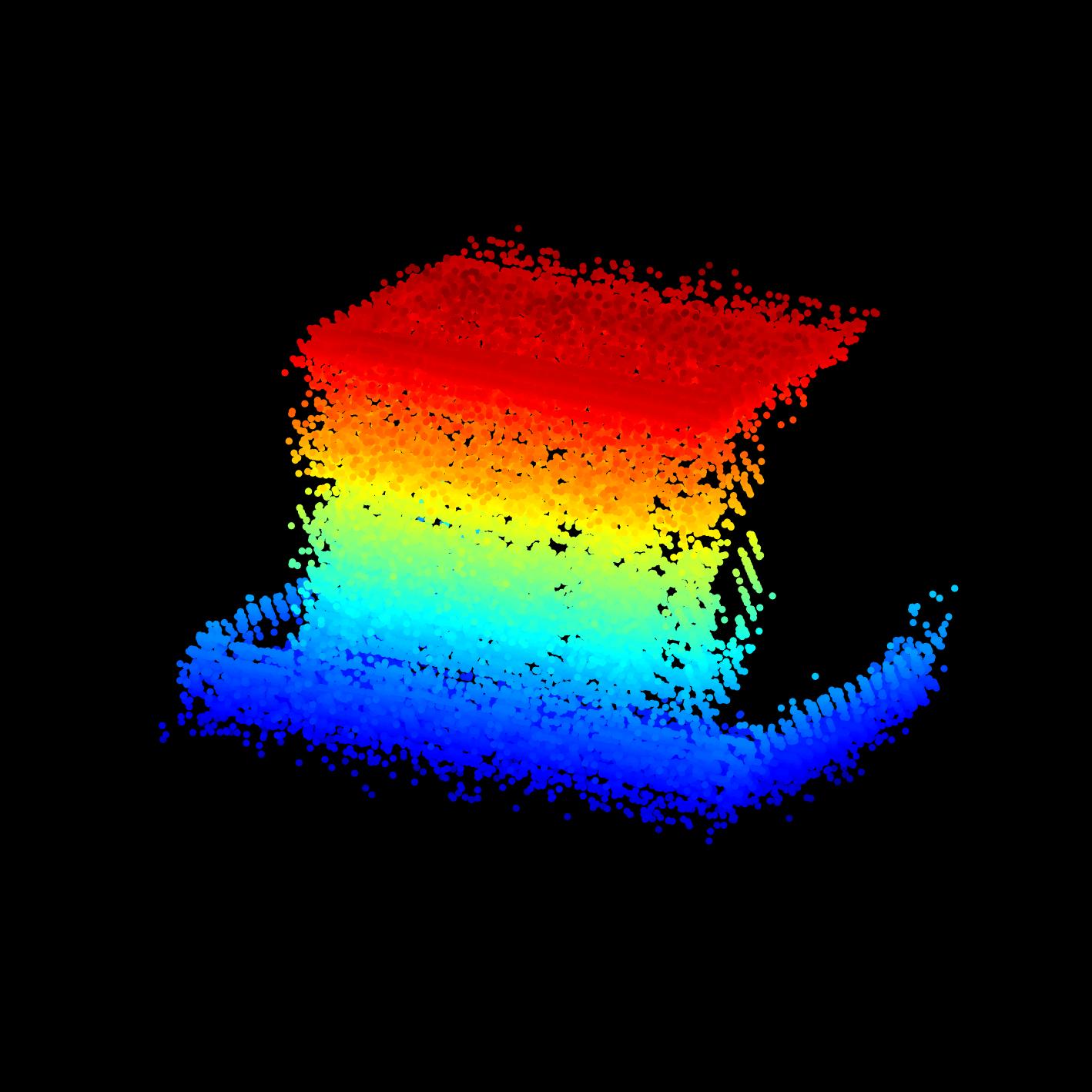}
        \caption*{{\fontsize{8pt}{10pt}\selectfont\centering (c-4)}}
        \label{figure27_c_4}
    \end{subfigure}
    \begin{subfigure}[b]{0.11\linewidth}
        \includegraphics[width=\linewidth]{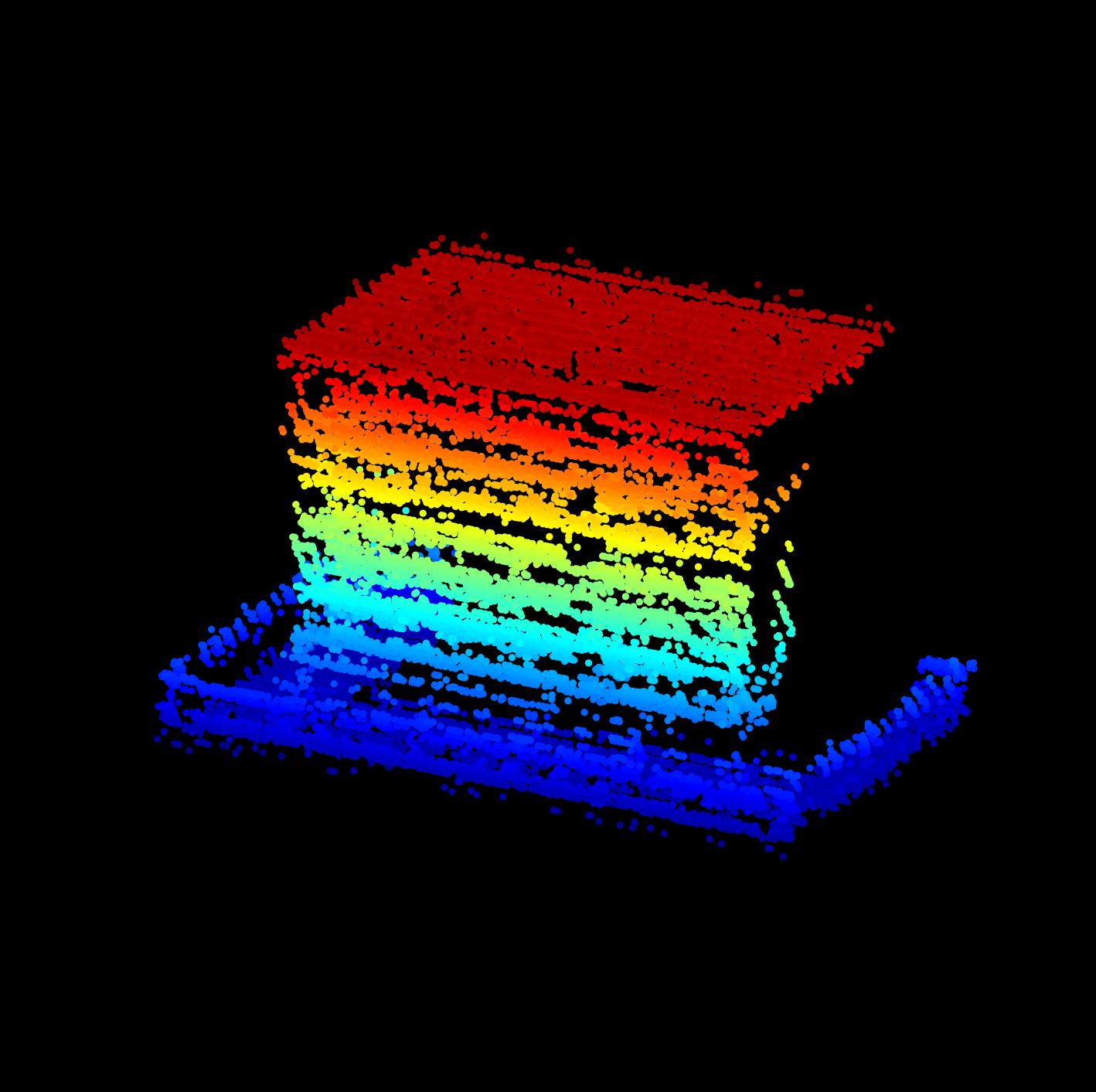}
        \caption*{{\fontsize{8pt}{10pt}\selectfont\centering (c-5)}}
        \label{figure27_c_5}
    \end{subfigure}
    \begin{subfigure}[b]{0.11\linewidth}
        \includegraphics[width=\linewidth]{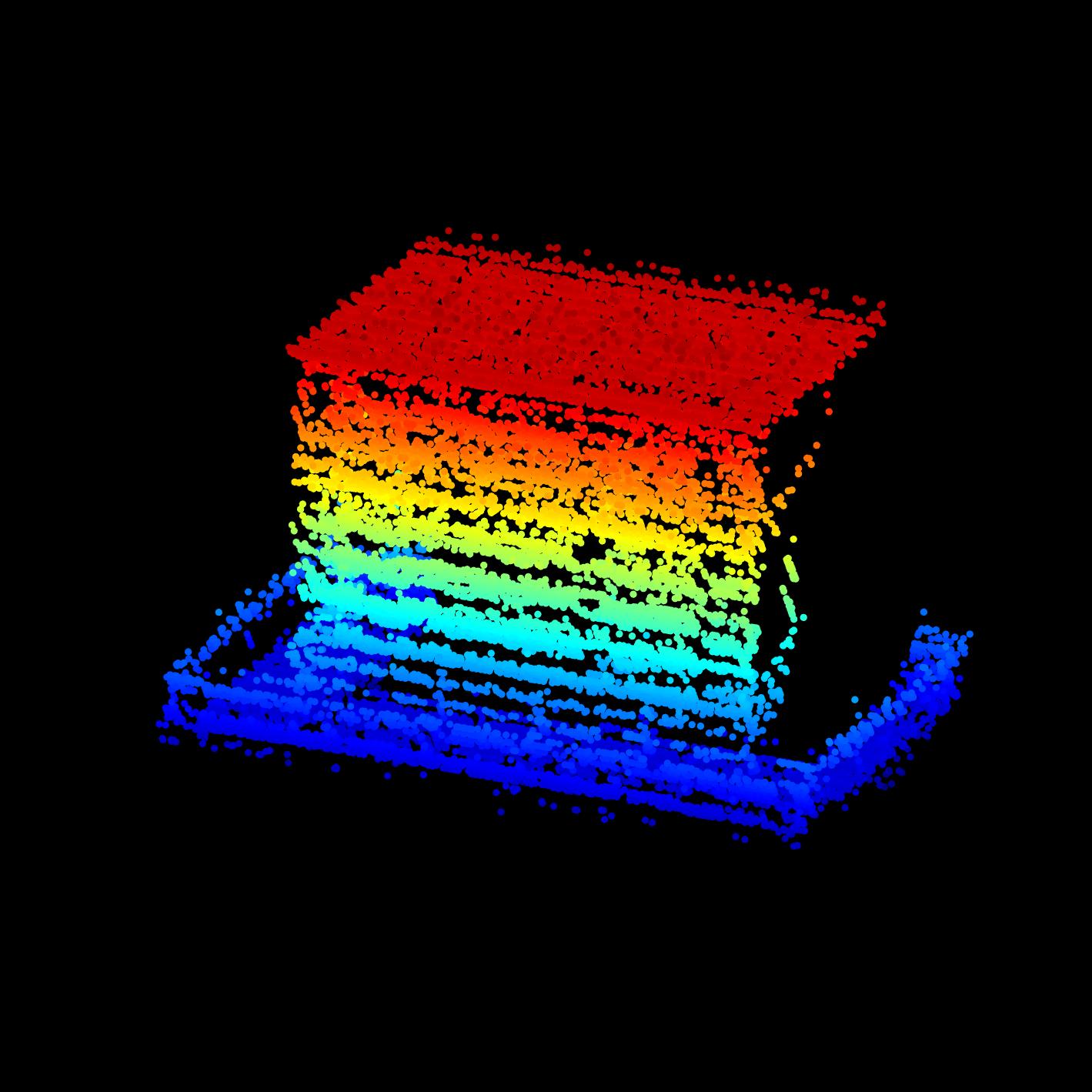}
        \caption*{{\fontsize{8pt}{10pt}\selectfont\centering (c-6)}}
        \label{figure27_c_6}
    \end{subfigure}
    \begin{subfigure}[b]{0.11\linewidth}
        \includegraphics[width=\linewidth]{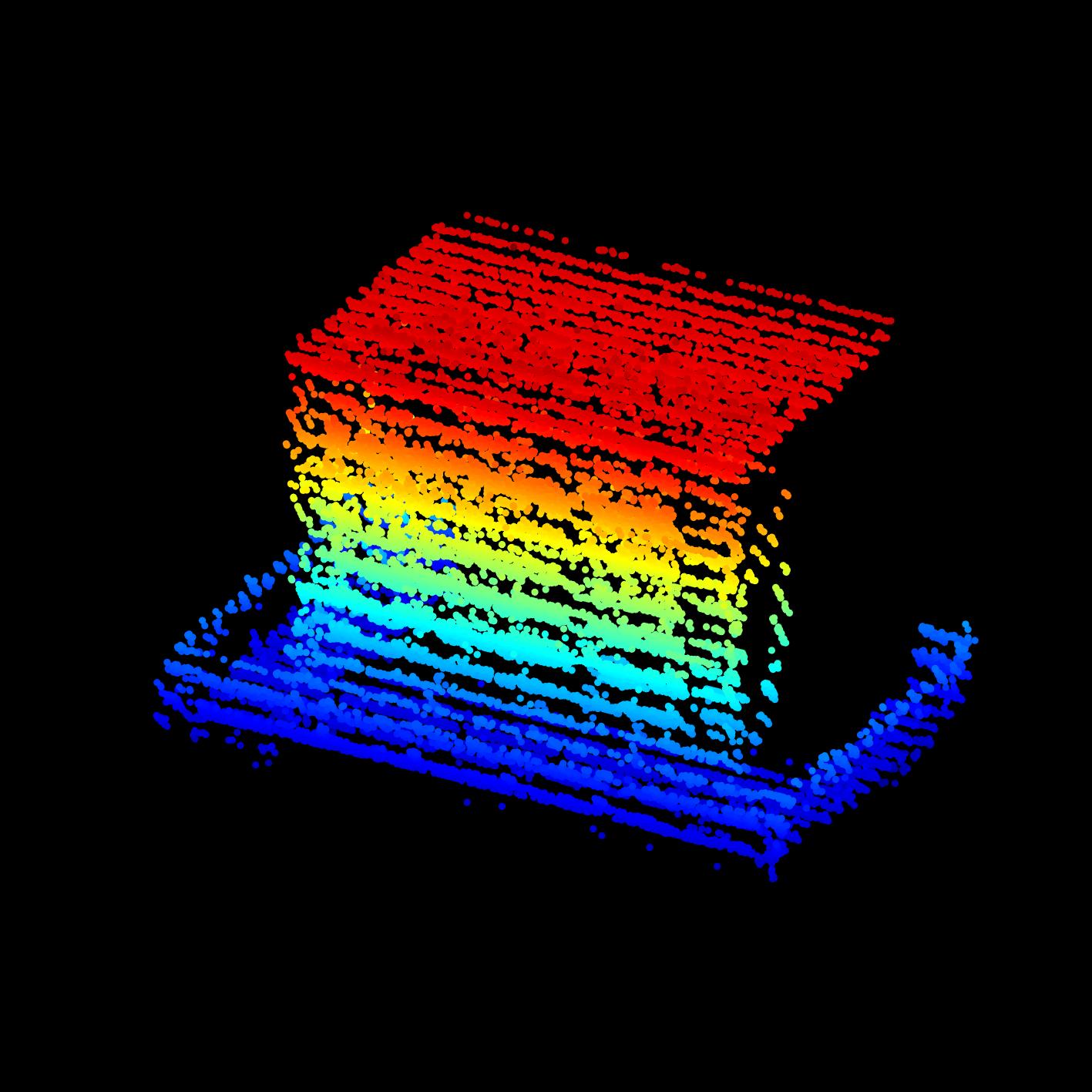}
        \caption*{{\fontsize{8pt}{10pt}\selectfont\centering (c-7)}}
        \label{figure27_c_7}
    \end{subfigure}
    \begin{subfigure}[b]{0.11\linewidth}
        \caption*{{\fontsize{8pt}{10pt}\selectfont\centering }}
        \includegraphics[width=\linewidth]{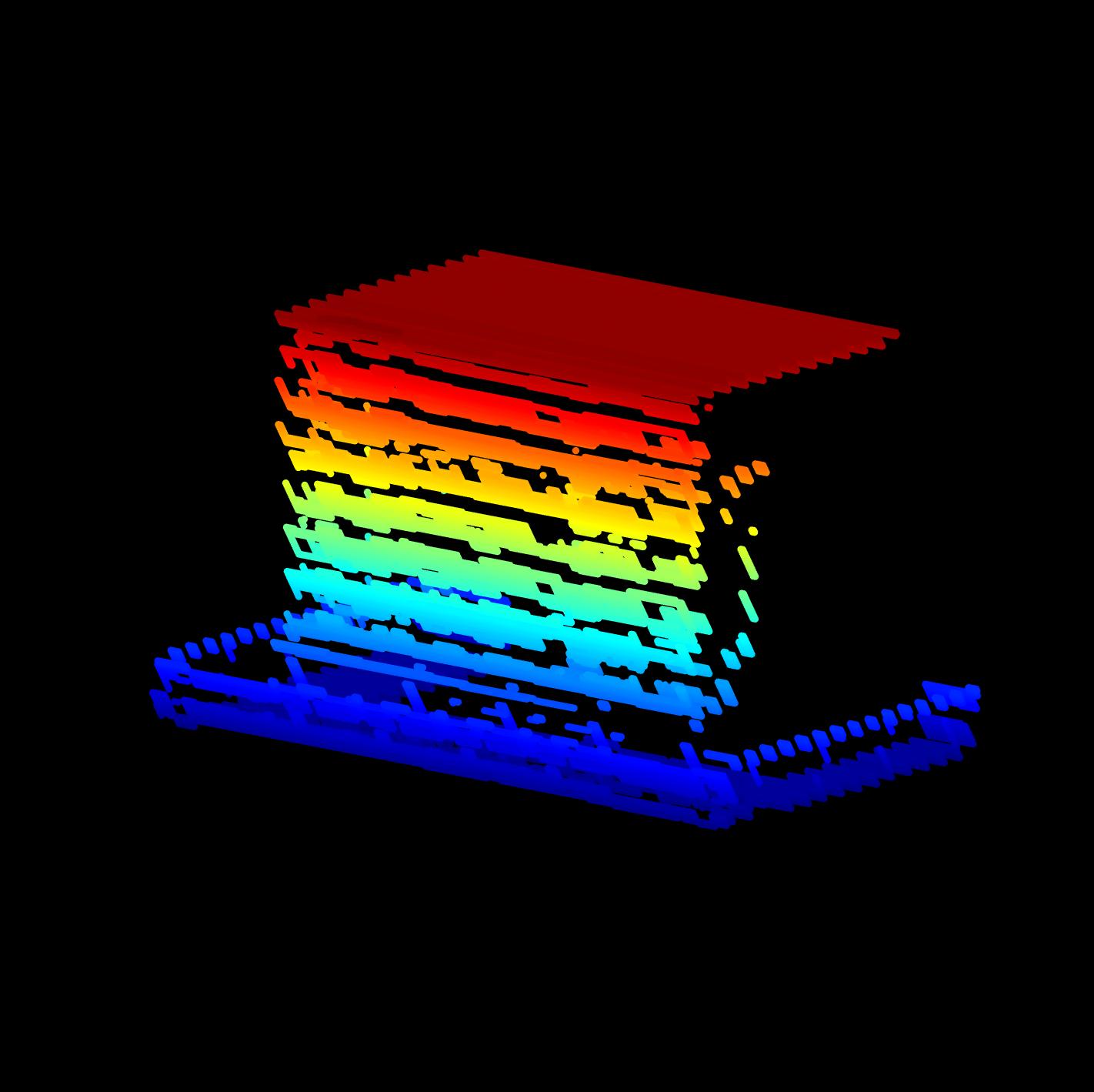}
        \caption*{{\fontsize{8pt}{10pt}\selectfont\centering (c-8)}}
        \label{figure27_c_8}
    \end{subfigure}

    % 第四行的八张图
    \begin{subfigure}[b]{0.11\linewidth}
        \includegraphics[width=\linewidth]{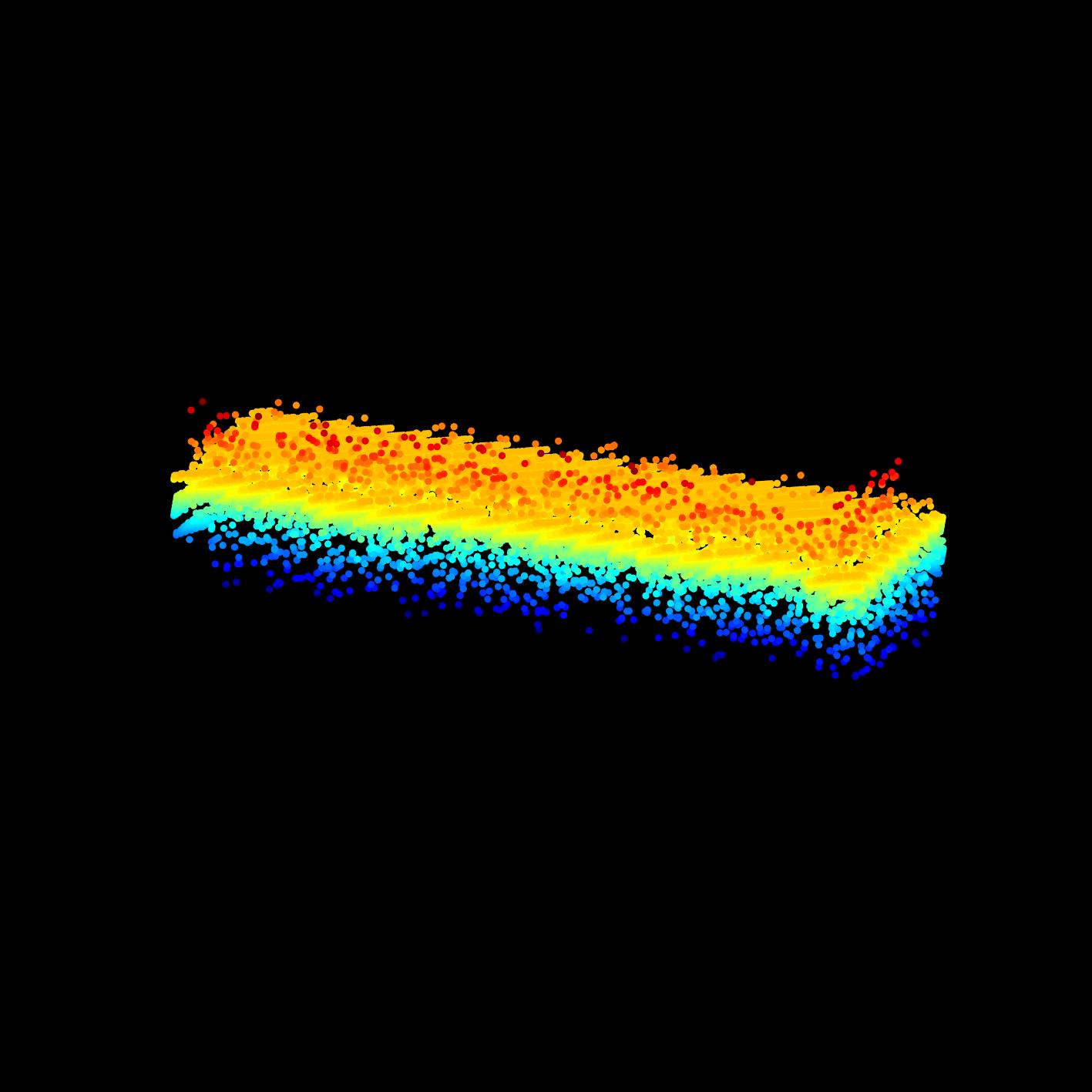}
        \caption*{{\fontsize{8pt}{10pt}\selectfont\centering (d-1)}}
        \label{figure27_d_1}
    \end{subfigure}
    \begin{subfigure}[b]{0.11\linewidth}
        \includegraphics[width=\linewidth]{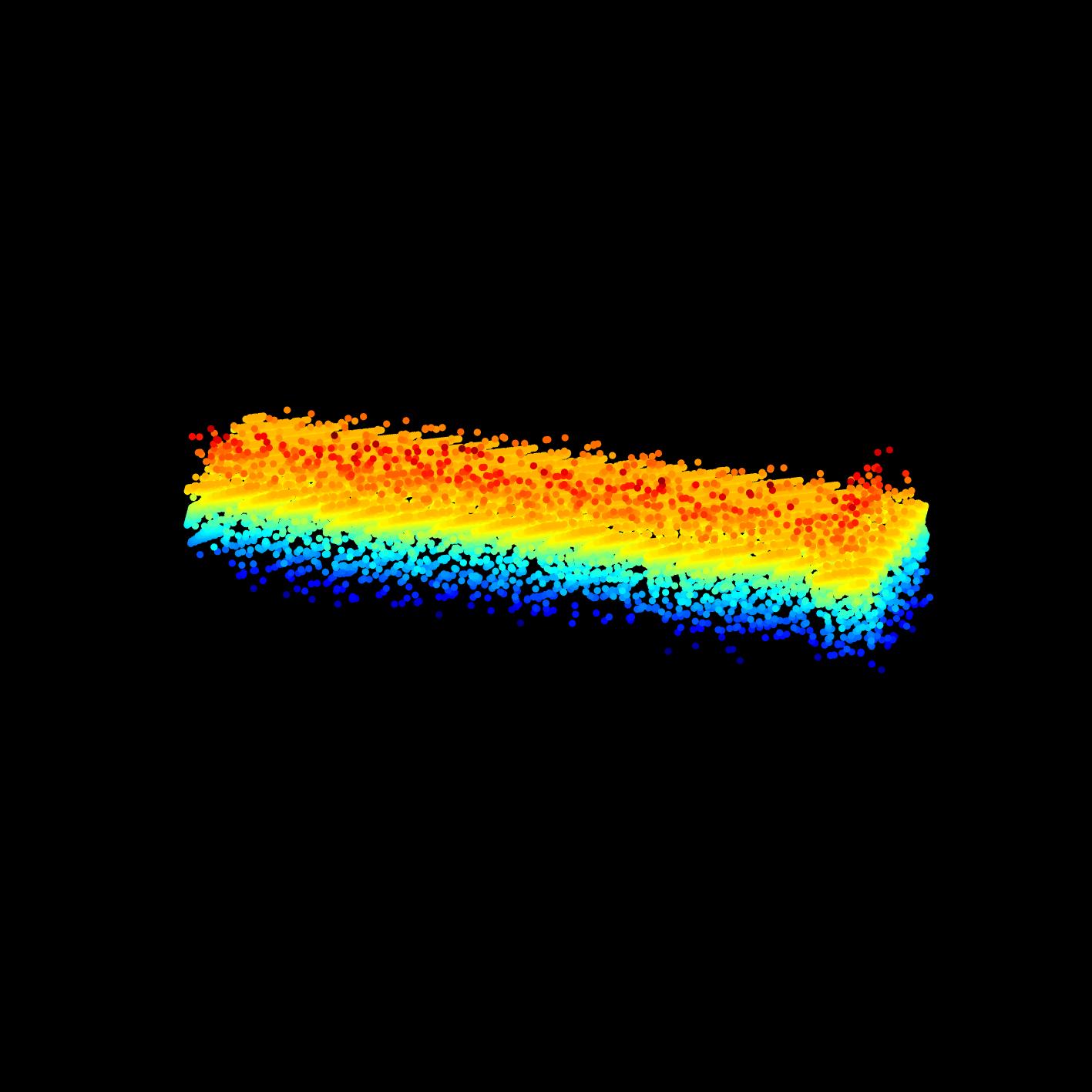}
        \caption*{{\fontsize{8pt}{10pt}\selectfont\centering (d-2)}}
        \label{figure27_d_2}
    \end{subfigure}
    \begin{subfigure}[b]{0.11\linewidth}
        \includegraphics[width=\linewidth]{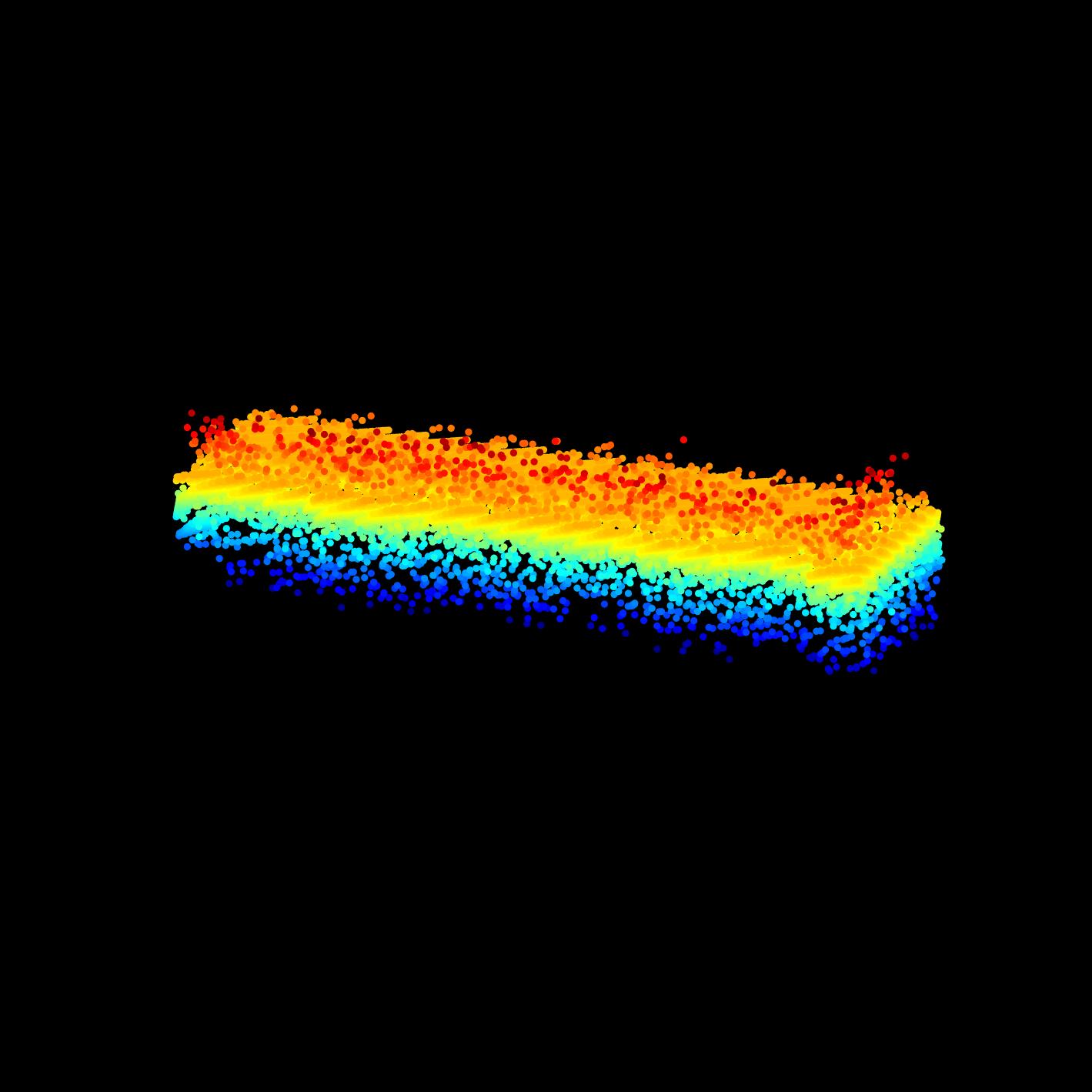}
        \caption*{{\fontsize{8pt}{10pt}\selectfont\centering (d-3)}}
        \label{figure27_d_3}
    \end{subfigure}
    \begin{subfigure}[b]{0.11\linewidth}
        \includegraphics[width=\linewidth]{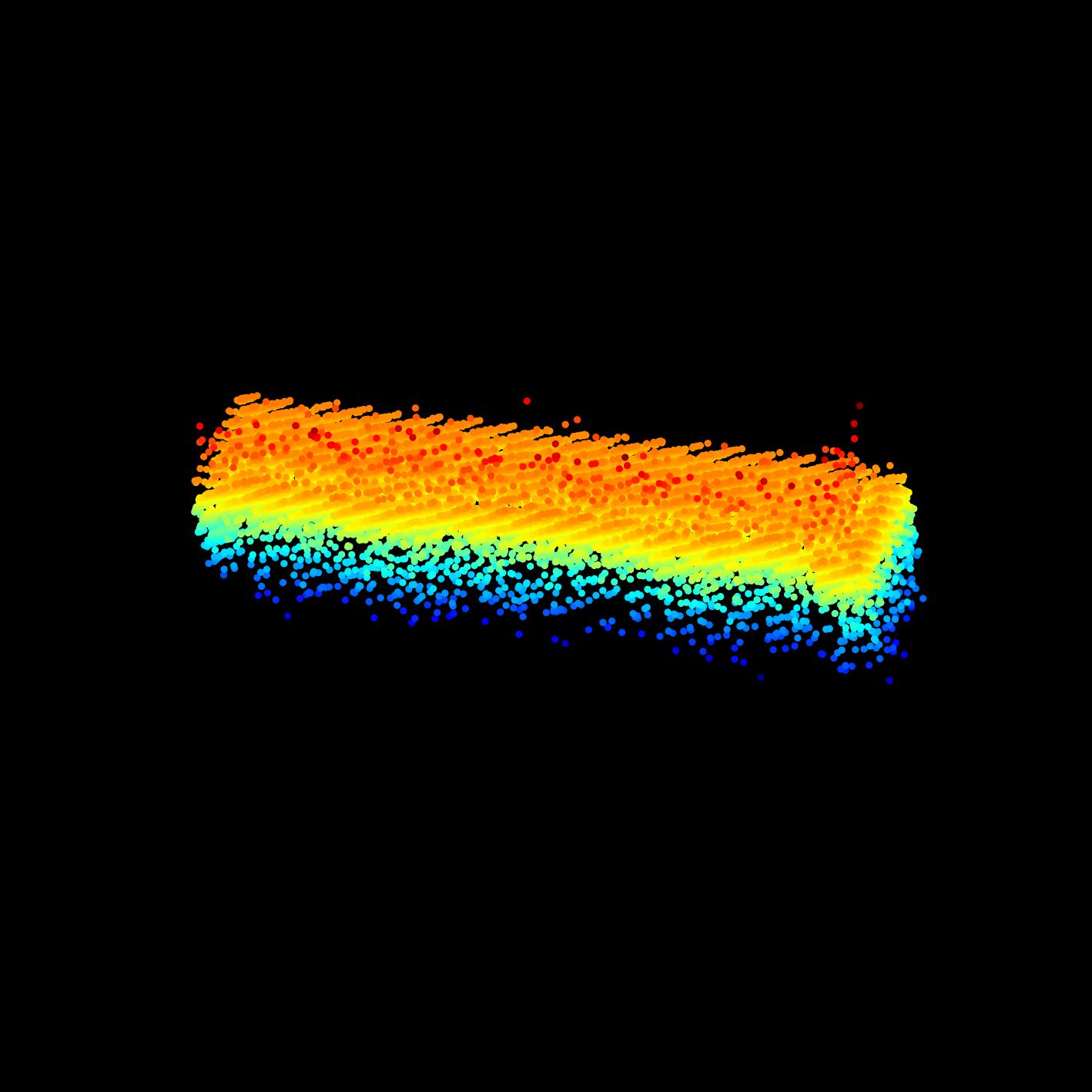}
        \caption*{{\fontsize{8pt}{10pt}\selectfont\centering (d-4)}}
        \label{figure27_d_4}
    \end{subfigure}
    \begin{subfigure}[b]{0.11\linewidth}
        \includegraphics[width=\linewidth]{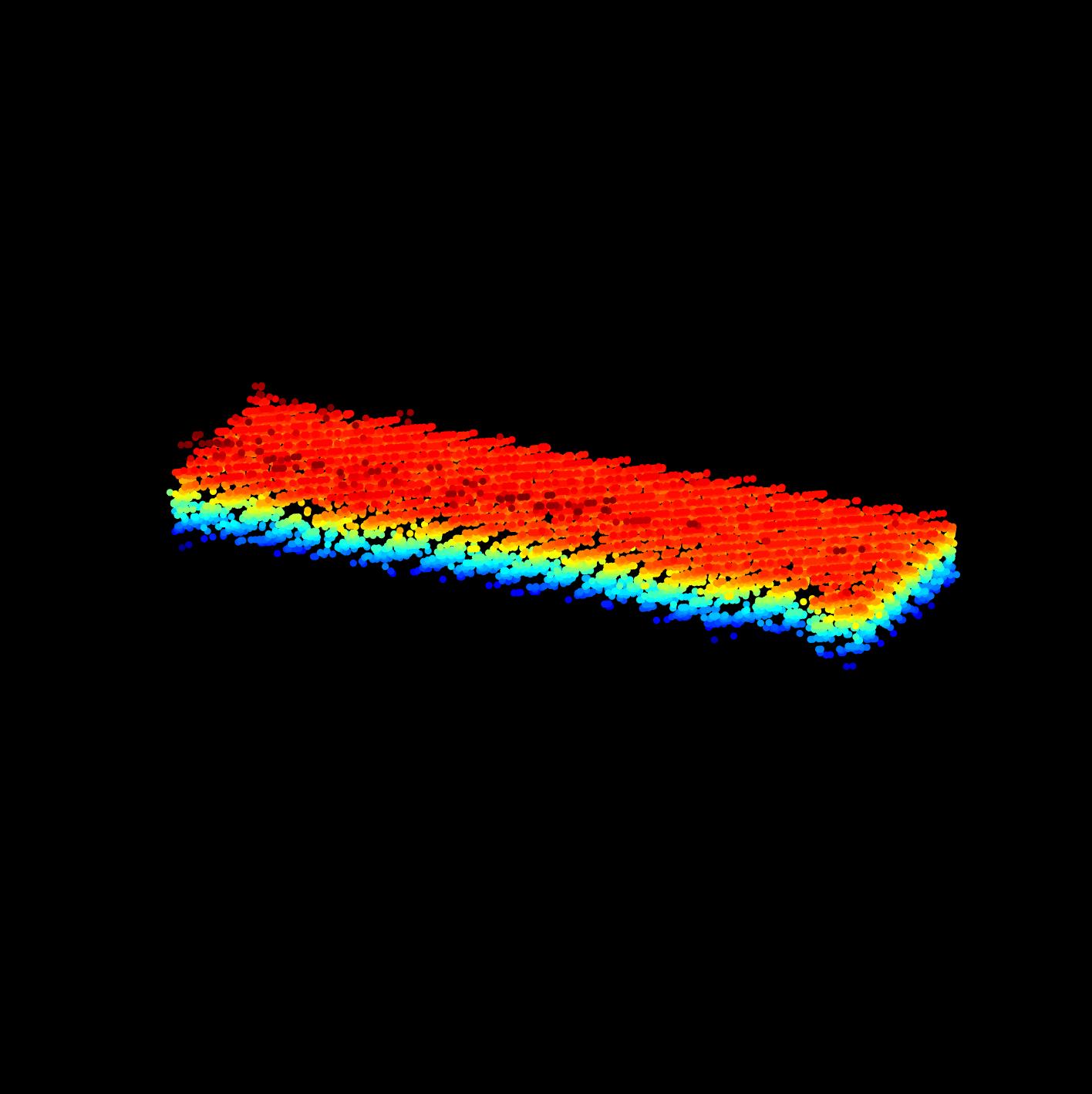}
        \caption*{{\fontsize{8pt}{10pt}\selectfont\centering (d-5)}}
        \label{figure27_d_5}
    \end{subfigure}
    \begin{subfigure}[b]{0.11\linewidth}
        \includegraphics[width=\linewidth]{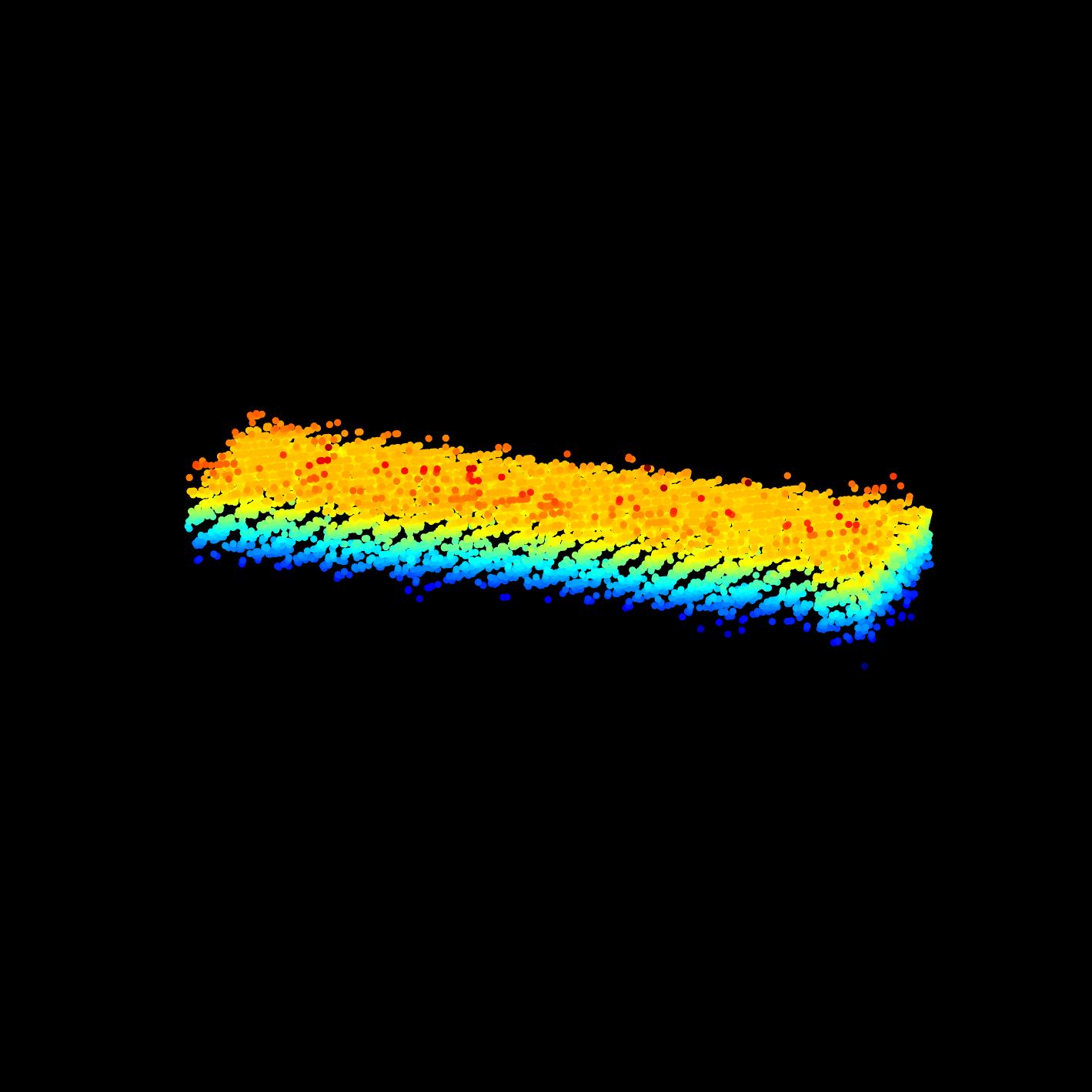}
        \caption*{{\fontsize{8pt}{10pt}\selectfont\centering (d-6)}}
        \label{figure27_d_6}
    \end{subfigure}
    \begin{subfigure}[b]{0.11\linewidth}
        \includegraphics[width=\linewidth]{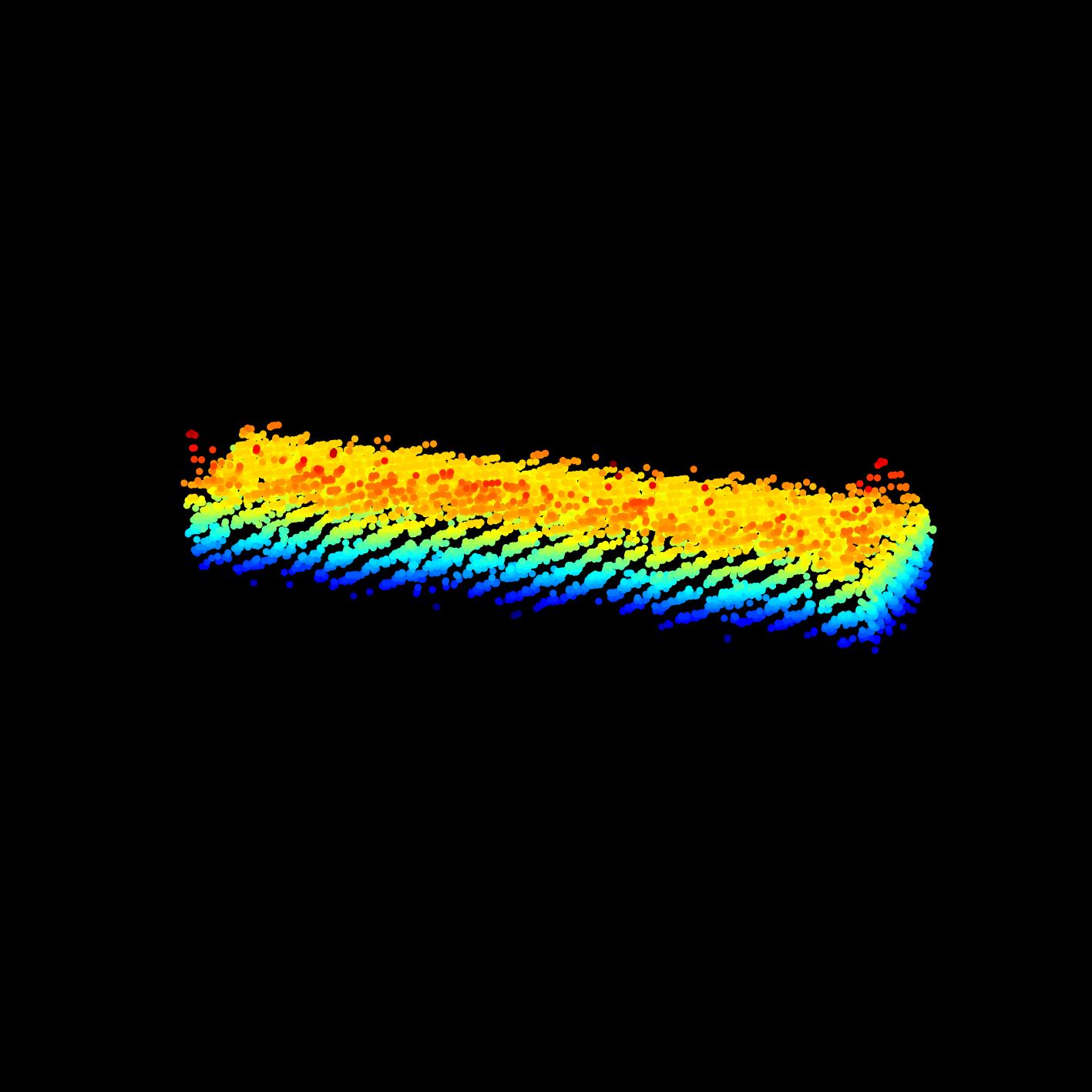}
        \caption*{{\fontsize{8pt}{10pt}\selectfont\centering (d-7)}}
        \label{figure27_d_7}
    \end{subfigure}
    \begin{subfigure}[b]{0.11\linewidth}
        \caption*{{\fontsize{8pt}{10pt}\selectfont\centering }}
        \includegraphics[width=\linewidth]{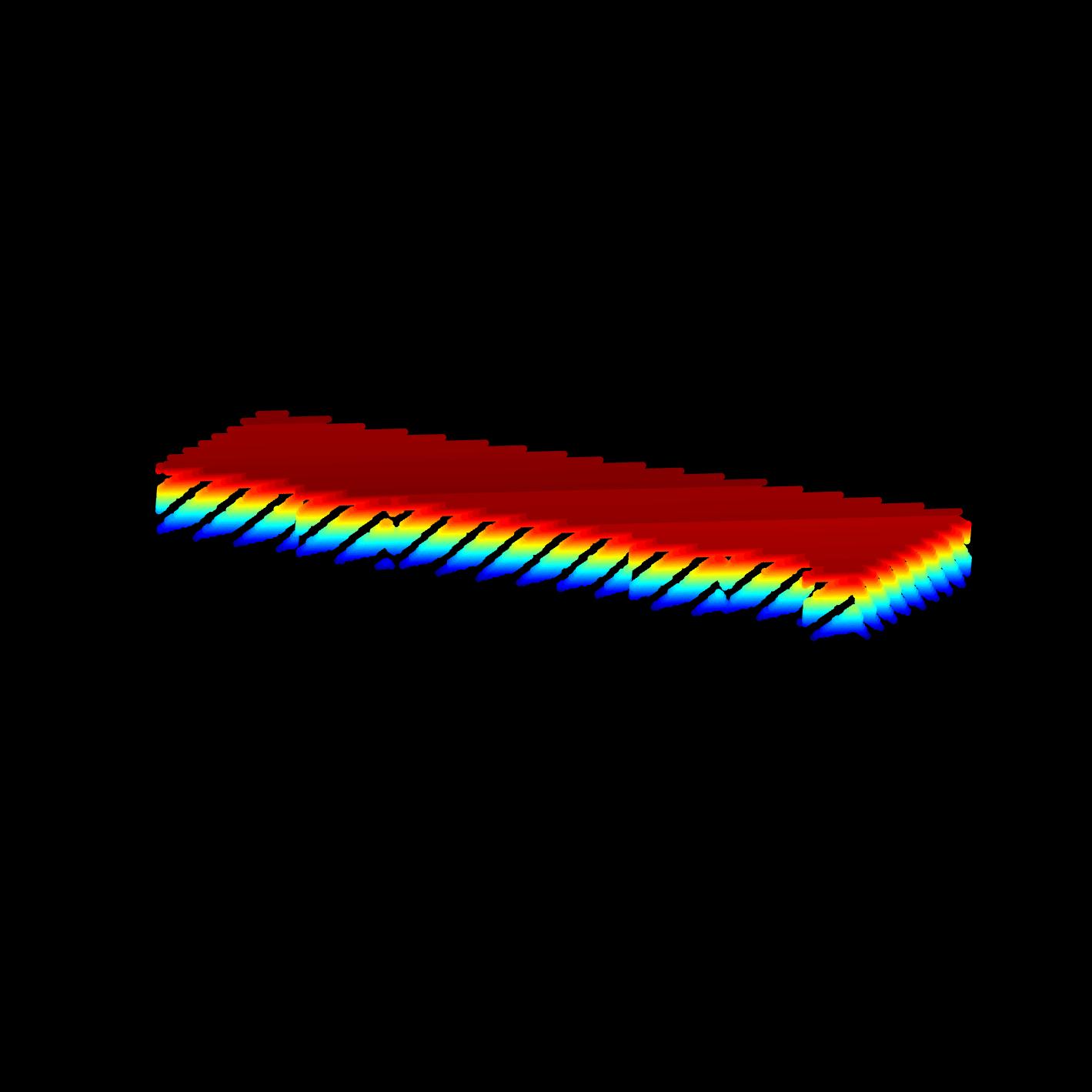}
        \caption*{{\fontsize{8pt}{10pt}\selectfont\centering (d-8)}}
        \label{figure27_d_8}
    \end{subfigure}

    \caption{Test 4 Results. The figure shows four different structure types, each represented in separate rows. From left to right, the columns display the outcomes of the FISTA-based method, SLIMMER, tomo-IRENet-Raw, tomo-IRENet-TV, tomo-LRNet-biU, tomo-LRENet-biU, tomo-LRENet-LSTM, and the ground truth, respectively.}
    \label{figure_27}
\end{figure}

In this test, we assess performance using complete 3D building models that include various architectural forms such as an L-shaped facade building, a complex planar structure, a low-rise box-type building, and a flat-type building. We have conducted these tests on 34 different building models, evaluating both qualitatively and quantitatively using the previously mentioned metrics. The quantitative results are listed in Table~\ref{table_3}. We also select four examples for qualitative comparisons, which are displayed in Fig.~\ref{figure_27}.

\begin{table}[h]
	\renewcommand{\arraystretch}{0.2} % Adjust row spacing
	\centering
	\captionsetup{justification=centering}
	\caption{Comparison of Various Methods on Different Metrics}
	\label{table_3}
	\footnotesize
	\hspace*{-4em}
	\begin{tabular}{@{}>{\centering\arraybackslash}m{0.19\textwidth} *{7}{>{\centering\arraybackslash}m{0.11\textwidth}} @{}}
	\toprule
	\textbf{Metric} & \textbf{FISTA} & \textbf{SLIMMER} & \textbf{tomo-IRENet-Raw} & \textbf{tomo-IRENet-TV} & \textbf{tomo-IRENet-U} & \textbf{tomo-LRENet-biU} & \textbf{tomo-LRENet-LSTM} \\
	\midrule
	\multicolumn{8}{c}{\textbf{3D Image Accuracy Metrics}} \\
	\midrule
	Root Mean Square Error (RMSE) & 0.0138 & 0.0084 & 0.0078 & 0.0245 & 0.0077 & 0.0039 & \textbf{0.0016} \\
	\midrule
	Peak Signal-to-Noise Ratio (PSNR) & 13.5132 & 13.7947 & 14.0860 & 12.2530 & 16.2224 & 17.5854 & \textbf{17.7281} \\
	\midrule
	\multicolumn{8}{c}{\textbf{3D Point Cloud Geometric Spatial Structure Metrics}} \\
	\midrule
	Scatterer Correspondence (Precision) & 89.55\% & 92.10\% & 87.32\% & 69.58\% & 95.78\% & 97.56\% & \textbf{98.74}\% \\
	\midrule
	Reconstruction Completeness (Recall) & 89.49\% & \textbf{90.67}\% & 89.20\% & 84.80\% & 83.84\% & 86.00\% & 86.05\% \\
	\midrule
	Average Euclidean Distance ($D_{pcm}$) & 0.28 & 0.21 & 0.34 & 0.80 & 0.18 & 0.11 & \textbf{0.09} \\
	\midrule
	Spatial Distribution Consistency ($V$) & 0.63 & 0.58 & 0.93 & 1.06 & 0.44 & 0.37 & \textbf{0.53} \\
	\midrule
	\multicolumn{8}{c}{\textbf{Reconstruction Efficiency Metric}} \\
	\midrule
	Average Reconstruction Time ($T_{ag}$) & 175.82s & 2497.31s & \textbf{1.63s} & 6.05s & 55.36s & 20.08s & 8.27s \\
	\bottomrule
	\end{tabular}
	\end{table}

We begin with the qualitative results. Starting from the first row down to the last, the results correspond to an L-shaped facade building, a complex planar structure, a low-rise box-type building, and a flat-type building, respectively. Among these, the results from tomo-IRENet-TV consistently appear more complete, depicting a denser point cloud for the building surfaces. It also tends to produce slightly fewer outliers compared to the other three methods on its left.

However, this could be due to two potential reasons. Firstly, incorporating shallow spatial feature modeling through the TV term may help in reducing outliers and contribute to a smoother outcome. Secondly, this approach might compromise the resolution, leading to a visually denser result. This latter implication may negatively affect the reconstruction precision, as seen in the subsequent quantitative results.

Excluding the tomo-IRENet-TV method, we can broadly classify the results into two parts: those on the left and those on the right. The right side, which takes spatial features into account, yields visually denser results with the reconstructed scatterers gathered more closely. There are also fewer outliers present in the rooftop part of the buildings. For a more detailed analysis of their performance differences, we delve into the quantitative results.

 It is clear from the findings that tomo-IRENet-TV's reconstruction precision is the poorest, even when not compared to traditional methods. This subpar performance is likely due to the harmful effects of limited resolution, as shown in previous tests. The limited resolution results in numerous false and position-mismatched scatters, deteriorating reconstruction precision.

Aside from this method, the 3D image accuracy metrics reveal that the other three new methods in the study (tomo-IRENet-U, tomo-LRENet-biU, and tomo-LRENet-LSTM) all have superior accuracy. Tomo-LRENet-LSTM ranks first, followed by tomo-LRENet-biU, and lastly, tomo-IRENet-U. The RMSE reduced considerably from tomo-IRENet-Raw (0.0078) to tomo-LRENet-LSTM (0.0016), while the PSNR increased significantly from tomo-IRENet-Raw (14.086) to tomo-LRENet-LSTM (17.7281).

Looking at the 3D point cloud geometric spatial structure metrics, we find that these three methods have the same positive impacts on average Euclidean distance and spatial distribution consistency. This indicates that the introduction of deep spatial feature modeling is beneficial for more consistent and matched 3D mapping results in terms of spatial structure.

When examining scatterer correspondence (precision) and reconstruction completeness (recall), it's interesting to note that compared to other methods, precision increased by at least 3\% to 6\%, while recall decreased by around 4\%. This suggests that other methods increase recall at the expense of precision reduction, resulting in more outliers. In contrast, the three new methods achieve a more balanced outcome.

Regarding reconstruction efficiency, tomo-IRENet-Raw is the fastest due to its simplified neural network architecture. Conversely, tomo-IRENet-U is the slowest, taking 55 seconds, due to the adoption of multiple U-Net structures. However, it is still at least three times faster than traditional iterative methods, such as the FISTA method. The two methods that adopt the light reconstruction and enhancement framework significantly reduce time cost to 20.08 and 8.27 seconds respectively. Among these, tomo-LRENet-LSTM offers the best balance between efficiency and precision. Compared to tomo-IRENet-Raw, which also takes less than 10 seconds, its precision is significantly higher across multiple metrics.

\vspace{0.5em}
\textit{D.5 results of test Object 5 (uniform urban structure applicability test)}
\vspace{0.5em}

\begin{figure}[h]
    \centering
    % 第一行四张图
    \begin{subfigure}[b]{0.24\linewidth}
        \caption*{{\fontsize{8pt}{10pt}\selectfont\centering  FISTA-based}}
        \includegraphics[width=\linewidth]{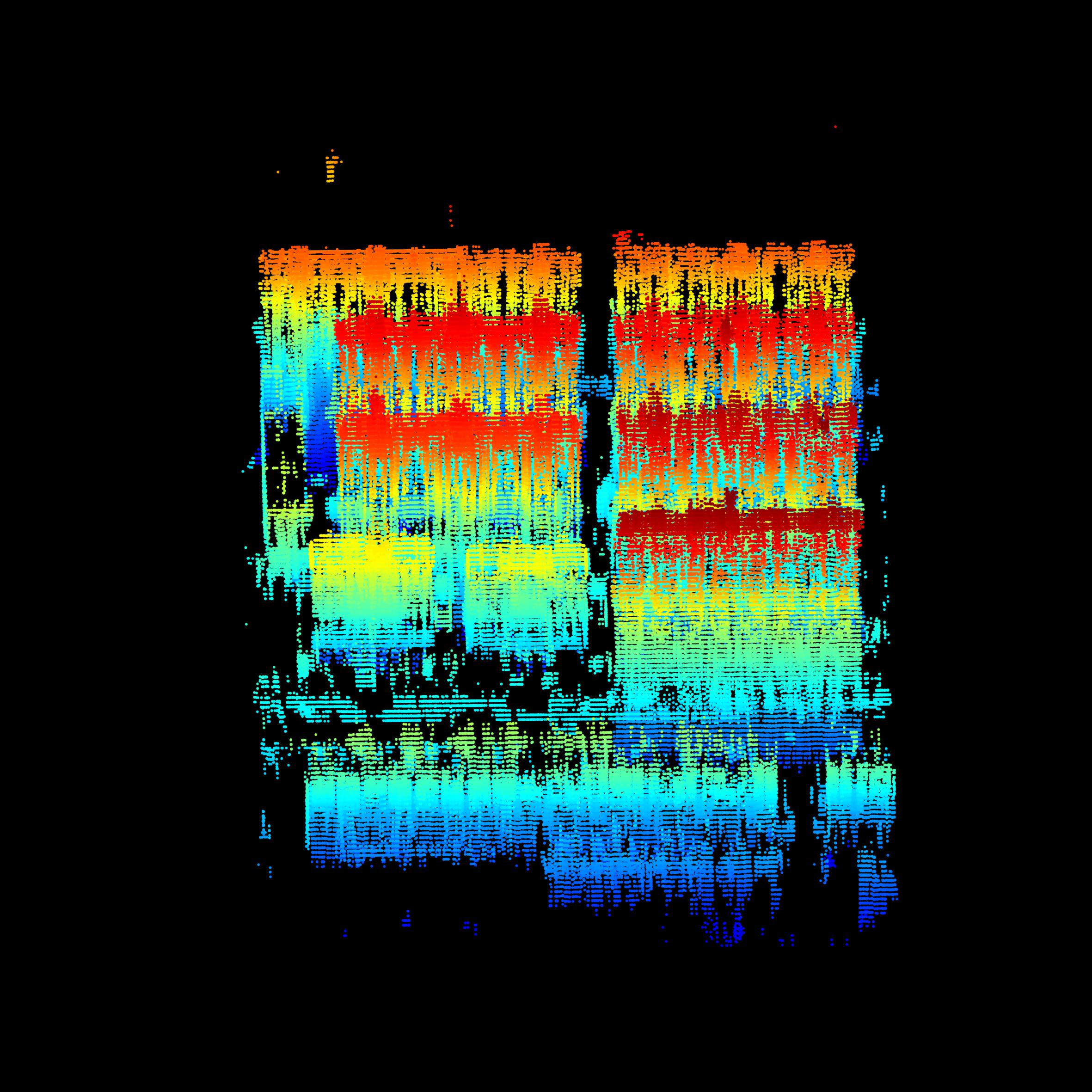}
        \caption{}
        \label{figure28_1}
    \end{subfigure}
    \begin{subfigure}[b]{0.24\linewidth}
        \caption*{{\fontsize{8pt}{10pt}\selectfont\centering SLIMMER}}
        \includegraphics[width=\linewidth]{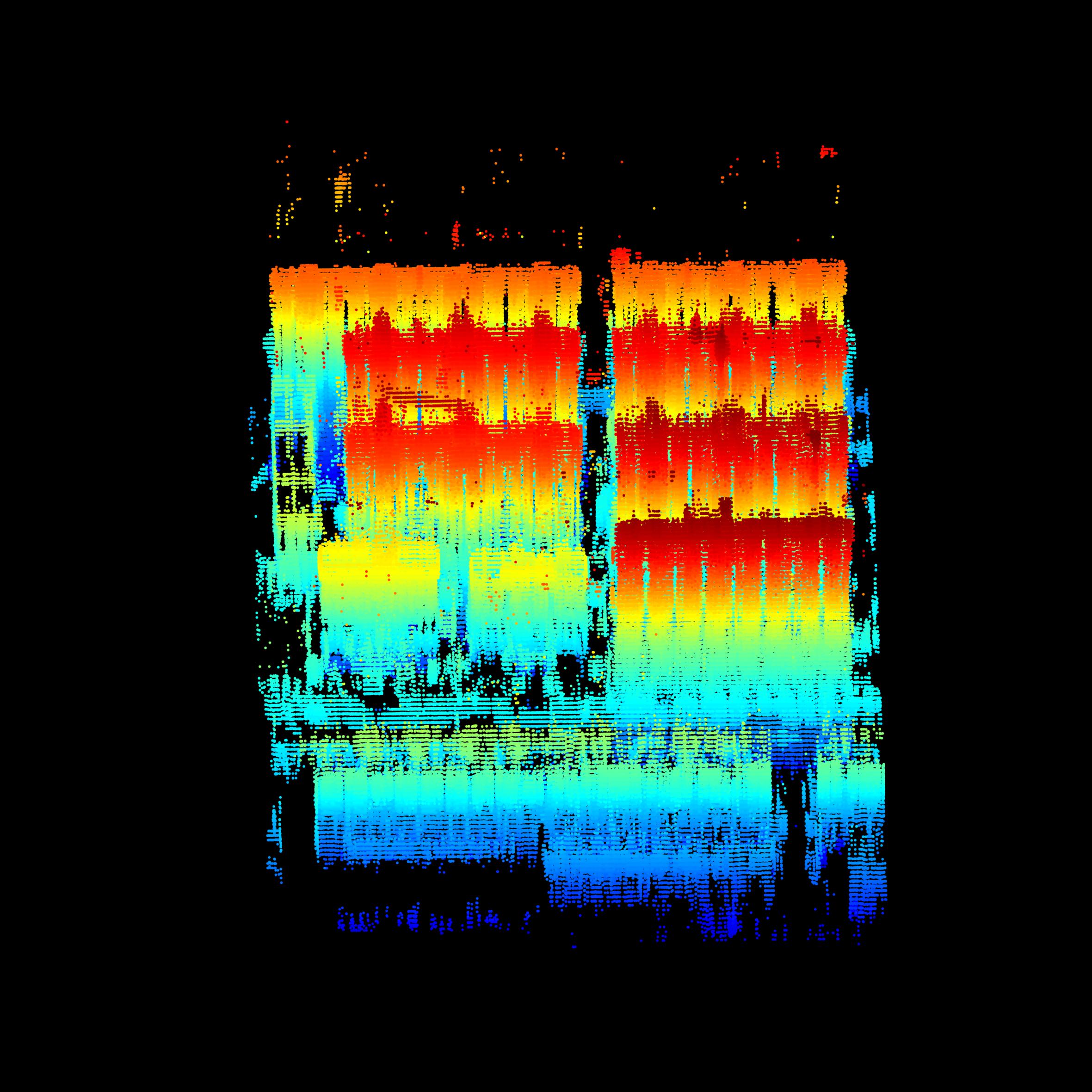}
        \caption{}
        \label{figure28_2}
    \end{subfigure}
    \begin{subfigure}[b]{0.24\linewidth}
        \caption*{{\fontsize{8pt}{10pt}\selectfont\centering tomo-IRENet-Raw}}
        \includegraphics[width=\linewidth]{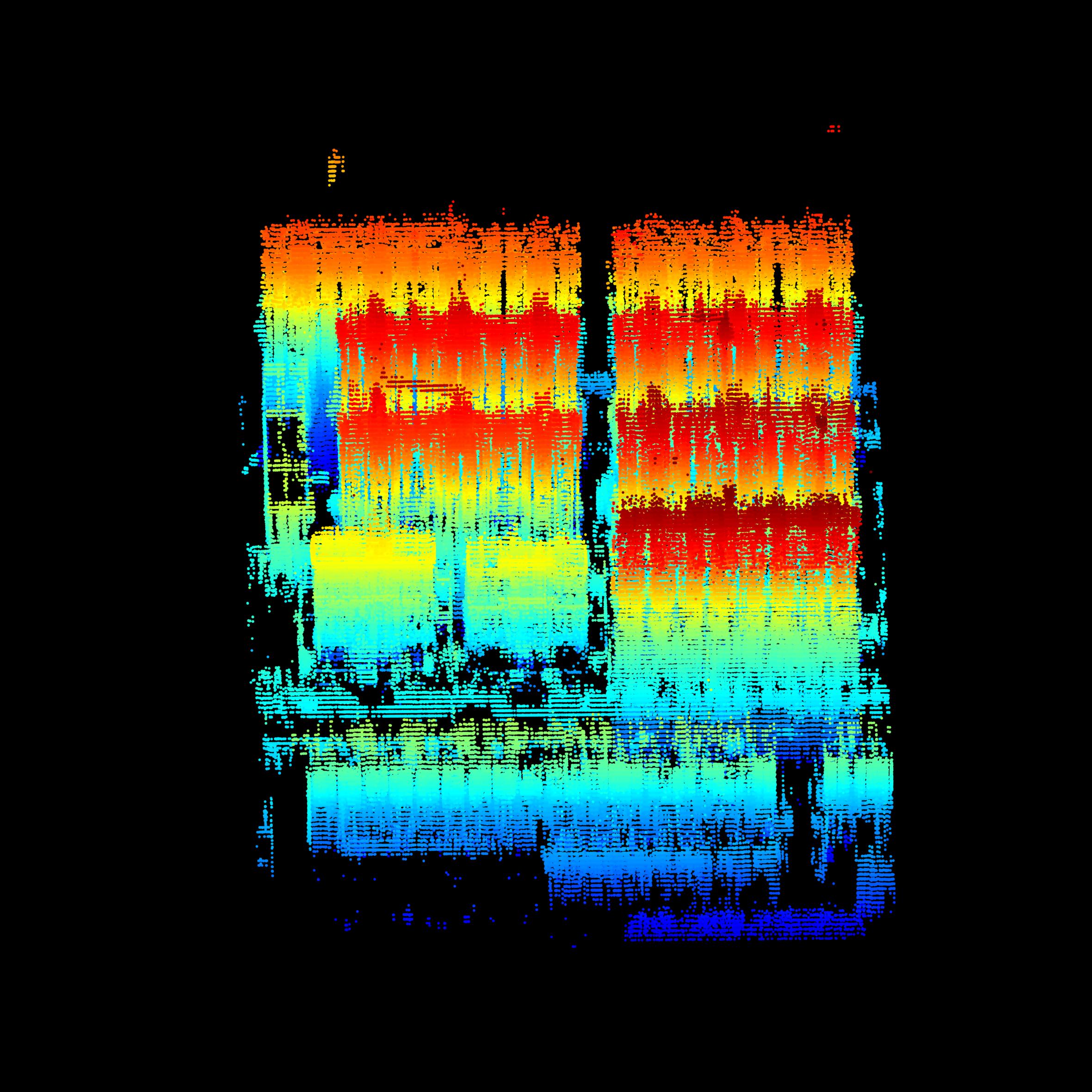}
        \caption{}
        \label{figure28_3}
    \end{subfigure}
    \begin{subfigure}[b]{0.24\linewidth}
        \caption*{{\fontsize{8pt}{10pt}\selectfont\centering  tomo-IRENet-TV}}
        \includegraphics[width=\linewidth]{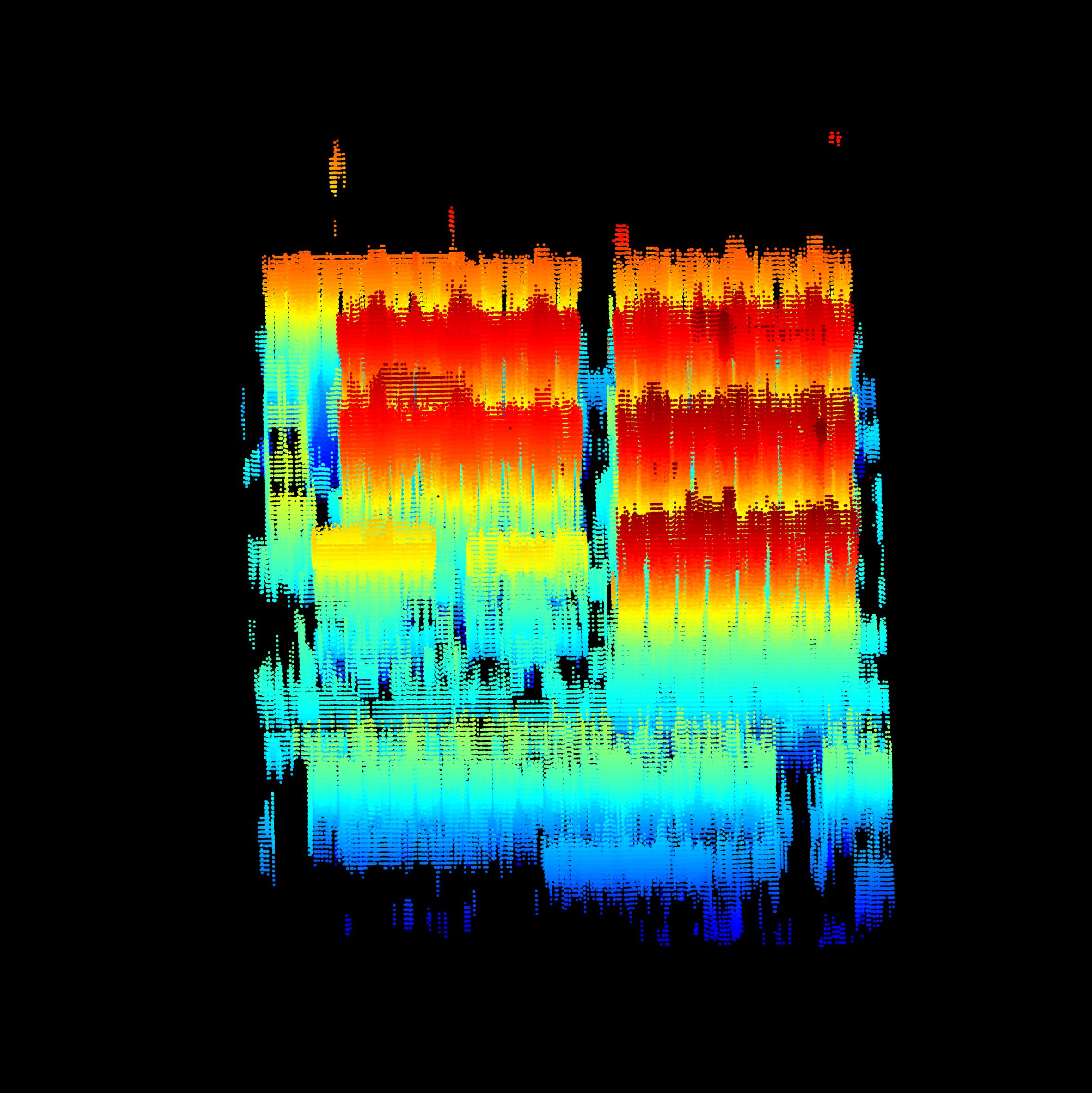}
        \caption{}
        \label{figure28_4}
    \end{subfigure}

    % 第二行三张图
    \begin{subfigure}[b]{0.24\linewidth}
        \caption*{{\fontsize{8pt}{10pt}\selectfont\centering tomo-IRENet-U}}
        \includegraphics[width=\linewidth]{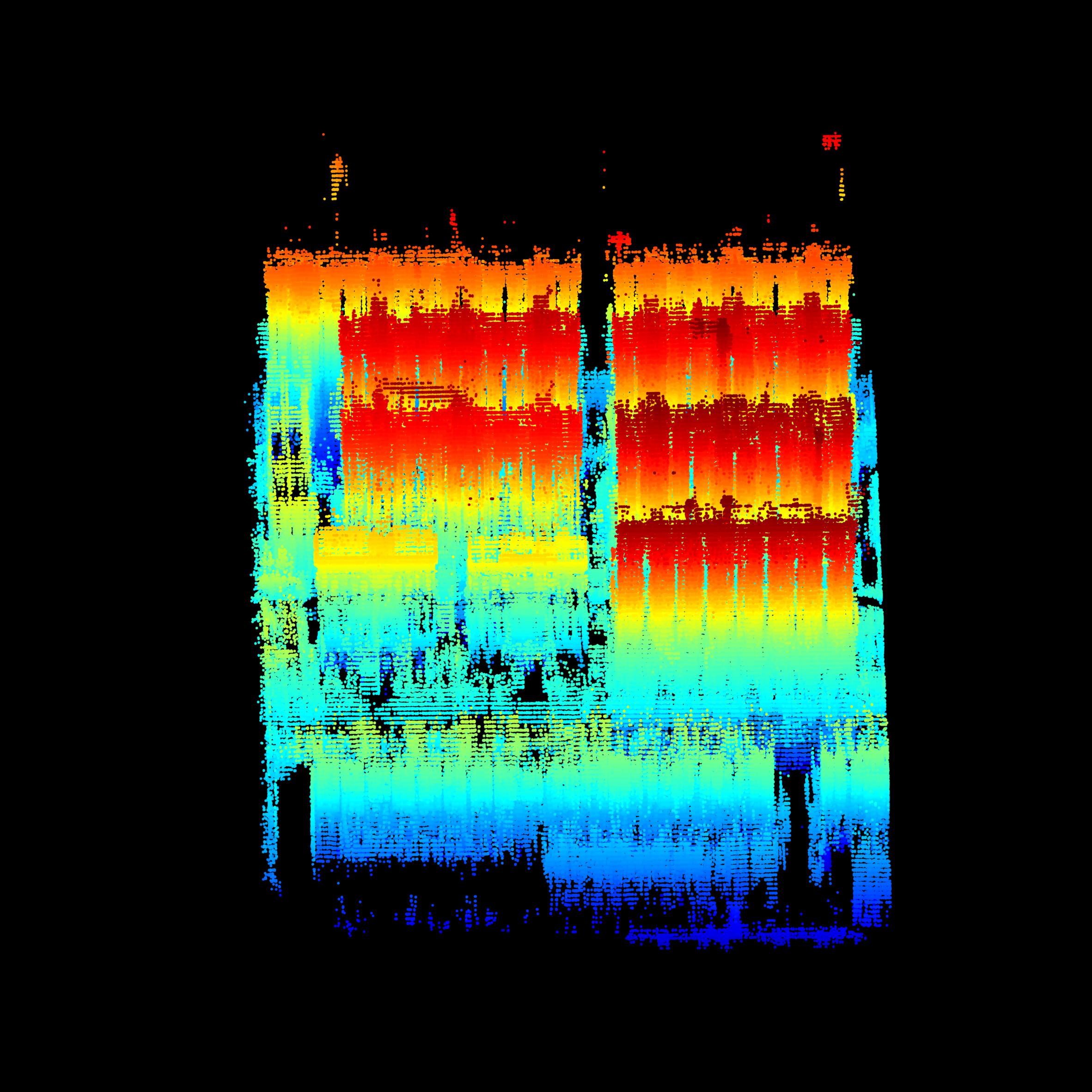}
        \caption{}
        \label{figure28_5}
    \end{subfigure}
    \begin{subfigure}[b]{0.24\linewidth}
        \caption*{{\fontsize{8pt}{10pt}\selectfont\centering tomo-LRENet-biU}}
        \includegraphics[width=\linewidth]{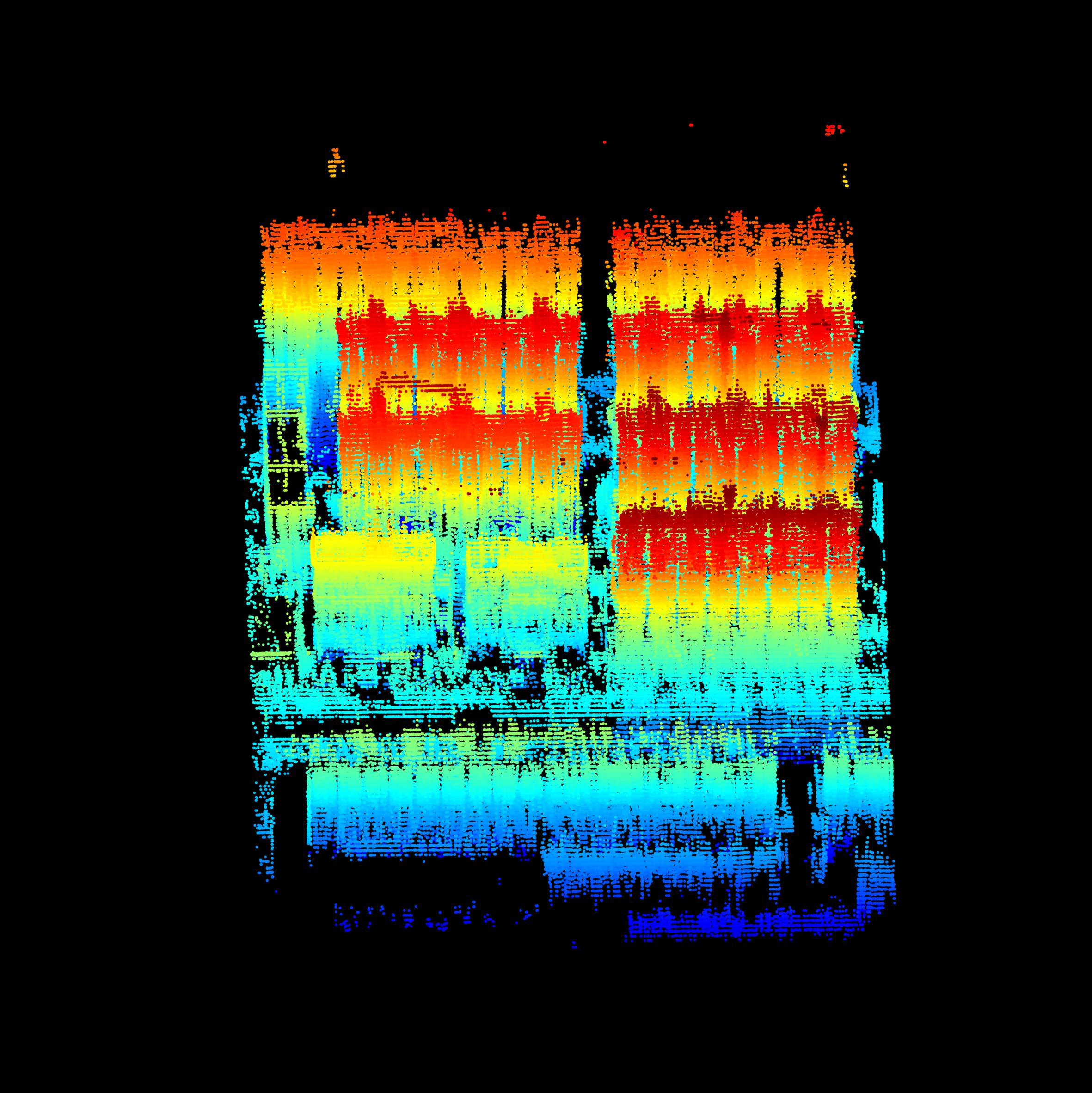}
        \caption{}
        \label{figure28_6}
    \end{subfigure}
    \begin{subfigure}[b]{0.24\linewidth}
        \caption*{{\fontsize{8pt}{10pt}\selectfont\centering  tomo-LRENet-LSTM}}
        \includegraphics[width=\linewidth]{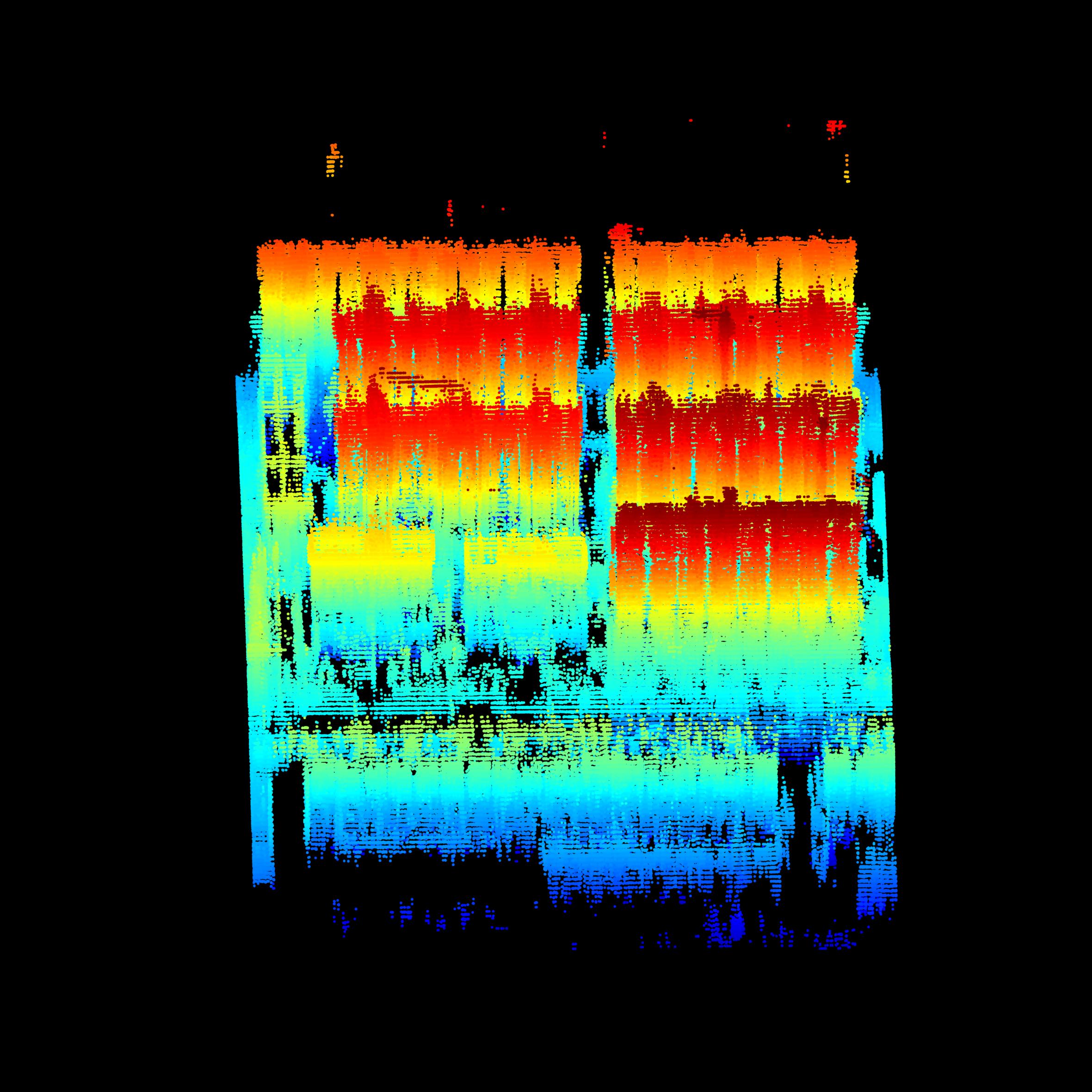}
        \caption{}
        \label{figure28_7}
    \end{subfigure}
    \caption{Results of test5. (a) Result of FISTA-based method. (b) Result of SLIMMER. (c) Result of tomo-IRENet-Raw. (d) Result of tomo-IRENet-TV. (e) Result of tomo-LRENet-U. (f) Result of tomo-LRENet-biU. (g) Result of tomo-LRENet-LSTM.}
    \label{figure_28}
\end{figure}

Starting from this point, we utilize public measured data. Initially, we employ the simpler Yuncheng data, which comprises a uniform urban structure of residential buildings. The results are depicted in Fig.~\ref{figure_28}.

Regarding outliers, SLIMMER's results contain the most, especially for areas at greater distances. Apart from SLIMMER, the remaining methods achieve similar levels, with the results from tomo-IRENET-TV and tomo-LRENET-LSTM being slightly worse.

Regarding wall surface completeness, FISTA's result is the poorest, followed by tomo-IRENET-Raw. The rest of the methods are on par. However, it's worth noting that FISTA and tomo-IRENet-TV have their drawbacks. FISTA improves completeness at the expense of more outliers, while tomo-IRENet-TV does so at the cost of degraded resolutions.

When evaluating the contour's smoothness, specifically the rooftops of buildings, tomo-LRENet-LSTM's result has the least jittered rooftop among tomo-IRENet-U, tomo-LRENet-biU, and tomo-LRENet-LSTM. The other two methods are comparable.

Taking all these various aspects into account, it becomes clear that tomo-LRENet-LSTM's performance offers the most balanced outcome. The results it delivers are preferred, making it the method of choice considering the balance it provides between the number of outliers, wall surface completeness, and contour smoothness.

\vspace{0.5em}
\textit{D.6 Results of test object 6 (diverse urban landscape challenge)}
\vspace{0.5em}

In this final test, we use Emei data to assess the performance of various methods across a diverse urban landscape. This landscape has significantly more variations than the previous case, including differences in height, orientations, and types (such as buses, trees, tower cranes, commercial centers, and residential apartment blocks). This diversity allows us to evaluate more detailed differences in the performance of the methods. 

\begin{figure*}[h!]
	\centering
	% 第一行三张图
	\begin{subfigure}[b]{0.32\linewidth}
		\caption*{{\fontsize{8pt}{10pt}\selectfont\centering  FISTA-based}}
		\includegraphics[width=\linewidth]{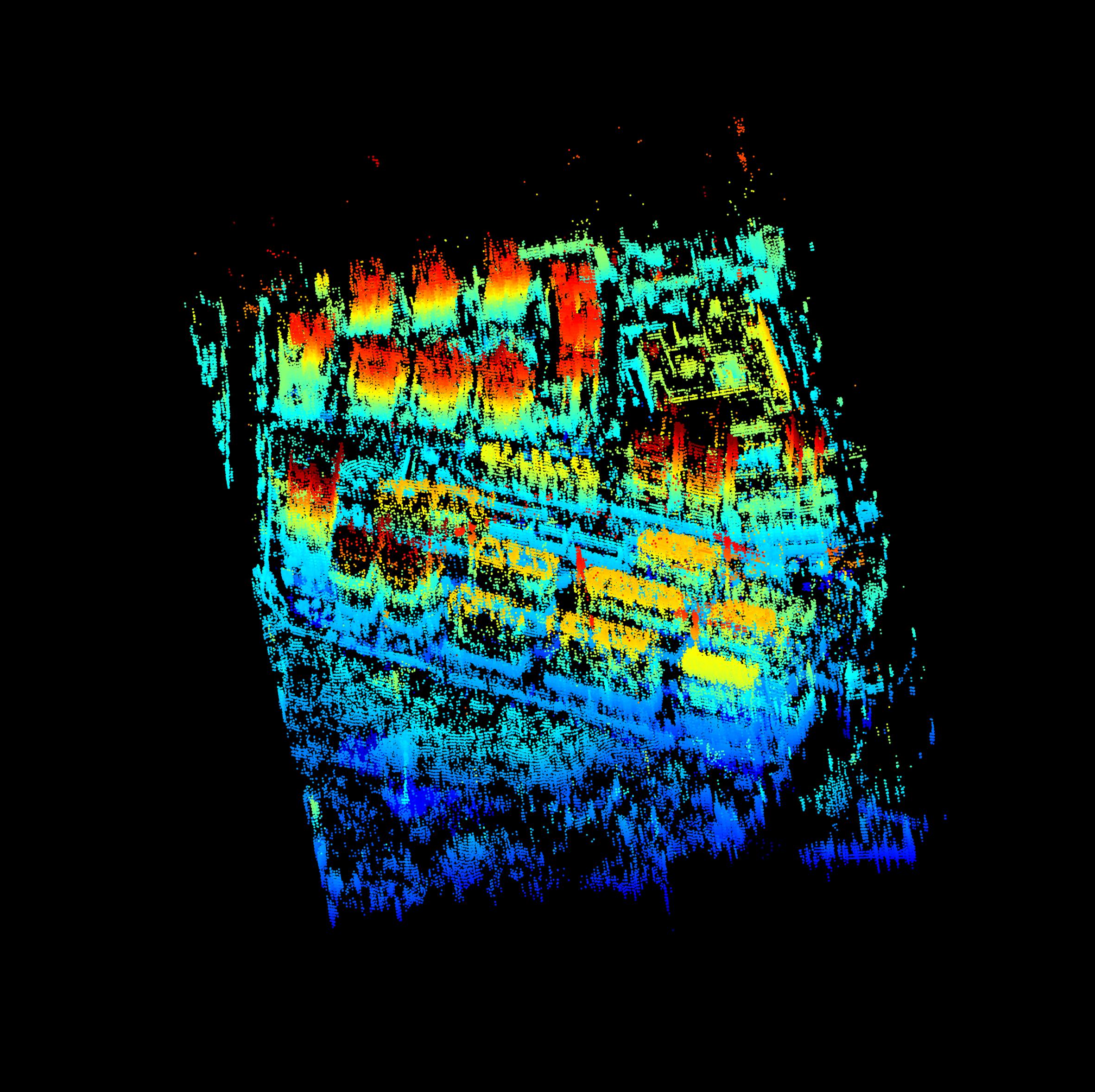}
		\caption{}
		\label{figure29_1}
	\end{subfigure}
	\begin{subfigure}[b]{0.32\linewidth}
		\caption*{{\fontsize{8pt}{10pt}\selectfont\centering SLIMMER}}
		\includegraphics[width=\linewidth]{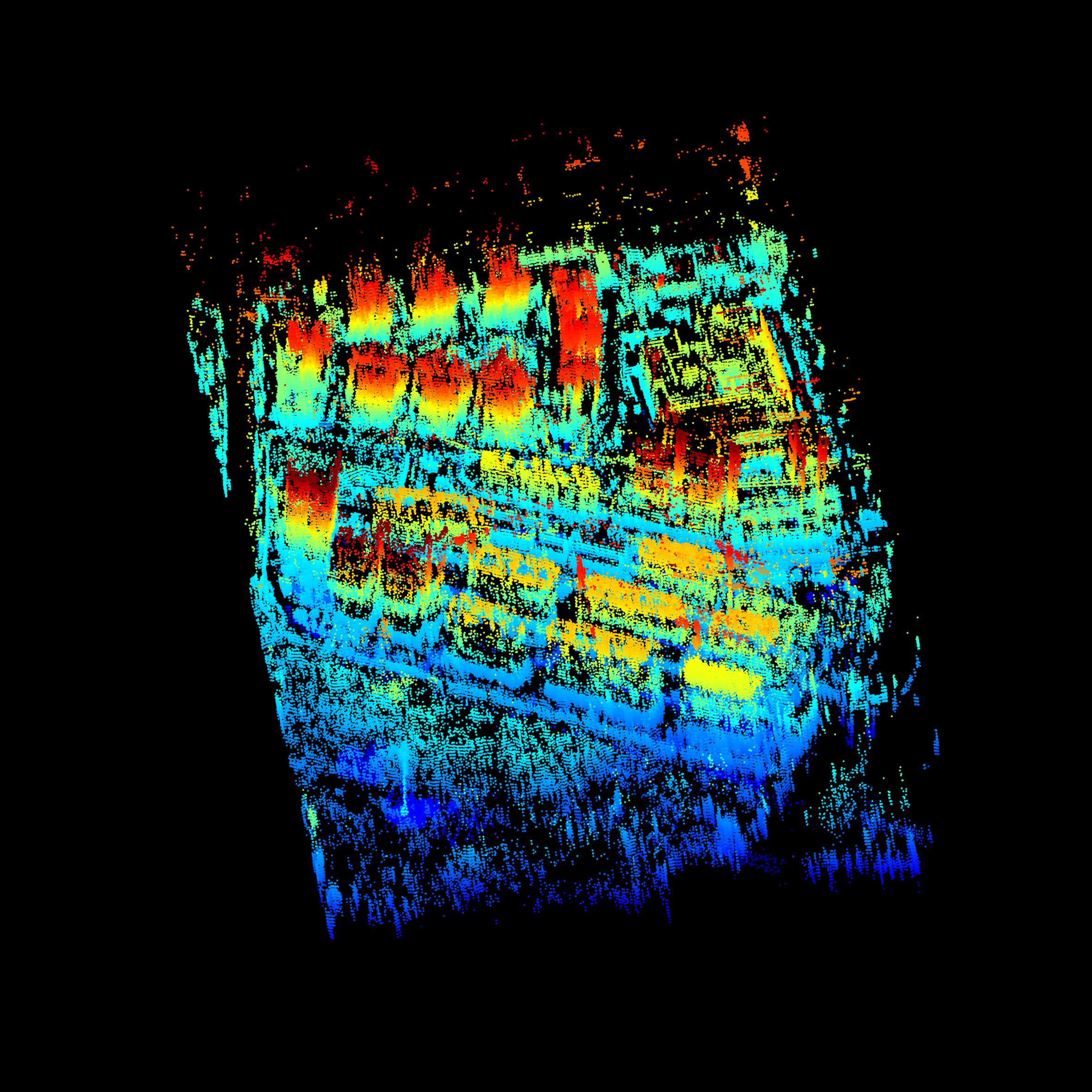}
		\caption{}
		\label{figure29_2}
	\end{subfigure}
	\begin{subfigure}[b]{0.32\linewidth}
		\caption*{{\fontsize{8pt}{10pt}\selectfont\centering tomo-IRENet-Raw}}
		\includegraphics[width=\linewidth]{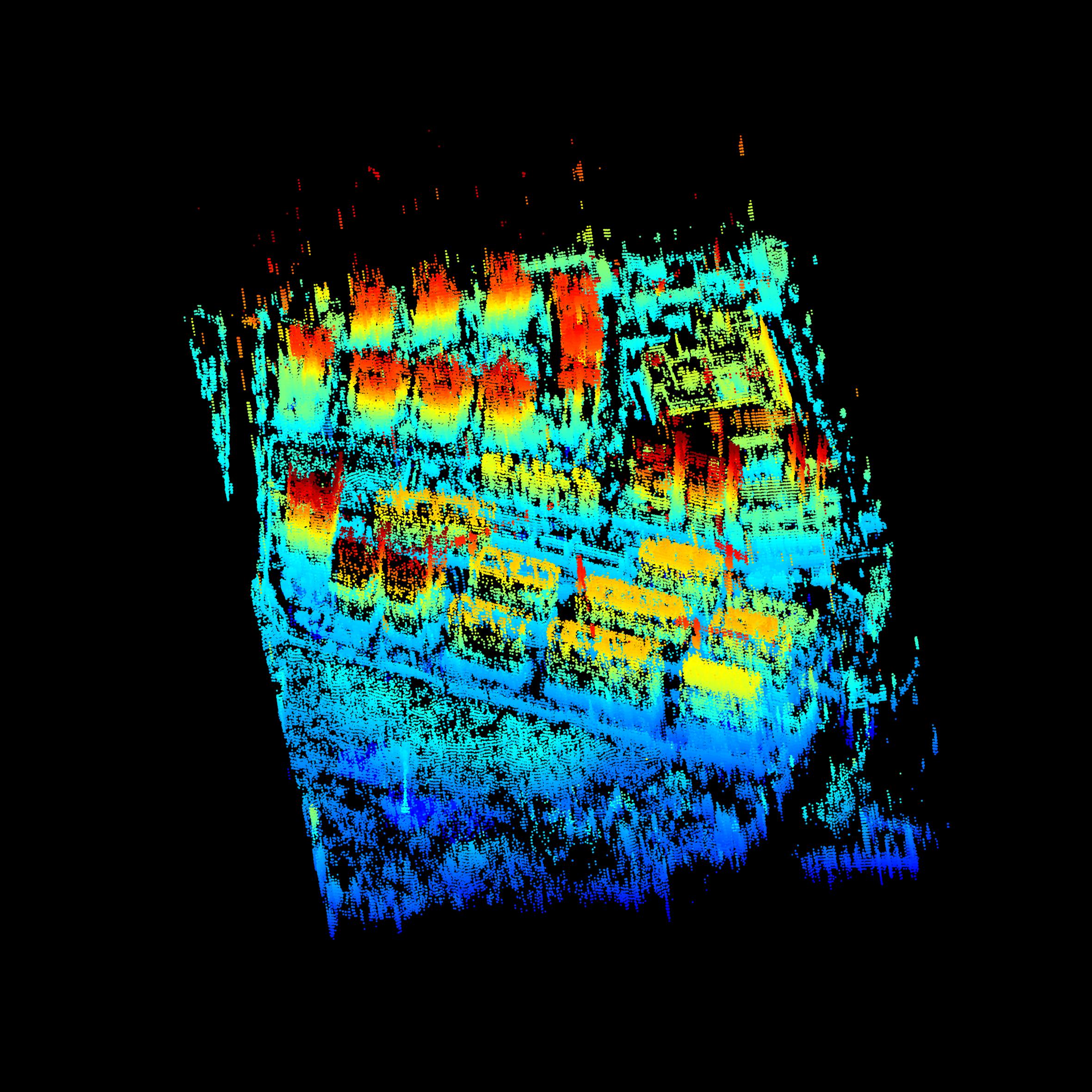}
		\caption{}
		\label{figure29_3}
	\end{subfigure}

	% 第二行两张图
	\begin{subfigure}[b]{0.32\linewidth}
		\caption*{{\fontsize{8pt}{10pt}\selectfont\centering  tomo-IRENet-TV}}
		\includegraphics[width=\linewidth]{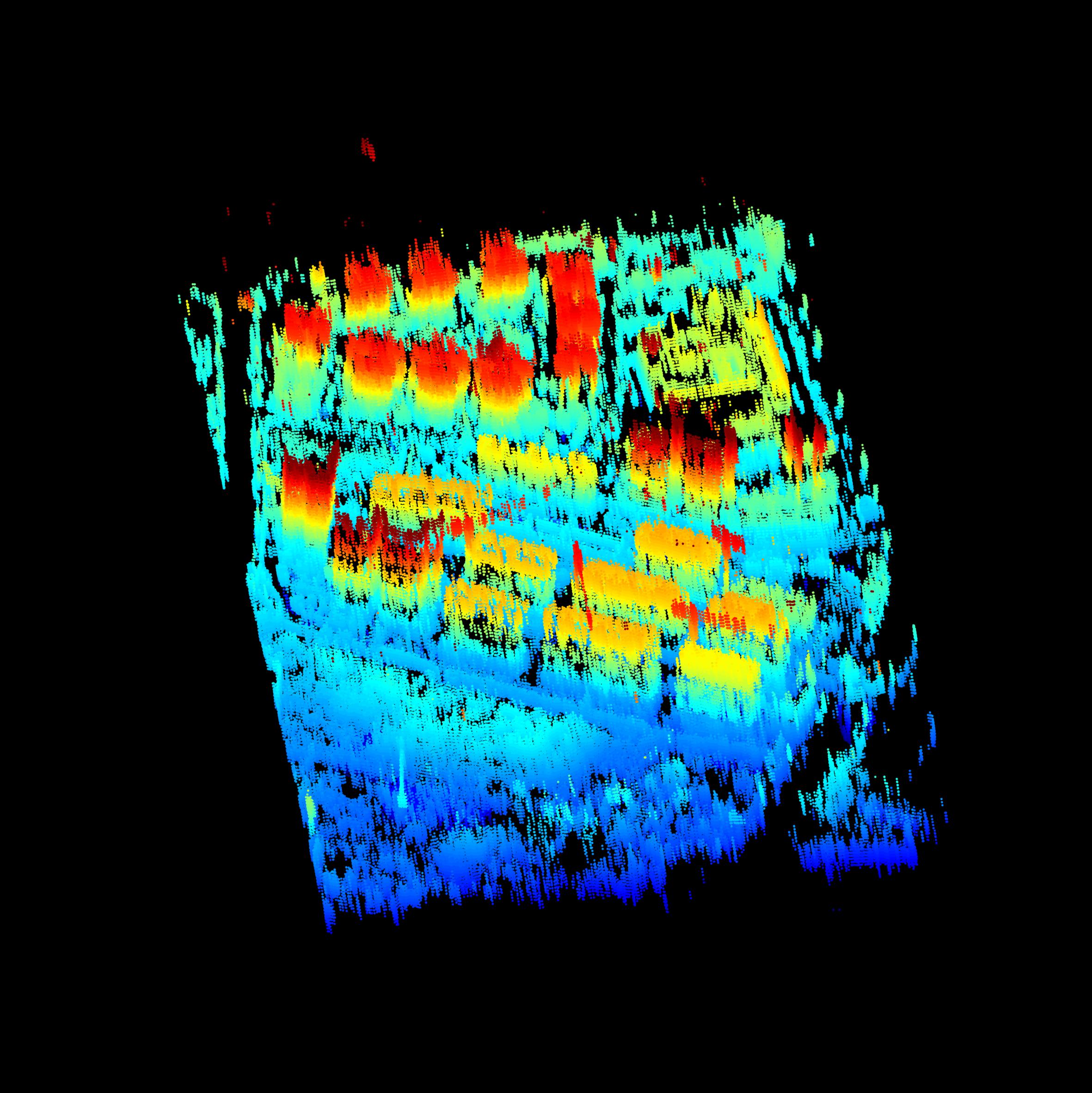}
		\caption{}
		\label{figure29_4}
	\end{subfigure}
	\hspace{20pt}
	\begin{subfigure}[b]{0.32\linewidth}
		\caption*{{\fontsize{8pt}{10pt}\selectfont\centering tomo-IRENet-U}}
		\includegraphics[width=\linewidth]{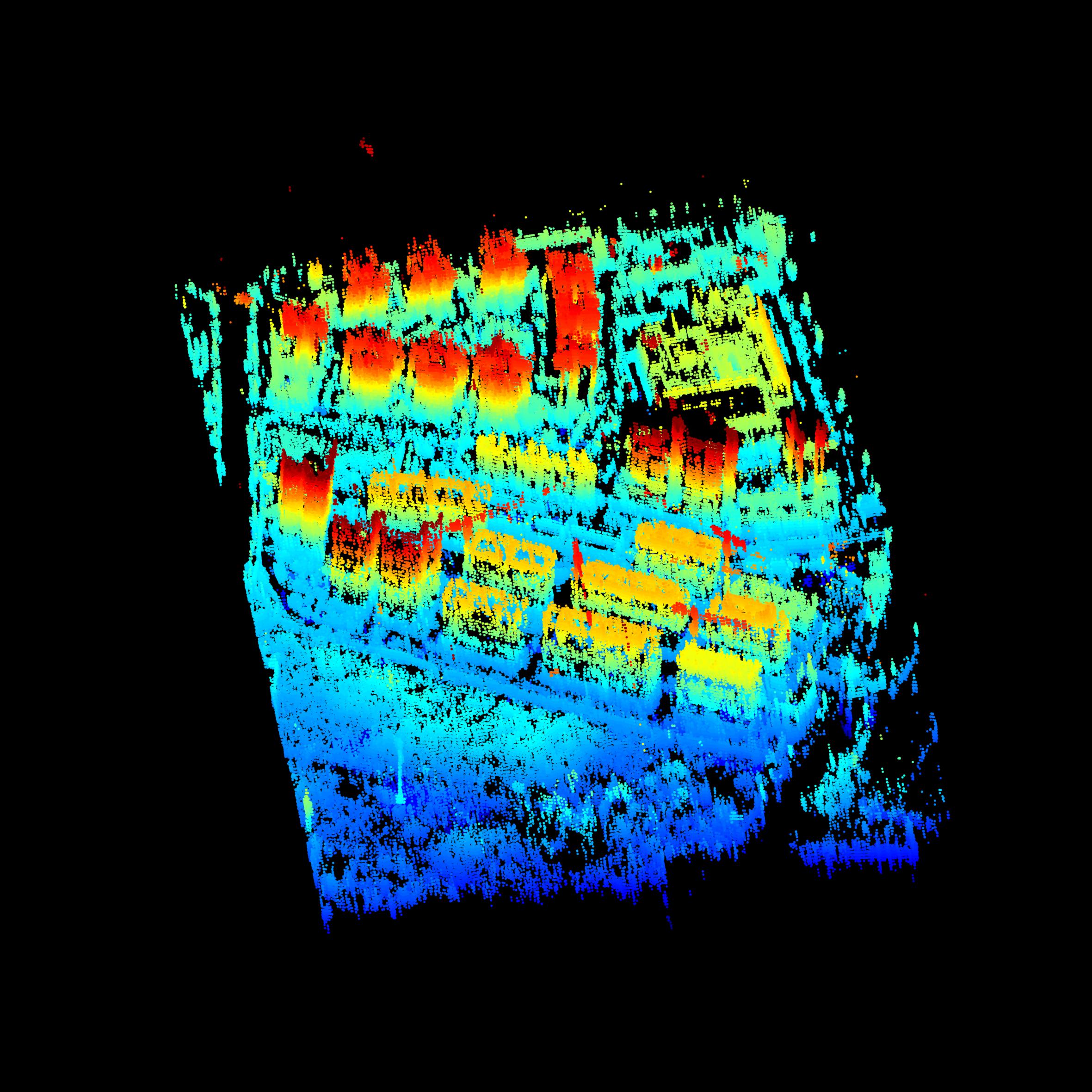}
		\caption{}
		\label{figure29_5}
	\end{subfigure}

	% 第三行两张图
	\begin{subfigure}[b]{0.32\linewidth}
		\caption*{{\fontsize{8pt}{10pt}\selectfont\centering tomo-LRENet-biU}}
		\includegraphics[width=\linewidth]{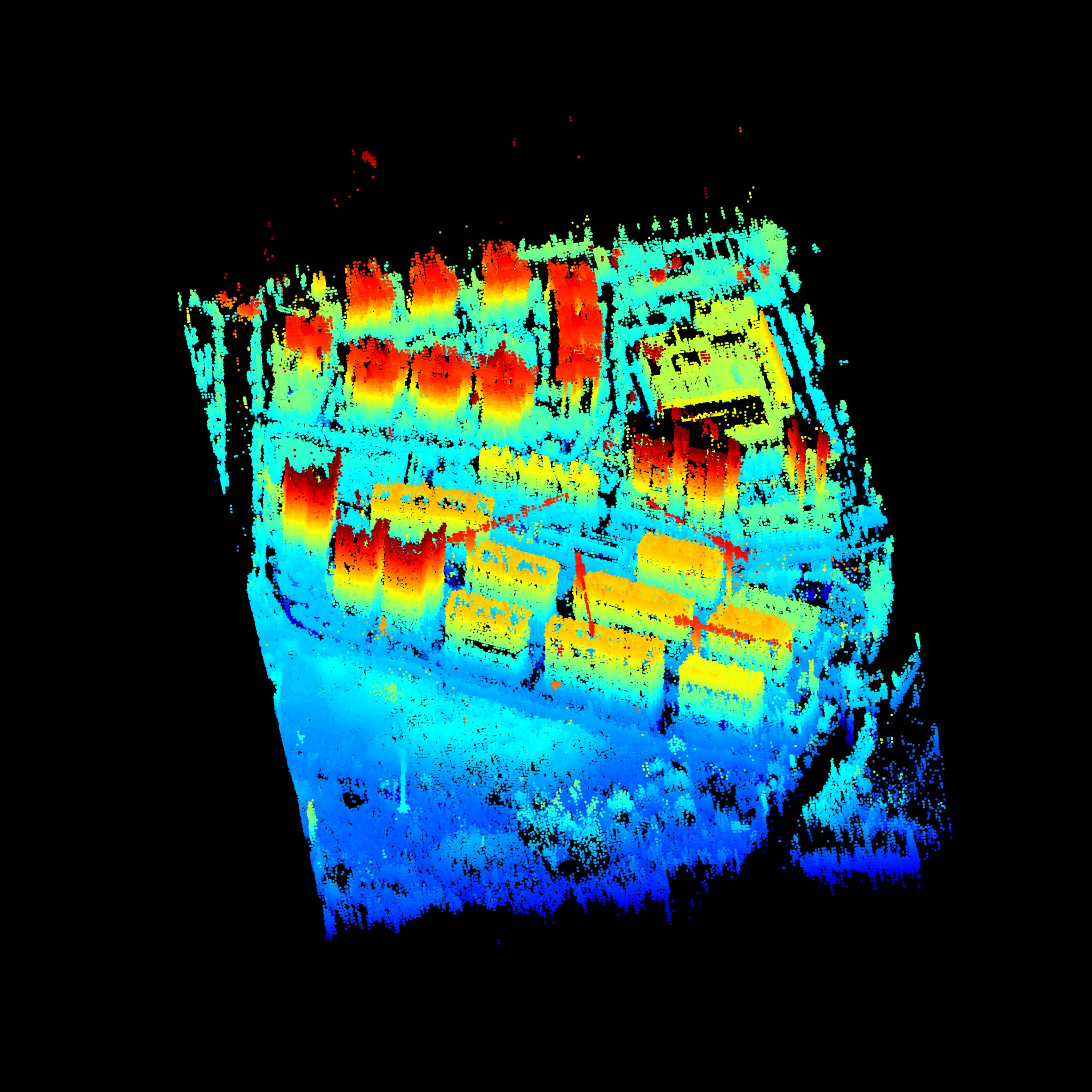}
		\caption{}
		\label{figure29_6}
	\end{subfigure}
	\hspace{20pt}
	\begin{subfigure}[b]{0.32\linewidth}
		\caption*{{\fontsize{8pt}{10pt}\selectfont\centering  tomo-LRENet-LSTM}}
		\includegraphics[width=\linewidth]{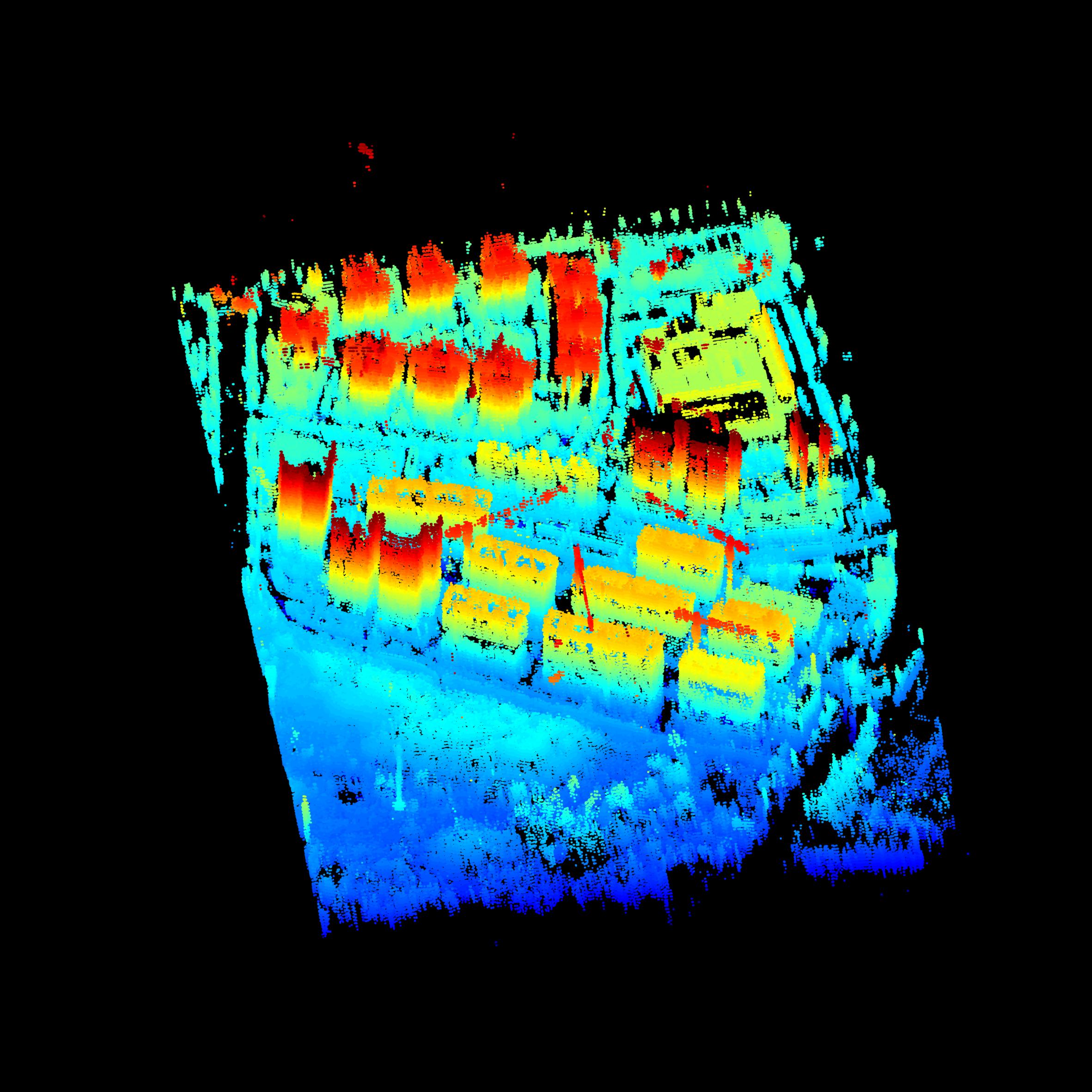}
		\caption{}
		\label{figure29_7}
	\end{subfigure}

	\caption{Results of test 6. (a) Result of FISTA-based method. (b) Result of SLIMMER. (c) Result of tomo-IRENet-Raw. (d) Result of tomo-IRENet-TV. (e) Result of tomo-LRENet-U. (f) Result of tomo-LRENet-biU. (g) Result of tomo-LRENet-LSTM.}
	\label{figure_29}
\end{figure*}

The results are displayed in Fig.~\ref{figure_29} at the following pages. The first row of results, which does not consider spatial features, contains more outliers compared to the other two rows that do. This is especially noticeable in the upper part of the scene, as primarily observed in the SLIMMER result. There are also noticeable gaps on the building surfaces, including walls and rooftops, especially in the lower part of the scene where three high-rise and six low-rise buildings are located. Considering these factors, the bottom four methods yield superior results, with the final two methods, tomo-LRENet-biU and tomo-LRENet-LSTM, performing the best. Besides the outlier and regular shaped surface aspects, the performance differnece also reflect from the line structure. As the lower part of the scene (the six low-rise buidlings) is under construction, there are four high-rise tower cranes existing, where their long arms are typical straight line structures. From the reconstructed arms, the results of tomo-LRENet-biU and tomo-LRENet-LSTM are superior in continuity compared to the other two methods among the bottom four methods. 

\begin{figure*}[h!]
	\centering
	% 第一排的四张图
	\begin{subfigure}[b]{0.24\linewidth}
		\caption*{{\fontsize{8pt}{10pt}\selectfont\centering scene}}
		\includegraphics[width=\linewidth]{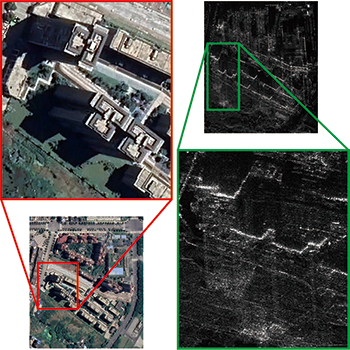}
		\caption{}
		\label{figure30_1}
	\end{subfigure}
	\begin{subfigure}[b]{0.24\linewidth}
		\caption*{{\fontsize{8pt}{10pt}\selectfont\centering FISTA-based}}
		\includegraphics[width=\linewidth]{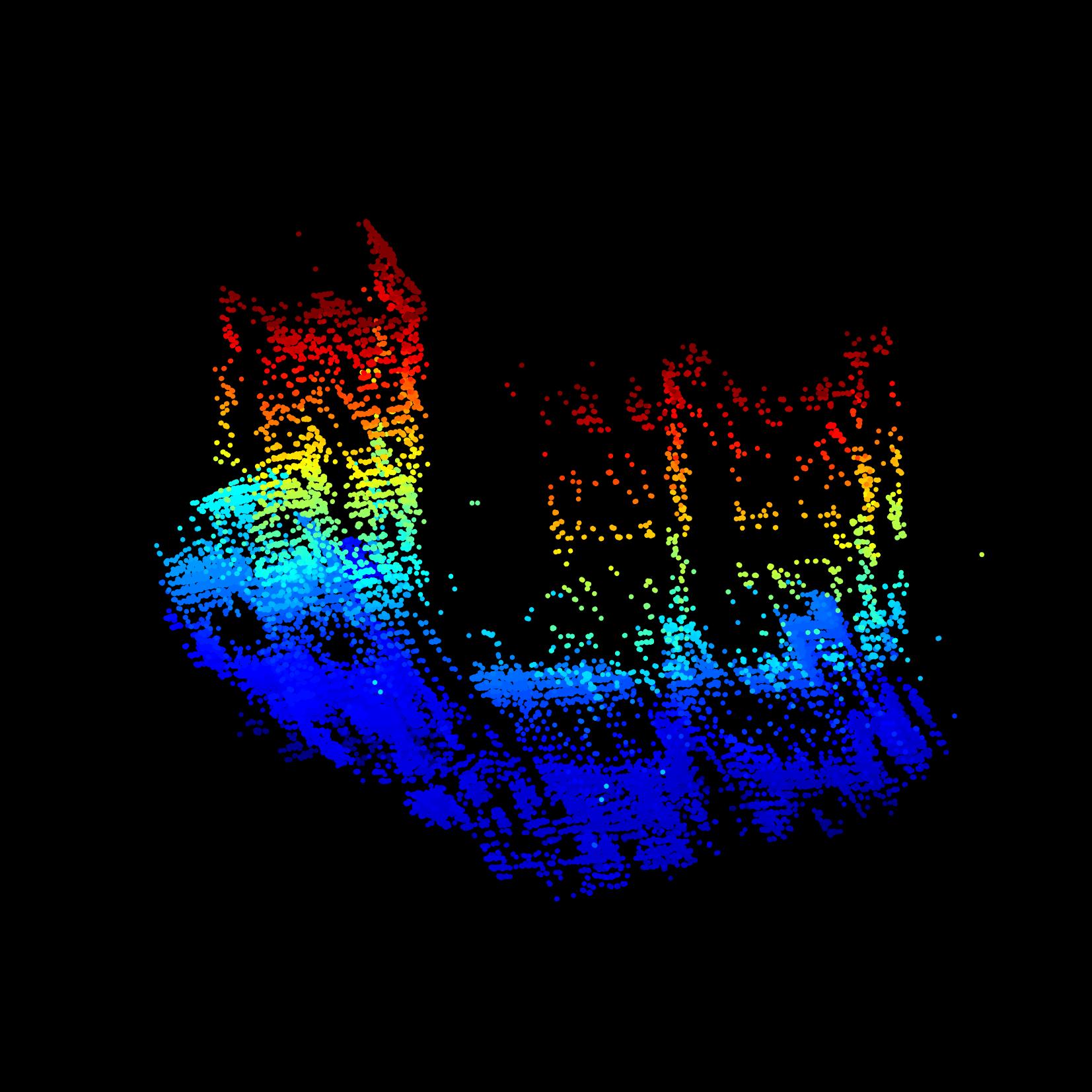}
		\caption{}
		\label{figure30_2}
	\end{subfigure}
	\begin{subfigure}[b]{0.24\linewidth}
		\caption*{{\fontsize{8pt}{10pt}\selectfont\centering SLIMMER}}
		\includegraphics[width=\linewidth]{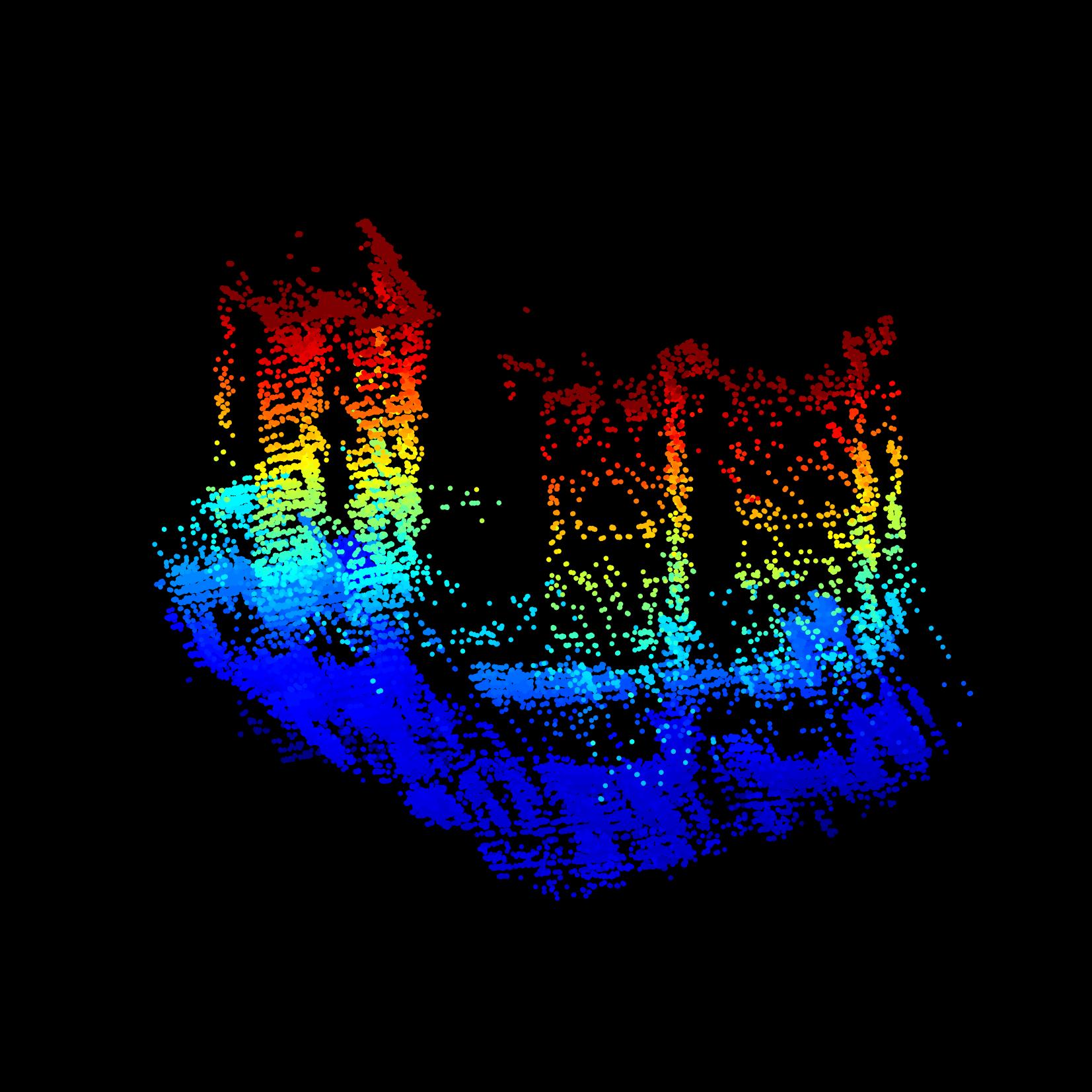}
		\caption{}
		\label{figure30_3}
	\end{subfigure}
	\begin{subfigure}[b]{0.24\linewidth}
		\caption*{{\fontsize{8pt}{10pt}\selectfont\centering tomo-IRENet-Raw}}
		\includegraphics[width=\linewidth]{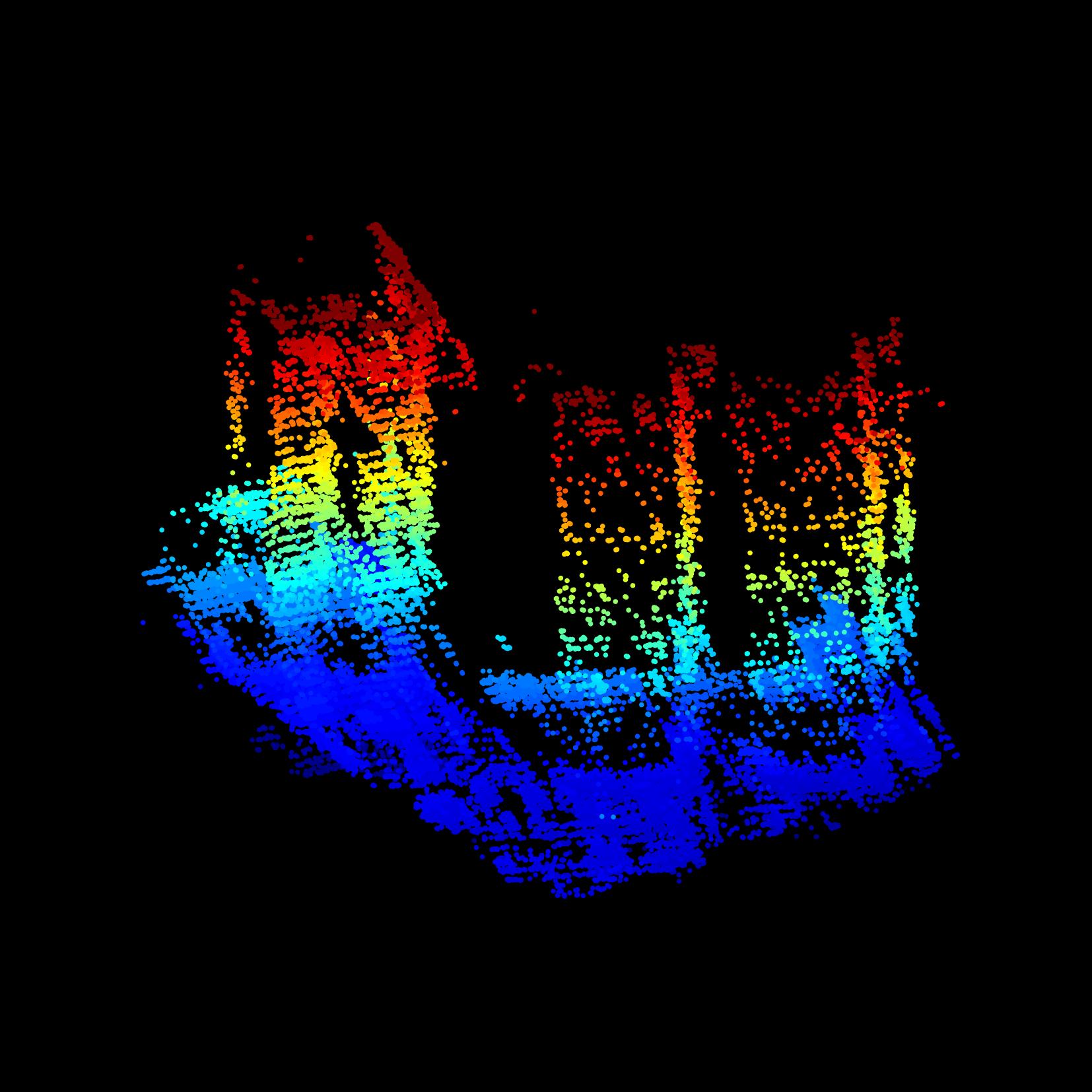}
		\caption{}
		\label{figure30_4}
	\end{subfigure}
	
	% 第二排的四张图
	\begin{subfigure}[b]{0.24\linewidth}
		\caption*{{\fontsize{8pt}{10pt}\selectfont\centering tomo-IRENet-TV}}
		\includegraphics[width=\linewidth]{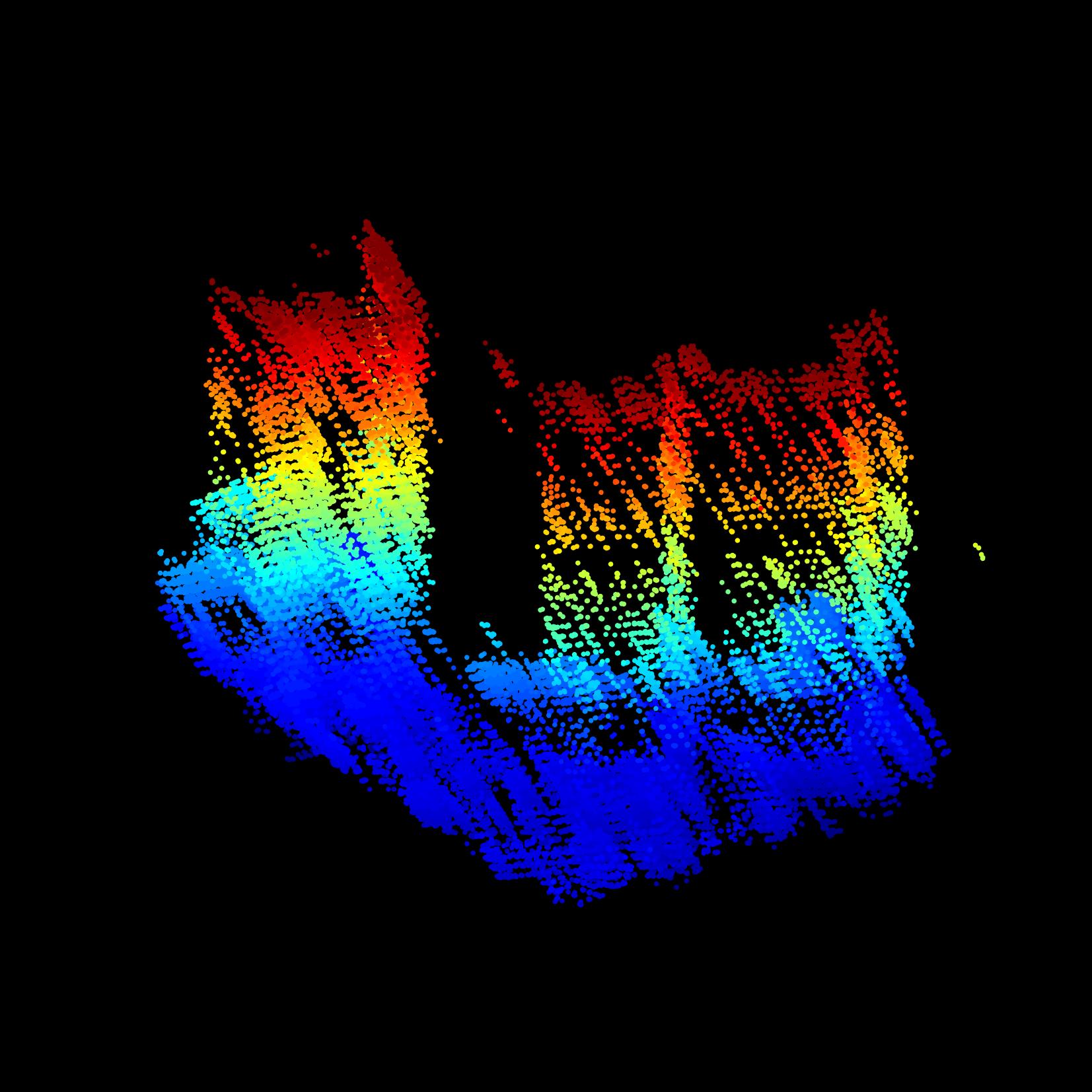}
		\caption{}
		\label{figure30_5}
	\end{subfigure}
	\begin{subfigure}[b]{0.24\linewidth}
		\caption*{{\fontsize{8pt}{10pt}\selectfont\centering tomo-IRENet-U}}
		\includegraphics[width=\linewidth]{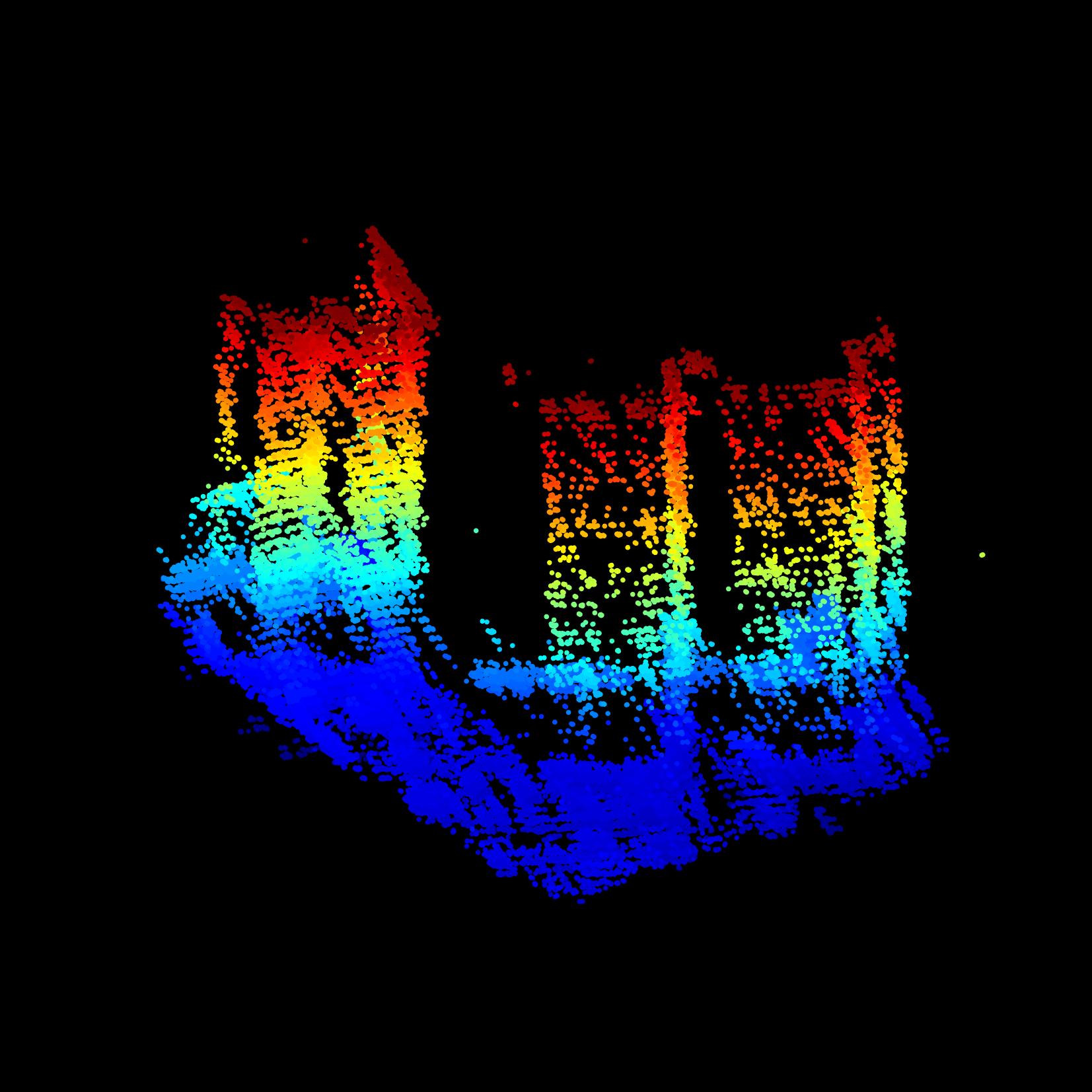}
		\caption{}
		\label{figure30_6}
	\end{subfigure}
	\begin{subfigure}[b]{0.24\linewidth}
		\caption*{{\fontsize{8pt}{10pt}\selectfont\centering tomo-LRENet-biU}}
		\includegraphics[width=\linewidth]{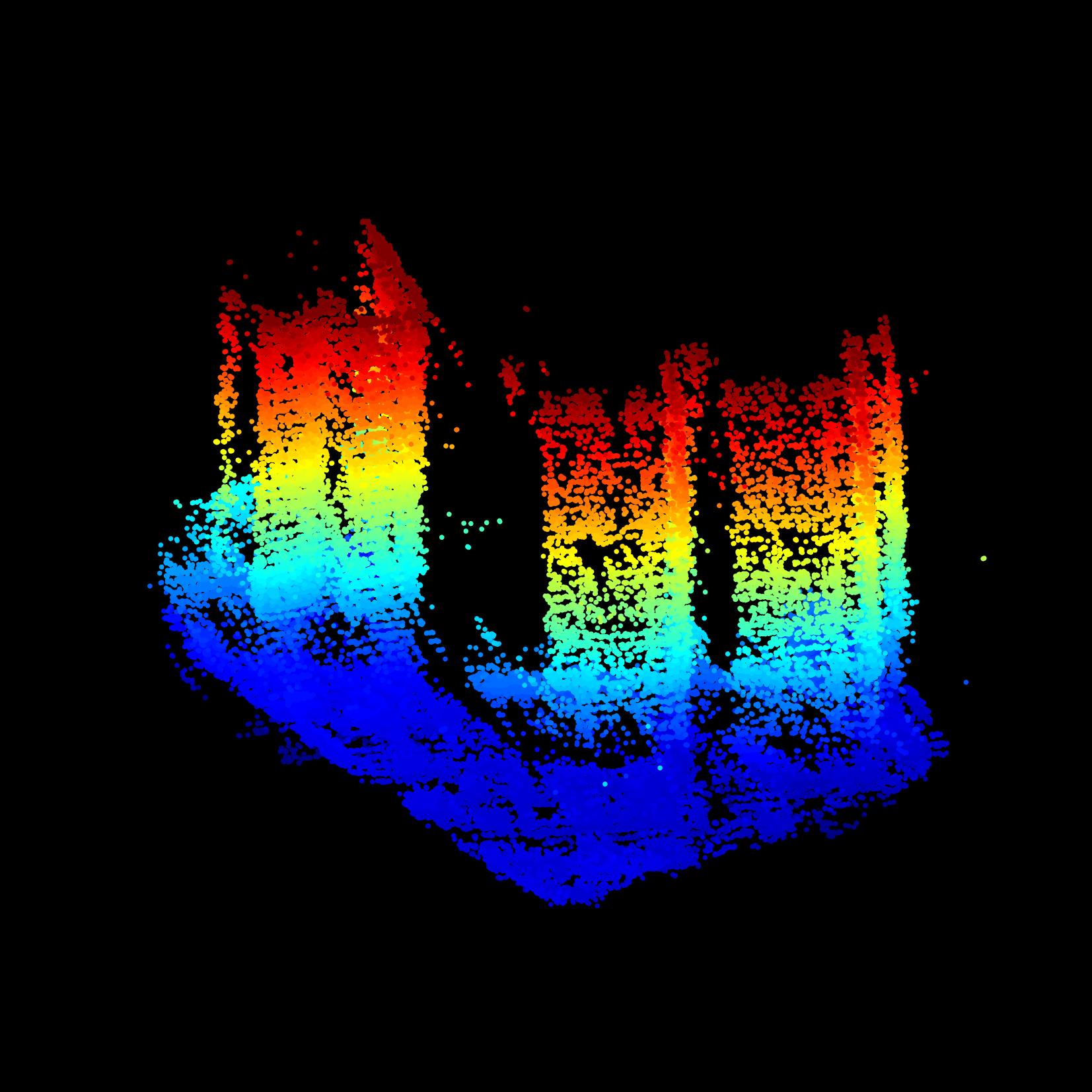}
		\caption{}
		\label{figure30_7}
	\end{subfigure}
	\begin{subfigure}[b]{0.24\linewidth}
		\caption*{{\fontsize{8pt}{10pt}\selectfont\centering tomo-LRENet-LSTM}}
		\includegraphics[width=\linewidth]{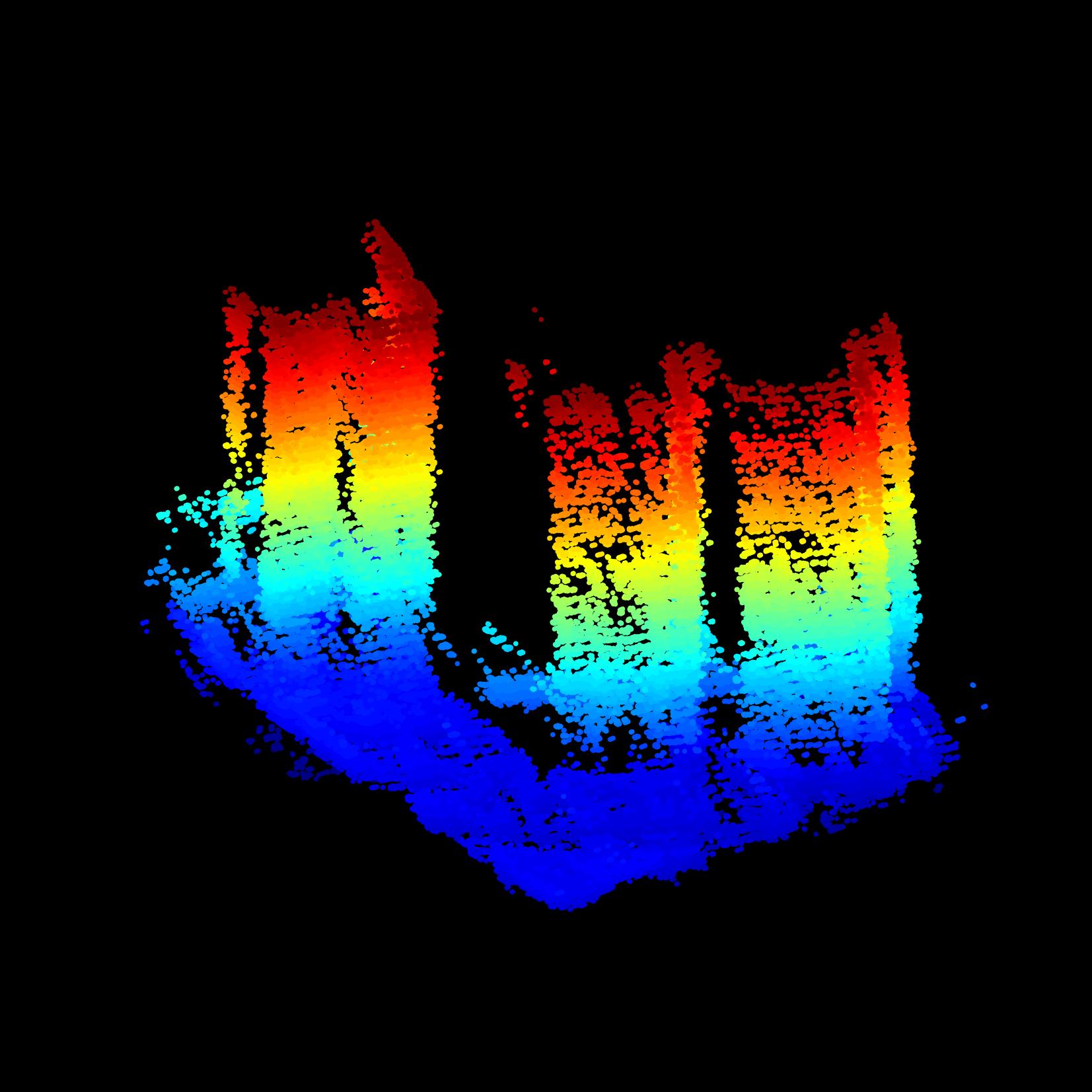}
		\caption{}
		\label{figure30_8}
	\end{subfigure}

	\begin{subfigure}[b]{0.24\linewidth}
		\caption*{{\fontsize{8pt}{10pt}\selectfont\centering scene}}
		\includegraphics[width=\linewidth]{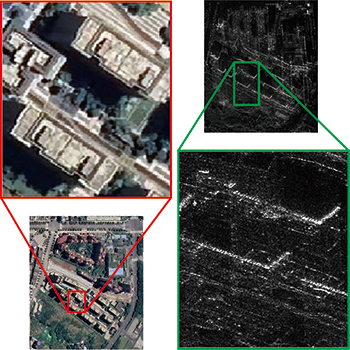}
		\caption{}
		\label{figure30_2_1}
	\end{subfigure}
	\begin{subfigure}[b]{0.24\linewidth}
		\caption*{{\fontsize{8pt}{10pt}\selectfont\centering FISTA-based}}
		\includegraphics[width=\linewidth]{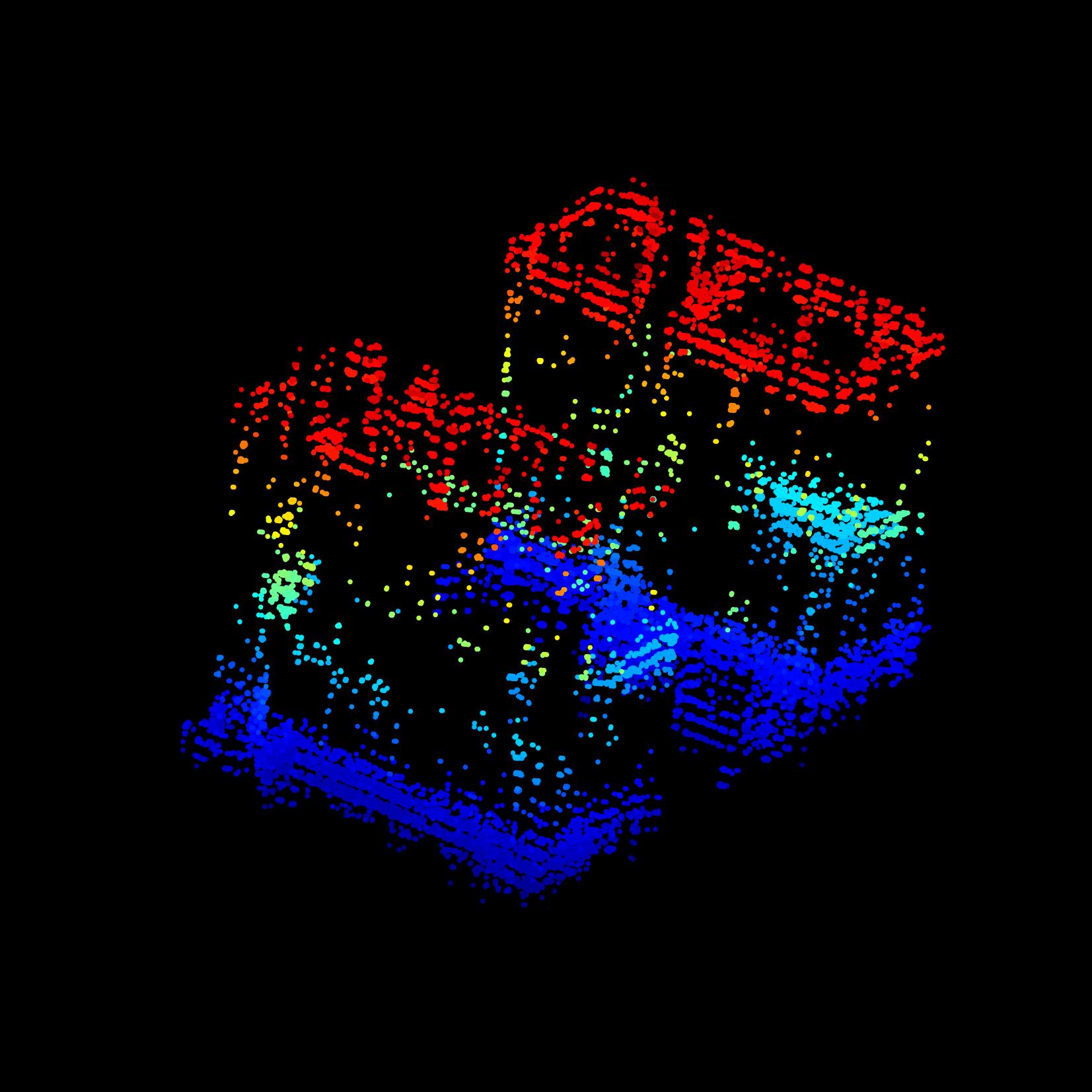}
		\caption{}
		\label{figure30_2_2}
	\end{subfigure}
	\begin{subfigure}[b]{0.24\linewidth}
		\caption*{{\fontsize{8pt}{10pt}\selectfont\centering SLIMMER}}
		\includegraphics[width=\linewidth]{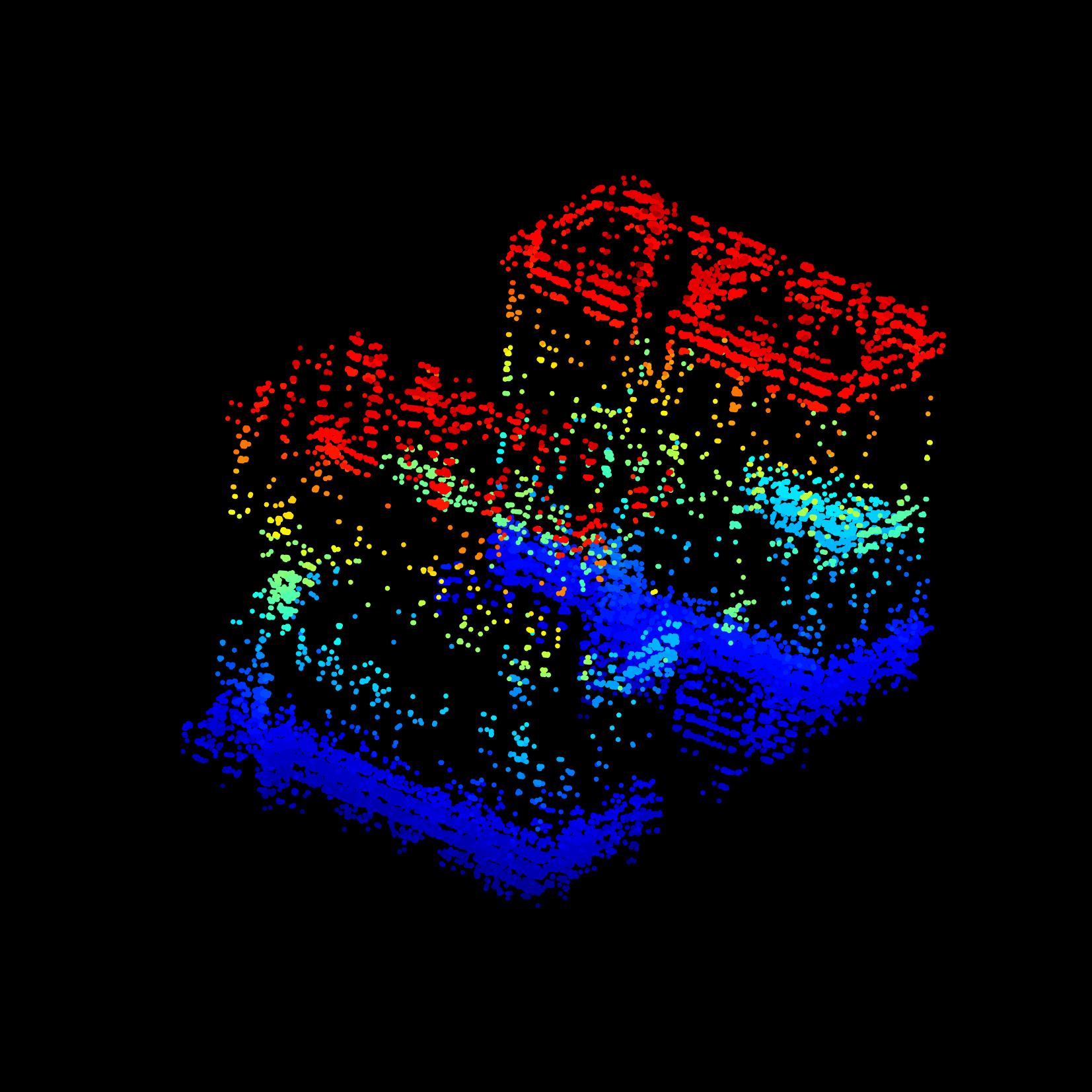}
		\caption{}
		\label{figure30_2_3}
	\end{subfigure}
	\begin{subfigure}[b]{0.24\linewidth}
		\caption*{{\fontsize{8pt}{10pt}\selectfont\centering tomo-IRENet-Raw}}
		\includegraphics[width=\linewidth]{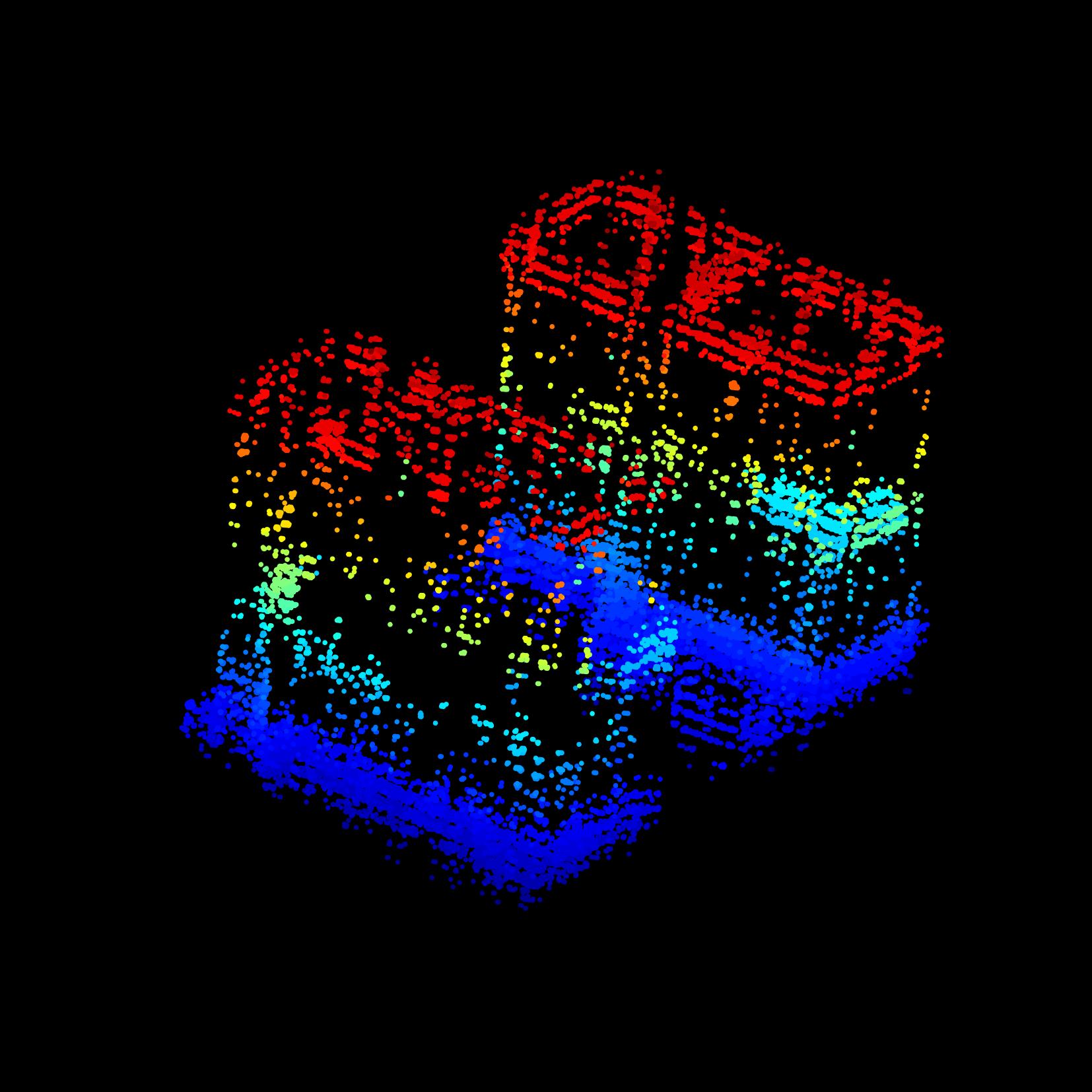}
		\caption{}
		\label{figure30_2_4}
	\end{subfigure}
	
	% 第二排的四张图
	\begin{subfigure}[b]{0.24\linewidth}
		\caption*{{\fontsize{8pt}{10pt}\selectfont\centering tomo-IRENet-TV}}
		\includegraphics[width=\linewidth]{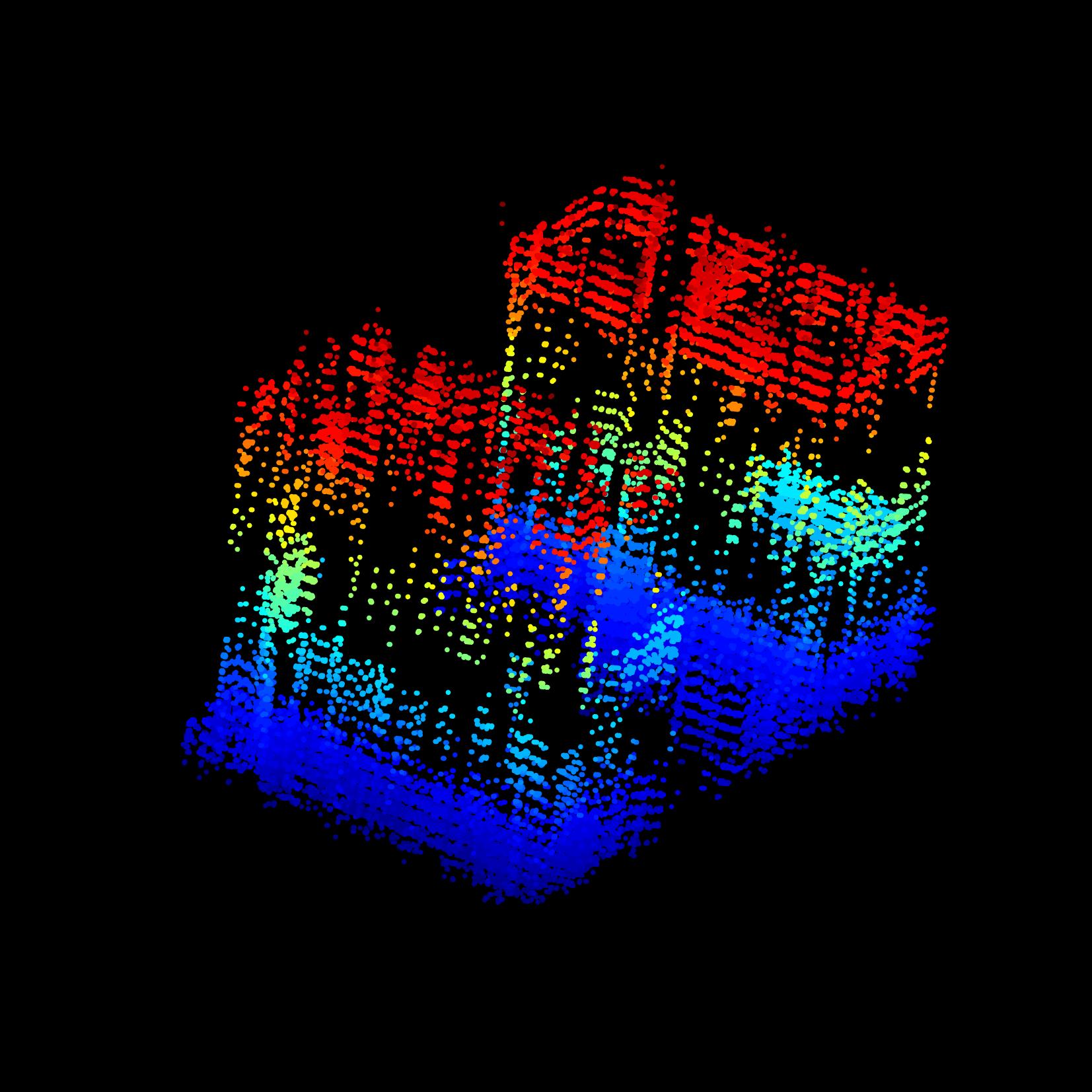}
		\caption{}
		\label{figure30_2_5}
	\end{subfigure}
	\begin{subfigure}[b]{0.24\linewidth}
		\caption*{{\fontsize{8pt}{10pt}\selectfont\centering tomo-IRENet-U}}
		\includegraphics[width=\linewidth]{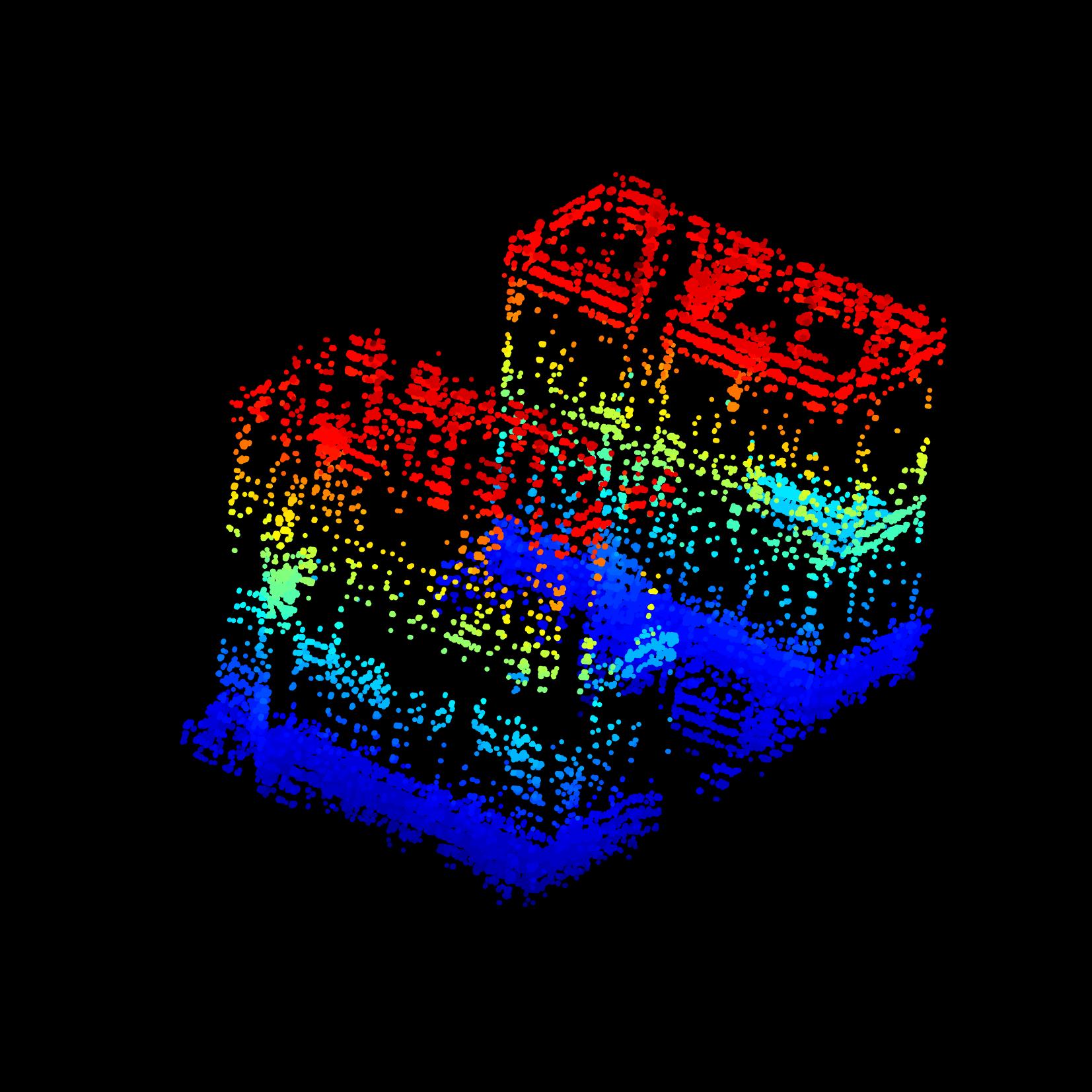}
		\caption{}
		\label{figure0_2_6}
	\end{subfigure}
	\begin{subfigure}[b]{0.24\linewidth}
		\caption*{{\fontsize{8pt}{10pt}\selectfont\centering tomo-LRENet-biU}}
		\includegraphics[width=\linewidth]{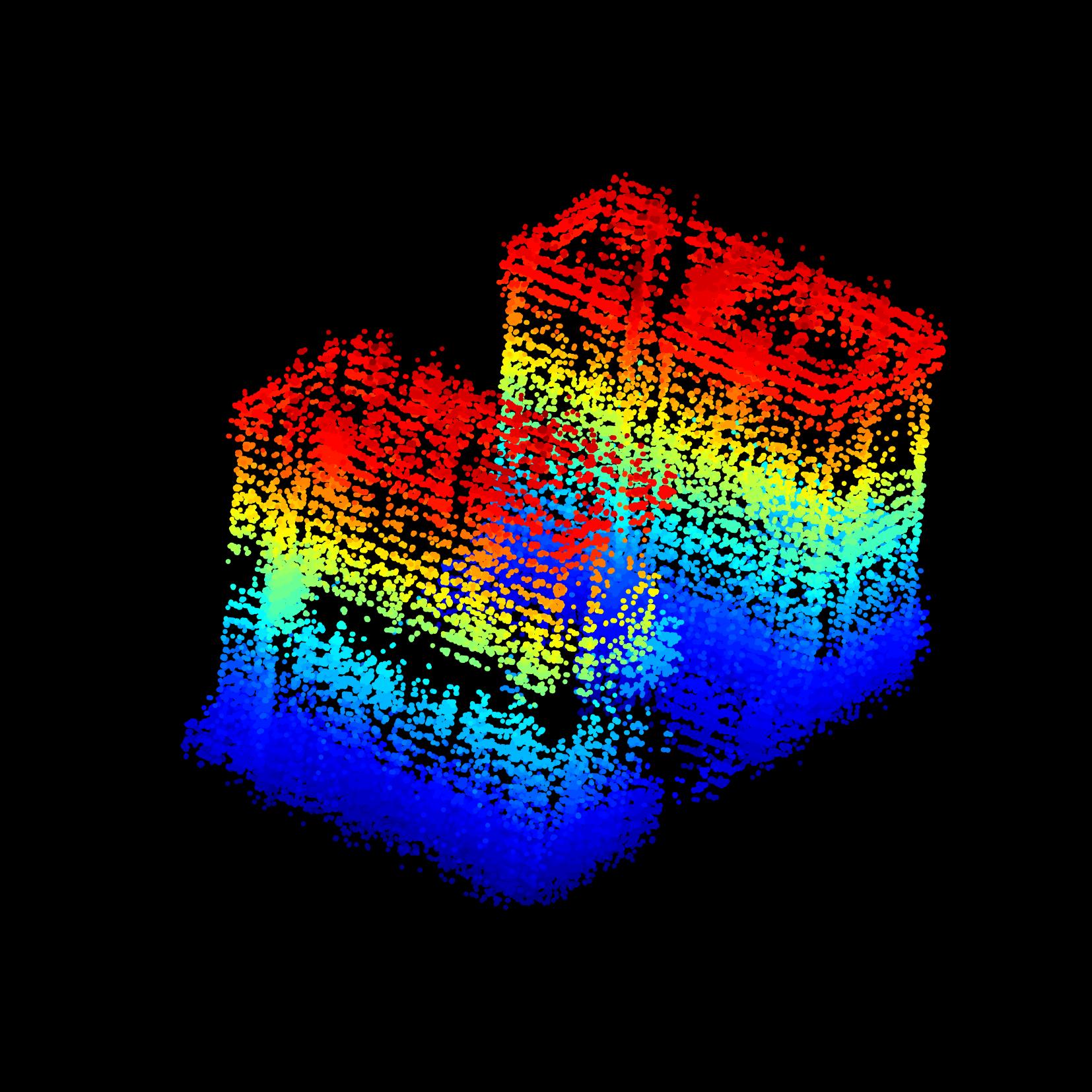}
		\caption{}
		\label{figure30_2_7}
	\end{subfigure}
	\begin{subfigure}[b]{0.24\linewidth}
		\caption*{{\fontsize{8pt}{10pt}\selectfont\centering tomo-LRENet-LSTM}}
		\includegraphics[width=\linewidth]{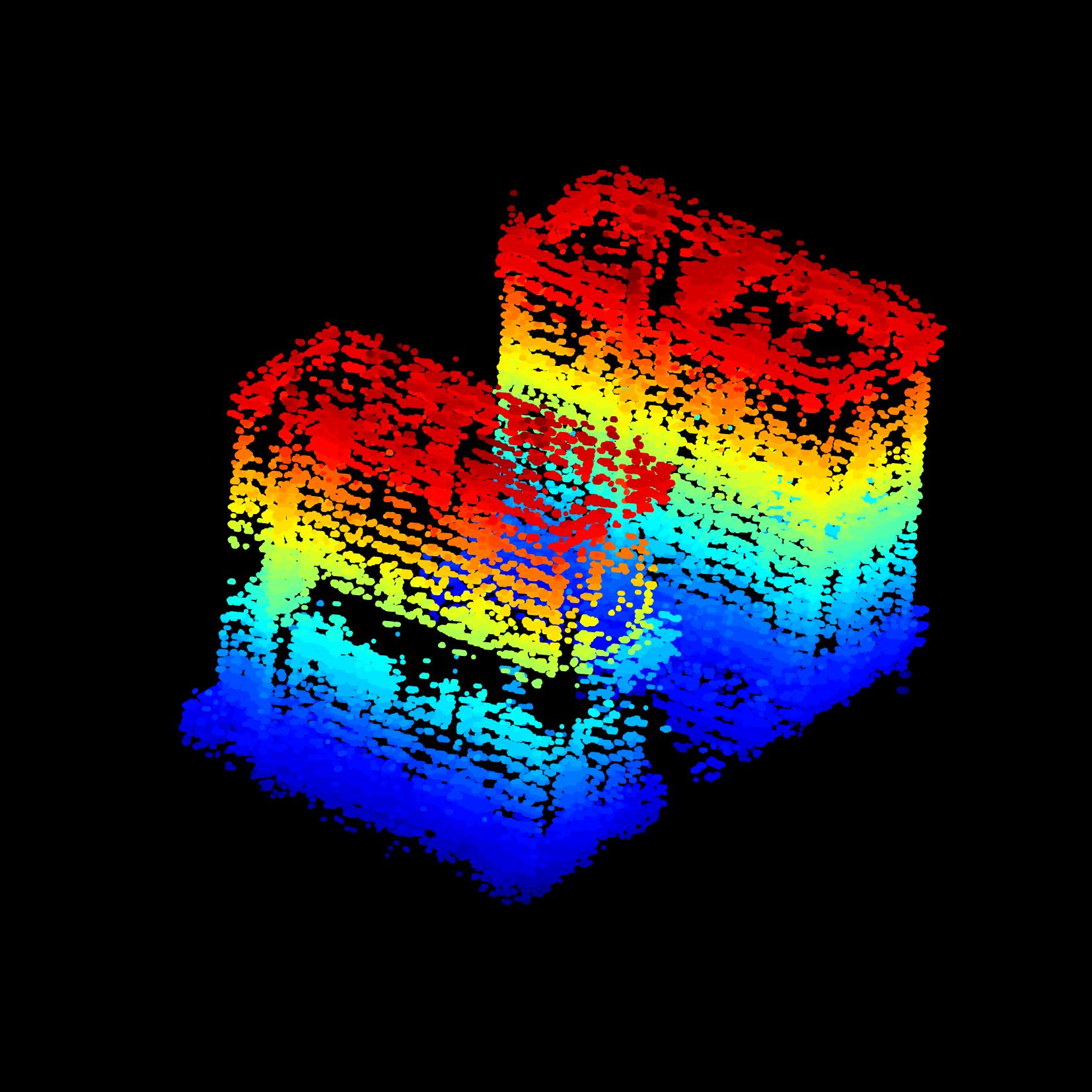}
		\caption{}
		\label{figure30_2_8}
	\end{subfigure}

	\caption{Detailed results of test 6, comparing reconstruction methods across two scenes. Scene 1 features a group of three regular-shaped residential buildings, while scene 2 includes two low-rise buildings with different orientations. (a) and (i) show optical and radar images for Scenes 1 and 2, respectively. (b-h) detail the reconstruction results for Scene 1 and (j-p) for Scene 2, from the FISTA-based method, SLIMMER, tomo-IRENet-Raw, tomo-IRENet-TV, tomo-IRENet-U, tomo-LRENet-biU, and tomo-LRENet-LSTM.} 
	\label{figure_30}
\end{figure*}

\begin{figure*}[h!]
	\centering
	% 第一排的四张图
	\begin{subfigure}[b]{0.24\linewidth}
		\caption*{{\fontsize{8pt}{10pt}\selectfont\centering scene}}
		\includegraphics[width=\linewidth]{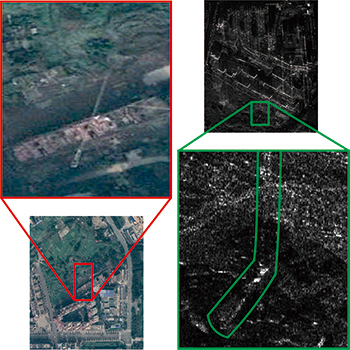}
		\caption{}
		\label{figure31_1_1}
	\end{subfigure}
	\begin{subfigure}[b]{0.24\linewidth}
		\caption*{{\fontsize{8pt}{10pt}\selectfont\centering FISTA-based}}
		\includegraphics[width=\linewidth]{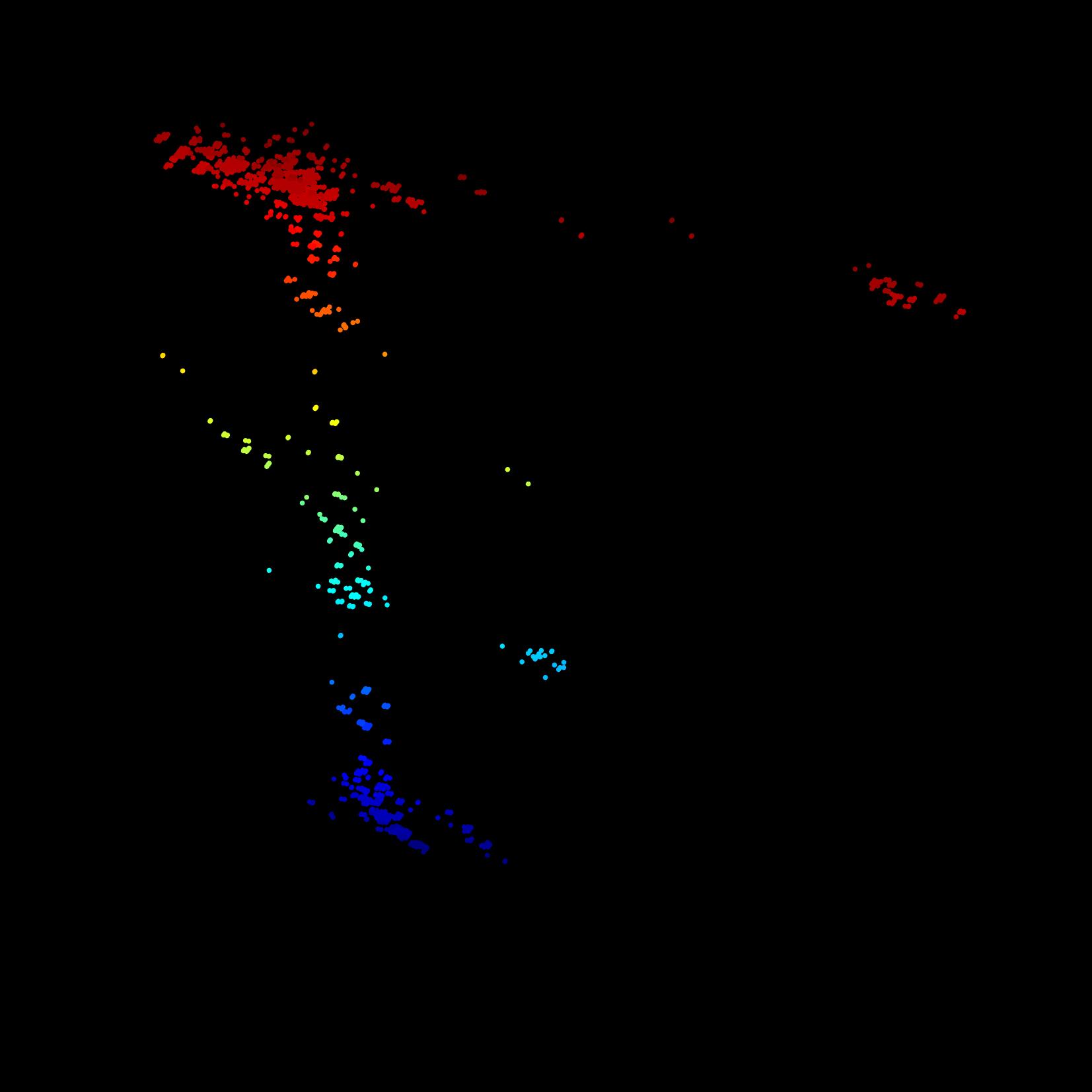}
		\caption{}
		\label{figure31_1_2}
	\end{subfigure}
	\begin{subfigure}[b]{0.24\linewidth}
		\caption*{{\fontsize{8pt}{10pt}\selectfont\centering SLIMMER}}
		\includegraphics[width=\linewidth]{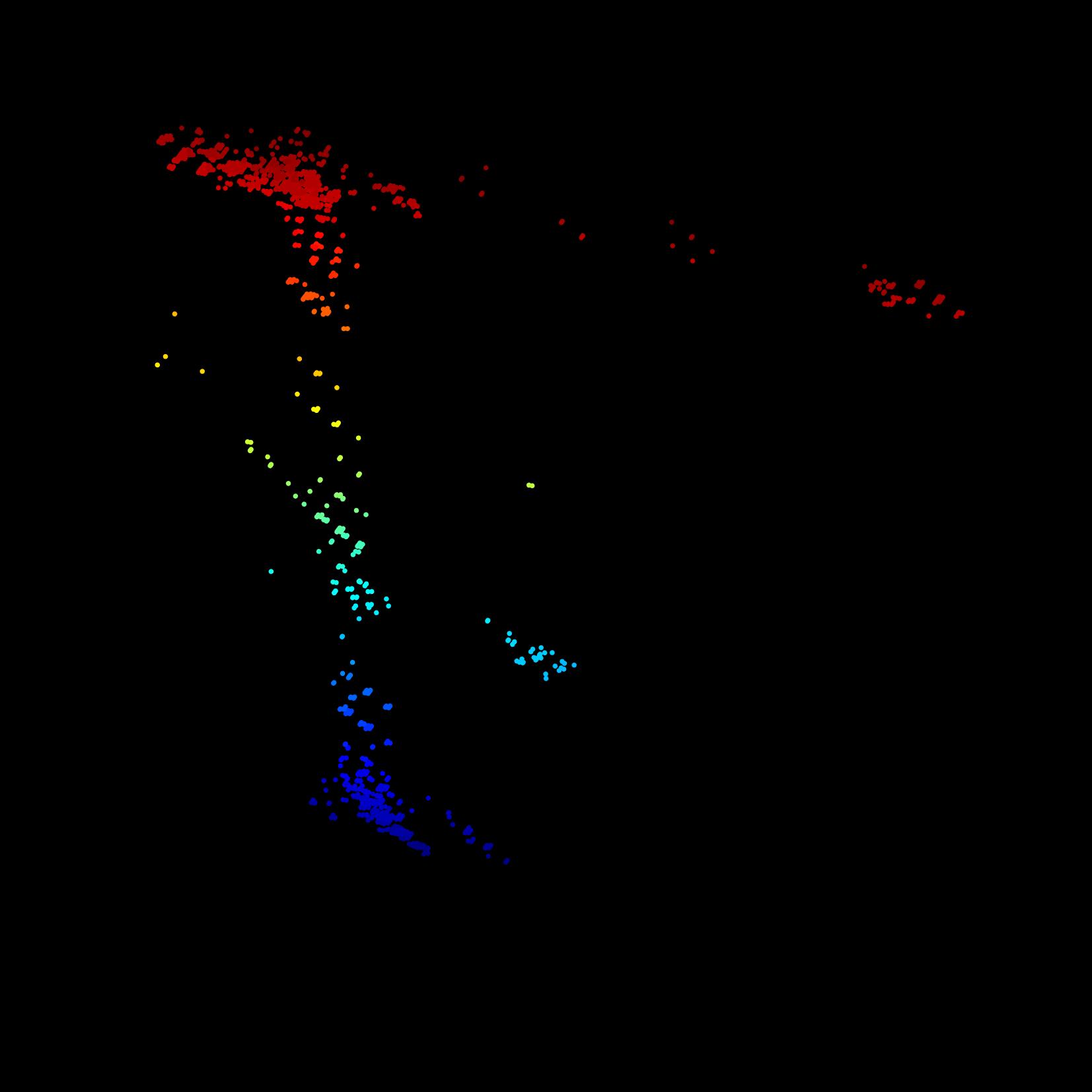}
		\caption{}
		\label{figure31_1_3}
	\end{subfigure}
	\begin{subfigure}[b]{0.24\linewidth}
		\caption*{{\fontsize{8pt}{10pt}\selectfont\centering tomo-IRENet-Raw}}
		\includegraphics[width=\linewidth]{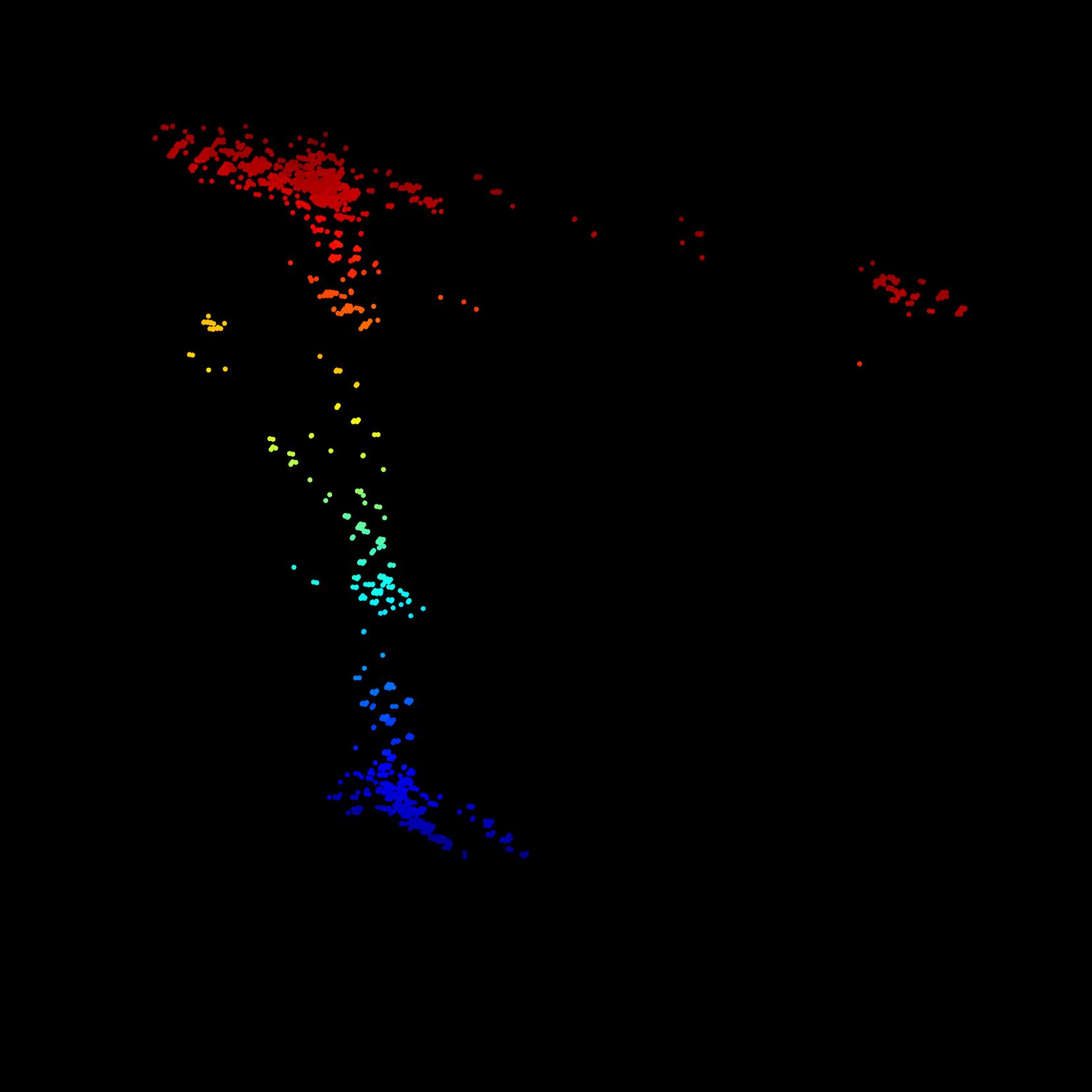}
		\caption{}
		\label{figure31_1_4}
	\end{subfigure}
	
	% 第二排的四张图
	\begin{subfigure}[b]{0.24\linewidth}
		\caption*{{\fontsize{8pt}{10pt}\selectfont\centering tomo-IRENet-TV}}
		\includegraphics[width=\linewidth]{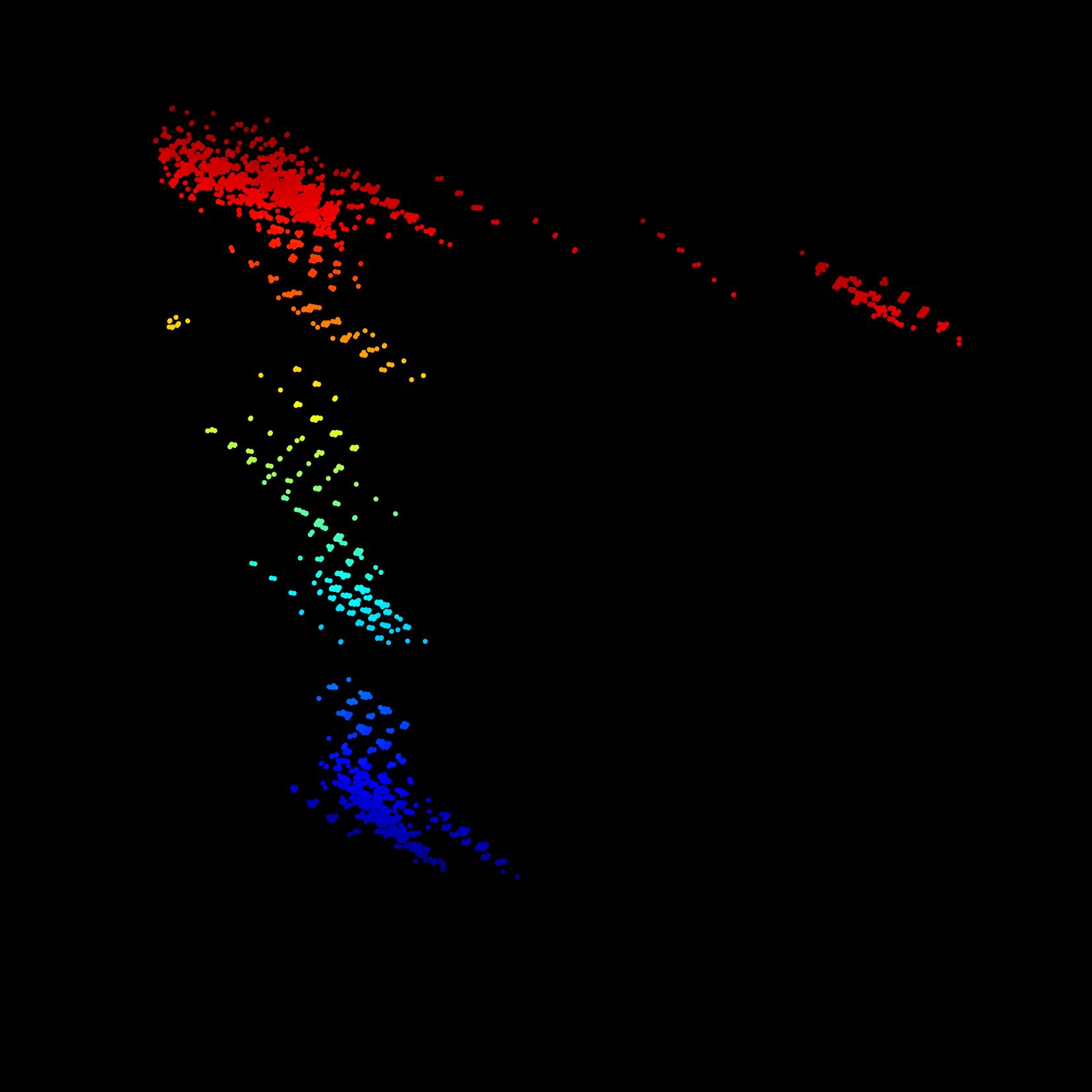}
		\caption{}
		\label{figure31_1_5}
	\end{subfigure}
	\begin{subfigure}[b]{0.24\linewidth}
		\caption*{{\fontsize{8pt}{10pt}\selectfont\centering tomo-IRENet-U}}
		\includegraphics[width=\linewidth]{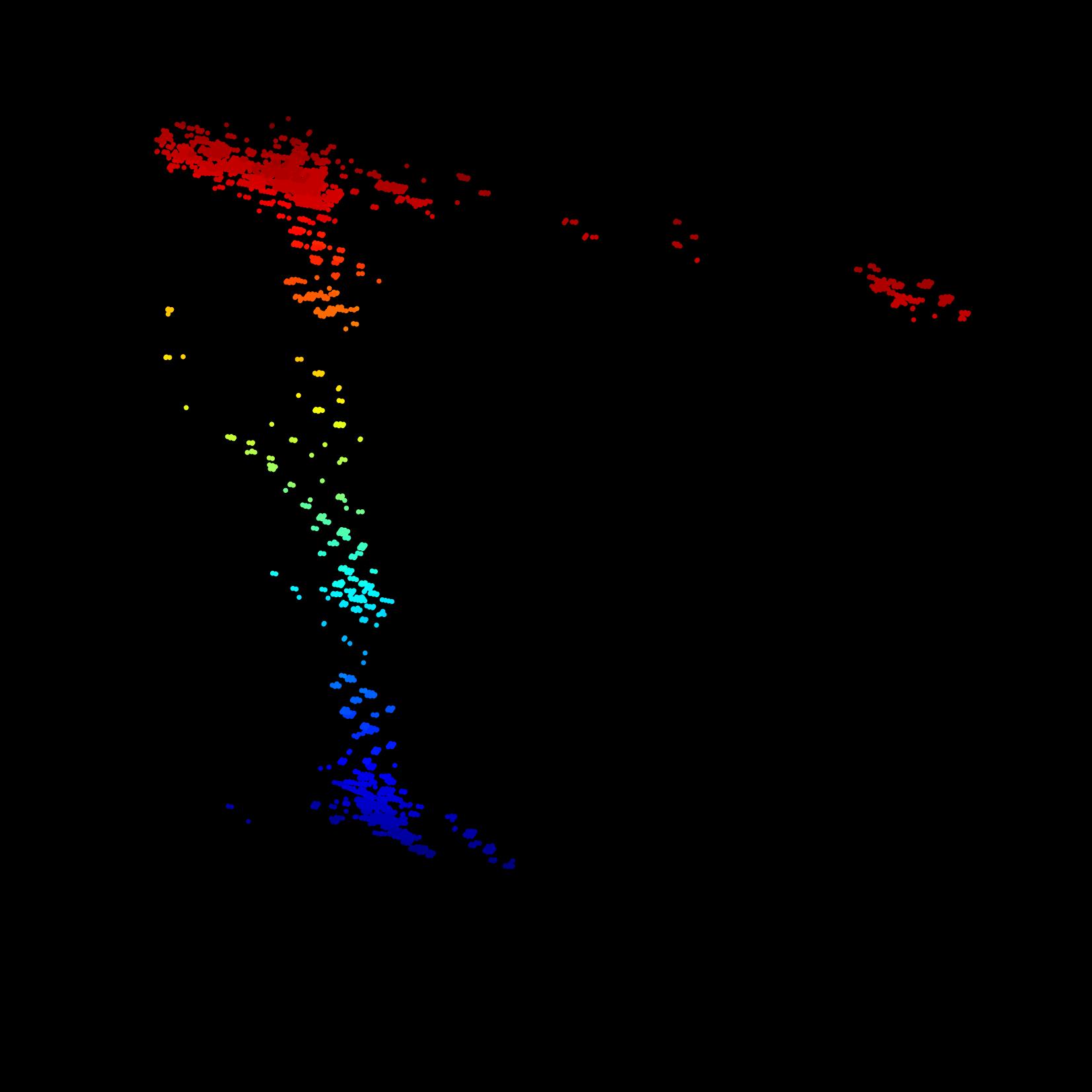}
		\caption{}
		\label{figure31_1_6}
	\end{subfigure}
	\begin{subfigure}[b]{0.24\linewidth}
		\caption*{{\fontsize{8pt}{10pt}\selectfont\centering tomo-LRENet-biU}}
		\includegraphics[width=\linewidth]{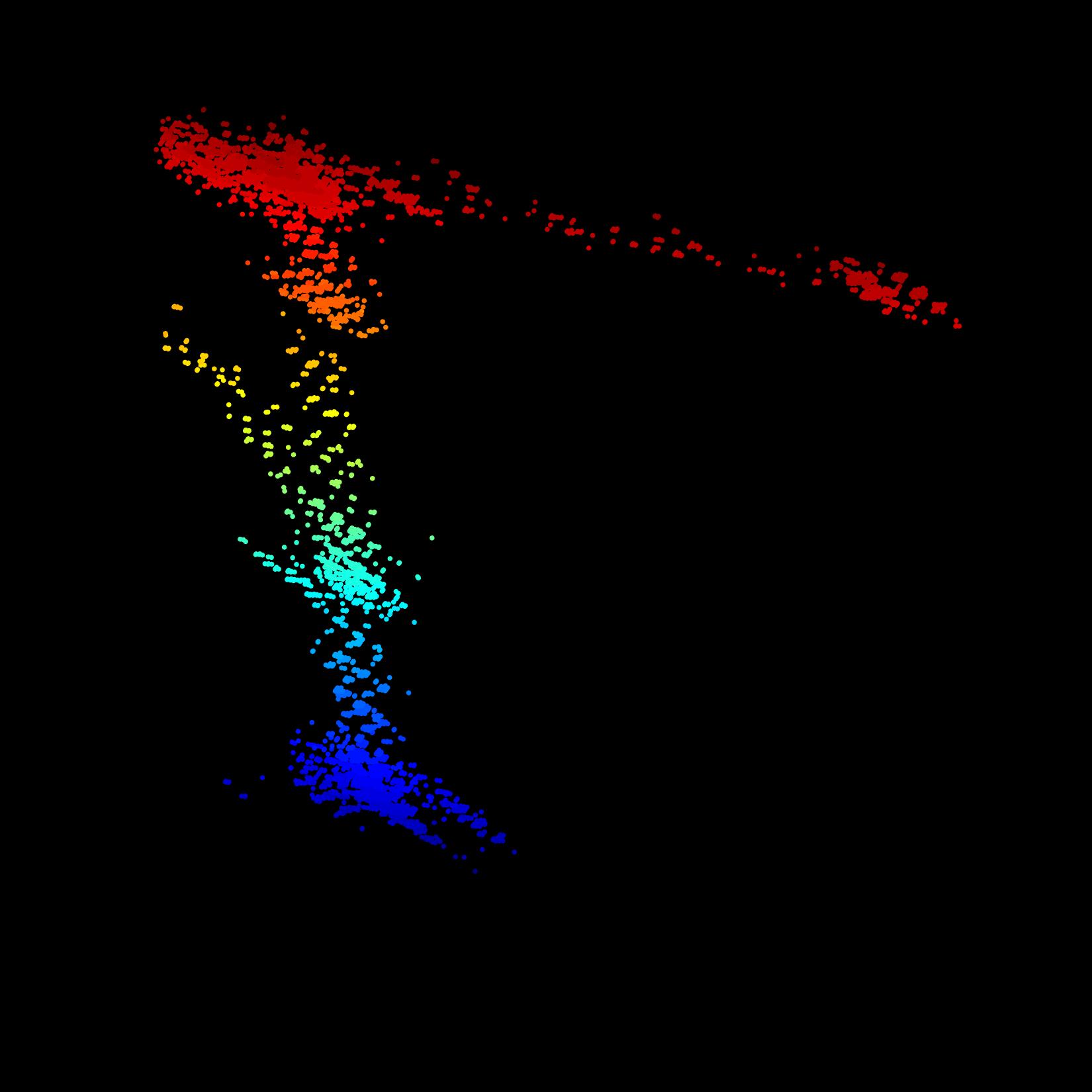}
		\caption{}
		\label{figure31_1_7}
	\end{subfigure}
	\begin{subfigure}[b]{0.24\linewidth}
		\caption*{{\fontsize{8pt}{10pt}\selectfont\centering tomo-LRENet-LSTM}}
		\includegraphics[width=\linewidth]{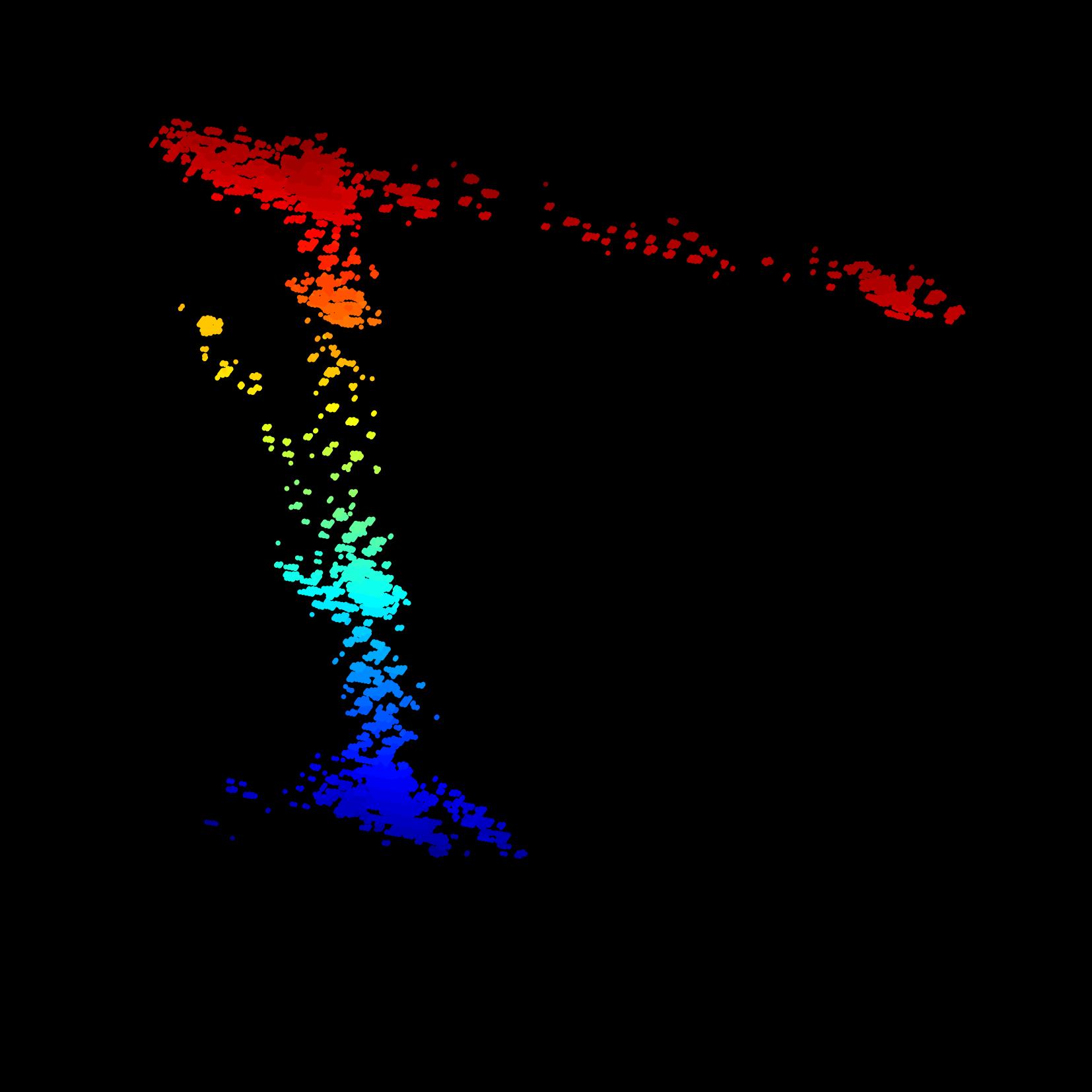}
		\caption{}
		\label{figure31_1_8}
	\end{subfigure}

	\begin{subfigure}[b]{0.24\linewidth}
		\caption*{{\fontsize{8pt}{10pt}\selectfont\centering scene}}
		\includegraphics[width=\linewidth]{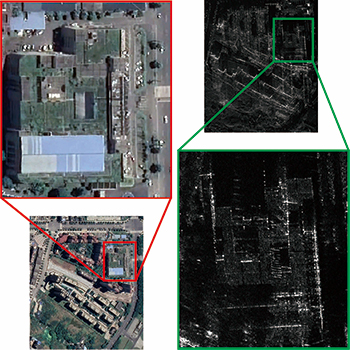}
		\caption{}
		\label{figure31_2_1}
	\end{subfigure}
	\begin{subfigure}[b]{0.24\linewidth}
		\caption*{{\fontsize{8pt}{10pt}\selectfont\centering FISTA-based}}
		\includegraphics[width=\linewidth]{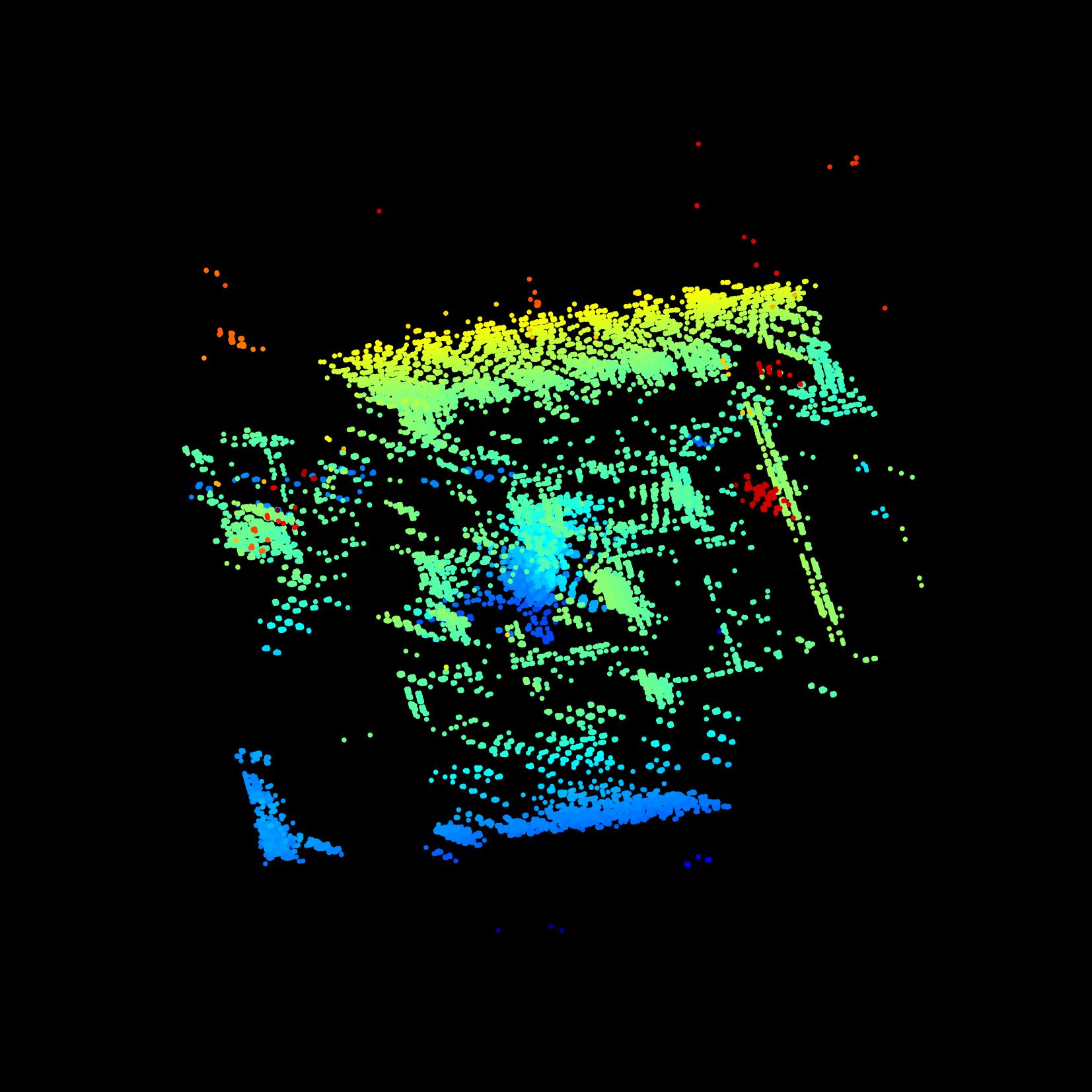}
		\caption{}
		\label{figure31_2_2}
	\end{subfigure}
	\begin{subfigure}[b]{0.24\linewidth}
		\caption*{{\fontsize{8pt}{10pt}\selectfont\centering SLIMMER}}
		\includegraphics[width=\linewidth]{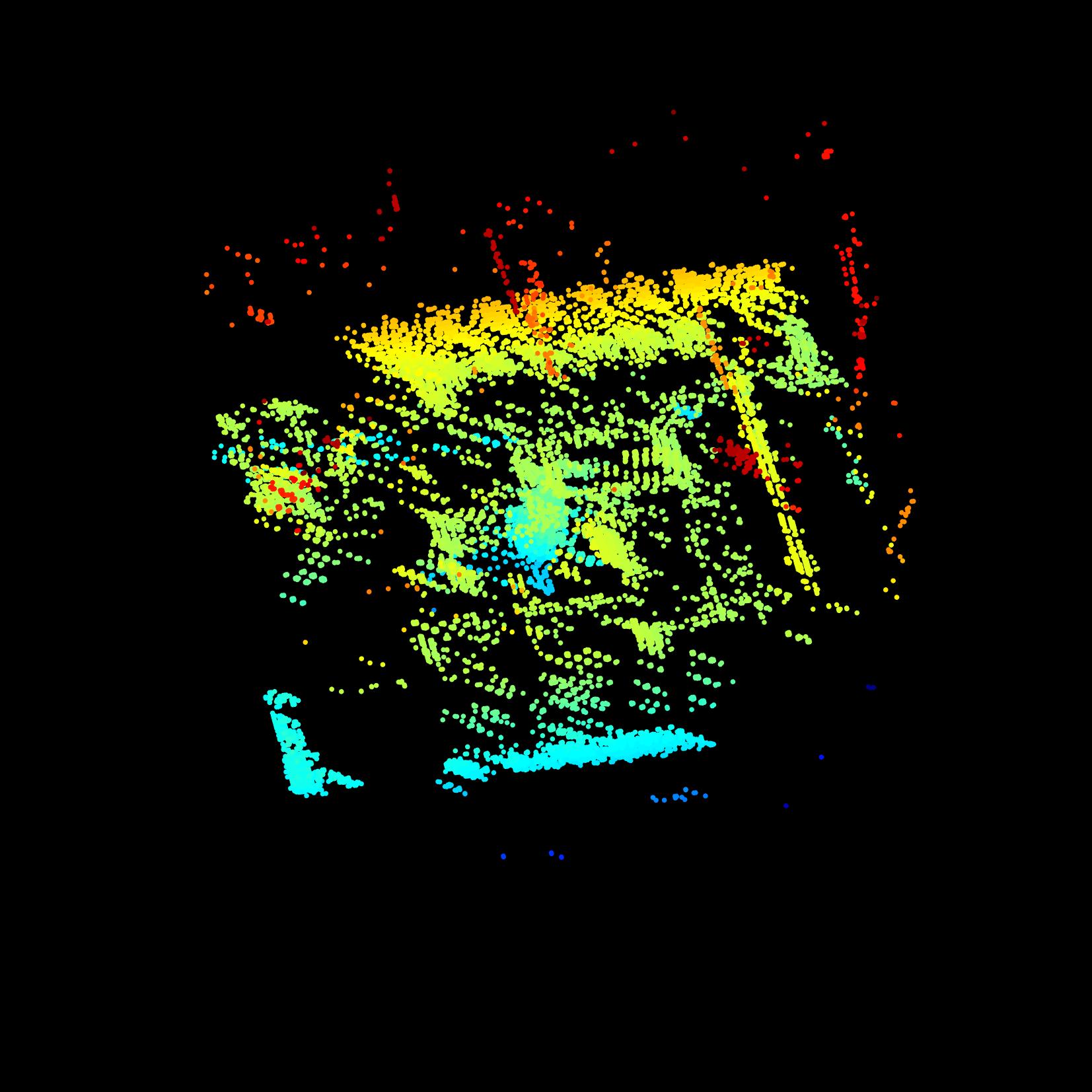}
		\caption{}
		\label{figure31_2_3}
	\end{subfigure}
	\begin{subfigure}[b]{0.24\linewidth}
		\caption*{{\fontsize{8pt}{10pt}\selectfont\centering tomo-IRENet-Raw}}
		\includegraphics[width=\linewidth]{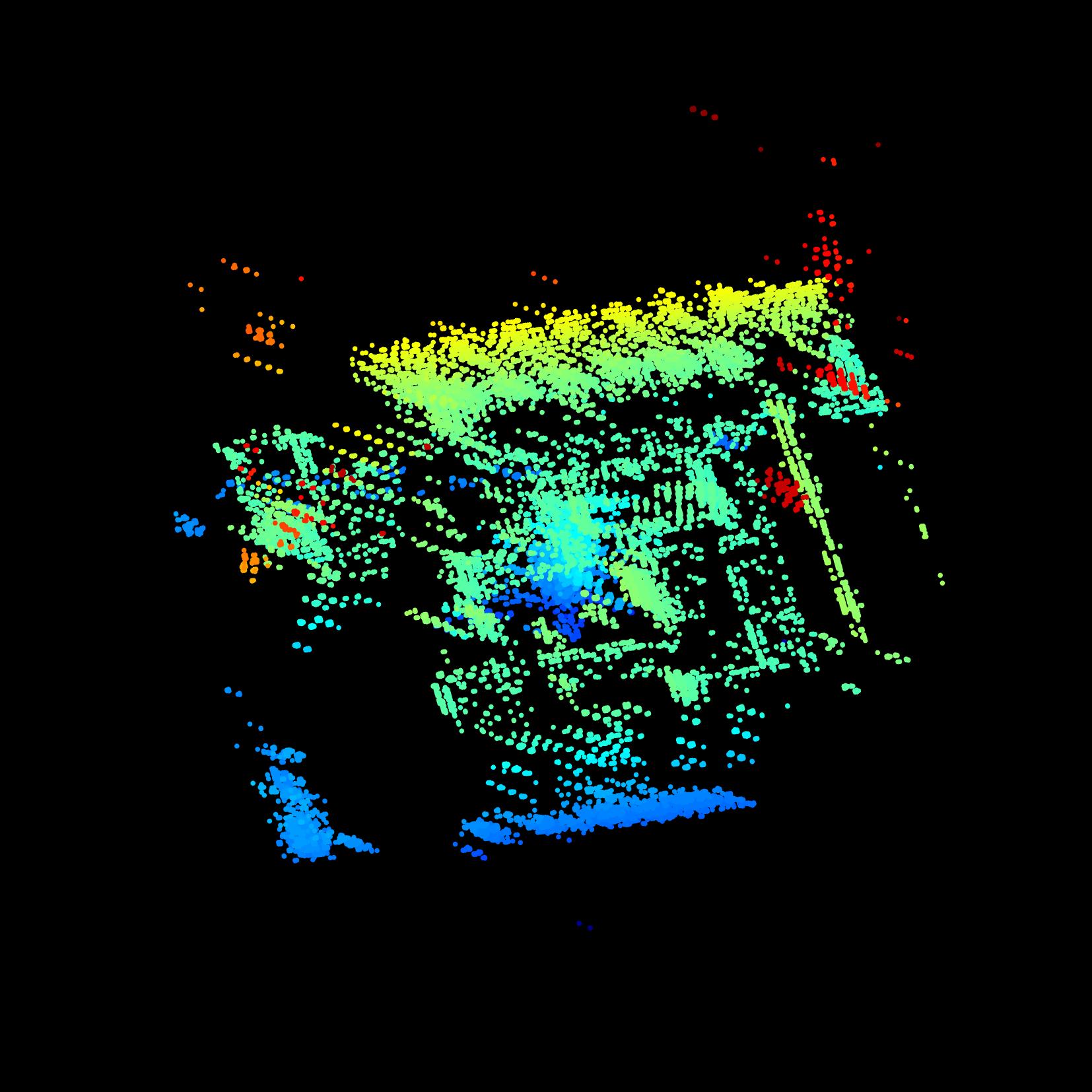}
		\caption{}
		\label{figure31_2_4}
	\end{subfigure}
	
	% 第二排的四张图
	\begin{subfigure}[b]{0.24\linewidth}
		\caption*{{\fontsize{8pt}{10pt}\selectfont\centering tomo-IRENet-TV}}
		\includegraphics[width=\linewidth]{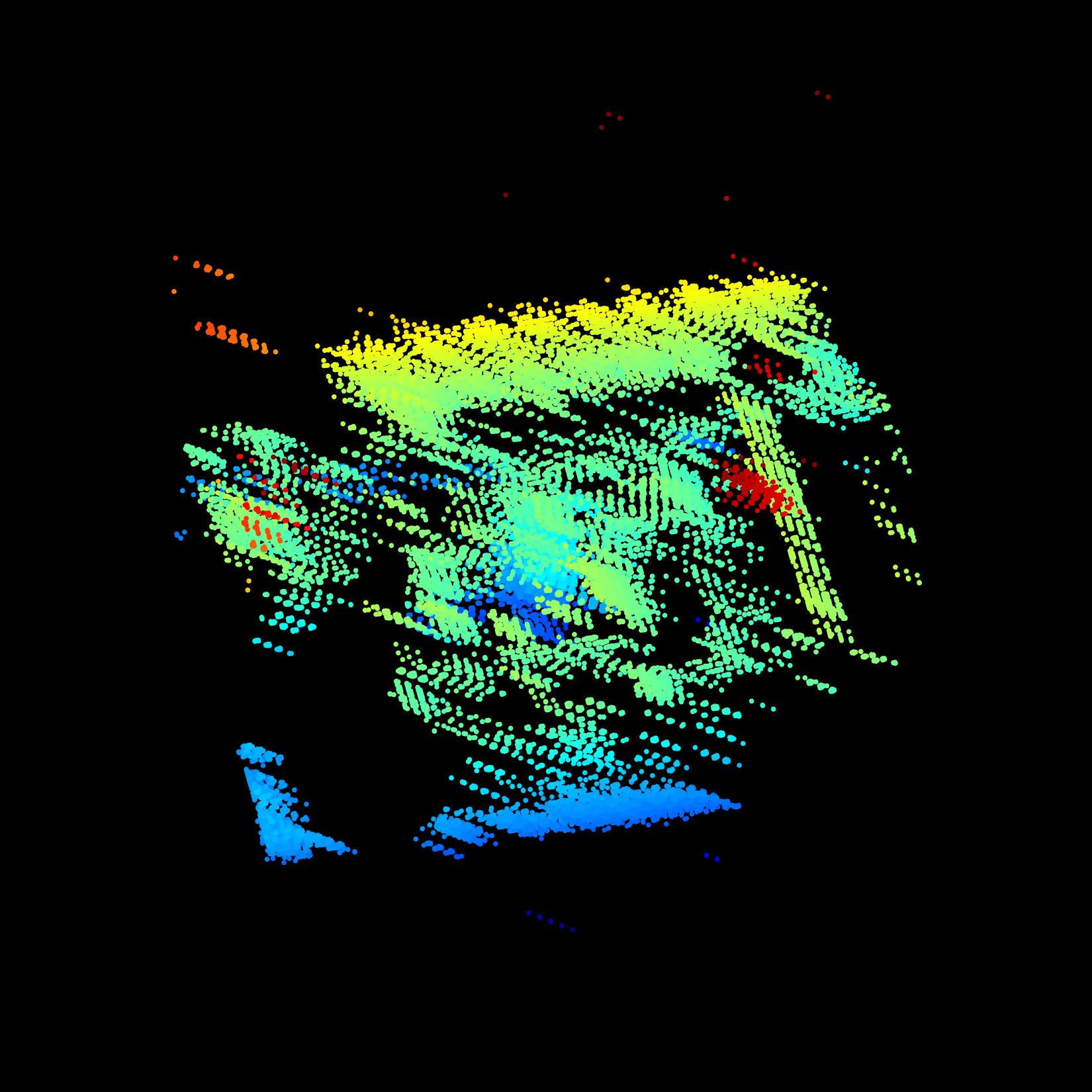}
		\caption{}
		\label{figure31_2_5}
	\end{subfigure}
	\begin{subfigure}[b]{0.24\linewidth}
		\caption*{{\fontsize{8pt}{10pt}\selectfont\centering tomo-IRENet-U}}
		\includegraphics[width=\linewidth]{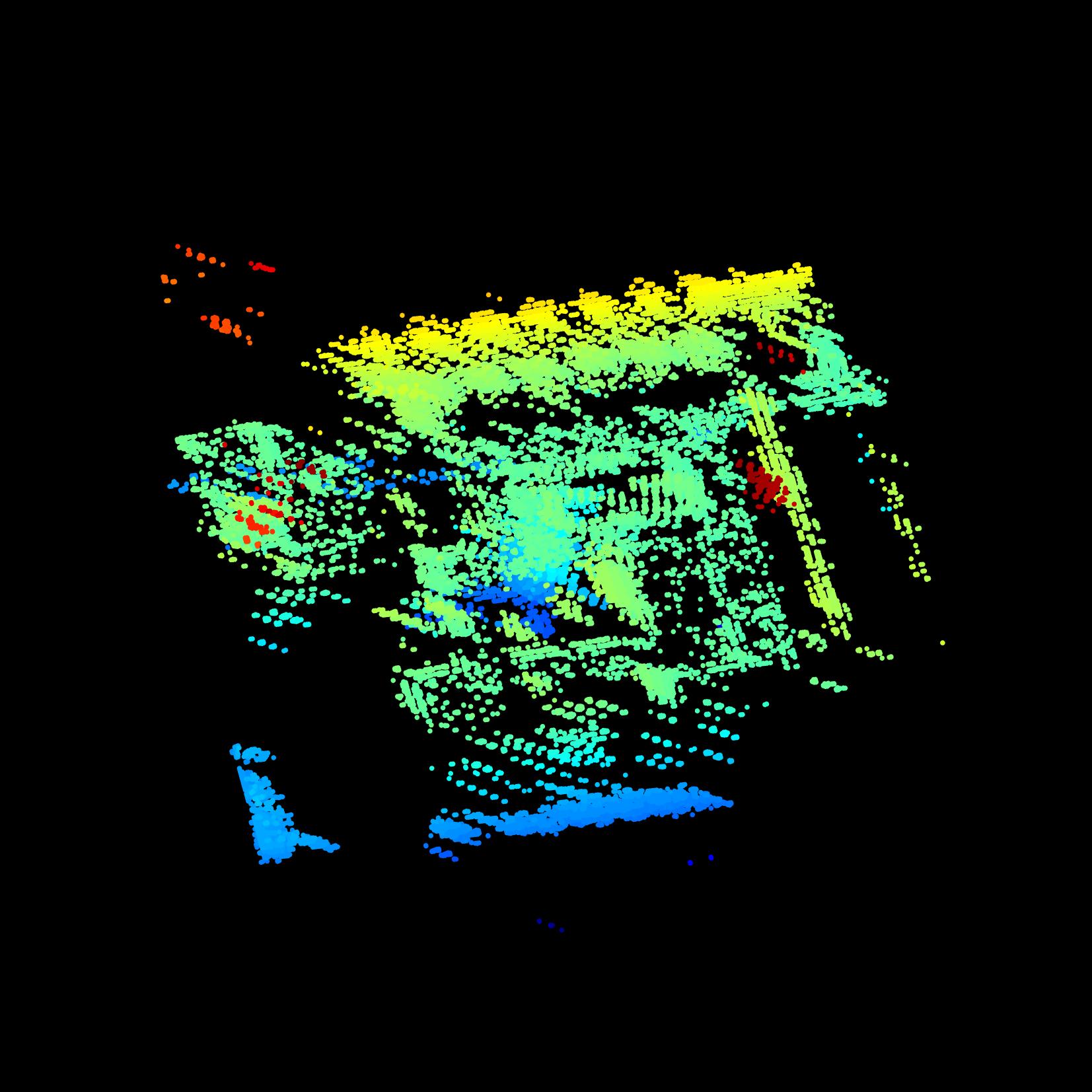}
		\caption{}
		\label{figure31_2_6}
	\end{subfigure}
	\begin{subfigure}[b]{0.24\linewidth}
		\caption*{{\fontsize{8pt}{10pt}\selectfont\centering tomo-LRENet-biU}}
		\includegraphics[width=\linewidth]{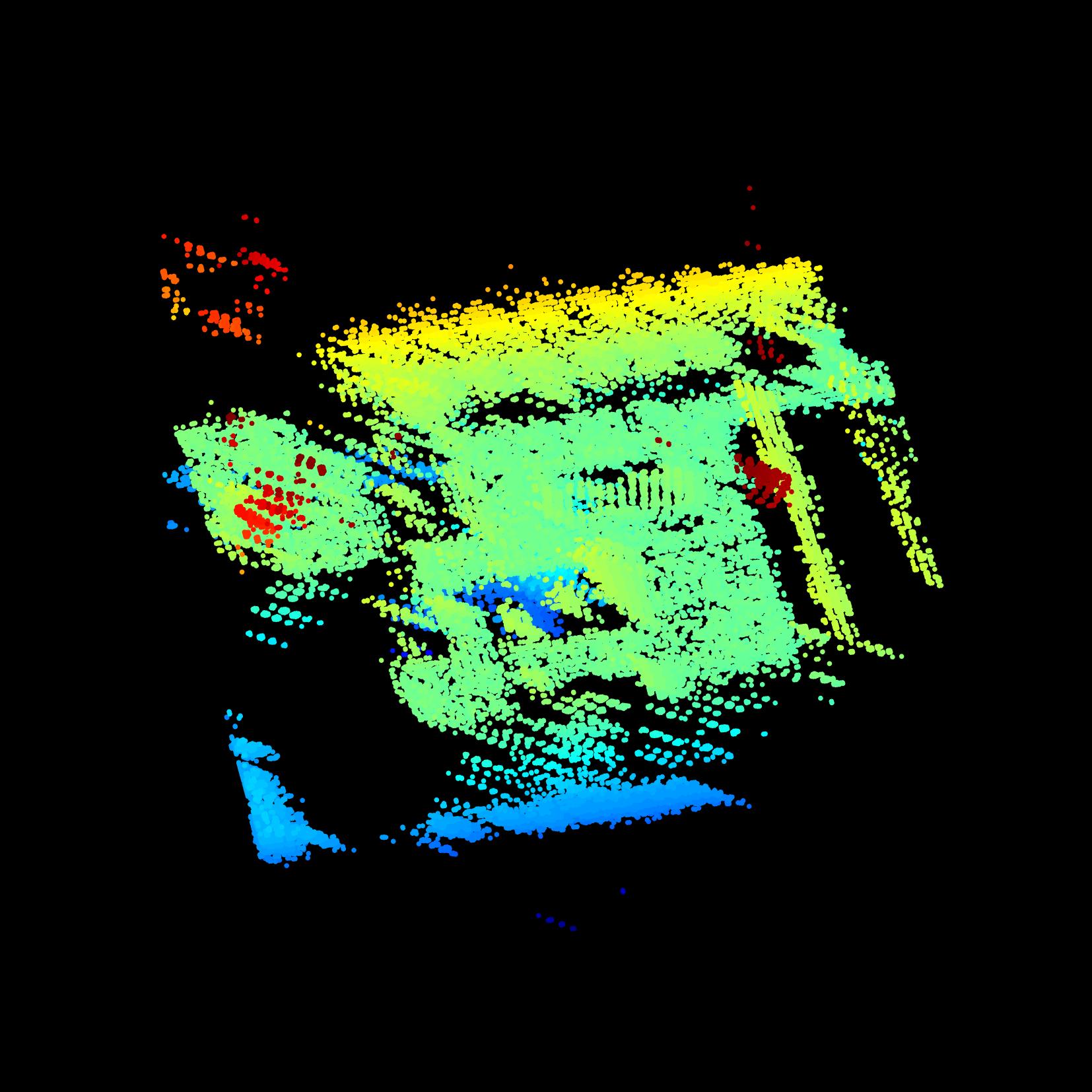}
		\caption{}
		\label{figure31_2_7}
	\end{subfigure}
	\begin{subfigure}[b]{0.24\linewidth}
		\caption*{{\fontsize{8pt}{10pt}\selectfont\centering tomo-LRENet-LSTM}}
		\includegraphics[width=\linewidth]{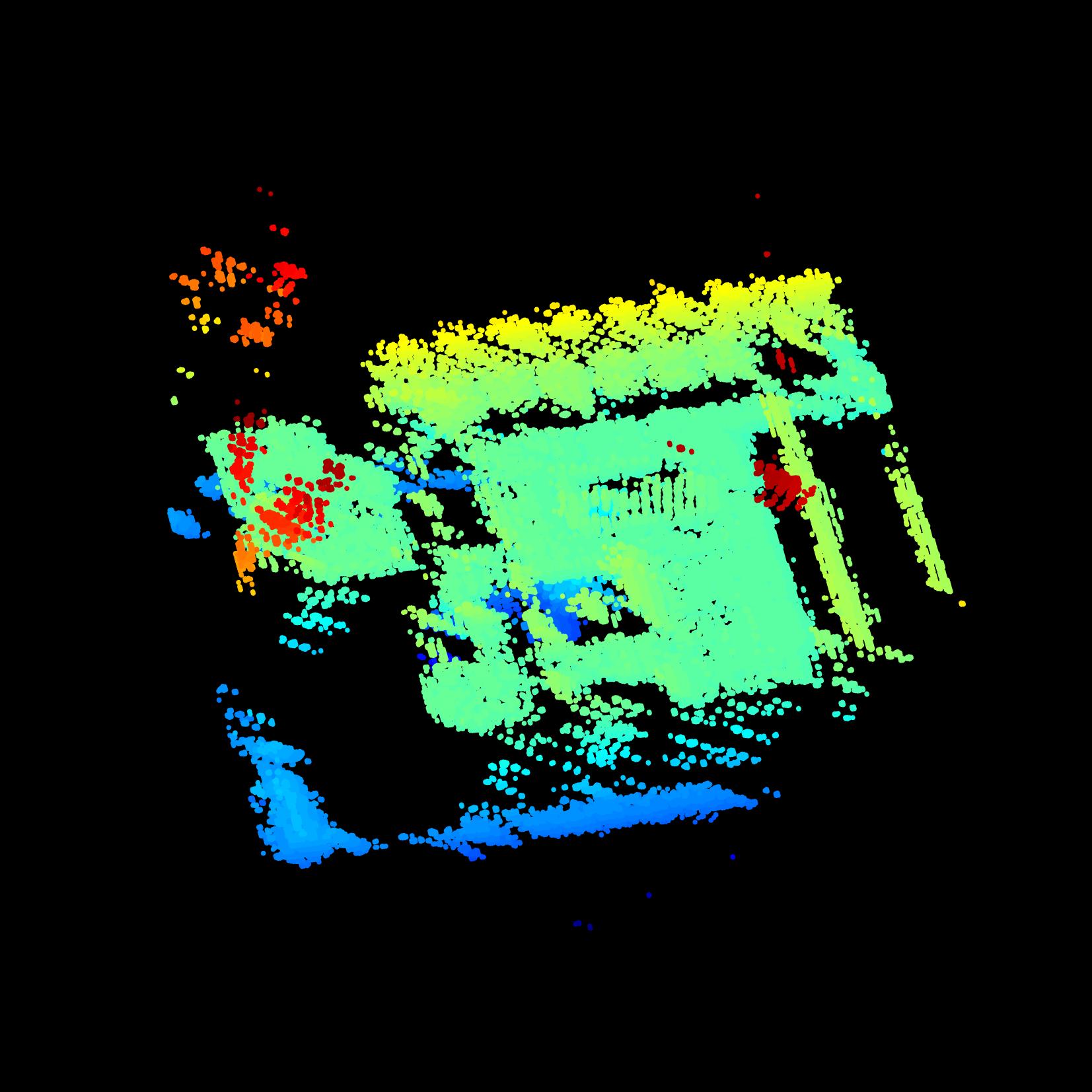}
		\caption{}
		\label{figure31_2_8}
	\end{subfigure}

	\caption{Detailed results of test 6, comparing reconstruction methods across another two scenes. Scene 1 features a high-rise tower crane , while scene 2 includes a commercial center with a flat rooftop. (a) and (i) show optical and radar images for Scenes 1 and 2, respectively. (b-h) detail the reconstruction results for Scene 1 and (j-p) for Scene 2, from the FISTA-based method, SLIMMER, tomo-IRENet-Raw, tomo-IRENet-TV, tomo-IRENet-U, tomo-LRENet-biU, and tomo-LRENet-LSTM.}
	\label{figure_31}
\end{figure*}

To effectively highlight their performance differences, we have selected four distinct types of building structures for a detailed comparison. These selected areas have been magnified for a more detailed examination. The results are presented in Fig.~\ref{figure_30} and Fig.~\ref{figure_31} at the following pages. 

The selected structures encompass a set of regularly shaped high-rise residential buildings with vertical wall surfaces, as shown in the first two rows of Fig.~\ref{figure_30}. They also include two low-rise buildings of varying orientations that feature not only vertical walls but also rooftops, as depicted in the bottom two rows of Fig.~\ref{figure_30}. In addition, the selection includes a high-rise tower crane, prominently featured in the first half of Fig.~\ref{figure_31}, with its support structure and long arms. Lastly, the structures comprise a commercial center with a large, flat rooftop, as seen in the bottom half of Fig.~\ref{figure_31}. For each set of results, we provide the optical image and the 2D radar image,  which is located in the upper-left corner of the figure. Targets are marked in a red box.

\clearpage
In these detailed and enlarged results comparisons, we can visually see that, by adopting spatial feature considerations, the second row's four methods generally have more reconstructed scatterers. For methods not considered, we often observe larger holes distributed on both wall and rooftop surfaces. In contrast, the new methods significantly reduce this issue.

In the context of these new methodologies, it is noteworthy to mention that the results procured from Tomo-IRENet-TV appear visually defocused. This defocusing is primarily due to the method's inferior resolution. Upon a comparative analysis between the outcomes of Tomo-IRENet-U and Tomo-LRENet-biU, it is noticeable that the latter method manages to reveal a greater quantity of scatterers both on the surface and the line. This observation underscores the potential advantages of enhancing spatial features in two parallel directions as opposed to a singular one. Furthermore, the results from Tomo-LRENet-LSTM stand out as superior when measured against all other methods. This is especially true in terms of the completeness and focus of the surface and line. Therefore, it can be concluded that Tomo-LRENet-LSTM, due to its exceptional performance, appears to be the most effective of all the methods examined.

Lastly, we present the reconstruction time for different methods applied to the measured data (tests 5 and 6) in Table~\ref{table_4}. As with test 4, the traditional methods are time-consuming, taking over 100 minutes. Among the remaining methods, tomo-IRENet-Raw is the most efficient, while tomo-IRENet-U is the most time-consuming. Tomo-LRENet-LSTM, which has the best precision performance, is the most efficient of the remaining three methods. It performs at a similar level to tomo-IRENet-Raw, which takes less than 5 minutes.

\begin{table}[!h]
    \centering
    \renewcommand{\arraystretch}{0.8} % Adjust row spacing

    \caption{Comparison of Reconstruction Time on Different Methods}
    \label{table_4}
    \footnotesize
	\hspace*{-4em}
    \begin{tabular}{@{}>{\centering\arraybackslash}m{0.19\textwidth} *{7}{>{\centering\arraybackslash}m{0.11\textwidth}} @{}}
    \toprule
    \textbf{Scene} & \textbf{FISTA} & \textbf{SLIMMER} & \textbf{tomo-IRENet-Raw} & \textbf{tomo-IRENet-TV} & \textbf{tomo-IRENet-U} & \textbf{tomo-LRENet-biU} & \textbf{tomo-LRENet-LSTM} \\
    \midrule
    Yuncheng scene \\ (test 5), unit: min & $\geq100$ & $\geq100$ & \textbf{1.99} & 5.36 & 13.43 & 6.76 & 2.84 \\
    \midrule
    Emei scene\\ (test 6), unit: min & $\geq100$ & $\geq100$ & \textbf{2.31} & 7.19 & 16.39 & 7.70 & 4.71 \\
    \bottomrule
    \end{tabular}
\end{table}

\section{Phase Five: Result Discussion}

In the previous phase of the study, we conducted a comprehensive evaluation of various methods' performance using diverse evaluation objects and metrics, yielding certain results. Based on these results, in this phase, we discuss and analyze them. Based on this analysis, we provide our considerations for the next stage of the study in the future.

\subsection{ Method performance}

Taking into account spatial features generally improves the quality of reconstruction. This results in sharper edges, like the arm of the tower crane, and more accurately shaped geometrical features, like vertical walls and flat rooftops. The reconstruction has fewer outliers and holes. Among the methods tested, the performance of tomo-IRENet-TV is below expectations set during the design phase. We will discuss this later. The other three models, tomo-IRENet-U, tomo-LRENet-biU, and tomo-LRENet-LSTM, mainly differ in the deep regularization modules they use.
    
The single U-Net structure used by tomo-IRENet-U, which is chosen to reduce computational and storage demands, is essentially part of the complete deep regularization module (Bi-Parallel-UNet Fusion and Sequential-UNet-LSTM Fusion). Compared to these more complex designs, the simplified structure yields less improvement in reconstruction performance. Additionally, it's worth noting that the sparse feature regularization component is left out, only including leaves in the U-Net structure during the design phase. We have also conducted experiments combining both the sparse feature regularization (as in tomo-LRENet-biU and tomo-LRENet-LSTM) and the U-Net structure. However, these modifications made little difference in the quality of the reconstruction. In the iterative reconstruction with enhancement regularization framework, the enhancement operator can only be effectively handled by the deep module alone, rather than through well-balanced hybrid approaches.
    
The Bi-Parallel-UNet Fusion and Sequential-UNet-LSTM Fusion modules employed in tomo-LRENet-biU and tomo-LRENet-LSTM illustrate different strategies for intra-slice and inter-slice modeling to assemble scene tensors from slices. Our experimental results indicate that tomo-LRENet-LSTM offers superior reconstruction quality in all tests compared to tomo-LRENet-biU. This implies that the LSTM's capability to capture and maintain long-term dependencies in data sequences can be crucial for interpreting complex spatial relationships in TomoSAR. The effectiveness of both tomo-LRENet-biU and tomo-LRENet-LSTM also implies that the light reconstruction and enhancement regularization framework strikes a good balance between quality and efficiency. This is significant for our aim of enhancing the accuracy of current deep learning-based methods without compromising efficiency.
    
\begin{table*}[!t]
    \centering
    \begin{threeparttable}
    \caption{Summary of the various methods in the study}
    \label{table_5} % 
    \setlength{\tabcolsep}{0pt}
    % 增加行间距
    \renewcommand{\arraystretch}{1.0}
    \footnotesize
    \begin{tabular}{>{\centering\arraybackslash}p{0.15\textwidth} 
                    >{\centering\arraybackslash}p{0.11\textwidth} 
                    >{\centering\arraybackslash}p{0.11\textwidth} 
                    >{\centering\arraybackslash}p{0.11\textwidth} 
                    >{\centering\arraybackslash}p{0.11\textwidth} 
                    >{\centering\arraybackslash}p{0.11\textwidth} 
                    >{\centering\arraybackslash}p{0.11\textwidth} 
                    >{\centering\arraybackslash}p{0.11\textwidth}}
    \toprule
     & \textbf{FISTA-based} & \textbf{SLIMMER} & \textbf{tomo-IRENet-Raw} & \textbf{tomo-IRENet-TV} & \textbf{tomo-IRENet-U} & \textbf{tomo-LRENet-biU} & \textbf{tomo-LRENet-LSTM} \\
    \midrule
    \textbf{Regularization Feature} & Sparse feature & Sparse feature & Sparse feature & Sparse feature + spatial feature & Sparse feature + spatial feature & Sparse feature + spatial feature & Sparse feature + spatial feature \\
    \midrule
    \textbf{Regularization Framework} & Iterative reconstruction with enhancement & Iterative reconstruction with enhancement & Iterative reconstruction with enhancement & Iterative reconstruction with enhancement & Iterative reconstruction with enhancement & Light reconstruction and enhancement & Light reconstruction and enhancement \\
    \midrule
    \textbf{Regularization Module} & Shallow modeling / Norm-based regularization in the intensity domain & Shallow modeling / Norm-based regularization in the intensity domain & Shallow modeling / Norm-based regularization in the intensity domain & Shallow modeling / Norm-based regularization in the gradient domain & Deep modeling / Deep U-Net based structure & Deep modeling / Bi-Parallel-UNet Fusion & Deep modeling / Sequential-UNet-LSTM \\
    \midrule
    \textbf{Resolving Ability} & $\uparrow$ & $\uparrow\uparrow$ & $\uparrow$ & $\downarrow$ & $\uparrow\uparrow$ & $\uparrow\uparrow\uparrow$ & $\uparrow\uparrow\uparrow$ \\
    \midrule
    \textbf{Suppression of Outliers} & $\uparrow$ & $\uparrow$ & $\uparrow$ & $\uparrow\uparrow$ & $\uparrow\uparrow$ & $\uparrow\uparrow$ & $\uparrow\uparrow\uparrow$ \\
    \midrule
    \textbf{Revealing of Spatial Feature} & $\uparrow$ & $\uparrow$ & $\uparrow$ & - & $\uparrow\uparrow$ & $\uparrow\uparrow\uparrow$ & $\uparrow\uparrow\uparrow\uparrow$ \\
    \midrule
    \textbf{Reconstruction Efficiency} & $\uparrow$ & $\uparrow$ & $\uparrow\uparrow\uparrow\uparrow\uparrow$ & $\uparrow\uparrow\uparrow$ & $\uparrow\uparrow$ & $\uparrow\uparrow\uparrow$ & $\uparrow\uparrow\uparrow\uparrow$ \\
    \bottomrule
    \end{tabular}

    \begin{tablenotes}
        \item[1] The symbols $\uparrow$ and $\downarrow$ are used to show relative comparisons between methods within the same aspect, indicating different levels of preference. The specific number of arrows isn't crucial, what matters is their relative count. It's key to note that comparisons are made within rows, not across columns.
    \end{tablenotes}
    \end{threeparttable}
\end{table*}

Lastly, let's discuss the unexpected performance of tomo-IRENet-TV. Traditionally, the addition of TV regularized terms for spatial feature modeling would enhance performance. However, when deep neural networks are used for solving, the performance diminishes. Here, "decreasing" refers to quantitative reconstruction precision, not the qualitative visual reconstruction performance. This could suggest that individual regularization terms may not be as beneficial as a single, hybrid regularization term, as shown by other deep modeling methods of regularization for enhancement. Whether to adopt separate or hybrid regularizations, particularly in the context of deep learning for tomoSAR reconstruction, warrants further, more comprehensive study in the future.

At last, we sumamrize the performance of each methods in the study in Table~\ref{table_5}. 

\clearpage
\subsection{ Limitations and future work}

\vspace{0.5em}
\textit{B.1 Target loss}
\vspace{0.5em}

During our study, we observed a common occurrence of target loss during the iterative reconstruction process  \cite{wang2023mada,qian2022gamma}. When target loss is severe, the subsequent enhancement operator cannot effectively recover the missing information. In terms of spatial features, lost targets manifest as holes that cannot be reconstructed, while the enhancement operator can only reduce outliers rather than fully restore the missing regions. This issue requires further investigation, particularly concerning the threshold settings applied to the image during the iterative process. Some recent studies suggest that leveraging historical iteration images or the corresponding measurement signals can help in determining a more appropriate threshold, potentially mitigating target loss \cite{wang2023mada,qian2022gamma}. However, this remains an open problem and warrants further research.

\vspace{0.5em}
\textit{B.2 Generalization ability across different systems}
\vspace{0.5em}

As demonstrated in our experiments with measured data, the proposed reconstruction neural network consistently performs well across two different SAR systems with varying parameters such as wavelength and channel numbers. This robustness is due to the physical-aware design of our network, where the forward measurement model is integrated via a deep unrolling framework.

However, one assumption made in our study, which is consistent with other works in the literatures \cite{qian2022basis, qian2022gamma, wang2023atasi}, is the absence of baseline deviations. In real-world scenarios, this assumption may not always hold \cite{wang2023mada}, potentially leading to model mismatch issues. We have previously conducted an in-depth analysis of this problem within the same deep unrolling framework \cite{zeng2021robust}. In general, baseline deviations can result in accumulated errors, such as increased noise patterns in the imaging results. To mitigate these effects, we propose incorporating an additional image enhancement network to reduce noise interference, a strategy that aligns with similar recent work. Furthermore, a recent study suggested introducing a compensation step or module during the reconstruction phase to address these deviations \cite{wang2023mada}. While addressing baseline deviations is not the primary focus of this study, we refer readers to the aforementioned works for further insights and potential solutions.

Beyond this, another aspect of generalization concerns the adoption of supervised learning using paired training samples, as in our study and many others \cite{qian2022basis, qian2022gamma, wang2023atasi,wang2023mada}. Although the deep unrolling framework provides some interpretability and is grounded in a physical measurement model, the reconstruction performance may still be potentially affected by variations in system or scene configurations, which limits generalization. To address this, future work could explore alternative frameworks such as unsupervised or self-supervised learning \cite{zeng2024unsupervised, wang2021single}, reinforcement learning, or methods like PnP \cite{wei2022tfpnp} to enhance generalization across diverse environments.

\vspace{0.5em}
\textit{B.3 Scene/layover variability}
\vspace{0.5em}

Our current approach addresses scene and layover variability by leveraging a variety of 3D building models. The reconstruction network processes adjacent slices, focusing on individual buildings, which provides a relatively localized perspective. The reconstruction of the entire scene is then achieved by assembling these sub-area reconstructions. Consequently, the current dataset does not encompass the full diversity that could arise from different combinations of building models, layouts, building densities, inter-building distances, or interactions between multiple buildings. Incorporating these additional variations would significantly enhance the representation of diverse layover scenarios encountered in real-world environments.
To address this limitation, we are actively working on expanding the scope of the reconstruction network to move beyond its current localized perspective towards a more comprehensive non-local approach. Prior to the advent of deep learning, non-local image processing techniques effectively leveraged the similarity and correlation of patches across the entire image, rather than focusing solely on local regions \cite{shi2019nonlocal, ren2022coprime}. Drawing inspiration from these, we are developing a reconstruction network that incorporates non-local information, enabling it to capture complex scene interactions and better generalize to diverse scenarios. Additionally, we are constructing a new dataset that reflects the actual layout of a real-world area with complex building arrangements, such as our school campus, to better model realistic urban environments.

\vspace{0.5em}
\textit{B.4 Generalization ability across different noise scenarios}
\vspace{0.5em}

Noise levels play a crucial role in TomoSAR reconstruction methods. This is because most deep learning models, including ours, are typically trained using datasets that follow a specific noise distribution \cite{qian2022basis, qian2022gamma, wang2023atasi,wang2023mada}. Consequently, when the noise conditions in the test data deviate from those in the training data, a noticeable drop in reconstruction performance can occur. In this study, achieving robustness against out-of-distribution noise levels was not our primary focus. Thus, when constructing our dataset, we chose to fix the noise level at a single SNR value rather than incorporating a range of SNR values. Despite this limitation, all the methods compared in this study were trained on the same dataset, and the proposed network still demonstrates improvements due to the incorporation of spatial feature regularization.
Nevertheless, we acknowledge that noise robustness is a critical issue that warrants further exploration, as improving robustness can significantly enhance the model’s performance and generalization ability. To address this, we are currently developing a more robust reconstruction network. Recent studies have shown that incorporating equivariance principles into the reconstruction process, particularly in neural networks, can improve the model’s generalization to unseen noise levels \cite{terris2024equivariant, chen2023imaging, chen2022robust, fu2024rotation}. We are exploring the integration of such equivariance principles into the TomoSAR reconstruction task and expect this direction to yield promising results in enhancing noise robustness.

\vspace{0.5em}
\textit{B.5 Reconstruction efficiency}
\vspace{0.5em}

Besides quality and generalization ability, efficiency should also be a future focus. Both this study and previous work show that reconstruction networks typically process the entire scene, including uninteresting objects like non-buildings in urban areas, which adds unnecessary storage and processing burdens. With our flexible computational framework integrating an image enhancement module in the reconstruction process, we could further reduce the solution space by adding a target detection module. 
Some related preliminary works can be found in \cite{jiao2023preliminary, li2023integrated, zhan2022target}.

\section*{ACKNOWLEDGMENT}

We would like to extend our special gratitude to Prof. Qiu and Prof. Ding at the Aerospace Information Research Institute, Chinese Academy of Sciences, as well as other researchers who contributed to making the SARMV3D dataset publicly available \cite{qiu2024microwave3d, xiaolan2021sarmv3d, xiaolan2022key}.

\newpage
\appendix

\section*{Derivation of the solver to the $l_1/ \text{TV}$ hybrid optimization problem in (\ref{equation_9})}\label{appendix_a}

Before we proceed with the formal derivation of the solver, let's clarify some concepts and operators for better understanding.

\begin{enumerate}
    \item \textit{Definition 1}: The tensor-to-matrix operator $\mathcal{M}\left(\cdot\right)$ maps a tensor $\mathcal{X}$ of size $N_z \times N_x \times N_y$ into a matrix $\mathbf{X}$ of size $N_z \times \left(N_xNy\right)$. It extracts the fibers $\mathbf{X}_{ij} = \mathcal{X}\left[:, i, j\right]$, and stack the fibers side by side to form the matrix $\mathbf{X}$ as the concatenation of columns: $\mathbf{X}=\left[\mathbf{X}_{11}, \mathbf{X}_{12}, \cdots, \mathbf{X}_{N_xN_y}\right]$. And the matrix-to-tensor operator $\mathcal{M}^\dagger\left(\cdot\right)$ is the adjoint operator, mapping a matrix $\mathbf{X}$ into a tensor $\mathcal{X}$ through reversing the steps.
    \item \textit{Remark 1}: The forward measurement operator $\mathcal{A}\left(\mathcal{X}\right)=\mathcal{M}^\dagger\left(\mathbf{A}\cdot\mathcal{M}\left(\mathcal{X}\right)\right)$, and its adjoint operator, the backwoard operator$\mathcal{A}^\dagger\left(\mathcal{X}\right)=\mathcal{M}^\dagger\left(\mathbf{A}^H\cdot\mathcal{M}\left(\mathcal{X}\right)\right)$. Here, $\mathbf{A}^H$ is the Hermitian transpose of $\mathbf{A}$.
    \item \textit{Definition 2}: The slices-to-matrix operator $\text{fold}\left(\cdot\right)$ transforms slices of a tensor $\mathcal{X}$ which follow a particular dimension, into a matrix. For instance, when considering horizontal slices along the first dimension, the horizontal slices-to-matrix operator  $\text{fold}_1\left(\cdot\right)$ starts by converting each slice $\mathcal{X[k,:,:]}$ into column vectors $\mathbf{x}_k$. Then, it stacks these column vectors to form a matrix $\mathbf{X}=[\mathbf{x}_1,\mathbf{x}_2,\cdots,\mathbf{x}_{N_z}]$, where $N_z$ is the size of the first dimension. In a similar fashion, we can also have lateral and frontal slices-to-matrix operators $\text{fold}_2\left(\cdot\right)$ and $\text{fold}_3\left(\cdot\right)$, respectively. And $\text{unfold}_1\left(\cdot\right)$, $\text{unfold}_2\left(\cdot\right)$  and $\text{unfold}_3\left(\cdot\right)$ rerepresent corresponding adjoint operators, respectively. 
\end{enumerate}

Given the problem (\ref{equation_9}) as follows.
\begin{equation}
\label{equation_a_1}
\hat{\mathcal{X}} = \mathop{\arg\min_{\mathcal{X}}} \Bigg\{\frac{1}{2} \| \mathcal{Y} - \mathcal{A}(\mathcal{X}) \|_F^2
+ \lambda_{1} \| \mathcal{X} \|_1 + \lambda_{2} g_{TV}(\mathcal{X}) \Bigg\}
\tag{A.1}
\end{equation}

In general, this problem can be solved through the proximal algorithm. The proximal algorithm involves iterating over two main steps:

1. Gradient descent step for the first measurement consistency term ${\left(\frac{1}{2} \| \mathcal{Y} - \mathcal{A}(\mathcal{X}) \|_F^2\right)}$. Computing its gradient and perform a gradient descent step:
\begin{equation}
\label{equation_a_2}
\begin{split}
\mathcal{Z}^{(k+1)} &= \mathcal{X}^{(k)} - \alpha \nabla_{\mathcal{X}^{\left(k\right)}}{\left(\frac{1}{2} \| \mathcal{Y} - \mathcal{A}(\mathcal{X}) \|_F^2\right)} \\
&= \mathcal{X}^{(k)} - \alpha \mathcal{A}^\dagger\left(\mathcal{A}\left(\mathcal{X}^{(k)}\right) - \mathcal{Y}\right)
\end{split}\tag{A.2}
\end{equation}
where $\alpha$ is the step size of the gradient descent step.

2. Proxiaml projection step for the left two feature consistency terms $\left(\lambda_{1} \| \mathcal{X} \|_1 + \lambda_{2} g_{TV}\left(\mathcal{X}\right) \right)$. This corresponds to solving the following new optimization problem.
\begin{equation}
\label{equation_a_3}
\hat{\mathcal{X}} = \mathop{\arg\min_{\mathcal{X}}} \Bigg\{\frac{1}{2\alpha}\|\mathcal{X} - \mathcal{Z}^{(k+1)}\|_F^2
+ \lambda_{1}\|\mathcal{X}\|_1 + \lambda_{2}\text{g}_\text{TV}(\mathcal{X}) \Bigg\}
\tag{A.3}
\end{equation}

For the convience of the following derivations, we altert this problem into a equivlanet new form with a more definitie version of the regularization term  regarding the total-variation operator of the tensor $\mathcal{X}$. Let's define $\text{g}_{TV}$ as $\|\mathbf{D}_1\mathbf{X}_1\|_1+\|\mathbf{D}_2\mathbf{X}_2\|_1+\|\mathbf{D}_3\mathbf{X}_3\|_1$. Here, $\mathbf{X}_1=\text{fold}_1\left(\mathcal{X}\right)$, $\mathbf{X}_2=\text{fold}_2\left(\mathcal{X}\right)$ and $\mathbf{X}_3=\text{fold}_3\left(\mathcal{X}\right)$. $\mathbf{D}_1,$ $\mathbf{D}_2$, and $\mathbf{D}_3$ represent the horizontal, lateral, and frontal forward first-order differential matrices, respectively. They are of the same size as their corresponding matrices $\mathbf{X}_1$, $\mathbf{X}_2$ and $\mathbf{X}_3$ of the tensor $\mathcal{X}$. The equaivalent new form of the optimization problem is formed as: 
\begin{equation}
\label{equation_a_4}
\hat{\mathbf{X}}= \mathbf{\arg\min_{\mathbf{X}_i}} \Bigg\{\frac{1}{2\alpha}\|\mathbf{X}_i - \mathbf{Z}_i^{(k+1)}\|_F^2 
 + \lambda_{1}\|\mathbf{X}_i\|_1 + \lambda_{2}\|\mathbf{D}_i\mathbf{X}_i\|_1  \Bigg\},i=1,2,3
\tag{A.4}
\end{equation}

Here, $\mathbf{Z}_1=\text{fold}_1\left(\mathcal{Z}^{(k+1)}\right)$, $\mathbf{Z}_2=\text{fold}_2\left(\mathcal{Z}^{(k+1)}\right)$ and $\mathbf{Z}_3=\text{fold}_3\left(\mathcal{Z}^{(k+1)}\right)$. In this problem, there is a coupling between $\| \mathbf{X}_i \|_1$ and $\|\mathbf{D}_i\mathbf{X}_i \|_1$ which complicates the solving process. To decouple it, we introduce an additional variable under the Bregman Split methodology. The problem is then reformed as follows:
\begin{equation}
\label{equation_a_5}
\hat{\mathbf{X}}_i, \hat{\mathbf{V}}_i, \hat{\mathbf{B}}_i = 
\mathop{{\arg\min}_{\mathbf{X}_i, \mathbf{V}_i,\mathbf{B}_i}}\\
\Bigg\{
\frac{1}{2\alpha}\|\mathbf{X}_i - \mathbf{Z}_i^{(k)}\|_F^2 + \lambda_{1}\|\mathbf{X}_i\|_1 + \lambda_{2}\|\mathbf{V}_i\|_1 
+ \frac{\mu}{2}\|\mathbf{D}_i\mathbf{X}_i - \mathbf{V}_i + \mathbf{B}_i\}\|_F^2\Bigg\}, i=1,2,3\
\tag{A.5}
\end{equation}

Here, $\mathbf{V}_i$ is the alternative variable for $\mathbf{D}_i\mathbf{X}_i$. $\frac{\mu}{2}\|\mathbf{D}_i\mathbf{X}_i - \mathbf{V}_i + \mathbf{B}_i\}\|_F^2$ is introduced to ensure that $\mathbf{D}_i\mathbf{X}_i$ approximates $\mathbf{V}_i$. $\mu$ is a penalty factor and $\mathbf{B}_i$ is a dual variable that is updated during each iteration to reduce the residual difference between $\mathbf{D}_i\mathbf{X}_i$ and $\mathbf{V}_i$. Once decoupled, this problem can be addressed through alternating iteration between solving two sub-problems and updating the dual variable, which can be expressed as follows:

\textit{1. Update $\mathbf{X}_i$ by solving sub-problem 1:}
\begin{equation}
\label{equation_a_6}
\hat{\mathbf{X}}_i= \mathbf{\arg\min_{\mathbf{X}_i}} \Bigg\{
\frac{1}{2\alpha}\|\mathbf{X}_i - \mathbf{Z}_i^{(k)}\|_F^2 + \lambda_{1}\|\mathbf{X}_i\|_1 + \frac{\mu}{2}\|\mathbf{D}_i\mathbf{X}_i - \mathbf{V}_i^{(k)} + \mathbf{B}_i^{(k)}\|_F^2  \Bigg\},i=1,2,3
\tag{A.6}
\end{equation}

\textit{2. Update $\mathbf{V}_i$ by solving sub-problem 2:}
\begin{equation}
\label{equation_a_7}
\hat{\mathbf{V}}_i= \mathbf{\arg\min_{\mathbf{V}_i}} \Bigg\{
\frac{\mu}{2}\|\mathbf{V}_i-\left(\mathbf{D}_i\mathbf{X}_i^{(k+1)} + \mathbf{B}_i^{(k)}\right)\|_F^2 + \lambda_{2}\|\mathbf{V}_i\|_1   \Bigg\},i=1,2,3
\tag{A.7}
\end{equation}

\textit{3. Update $\mathbf{B}_i$ by:}
\begin{equation}
\label{equation_a_8}
\mathbf{B}_i^{(k+1)} = \mathbf{B}_i^{(k)} + \mathbf{D}_i\mathbf{X}_i^{(k+1)} - \mathbf{V}_i^{(k+1)}, i=1,2,3\tag{A.8}
\end{equation}

Next, we delve deeper into the process of solving the two subproblems.

For sub-problem 1, by combining the quadratic term, it can be reformulated as:
\begin{equation}
\label{equation_a_9}
\begin{split}
\hat{\mathbf{V}}_i= \mathbf{\arg\min_{\mathbf{V}_i}} \Bigg\{
\frac{1}{2}\mathbf{X}_i^H \mathbf{Q}_i \mathbf{X}_i - \mathbf{X}_i^H \mathbf{p}^{\left(k\right)}_i + \lambda_1\|\mathbf{X}_i\|_1    \Bigg\},i=1,2,3
\end{split}\tag{A.9}
\end{equation}

Where, $\mathbf{Q}_i = \frac{1}{\alpha}I + \mu\mathbf{D}_i^H\mathbf{D}_i,$ $\mathbf{p}^{\left(k\right)}_i = \frac{1}{\alpha}\mathbf{Z}_i^{(k)} + \mu\mathbf{D}_i^H(\mathbf{V}_i^{(k)} - \mathbf{B}_i^{(k)})$. This problem can also be solved using proximal algorithms, as follows.

\begin{algorithm}[b]
	\caption{Proximal Algorithm for Solving the \(l_1/ \text{TV}\) Hybrid Optimization Problem}
	\begin{algorithmic}[1]
	\State \textbf{Input:} Initial tensor \(\mathcal{X}^{(0)}\), \(\mathbf{X}_i^{(0)} = \text{fold}_i(\mathcal{X}^{(0)})\), \(\mathbf{V}_i^{(0)} = \mathbf{D}_i\mathbf{X}_i^{(0)}\), \(\mathbf{B}_i^{(0)}\) for \(i=1,2,3\).
	\State Parameters: \(\alpha\), \(\lambda_1\), \(\lambda_2\), \(\mu\), \(\tau_1\) and \(\tau_2\).
	\Repeat
		\State \textbf{Step 1:} Update \(\mathcal{Z}^{(k+1)}\) according to equation~(\ref{equation_a_2}).
		\State \textbf{Step 2:} Update \(\mathbf{Z}_i^{(k+1)} = \text{fold}_i(\mathcal{Z}^{(k+1)})\) for each \(i\).
		\State \textbf{Step 3:} For each \(i\), update \(\mathbf{X}_i^{(k+1)}\) according to equations~(\ref{equation_a_10}) and~(\ref{equation_a_11}).
		\State \textbf{Step 4:} For each \(i\), update \(\mathbf{V}_i^{(k+1)}\) according to equations~(\ref{equation_a_12}) and~(\ref{equation_a_13}).
		\State \textbf{Step 5:} For each \(i\), update \(\mathbf{B}_i^{(k+1)}\) according to equation~(\ref{equation_a_8}).
		\State \textbf{Step 6:} Update \(\mathcal{X}^{(k+1)}\) according to equation~(\ref{equation_a_14}).
	\Until the stopping criterion \(\frac{\|\mathcal{X}^{(k)}-\mathcal{X}^{(k-1)}\|_F^2}{\|\mathcal{X}^{(k)}\|_F^2} < \sigma\) is met.
	\State \textbf{Output:} Reconstructed scene image  \(\mathcal{X}^{(k)}\).
	\end{algorithmic}
	\end{algorithm}

\textit{1. Gradient Descent:}
\begin{equation}
\label{equation_a_10}
\begin{split}
\mathbf{X}_i' = &\mathbf{X}_i^{\left(k\right)} - \tau_1 \nabla_{\mathbf{X}_i^{\left(k\right)}} \left(\frac{1}{2}\left(\mathbf{X}_i^{\left(k\right)}\right)^H\mathbf{Q}_i \mathbf{X}_i^H - \mathbf{X}_i^H \mathbf{p}^{\left(k\right)}_i\right)\\=&\mathbf{X}_i^{\left(k\right)} -\tau_1 \left(\mathbf{Q}_i \mathbf{X}_i^{\left(k\right)} - \mathbf{p}_i^{\left(k\right)}\right),i=1,2,3
\end{split}\tag{A.10}
\end{equation}

Here, $\tau_1$ represents the step size.

\textit{2. Proximal Mapping for } $l_1$ \textit{ Regularization:}
\begin{equation}
\label{equation_a_11}
\begin{split}
\mathbf{X}_i^{(k+1)} = \text{prox}_{\lambda_1 \tau_1}(\mathbf{X}_i^{'}),i=1,2,3
\end{split}\tag{A.11}
\end{equation}

Here, $\text{prox}_{\lambda_1 \tau_1}\left(\mathbf{X}\right) = \text{sign}\left(\mathbf{X}\right)\odot \max\left(|\mathbf{X}| - \lambda_1 \tau_1, 0\right)$ is the element-wise soft-thresholding operator. $\odot$ represents the matrix hadamard product.  

Similarly, sub-problem 2 can be solved using proximal algorithms as follows.

\textit{1. Gradient Descent:}
\begin{equation}
\label{equation_a_12}
\begin{split}
\mathbf{V}_i' = & \mathbf{V}_i^{(k)} - \tau_2 \nabla_{\mathbf{V}_i^{(k)}} \left(\frac{\mu}{2}\|\mathbf{V}_i^{(k)}-\mathbf{D}_i\mathbf{X}_i^{(k+1)} - \mathbf{B}_i^{(k)}\|_F^2\right)\\
=& \mathbf{V}_i^{(k)} -\tau_2 \mu \left(\mathbf{V}_i^{(k)} - \mathbf{D}_i\mathbf{X}_i^{(k+1)} - \mathbf{B}_i^{(k)}\right),i=1,2,3
\end{split}\tag{A.12}
\end{equation}

Here, $\tau_2$ represents the step size.

\textit{2. Proximal Mapping for } $l_1$ \textit{Regularization:}
\begin{equation}
\label{equation_a_13}
\begin{split}
\mathbf{V}_i^{(k+1)} = \text{prox}_{\lambda_2 \tau_2}(\mathbf{V}_i^{'}),i=1,2,3
\end{split}\tag{A.13}
\end{equation}

Here, $\text{prox}_{\lambda_2 \tau_2}\left(\mathbf{X}\right) = \text{sign}\left(\mathbf{X}\right)\odot \max\left(|\mathbf{X}| - \lambda_2 \tau_2, 0\right)$ is the element-wise soft-thresholding operator.

Since we conduct proximal projections of (\ref{equation_a_2}) tensor-wise and (\ref{equation_a_3}) matrix-wise, we must transform the matrix back into the tensor form as follows.
\begin{equation}
\label{equation_a_14}
\mathcal{X}^{(k+1)} = \frac{1}{3} 
\Bigg\{
\text{unfold}\left(\mathbf{X}_1^{(k+1)}\right) + \text{unfold}\left(\mathbf{X}_2^{(k+1)}\right)
+\text{unfold}\left(\mathbf{X}_3^{(k+1)}\right) 
\Bigg\}\tag{A.14}
\end{equation}

In summary, the complete iteration steps can be summarized in Algorithm 1.

\section*{Detailed Flowchart of Simulation}\label{appendix_b}
 Generally, the simulation involves three phases: 1) raw material collection, 2) point cloud preparation, and 3) paired data generation.

\begin{enumerate}
    \item \textit{Phase of raw material collection:} Inspired by \cite{chen20193d}, this study uses 3D building models to effectively reveal spatial structures. These models offer detailed representations of buildings, capturing their shapes, sizes, and architectural features. To ensure diversity, the study includes various types of buildings, such as residential, commercial, high-rise, and other relevant structures.
    \item \textit{Phase of point cloud generation:} In this phase, ensuring the quality and diversity of the dataset involves further processing of the raw materials through point cloud sampling and point cloud augmentation.
    \begin{itemize}
        \item \textit{Point cloud sampling:} To align with the SAR sensor's relationship with the observing scene, which assumes point scattering, the raw 3D vortex-based models must be converted into 3D point cloud models through sampling. This process captures the electromagnetic wave's illumination and the occlusion of building components. For instance, walls facing the SAR sensor generate detectable points, while occluded walls do not. To achieve this, we use an enhanced-POVRay method for ray tracing-based point cloud sampling\cite{auer2009ray}, a highly efficient SAR image simulator, this method allows us to generate accurate and realistic point cloud samples.
        \item \textit{Point cloud augmentation:} To enhance the diversity of the input, including building sizes and spatial distribution in 3D space, the point cloud undergoes random rescaling and translation in various directions. This introduces additional variability, ensuring a wider range of building sizes and spatial arrangements within the dataset.
    \end{itemize}

    \item \textit{Phase of paired data generation:} In this phase, ensuring the effectiveness of supervised learning and evaluation requires generating paired data using the prepared point cloud. This involves creating two data sets: input data (ground truth of the reconstruction result) and output data (multi-baseline observation data).
    \begin{itemize}
        \item \textit{Input data generation:} To incorporate the TomoSAR imaging coordinate system, the discrete point cloud initially represented in the ground-range/azimuth/height coordinate system is converted to the slant-range/azimuth/elevation coordinate system. Next, to accommodate the voxel-based reconstruction results of the network, which are organized in a 3D grid, the irregular 3D point cloud is sampled to align with this grid structure. Furthermore, the amplitude and phase values of the points are assigned using a uniform distribution, following established methods described in the literature\cite{qian2022gamma}.
        \item \textit{Output data generation:} To efficiently incorporate the TomoSAR forward observation process and generate large amounts of data, the output data is directly produced using the forward observation matrix. This approach eliminates the need to calculate raw echoes and form 2D images with different baselines.
    \end{itemize}
    
\end{enumerate}

\begin{algorithm}[t]
	\caption*{\textbf{Flowchart 1:} Point Cloud Sampling Process} 
	\begin{algorithmic}[1]
	\label{flowchart_1}
	\State \textbf{Input:} 3D model (in .pov format), illumination angle
	\State \makebox[1.1cm][l]{\textbf{Step 1:}} Use Enhanced POV-Ray to obtain the point cloud with a 3D model and a certain illumination angle.
	\State \makebox[1.1cm][l]{\textbf{Step 2:}} Rotate the point cloud to the ground-range-azimuth coordinate system.
	\State \makebox[1.1cm][l]{\textbf{Step 3:}} Shift all negative coordinates to the positive range. For each point $x \in \mathbb{X}$, $y \in \mathbb{Y}$, $z \in \mathbb{Z}$. $x = x - \min(\mathbb{X})$, $y = y - \min(\mathbb{Y})$, $z = z - \min(\mathbb{Z})$.
	\State \makebox[1.1cm][l]{\textbf{Step 4:}} Normalize the coordinates proportionally. For each coordinate $x$, $y$, $z$, divide $x$, $y$, $z$ by $\max\{\max(\mathbb{X}), \max(\mathbb{Y}), \max(\mathbb{Z})\}$.
	\State \textbf{Output:} Point cloud coordinate set ($\mathbb{X}, \mathbb{Y}, \mathbb{Z}$).
	\end{algorithmic}
	\end{algorithm}

\vspace{0.5em}
\textit{A. Point cloud sampling}
\vspace{0.5em}

\newcounter{customalgcounter} 
\stepcounter{customalgcounter} 

The original form of a prepared 3D building model is a continuous, patch-based 3D mesh. Following the assumption of point scattering, we must discretize the 3D model through point cloud sampling to obtain isolated scatterers. We use the open-source Enhanced POV-Ray to generate a 3D point cloud in the slant range-azimuth-elevation coordinate system from the original 3D model. Its use of ray tracing accounts for obstruction between different parts of the buildings, resulting in a more realistic simulation\cite{auer2009ray}. After sampling, we rotate the point cloud to the ground range-azimuth-elevation coordinate system, allowing for the easy addition of random scaling and translation transformations to augment the original point cloud. Additionally, we normalize all point clouds in terms of their coordinates. The above process is summarized in Flowchart 1.

\vspace{0.5em}
\textit{B. Paired data generation}
\vspace{0.5em}

After obtaining the point cloud, the next phase is generating the input data (ground-truth data of the observed scene) and the output data (paired multi-baseline observation data).

To adapt to the TomoSAR imaging coordinate system, the 3D point cloud, initially expressed in the ground-range/azimuth/height coordinate system, is transformed into the slant-range/azimuth/elevation coordinate system. For each point, the slant range \( R \) and incident angle \( \theta \) are calculated based on the point's coordinates and the radar system center coordinates. The coordinates in the slant-range/azimuth/elevation system are determined as \((R \sin(\theta), R \cos(\theta))\). To align with the network's voxel-based reconstruction results, which are organized in a 3D grid, the uneven 3D point cloud is sampled to fit this grid structure. If multiple scatter points fall within one grid, only one is retained. Finally, the points' amplitude and phase values are assigned using a uniform distribution, as described in \cite{qian2022gamma}, with the scattering intensity of one scatter randomly set between 1 and 4. This process is summarized in Flowchart 2.

\begin{algorithm}[t]
	\caption*{\textbf{Flowchart 2:} Input Data Generation Process}
	\begin{algorithmic}[1]
	\label{flowchart_2}
	\State \textbf{Input:} normalized point cloud, system parameters.
	\Repeat
		\State \makebox[1.1cm][l]{\textbf{Step 1:}} Calculate the slant range and the illumination angle, and project it into the 3D grid.
		\State \makebox[1.1cm][l]{\textbf{Step 2:}} According to the slant range, add the slant-range phase.
		\State \makebox[1.1cm][l]{\textbf{Step 3:}} randomly add the intensity.
	\Until all the points in the cloud are calculated.
	\State \textbf{Output:} Input data.
	\end{algorithmic}
	\end{algorithm}

	\begin{algorithm}[t]
		\caption*{\textbf{Flowchart 3:} Output Data Generation Process}
		\begin{algorithmic}[1]
		\label{flowchart_3}
		\State \textbf{Input:} input data, scene parameters(system centers of different observations $R_{bi}$, scene center $R_c$, vertical baseline vector of different observations $\mathbf{b}$, wavelength $\lambda$), elevation grid vector $\mathbf{s}$.
		\State \makebox[1.1cm][l]{\textbf{Step 1:}} Calculate the reference slant range $R_0=\|{R_{b0}-R_c}\|_2$.
		\State \makebox[1.1cm][l]{\textbf{Step 2:}} Calcuate the forward observation matrix $\mathbf{A}=\left[ \begin{array}{cccc}
		  exp\left(j\frac{4\pi\mathbf{s}\left(0\right)}{\lambda R_0}\mathbf{b}\right) & \ldots & exp\left(j\frac{4\pi\mathbf{s}\left(N_z\right)}{\lambda R_0}\mathbf{b}\right)\\
		\end{array} \right]$, $N_z$ is the grid number in the elevation.
		\State \makebox[1.1cm][l]{\textbf{Step 3:}} Calculate output data $\mathbf{y}=\mathbf{Ax} + \mathbf{n}$, $\mathbf{n}$ is the random noise.
		\State \textbf{Output:} output data.
		
		\end{algorithmic}
		\end{algorithm}

For efficient incorporation of the tomoSAR forward observation process that generates large amounts of data, the output data is created directly using the forward observation matrix. This process is summarized in Flowchart 3.

\clearpage

\bibliographystyle{IEEEtran}
\bibliography{IEEEabrv, reference}

\end{document}